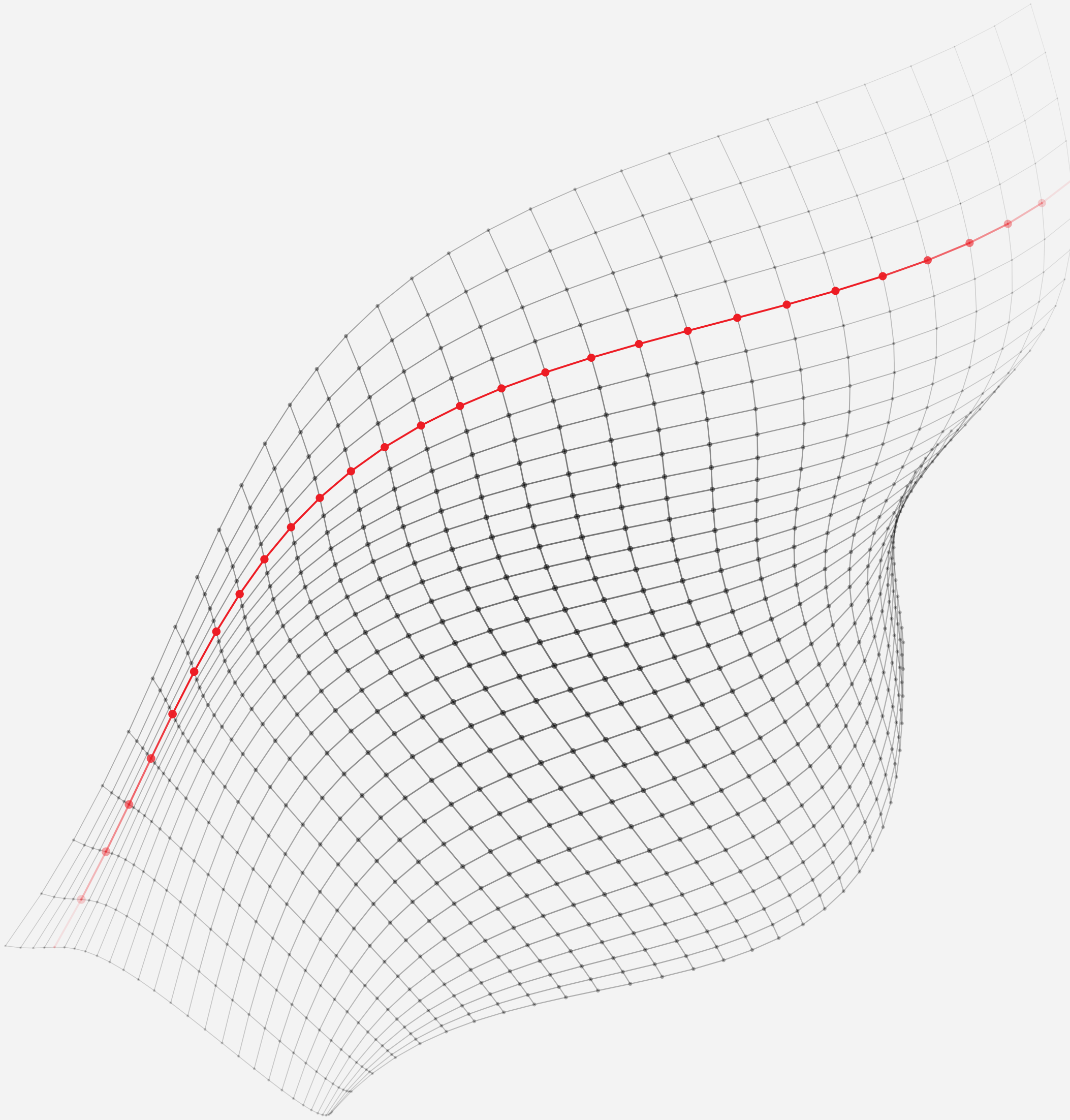

# QIR | Quantum Index Report 2026

MIT INITIATIVE ON THE DIGITAL ECONOMY

MIT DIGITAL

## How to cite this report

Ruane, J., Kiesow, E., Dukatz, C., Converso, L., “The Quantum Index Report 2026”, MIT Initiative on the Digital Economy, Massachusetts Institute of Technology, Cambridge, MA, October 2026.

## Interactive website and public data

The Quantum Index Report 2026 is accompanied by interactive tools available on our website (qir.mit.edu) and we share our raw data with the community, which is available to download from our website (qir.mit.edu/data).

Center for Quantum Networks
A National Science Foundation Engineering Research Center

This work was supported by the Engineering Research Centers Program of the National Science Foundation under NSF Cooperative Agreement No. 1941583. Any opinions, findings and conclusions or recommendations expressed in this material are those of the authors and do not necessarily reflect those of the National Science Foundation.

## Team

**Jonathan Ruane**, Principal Investigator and Editor-in-Chief
MIT Sloan School of Management
MIT Initiative on the Digital Economy

**Elif Kiesow**, Senior Researcher and Project Manager
MIT Initiative on the Digital Economy

**Carl Dukatz**
Accenture

**Laura Converso**
Accenture

This research is a collaboration between Accenture and the MIT Initiative on the Digital Economy (IDE) and was performed under the MIT and Accenture Convergence Initiative for Industry and Technology.

Acknowledgments: Special thanks to Prashant Shukla who contributed to this report while he was at Accenture.

# Introduction

## 2026 Quantum Index Report

The decisions that shape important technologies are often made long before their ultimate capabilities and long-term impact are clear. What we choose to fund, the people we train, the infrastructure we build and the institutions we establish all influence what becomes possible. The second quantum revolution is no longer only a scientific project. It is also becoming an industrial, economic and geopolitical one.

This second edition of the Quantum Index Report captures a field attracting greater resources, broader participation and closer strategic attention. Capital raised tracked in this report exceeded $10 billion in 2025, nearly four times the previous year’s total. Governments around the world are placing greater emphasis on commercialization, procurement and sovereign capability. Demand for quantum skills is growing, educational pathways are expanding and the range of commercially available quantum processors continues to widen. Understanding how these developments translate into scientific progress, practical capability and economic value requires careful measurement.

We created the Quantum Index Report to help people make these decisions in an informed way. Our ambition is to provide a shared foundation of evidence for researchers, entrepreneurs, business leaders, policymakers, educators and the wider public. Many of the people making important contributions to the expanding quantum landscape will bring expertise from outside quantum science. They need reliable information that is accessible without sacrificing the complexity that makes this field distinctive.

> The Quantum Index Report seeks to build an evidence-based understanding of the quantum ecosystem, tracking how the field is evolving and the forces shaping its future.

This year's findings underline the importance of that approach. Our data indicates that leadership varies across technologies and measures. The United States continues to lead in quantum computing research influence, while China leads in publication volume and across multiple measures of quantum communications research. Smaller nations also make contributions that exceed what their size might suggest. Understanding this landscape requires attention to the particular capabilities being developed and the conditions that support them.

This second edition builds on the foundation laid in 2025. The inaugural report's findings were cited or covered by the *Financial Times*, *The Washington Post*, *Scientific American*, the Federal Trade Commission, and the European Securities and Markets Authority. We have expanded the report to 11 chapters, so it keeps pace with a field that is broadening quickly. New analysis covers quantum networking applications, the organizations building the most influential patent portfolios and, in a new chapter, the growing role of quantum technologies in defense. We continue to aggregate data from academic, industry and policy sources and aim to provide nonpartisan insights. Where possible, the underlying data is available at www.qir.mit.edu.

Even with this broader scope, important limits remain. The field is still nascent, and some of the data we need is sparse, inconsistent, uneven across countries or does not yet exist. Definitions evolve, disclosure varies and important activity can be difficult to observe. We see improving this evidence base as an ongoing responsibility.

We are grateful to everyone who responded to last year's invitation to contribute, and we renew it. We welcome datasets, analysis, corrections, commentary and suggestions for what we should measure next. Please connect through the Get Involved section of our website (www.qir.mit.edu/get-involved) or by email (contact@qir.mit.edu).

My thanks go first to Elif Kiesow, our Senior Researcher, who led this edition from its first datasets to its final pages. I also thank our collaborators at Accenture, colleagues across MIT and the many contributors whose expertise and data make this work possible. This research was supported by the National Science Foundation, and I am grateful for its continued support. Together, our team's efforts serve a purpose I consider essential: enabling more people to take part thoughtfully in shaping a technology with potentially far-reaching consequences.

My hope is that readers will use this report to ask sharper questions, test assumptions and make better-informed decisions. The choices being made today will help determine how quantum technologies develop and whom they benefit tomorrow. Those choices deserve the strongest evidence we can provide.

*Jonathan Ruane*

Editor-in-Chief, Quantum Index Report
MIT Sloan School of Management and the MIT Initiative on the Digital Economy
Cambridge, MA, USA

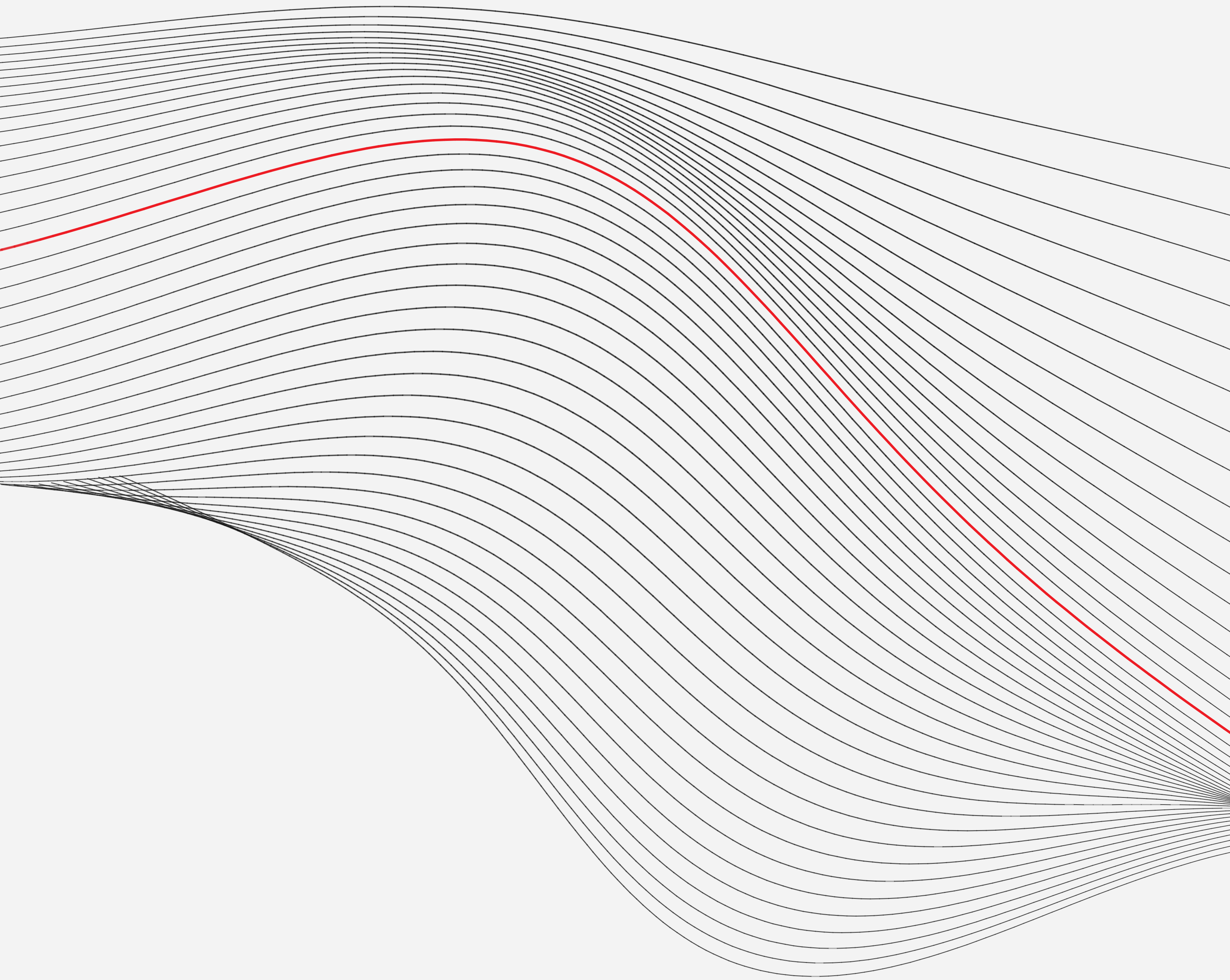

2026 Quantum Index Report

# Table of Contents

2026 Quantum Index Report

# Key Insights

**01** In 2025, China filed more quantum computing patents than **the United States, WIPO, the European Patent Office, and Japan combined**.

**02** Quantum funding nearly quadrupled in a single year to a **record $10.9 billion** in 2025; established players appear to be receiving financing priority ahead of startups.

**03** The U.S. maintains a **global lead in research influence and citation impact** in quantum computing, while China leads in the sheer number of publications.

**04** In quantum communications, China leads globally **across every dimension at once**: output, influence, and citation impact alike.

**05** Quantum technology has moved decisively into the national security and defense policy arena over the past two years. U.S. Department of Defense spending on quantum technology awards hit a **record $285.3 million** in 2025.

**06** **Sovereign capability** in quantum technologies was a focus of national quantum strategies in 2025–2026, together with commercialization and procurement.

07

Public opinion shifted toward more nervousness about quantum technology applications and **lower trust in the government's role to oversee** quantum technology, even as support for funding the research itself stayed resilient.

08

U.S. job postings requiring quantum skills have grown **more than fivefold since 2011** and hit an all-time high in early 2026.

09

Commercially available QPUs grew **from 40 to 56** in a year, with the U.S. retaining a clear lead with 23 and China moving up to second place.

10

Investment and advancement in **quantum repeaters, satellite infrastructure, and optical network switches** could benefit numerous quantum networking applications simultaneously.

11

**AI data centers** are emerging as an important new customer market for quantum companies, giving vendors a path to sell hardware and services into the much larger cloud, HPC, and AI infrastructure ecosystem.

12

The number of dedicated **quantum-titled master's programs** worldwide grew substantially over the past year, with several countries launching their first such degree.

## Recent Highlights

JULY 2025 – APRIL 2026

**JULY 2025**

The European Commission adopts its **Quantum Europe Strategy**.

**NOVEMBER 2025**

The Pentagon designates **Quantum and Battlefield Information Dominance** as one of six Critical Technology Areas, formally elevating quantum's defense priority status.

**DECEMBER 2025**

China formally launches its **National Venture Capital Guidance Fund**, a long-term financing mechanism explicitly covering quantum technology as a strategic industry.

**MARCH 2026**

Xanadu **begins trading** on Nasdaq, becoming the first pure-play photonic quantum computing company to go public.

**MARCH 2026**

The United Kingdom announces a major expansion of its **National Quantum Strategy** worth up to £2 billion, anchored by a world-first government procurement commitment for commercial-scale quantum computers.

**APRIL 2026**

The Czech Quantum Communication Infrastructure launches a completed quantum backbone linking Prague, Brno, and Ostrava under the **EuroQCI** program.

# Recent Highlights

JUNE 2026 — AUGUST 2026

**JUNE 2026**

Quantinuum makes its Nasdaq debut, with a valuation of $17.6 billion.

**JUNE 2026**

The U.S. Executive Order "Ushering in the Next Frontier of Quantum Innovation" is signed, overhauling the National Quantum Strategy toward commercialization, supply-chain resilience, and a government-wide workforce strategy.

**JULY 2026**

IQM begins trading on Nasdaq as the first European quantum computing company to be listed on a major U.S. stock exchange.

**JULY 2026**

D-Wave announces it is moving its stock listing to Nasdaq, capping a wave of quantum companies transitioning to U.S. public markets.

**AUGUST 2026**

Oracle partners with Quantinuum to accelerate hybrid quantum compute adoption in AI data centers.

**AUGUST 2026**

The White House releases its National Security Science & Technology Strategy, formally classifying quantum information technologies among "potentially transformative emerging technologies."

# Executive Summary
## 2026 Quantum Index Report

## Chapter 1 | **Patents**

Quantum computing patent activity continues to mature and globalize, with corporations and universities together accounting for the vast majority of filings, though the balance between them has shifted: university filings have now overtaken corporate filings for two consecutive years, a reversal from a decade ago when corporations filed several times more patents than academia. China has emerged as the dominant force in the country-level landscape, filing more patents than the next several top countries combined, while its two-track model of state-funded research feeding into commercialization vehicles has placed both a domestic specialist firm and a national research academy among the top three patent holders globally. Country patterns also diverge in whether a patent landscape is led by domestic players or by global incumbents extending their footprint internationally.

## Chapter 2 | **Academic Research**

Academic publishing data reveals a split between research volume and research influence, with China and the United States dominating global output but pursuing markedly different strategic priorities. In quantum computing, China leads narrowly in publication share while the U.S. holds a commanding lead in research impact and citation influence, suggesting a marginally smaller but more consistently cited body of American work. In quantum communications, however, China leads decisively across every measure: output, impact and citation share alike, reflecting a sustained national emphasis on secure networking. A handful of smaller research nations, including Germany, the UK, Switzerland and the Netherlands, consistently convert modest publication volume into outsized scientific influence.

## Chapter 3 | **Funding**

Quantum funding reached an unprecedented high, surpassing $10 billion in 2025. Divergence is also observed in how capital is entering the sector: rather than fueling new startups, the recent surge was driven overwhelmingly by post-IPO financing and late-stage growth rounds flowing to already-established players. This capped a decade defined by three distinct eras: steady early growth, a SPAC-fueled breakout phase, and a step-change phase that included the sector's only real correction before rebounding sharply. The 2025–2026 period was also marked by a wave of public-market entries, as several previously venture-backed quantum companies transitioned onto major exchanges, and by continued U.S. dominance in both total capital raised and the pace of recent growth, even as funding broadened across other geographies. The integration of quantum systems into AI data centers suggests that quantum companies are beginning to access a substantial new commercial market, with cloud providers and large enterprises becoming potential buyers of quantum hardware and services.

## Chapter 4 | **Quantum in Corporate Communications**

Mentions of quantum computing across corporate and media channels have grown substantially, suggesting that the topic is increasingly diffusing from a niche technical subject into mainstream business and financial discourse. News coverage has grown far faster than companies' own disclosures, suggesting the public narrative around quantum technology is currently being shaped more by external media framing than by corporate communication itself, though company documents and earnings-call mentions have also risen steadily and appear to be settling into a stable, recurring feature of business discourse rather than a passing trend. Leading firms such as IBM and IonQ dominate document volume, though several smaller, quantum-native companies posted much faster growth rates, narrowing the gap with larger incumbents.

## Chapter 5 | **Policy**

Over the past two years, national quantum strategies have moved toward more support for commercialization, deployment and sovereign technological capability. A wave of countries launched or substantially expanded national strategies during this period, with major economies like the United States and United Kingdom formalizing new executive-level commitments that emphasize domestic manufacturing, government procurement and industry partnerships. Nearly every country profiled also expanded international cooperation through bilateral and multilateral agreements, while several, including the U.S., China, Australia and Canada, increasingly linked quantum policy explicitly to defense and national-security objectives, reflecting a broader convergence between quantum industrial policy and strategic competition.

## Chapter 6 | **Workforce**

Government efforts to build a quantum-capable workforce have intensified across major economies, with the United States, United Kingdom, Australia and the European Union each pursuing distinct but complementary models spanning apprenticeships, dedicated training institutes, and coordinated talent-mobility programs. Labor market data shows that demand for quantum-specific skills has grown consistently and recently reached its highest level on record, a pattern distinct from the more cyclical, boom-and-bust trajectory of general "quantum" mentions in job postings. This divergence suggests the market is maturing beyond broad interest and buzzword adoption toward more concrete, skill-specific hiring needs.

## Chapter 7 | **Education**

Quantum education is broadening well beyond its traditional postgraduate, physics-heavy roots, with policies increasingly focused on introducing quantum concepts earlier, training teachers, and creating accessible pathways for engineers, technicians and industry professionals. The number of dedicated quantum-titled master's programs worldwide grew substantially over the past year, with several countries launching their first such degree, while enrollment data in adjacent fields like computer science and electrical engineering shows shifting, sometimes diverging trajectories. Major coordinated efforts in recent years illustrate a broader shift toward differentiated, system-level education strategies rather than reliance on PhD-level training alone.

## Chapter 8 | **Public Opinion**

Survey data on U.S. public attitudes reveals a population still largely unfamiliar with quantum technologies, with a majority reporting limited familiarity with both quantum computing and quantum networking. Emotional responses vary considerably by application: uses tied to tangible benefits, such as pharmaceutical discovery or secure communications, generate clear enthusiasm, while applications touching sensitive systems, like data security or infrastructure scaling, provoke far more ambivalence. Compared to 2024 survey results, in 2025, public trust in the government's role to effectively oversee organizations and companies using quantum technology decreased, even as support for funding quantum research, whether public or private, remained comparatively resilient.

## Chapter 9 | **Quantum Networking**

Quantum networking is advancing as an incremental, application-driven technology, with near-term commercial value emerging first in constrained applications like short-distance QKD and quantum digital signatures, while long-distance distributed quantum computing remains a more distant prospect. A wave of new testbeds launched across the U.S., Europe and Asia over the past two years is laying the groundwork for larger-scale networks, while examples emerge that build quantum networks as a hybrid layer atop existing classical telecommunications infrastructure rather than as an entirely separate system. The CQN-QED-C industry roadmap identifies a small set of enabling technologies, including quantum repeaters, satellite links and optical switching, as critical bottlenecks whose advancement would benefit numerous applications simultaneously.

## Chapter 10 | **Quantum Processor Benchmarking**

Benchmarking data shows that quantum hardware progress is fracturing into two distinct strategies, with some manufacturers chasing raw qubit scale and others prioritizing circuit quality and fidelity, sometimes releasing devices representing opposite extremes within months of each other. The number of commercially available quantum processors has grown substantially, from 40 to 56, with the United States retaining a clear lead with 23, and China solidifying its position as the closest competitor, while different underlying hardware modalities, including superconducting, trapped-ion, neutral-atom, and photonic systems, continue to show distinct strengths and trade-offs without a clear frontrunner emerging. Forward-looking roadmaps reveal ambitious long-term targets, though the most aggressive scale claims remain notably undated compared to more conservative, year-by-year commitments from other manufacturers.

## Chapter 11 | **Quantum in Defense**

Quantum technology has become an increasingly prominent component of defense and national-security policy, with major governments formally elevating it as a strategic priority and reorganizing institutional structures accordingly, including the Pentagon's designation of quantum as a critical technology area and NATO's first dedicated quantum strategy. U.S. defense spending on quantum-related work has grown substantially over the technology's history, with the majority of cumulative investment concentrated in just the past few years, reflecting an accelerating institutional commitment. Beyond computing, quantum sensing for PNT and secure communications are emerging as some of the more mature near-term military applications, with allied programs increasingly emphasizing hybrid integration with existing classical systems and expanding testing infrastructure and international cooperation frameworks such as AUKUS and NATO's Transatlantic Quantum Community.

# 1 | Patents

The rising trajectory of quantum patent filings serves as an indication of innovation and demonstrates the maturation of research efforts and intensifying geopolitical and institutional competition. Intellectual property assets have become pivotal to establishing competitive advantage as quantum technologies progress from experimental research to commercialization.

This report follows a standardized framework for analyzing global innovation trends by tracking patent data based on patent families, which consist of multiple patent applications covering the same fundamental invention. By accounting for cross-border filing variations and normalizing discrepancies between regional patent systems, patent families offer a more robust metric for evaluating technological advancement than individual patent counts. As a result, this approach enables researchers to evaluate the global scope of innovation, track international intellectual property protection, and detect geographic market diffusion.

The data for this chapter was sourced from Accenture Research and The Quantum Insider.

## 1.1 | Patents by entity

**Fig. 1.a** | Quantum computing patent families by origin, 1999–2025

MIT Quantum Index Report

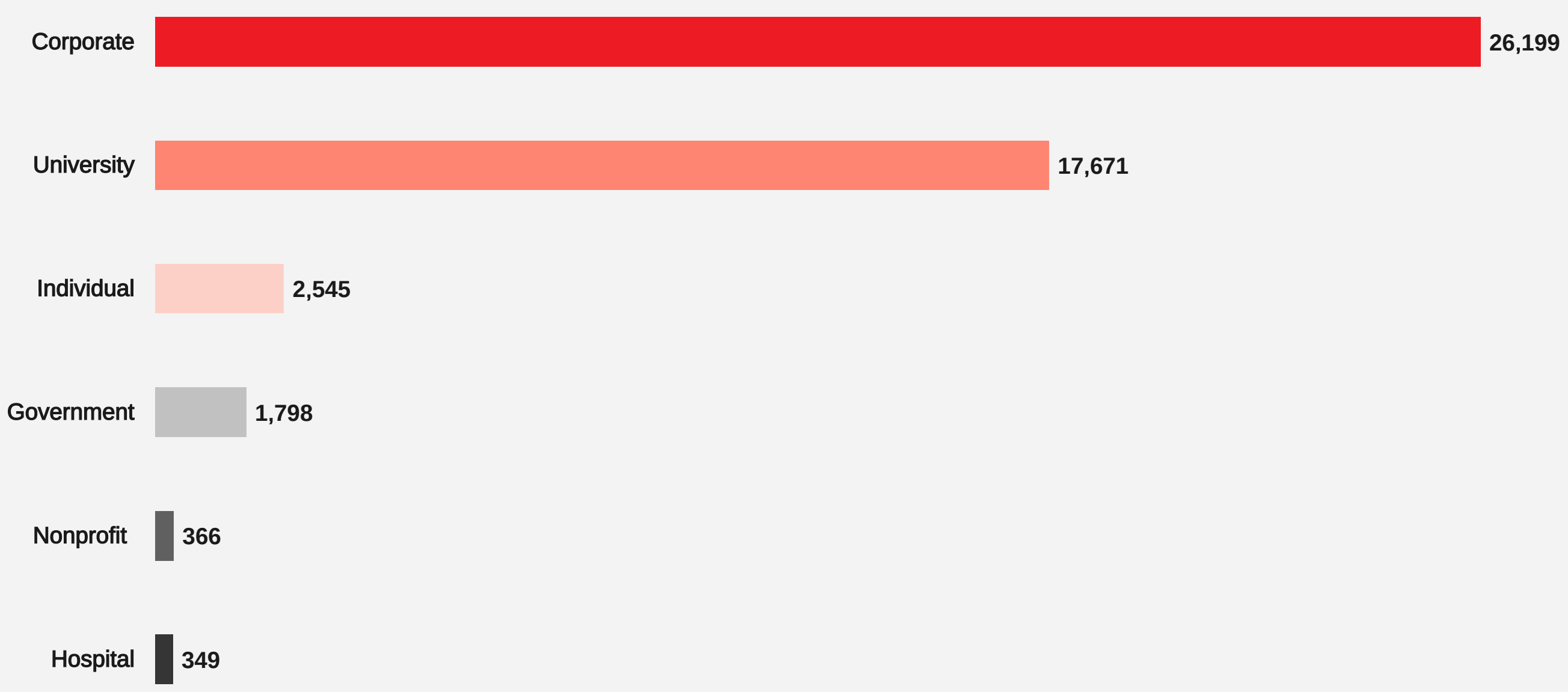

Corporations have emerged as the dominant force in quantum computing patent development, demonstrating significant investment in intellectual property protection. Corporate entities lead the landscape with 53.5% of total patents (26,199), followed by universities with 36.1% (17,671). Together, these two sectors account for nearly 90% of all quantum computing patents, underscoring a high concentration around these two categories. Individual inventors hold the third position with 5.2% (2,545), followed by government institutions at 3.7% (1,798). Nonprofit organizations contribute a modest 0.7% (366), while hospitals show the least participation at 0.7% (349), trailing only marginally behind nonprofits.

The disparity between the top two sectors and the remaining four highlights a pronounced divide in the patent landscape: corporate and academic institutions operate at a scale roughly seven to ten times larger than individual inventors and government bodies, while nonprofits and hospitals remain peripheral contributors. This distribution suggests that meaningful entry into quantum computing patent development still requires the substantial R&D infrastructure and capital that corporations and universities are uniquely positioned to provide.

**Fig. 1.b** | Quantum computing patent families by origin, 1999–2025

MIT Quantum Index Report

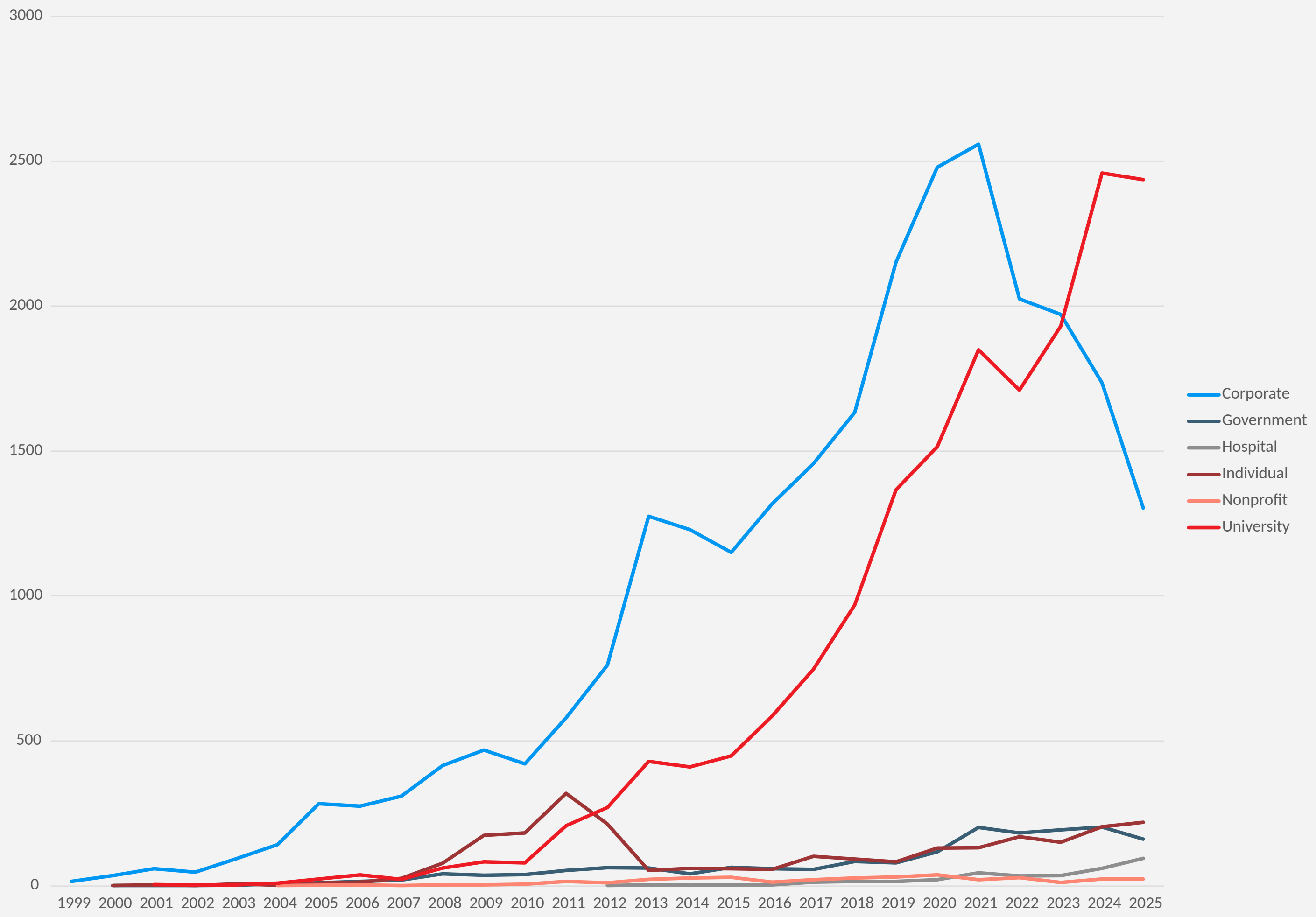

> While universities have recently led annual patent family filings, corporations remain the primary driver of total historical volume across the entire period.

The transition period from 2005 to 2012 marked a significant shift in patent activity across nearly every sector.

Beginning in 2005, corporate patents jumped notably, doubling from 142 filings in 2004 to 284, and then continuing a steady upward march through the decade to reach 762 by 2012, an over fivefold increase in just eight years. This early surge established corporations as the clear engine of quantum computing innovation during the field's formative years. Individual inventors began making more substantial contributions after 2008, when filings nearly tripled from 26 to 78, then surged again to 174 in 2009 before settling into the 180–320 range through 2011–2012, establishing individuals as a consistent, if secondary, presence in the landscape. Government participation also strengthened during this window, rising from just 20 patents in 2007 to 41 in 2008 and climbing further to 63 by 2012, while hospitals made their first tentative appearance in 2006 with a single filing, followed by a second isolated filing in 2012, underscoring how peripheral clinical involvement remained at this stage. Universities, meanwhile, grew from a mere 10 patents in 2004 to 270 by 2012, a twenty-sevenfold expansion that quietly laid the groundwork for academia's eventual dominance.

The years 2013 to 2020 represent a period of accelerated growth and sectoral diversification. Corporate filings climbed sharply, jumping 67% from 762 in 2012 to 1,275 in 2013, and continued an upward trajectory, with only a brief dip in 2014–2015, to reach 2,479 by 2020, more than triple the 2012 figure. Universities mirrored this expansion even more dramatically, growing from 270 patents in 2012 to 430 in 2013 alone, and eventually surpassing 1,500 annual filings by 2020, a nearly sixfold increase over the period that signaled academia's rapidly deepening involvement in quantum computing research. Government filings roughly doubled again, moving from 62 in 2013 to 118 in 2020, while nonprofit activity, though consistently small, peaked at 38 filings that same year. Hospitals began contributing more consistently from 2017 onward, moving from occasional single-digit filings to double digits by 2018 and reaching 21 by 2020, a modest but telling sign of clinical and applied interest entering the field for the first time on a sustained basis.

Government patent activity grew more than tenfold in under two decades, rising from 20 filings in 2007 to a peak of 202 in 2021, making it, alongside universities, one of the fastest-growing categories even though it remains a small share of the overall total.

The most recent period, from 2021 to 2025, reveals a notable inflection point in the overall structure of the patent landscape. Corporate patents peaked in 2021 at 2,559 filings before entering a decline, falling every year thereafter to just 1,304 by 2025, a drop of roughly 49% from their peak. Universities, by contrast, continued to expand through this window even as corporate activity retreated, reaching 2,459 filings in 2024 and 2,436 in 2025, meaning that for the first time, annual university patent activity exceeded corporate activity in two consecutive years, both 2024 and 2025. Hospitals also saw their strongest growth during this stretch, climbing from 34 filings in 2022 to 95 in 2025, nearly tripling in just three years and marking the fastest proportional growth of any sector in the most recent period. Individual inventors likewise reached their highest recorded levels in 2025 at 220 filings, suggesting renewed grassroots engagement even as corporate momentum cooled. Government filings, after peaking at 202 in 2021, have since moderated but remain well above pre-2020 levels.

The corporate-to-university ratio has essentially inverted, which might pave the way for increased cooperation in patent applications between companies and universities[1]. In 2013, corporations filed nearly 3 times as many patents as universities (1,275 vs. 430); by 2025, universities filed nearly 2 times as many patents as corporations (2,436 vs. 1,304), a significant reversal in relative position over just twelve years. Part of this shift may be compositional, as China's share of total filings has grown sharply (see 1.2) and some of its leading filers, such as the Chinese Academy of Sciences, sit at the boundary between state research institutions and commercial entities.[2] Another potential driver is a shift by corporations toward protecting certain research as trade secrets rather than filing patents, as the field matures and disclosure becomes more strategically costly. We were unable to test for either of these explanations in our data.

> Together, these patterns point to a maturing field whose visible innovation is increasingly academic in origin. Whether this reflects a genuine slowdown in corporate activity, a shift in how and where that activity is disclosed, or changes in who is doing the filing in the first place is a question the data cannot yet answer.

## 1.2 | Patents by country

**Fig. 1.c** | Quantum technology patents by country, 2014–2025

MIT Quantum Index Report

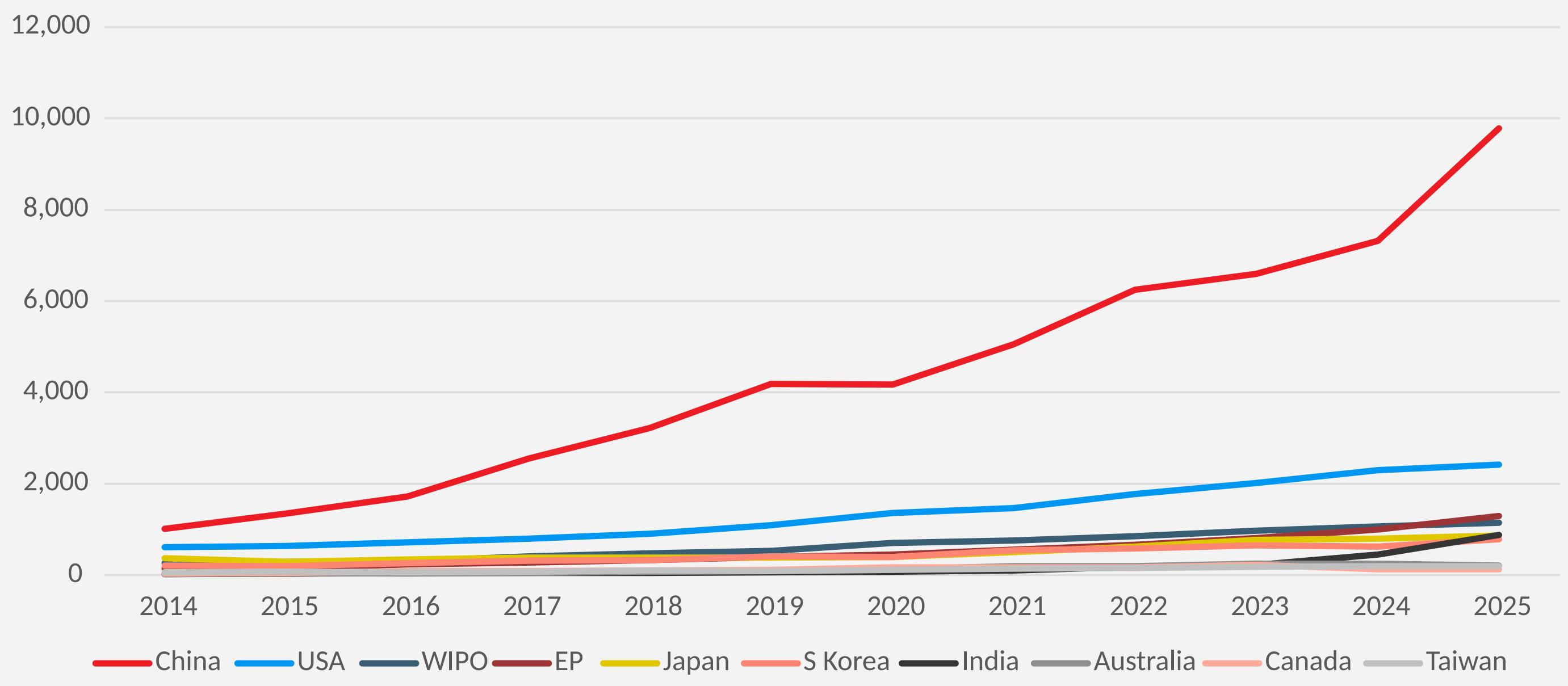

The global distribution of patent filings from 2014 to 2025 tells a story of increasingly lopsided growth, with China's rise reshaping the competitive landscape almost entirely on its own.

In 2014, China led the field with 1,013 patents, already ahead of the United States' 613, but the gap was still relatively contained, with China having roughly 1.65 times as many patents as the U.S. Over the following decade, that gap widened dramatically. China's patent numbers grew nearly tenfold, reaching 9,790 by 2025, while the U.S. grew at a comparatively modest pace, nearly quadrupling to 2,423. As a result, China now shows roughly four times as many patents annually as the United States, a decisive shift from the 2014 ratio and a clear signal of China's increasing importance in intellectual property across this space.[3]

**9,790**

In China, more quantum technology patents were filed than in the rest of the world's top five combined: China's 9,790 filings in 2025 exceed the combined total of the U.S., WIPO, EP and Japan, demonstrating a strong contrast.

The 2014–2018 period saw steady, broad-based growth across most major players. China's patents grew from 1,013 to 3,224, more than tripling in five years, while the U.S. grew more gradually from 613 to 913. WIPO and EP filings also expanded steadily, with EP more than doubling from 150 to 331 and WIPO growing from 265 to 476. Japan, by contrast, remained relatively flat during this window, dipping slightly in 2015 before recovering to 380 by 2018, suggesting a period of consolidation rather than expansion. South Korea showed modest but consistent gains, rising from 210 to 330, while countries like India and Taiwan grew from a low base but had not yet entered a phase of rapid acceleration.

The 2019–2022 period marks the beginning of China's most explosive growth phase. Patent numbers jumped from 4,184 in 2019 to 6,255 in 2022, an increase of 50% in just three years, while the global total grew more slowly by comparison, indicating China was capturing a rapidly increasing share of overall patent activity.

The U.S. also grew meaningfully during this stretch, climbing from 1,095 to 1,772, but its growth rate remained well below China's. This period also saw South Korea briefly overtake Japan for the first time, a notable reversal after years of Japan holding a consistent lead. India's growth began accelerating sharply as well, more than doubling from 79 filings in 2020 to 181 in 2022, foreshadowing the dramatic gains still to come. During this period, Canada and Australia both showed an early high around 2021, with Canada at 176 and Australia climbing steadily toward 198, before both showed a small decline in the subsequent year.

The most recent period, 2023–2025, has been defined by China's continued acceleration even as several other filers plateaued or pulled back. China surged from 6,603 in 2023 to 9,790 in 2025, adding nearly 3,200 patents in just two years. India's growth during this window has been noticeable, nearly doubling from 226 in 2023 to 449 in 2024, then almost doubling again to 883 in 2025, making it the fastest-growing location by a wide margin. Japan and South Korea have continued their close competition, with Japan retaining a narrow lead, including in 2025 (874 vs. 792), reversing South Korea's brief lead from the previous period. Meanwhile, Canada and Australia both show signs of contraction, with Canada falling from a peak of 220 in 2023 to just 127 by 2025, and Australia dropping from 246 to 210 over the same period, the only two countries to show declines from their respective peaks.

**2×**

▶ In the 2023–2025 period, India was the fastest-growing country, nearly doubling from 226 in 2023 to 449 in 2024, then almost doubling again to 883 in 2025, while Canada and Australia were the only two players to shrink from their peak levels.

Looking at the landscape as a whole, global patent totals for the top 10 grew from 2,773 in 2014 to 17,749 in 2025, marking a more than sixfold increase over the eleven-year span, but this growth was not evenly distributed.

China alone accounted for roughly 37% of the 2014 total and had grown to account for over 55% of the 2025 total. The U.S. share, by contrast, actually shrank over the same period, from about 22% of the global top 10 total in 2014 down to roughly 14% in 2025, despite growing in absolute terms, illustrating that China's expansion has outpaced even the next-largest filer.

## 1.2.1 | Leading organizations in quantum patents

**Fig. 1.d** | Top 3 quantum technology patent filing organizations by country (Patent Landscape 2024–2025)

MIT Quantum Index Report

| Country | 1ST PLACE | 2ND PLACE | 3RD PLACE |
|---|---|---|---|
| **Australia** | 1 Alphabet | 2 IBM | 3 Terra Quantum |
| **Canada** | 1 Alphabet | 2 Terra Quantum | 3 Northrop Grumman |
| **China** | 1 Origin Quantum | 2 Chinese Academy of Sciences | 3 State Grid Corp |
| **India** | 1 Vellore Institute of Technology | 2 Saveetha Institute of Medical and Technical Sciences | 3 IBM |
| **Japan** | 1 IBM | 2 NEC Corp | 3 Fujitsu |
| **S Korea** | 1 Samsung | 2 National Research Council of Science and Technology | 3 Korea University |
| **Taiwan** | 1 Fujifilm | 2 Shin-Etsu Chemical | 3 Photonic |
| **USA** | 1 IBM | 2 Alphabet | 3 IonQ |

**Note:** There is an approximately 18-month lag in patent publication data; therefore, recent years' data may be subject to change.

IBM and Alphabet show a similar cross-border pattern, with IBM leading in the United States and Japan, and Alphabet leading in both Australia and Canada. China stands out as a major market where a domestic specialist, Origin Quantum Computing Technology (Hefei), leads, consistent with China's broader push for self-reliance in strategic technologies and quantum in particular. India is the only country where the patent filings in this recent period were led by an academic institution, Vellore Institute of Technology, rather than a corporation, which might suggest a market where research output is translating into practice-oriented IP.

It is also worth noting that IBM is listed most frequently in the top 3, appearing 4 times (Australia #2, India #3, Japan #1, USA #1), followed by Alphabet with 3 appearances (Australia #1, Canada #1, USA #2), and Terra Quantum with 2 (Australia #3, Canada #2).

Australia, Canada, India, and Japan have IBM, Alphabet, or both somewhere in their top 3, whereas China, South Korea and Taiwan do not have an American company appear in their top 3 rankings, which are instead mostly populated by domestic or regional players. This shows that quantum/IP landscapes in East Asia reflect strong national R&D ecosystems and a degree of patent self-sufficiency, versus other regions where global tech incumbents extend their patent footprint internationally rather than being crowded out by local competitors.

In the U.S., Japan, Australia and Canada, every single company in the top 3 is a corporation. But in India, China and South Korea, academic and research institutions crack the top 3, suggesting more university-driven quantum patent activity than commercially driven, unlike the more corporate-dominated landscapes elsewhere.

**Fig. 1.e** | Global leaders in quantum technology patents, 2024–2025

MIT Quantum Index Report

**1** — **Origin Quantum**
Specialized quantum firm, China
1ST PLACE

**2** — **IBM**
Large diversified multinational, United States
2ND PLACE

**3** — **Chinese Academy of Sciences**
Research institution, China
3RD PLACE

**Note:** There is an approximately 18-month lag in patent publication data; therefore, recent years' data may be subject to change.

The global top three show that quantum patent leadership is currently shared by a mix of a specialized quantum firm, a large diversified multinational and a research institution, indicating there are different organizational models at the frontier of this technology. Origin Quantum leads globally, followed by IBM, with the Chinese Academy of Sciences in third. With Origin Quantum in the top spot worldwide and the Chinese Academy of Sciences in third place, China holds two of the top three global positions. It might shed some light on China's deliberate two-track strategy: state-funded fundamental research feeding into commercialization vehicles, such as Origin Quantum, which was incubated out of the Chinese Academy of Sciences[4]. IBM's second-place global ranking is notable given its scale and breadth as a diversified technology company rather than a firm solely focused on quantum research.

Its consistent lead in both the U.S. and Japan further points to a substantial and sustained quantum patent success running across multiple markets simultaneously.

For a more granular analysis, an extensive research report[5] by the European Quantum Industry Consortium (QuIC) looks into patent applications between 2015 and 2023, and maps the leading organizations for the four major qubit platforms for quantum computing applications based on the number of active patent families.

**Fig. 1.f** | Leading organizations by qubit modality (patent landscape ranking)

MIT Quantum Index Report

### Superconducting

Qubits made from superconducting circuits cooled close to absolute zero

|  |  | |
|---|---|---|
| 1 | **IBM** | 1ST |
| 2 | D-Wave | 2ND |
| 3 | Google | 3RD |

### Spin

Qubits stored in the spin of single electrons or nuclei in silicon

|  |  | |
|---|---|---|
| 1 | **IBM** | 1ST |
| 2 | D-Wave | 2ND |
| 3 | Origin Quantum | 3RD |

### Trapped ions

Charged atoms held in place by electric fields and controlled with lasers

|  |  | |
|---|---|---|
| 1 | **IonQ** | 1ST |
| 2 | Qudoor | 2ND |
| 3 | Quantinuum | 3RD |

### Cold atoms

Neutral atoms arranged in arrays and controlled with lasers

|  |  | |
|---|---|---|
| 1 | **Pasqal** | 1ST |
| 2 | Infleqtion | 2ND |
| 3 | Shanxi University | 3RD |

**Note:** There is an approximately 18-month lag in patent publication data; therefore, recent years' data may be subject to change.

IBM, D-Wave and Google form the top three in active patents on superconducting qubits; IBM and D-Wave also keep their relative positions in spin qubits where Origin Quantum ranks third. IonQ, Qudoor and Quantinuum are placed in the top three for most patents regarding trapped ions, while Pasqal, Infleqtion and Shanxi University form the top three in cold atoms.

## 1.3 | Future research

We aim to keep the community updated on the rapidly shifting quantum patent landscape. This involves tracking where patent activity is concentrating globally, as well as identifying new technology hubs and their areas of expertise. Moving forward, we plan to analyze these patents by specific quantum subfields and classification categories. We welcome any stakeholders interested in contributing data or insights to collaborate with us.

You can reach us at contact@qir.mit.edu

---

### Footnotes

[1] Matt Swayne, 'Penn State Files Three Patent Applications Related to Delta Sponsored Research' (The Quantum Insider, 1 June 2026) <https://thequantuminsider.com/2026/06/01/penn-state-files-three-patent-applications-related-to-delta-sponsored-research/> accessed 13 June 2026.

[2] This dynamic is also observable globally, for example, some of China's leading filers, such as the Chinese Academy of Sciences, sit at the boundary between state research institutions and commercial entities. See: Hideki Tomoshige and Phillip Singerman, 'Understanding China's Quest for Quantum Advancement' <https://www.csis.org/analysis/understanding-chinas-quest-quantum-advancement> accessed 26 August 2026.

[3] Although a more granular approach might demonstrate different results, a June 2026 report by Special Competitive Studies Project looks into patent impact and finds that regarding quantum error correction 55% of the most cited patents are U.S.-affiliated, compared to 18% for China. 'Quantifying the Competition: Assessing Leadership in Quantum Technologies' (SCSP 2026).

[4] Matt Swayne, 'China's Origin Quantum Ships First Quantum Computer' (*The Quantum Insider*, 3 February 2023) <https://thequantuminsider.com/2023/02/03/chinas-origin-quantum-ships-first-quantum-computer/> accessed 26 August 2026.

[5] 'A Portrait of the Global Patent Landscape in Quantum Technologies' (QuIC 2026).

# 2 | Academic Research

Academic publications offer an early look at scientific progress, building the base for wider commercial application of quantum technologies. By analyzing publication metrics, we intend to spot emerging trends, track how knowledge spreads, and better understand how global quantum ecosystems are evolving. This chapter draws on research by ASPI[1] to present publication data across quantum computing and quantum communications.

Quantitative metrics give us a clearer view of national capabilities in quantum technologies. While total publication counts show overall activity, reflecting the scale of a country's research infrastructure and talent pool, they tell only part of the story. Combining publication volume with h-index[2] scores and citation data[3] balances sheer output with actual impact and influence. This dual perspective makes it easier to separate quantity from quality, giving a clearer picture of what each nation contributes to the quantum knowledge base.

> The three-metric comparison reveals that publication volume alone does not predict research impact.

# 2.1 | Quantum computing

## 2.1.1 | Rankings by published research

**Fig. 2.a** | National share of quantum computing published research, 2021–2025

MIT Quantum Index Report

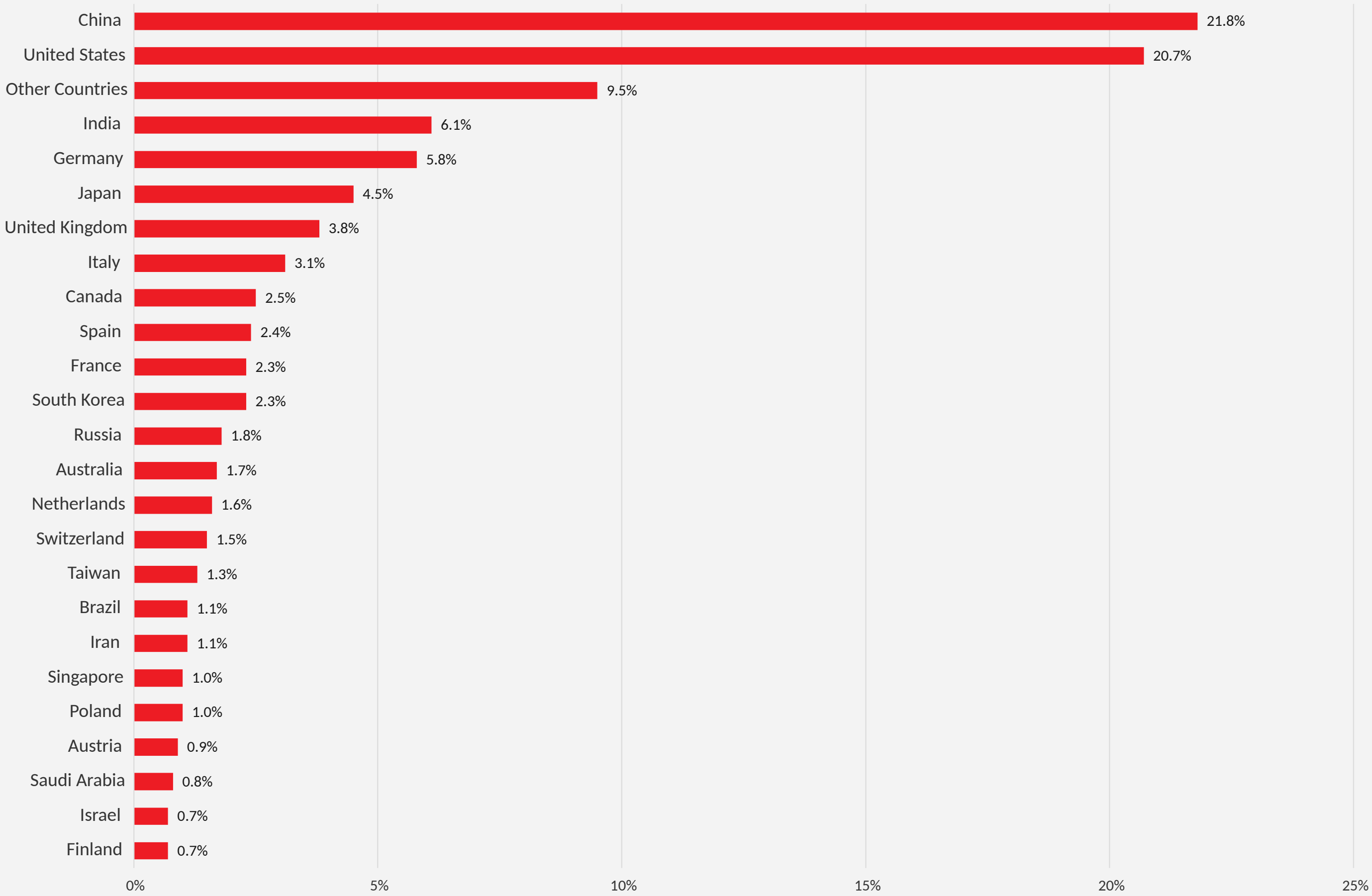

The global landscape of quantum computing publications over 2021–2025 reveals a highly concentrated distribution pattern, with China and the United States collectively accounting for 42.5% of all published research. China leads with 21.8% of publications, followed closely by the United States at 20.7%, signaling a close competition at the top.

India emerges as the leading individual contributor outside the top two, with 6.1% of global publications, followed by Germany at 5.8%. Together, the top five contributors (China, U.S., India, Germany and Japan at 4.5%) account for 58.9% of all quantum computing research, meaning just five countries generate nearly six in every ten papers published worldwide.

**42.5%**

China and the United States alone account for 42.5% of all quantum computing academic publications, nearly matching the combined output of every other country in the top 25.

The middle tier shows diversity: Japan (4.5%), the United Kingdom (3.8%), and Italy (3.1%) form a distinct band of established research producers, together contributing 11.4% of global output. Just behind them, Canada (2.5%), Spain (2.4%), South Korea (2.3%), France (2.3%), and Russia (1.8%) make up a secondary cluster contributing a nearly identical 11.3% combined, indicating a competitive mid-tier.

At the lower end, a long tail of smaller but active research nations rounds out the picture: Australia (1.7%), the Netherlands (1.6%), Switzerland (1.5%), and Taiwan (1.3%) form a modest fifth tier (6.1% combined), while Iran, Brazil, Poland, Singapore, Austria, Saudi Arabia, Finland, and Israel, each contributing between 0.7% and 1.1%, collectively add another 7.3%. Notably, this tail includes a geographically diverse mix spanning the Middle East (Iran, Israel, Saudi Arabia), Latin America (Brazil), and Europe (Poland, Finland, Austria), suggesting quantum computing research, while still top-heavy, is becoming a genuinely global pursuit rather than one confined to a handful of traditional research powers.

## 2.1.2 | Rankings by H-index

**Fig. 2.b** | National quantum computing published research ranked by H-index, 2021–2025

MIT Quantum Index Report

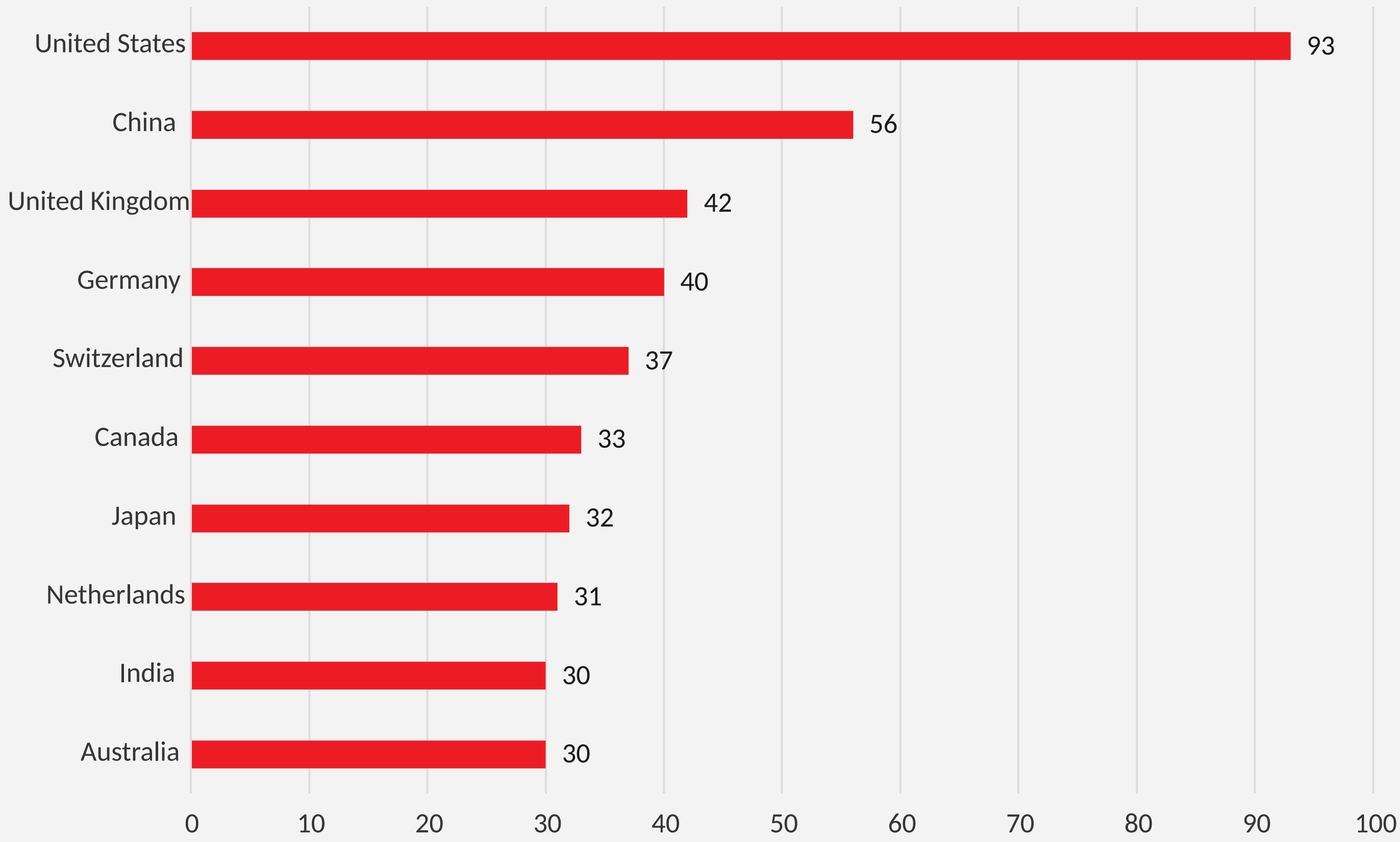

93

▶ U.S. quantum computing research quality is ranked highest in the world, with an h-index of 93 that leaves every other country well behind. China ranks second with an h-index of 56, meaning even though China rivals the U.S. in publication volume, its research impact trails by a much wider margin. The United Kingdom follows in third place with an h-index of 42.

This top three forms a clear leading group, with all other countries falling to 40 or below. Germany (40), Switzerland (37) and Canada (33) round out the upper-middle tier, followed closely by Japan at 32. The Netherlands (31) and a tie between Australia and India (30 each) complete the top ten, showing a tight cluster of countries within just a few points of one another at the lower end of the ranking.

Comparing this ranking to publication volume, the U.S. produces slightly fewer papers than China in raw output terms, yet its h-index is 66% higher, pointing to research that is cited and built upon far more consistently. Similarly, the UK and Germany, mid-tier producers by publication share, punch well above their weight on impact, ranking third and fourth globally despite trailing India in sheer volume of published research.

The h-index rankings reveal a quality gap that does not mirror the quantity rankings: the U.S.'s commanding lead (93) is nearly double China's (56) and more than double the UK's (42), suggesting that research influence in quantum computing is even more concentrated at the top than publication output alone would indicate.

## 2.1.3 | Rankings by most highly cited papers

**Fig. 2.c** | National share of top 10% most highly cited quantum computing publications, 2021–2025

MIT Quantum Index Report

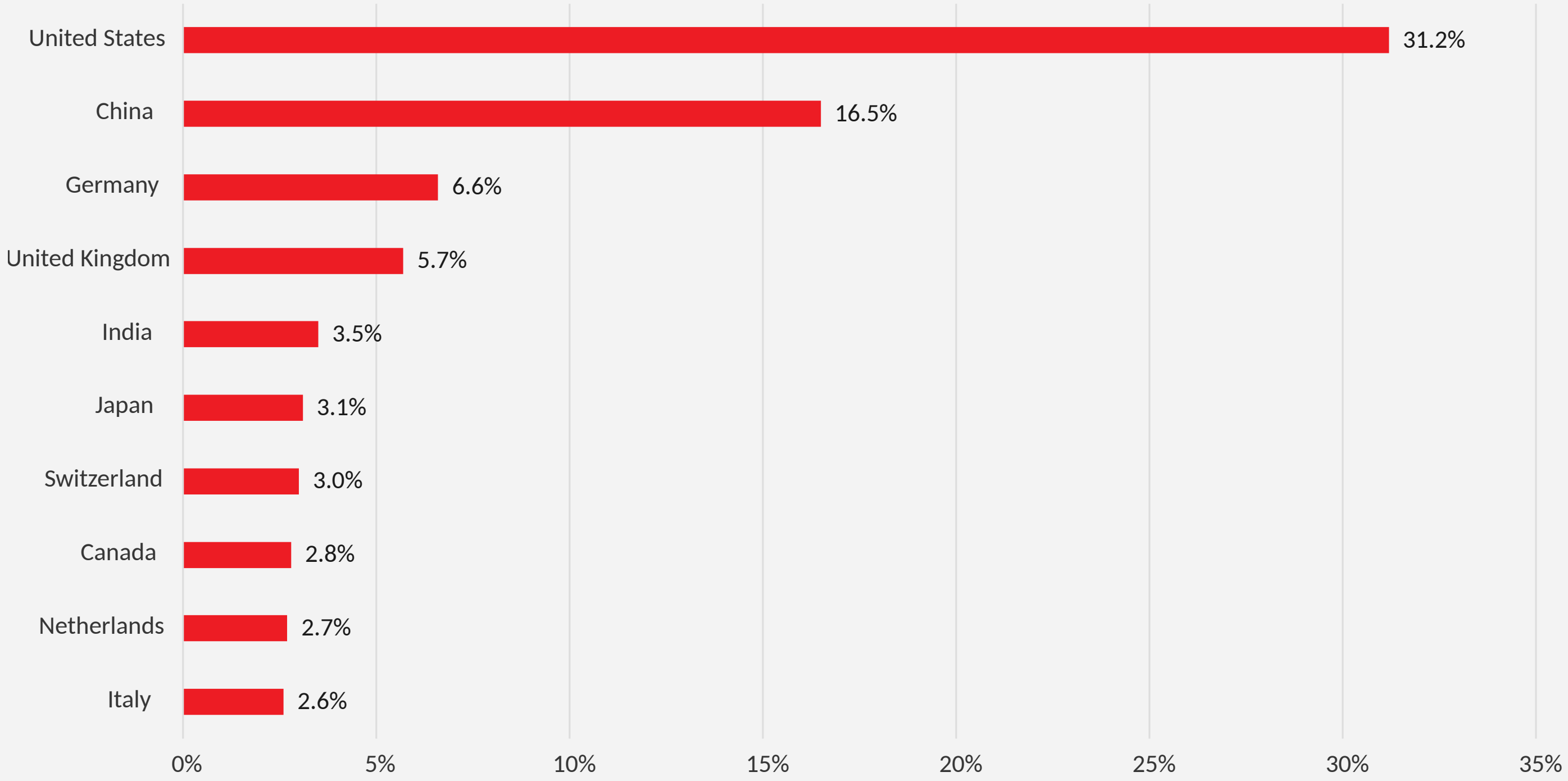

> Among the top 10% of the most highly cited papers in quantum computing, the United States holds the largest share by a wide margin, with 31.2% of all highly cited publications.

China follows in second place with 16.5%, just over half the U.S. share, reinforcing the pattern seen in the h-index rankings, where China's research volume is not matched by an equivalent share of high-impact output.

Germany ranks third with 6.6%, followed by the United Kingdom at 5.7%, together forming a clear second tier well behind the two leaders but still notably ahead of the rest of the field. India comes in fifth with 3.5%, edging out Japan (3.1%) and Switzerland (3.0%) in a closely clustered group.

The remainder of the top ten, Canada (2.8%), the Netherlands (2.7%) and Italy (2.6%), cluster within just 0.2 percentage points of one another, indicating that beyond the top four countries, the ability to produce highly cited research becomes far more evenly distributed among a broader set of nations.

> The U.S. and China alone are responsible for 47.7% of the most highly cited publications, meaning the two leading nations produce nearly half of the world's most influential quantum computing research, a concentration even sharper than their already-dominant 42.5% share of total publications by volume.

## 2.1.4 | Quantum computing rankings overall

**Fig. 2.d** | Quantum computing research rankings overall, 2021–2025

MIT Quantum Index Report

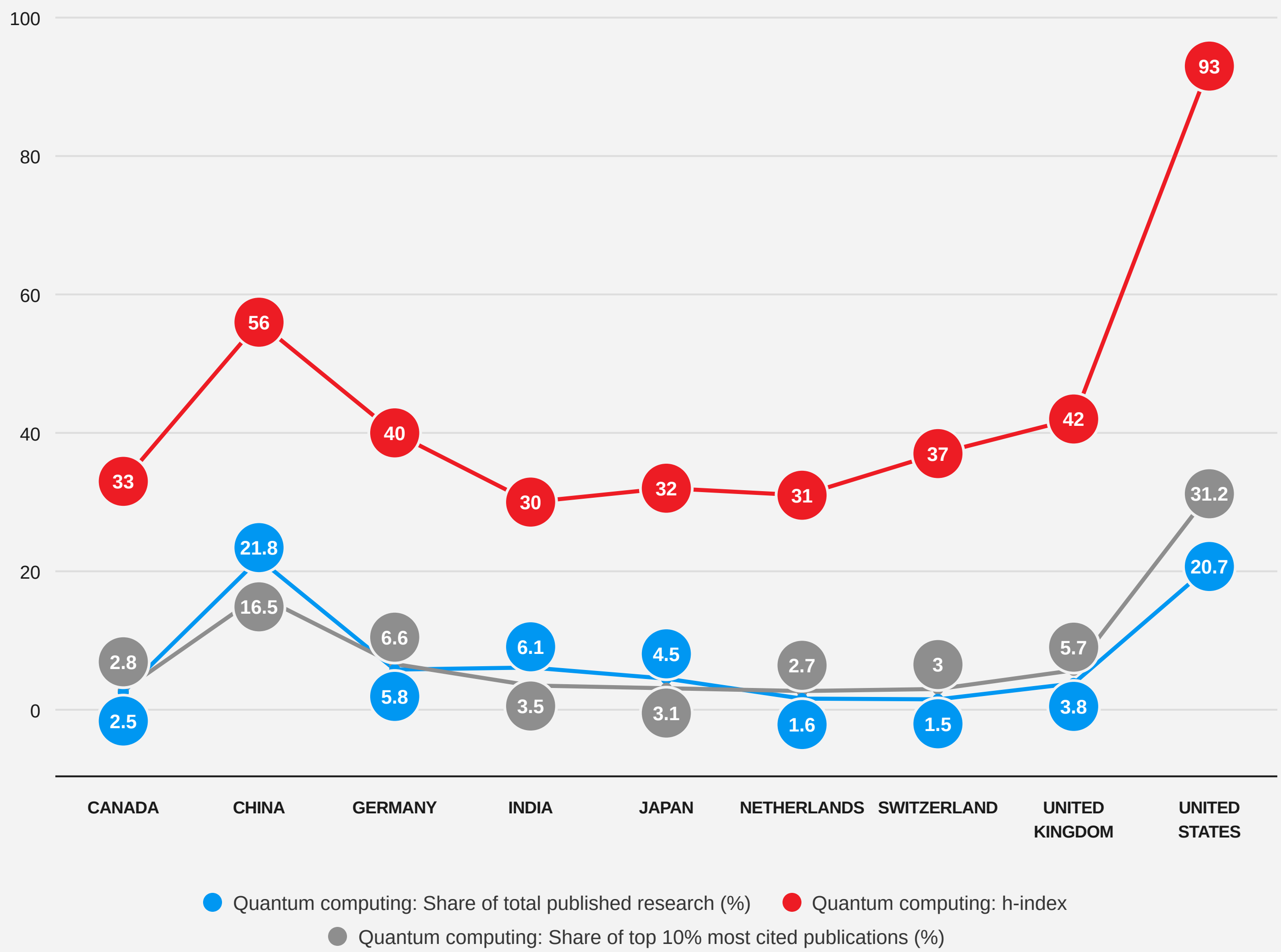

The three metrics together show a fuller picture of global quantum computing research and reveal that volume, quality and impact do not always move together.

The United States demonstrates remarkable consistency in leadership despite not leading in raw output. It ranks second in total publications (20.7%, just behind China), but takes the top position decisively in both the h-index (93) and highly cited publications (31.2%).

This pattern suggests the U.S. research system converts a slightly smaller volume of output into disproportionately higher quality and influence.

China presents the inverse case. It leads narrowly in total publications (21.8%), yet its performance drops sharply in the quality metrics, second in h-index (56, a 37-point gap from the U.S.) and second in highly cited papers. Germany and the United Kingdom emerge as the strongest "quality relative to volume" performers outside the top two. Both countries consistently outperform their publication share in the impact metrics, suggesting efficient, high-quality research systems.

India tells a different story, ranking higher in volume but lower in impact. Switzerland and the Netherlands stand out as small but disproportionately influential contributors. Both nations produce a small fraction of global output but consistently rank among the more cited and influential research bases, which could be interpreted as a sign of concentrated, high-caliber research programs rather than broad output. Japan and Canada occupy a more balanced middle ground, with rankings that stay relatively proportional across all three metrics.

> The three-metric comparison reveals that publication volume and research impact are not always aligned. The U.S. leads decisively on quality and impact despite a marginally smaller publication share than China, while China and India lead on volume but rank lower on influence, and a group of smaller, high-efficiency research nations (Germany, the UK, Switzerland, the Netherlands) consistently convert modest output into outsized global impact.

# 2.2 | Quantum communications

## 2.2.1 | Rankings by published research

**Fig. 2.e** | National share of quantum communications published research, 2021–2025

MIT Quantum Index Report

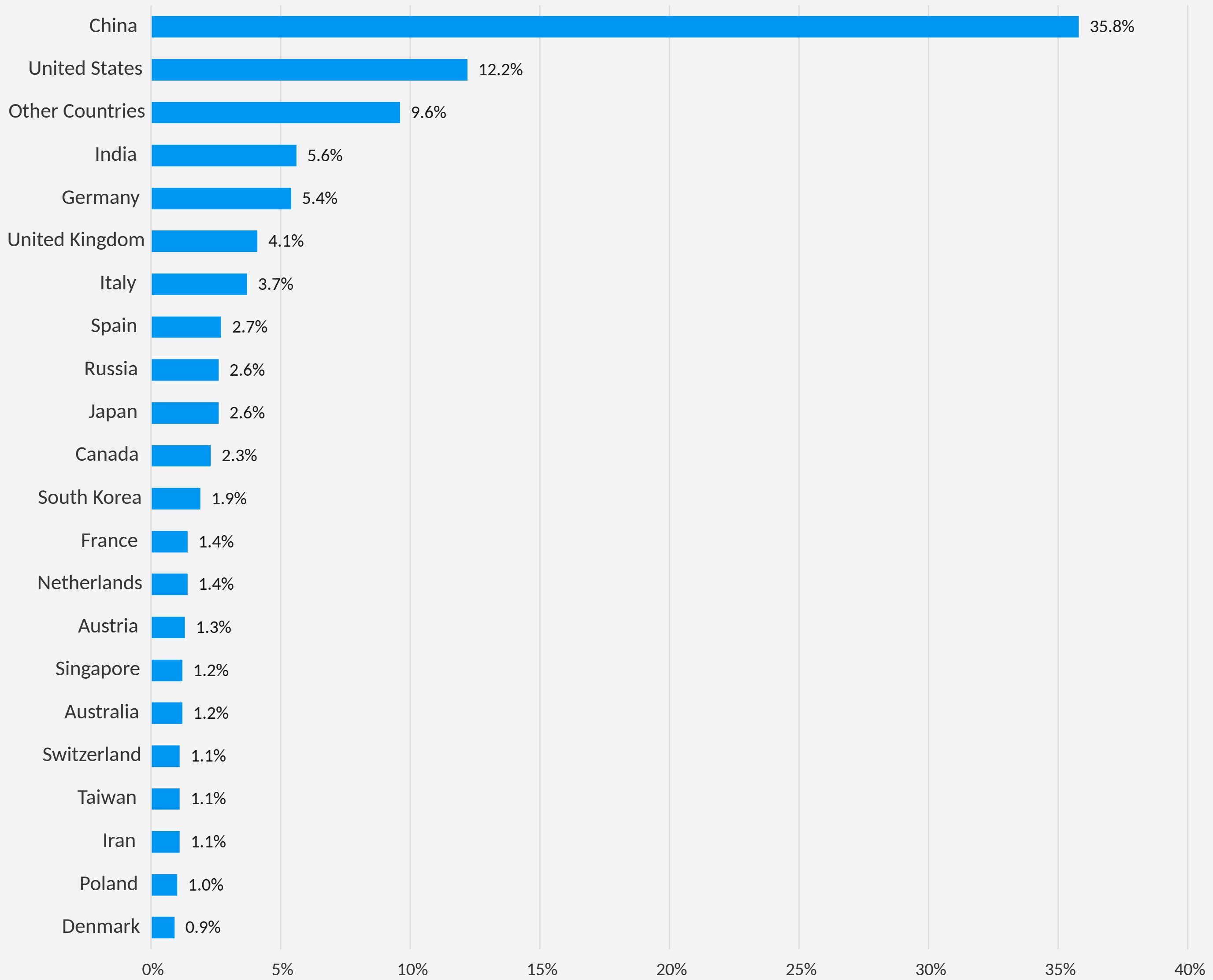

**35.8%**

China leads quantum communications research to a degree far above its already-strong position in quantum computing, holding 35.8% of global publications, nearly three times the share of the United States, which ranks a distant second at 12.2%.

India (5.6%) and Germany (5.4%) form a closely matched third and fourth, together contributing 11%. Combined, the top four countries, China, the U.S., India, and Germany, account for 59% of all quantum communications publications, underscoring the concentration in this research domain.

A secondary tier follows with the United Kingdom (4.1%) and Italy (3.7%), together adding 7.8%, followed by Spain (2.7%), Japan (2.6%), and Russia (2.6%), a closely clustered trio contributing 7.9% combined. Canada (2.3%) and South Korea (1.9%) round out the upper-middle tier, each still producing a meaningful share of global research despite trailing the leaders significantly.

The lower tier reveals a long list of contributors: the Netherlands and France tie at 1.4%, Austria, Australia and Singapore sit near 1.2–1.3%; and Iran, Taiwan, and Switzerland all cluster tightly around 1.1%, followed by Poland (1.0%) and Denmark (0.9%) ranking at the end.

Comparing this to the quantum computing landscape, China's quantum communications share (35.8%) is 64% higher than its quantum computing share (21.8%), while the U.S. shows the opposite pattern, with its quantum communications share (12.2%) roughly 41% lower than its quantum computing share (20.7%). This divergence suggests quantum communications is a field where Chinese research investment and output have scaled far more aggressively than in quantum computing, while the U.S. maintains a comparatively smaller presence, though still securing second place in both domains.

## 2.2.2 | Rankings by H-index

**Fig. 2.f** | National quantum communications published research ranked by H-index, 2021–2025

MIT Quantum Index Report

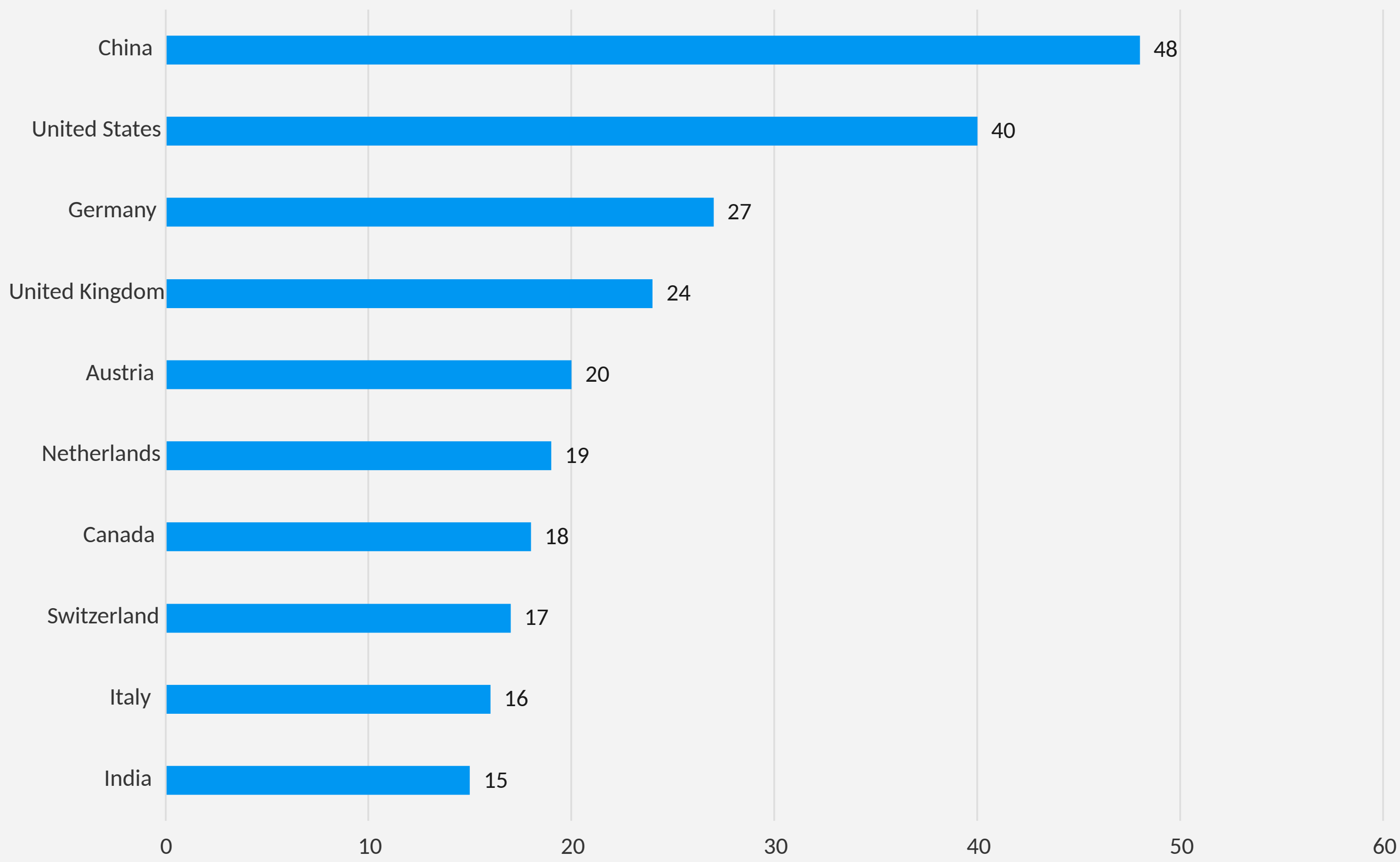

▶ China leads quantum communications research quality with an h-index of 48, followed by the United States at 40, a gap of 8 points that, while notable, is far narrower than the near-threefold lead China holds in publication volume. This suggests that while China's dominance in communications is most pronounced in sheer output, its research impact edge over the U.S. is comparatively modest.

Germany ranks a clear third with an h-index of 27, followed by the United Kingdom at 24, together forming a distinct second tier well behind the two leaders. Austria (20) and the Netherlands (19) follow each other closely, while Canada (18) and Switzerland (17) rank just below them.

Italy (16) and India (15) round out the top ten, closing out a ranking where the gap between third and tenth place spans just 12 points, compared with the 8-point gap between the top two alone, indicating a relatively compressed and competitive mid-to-lower tier.

In comparison to quantum computing rankings where the U.S. held a commanding h-index lead (93 vs. China's 56), in communications, China actually overtakes the U.S. on research quality (48 vs. 40), not just volume. China's leadership in quantum communications reflects genuinely higher-impact research in this specific subfield, unlike quantum computing where its volume lead was not matched by citation influence.

## 2.2.3 | Rankings by most highly cited papers

**Fig. 2.g** | National share of top 10% most highly cited quantum communications publications, 2021–2025

MIT Quantum Index Report

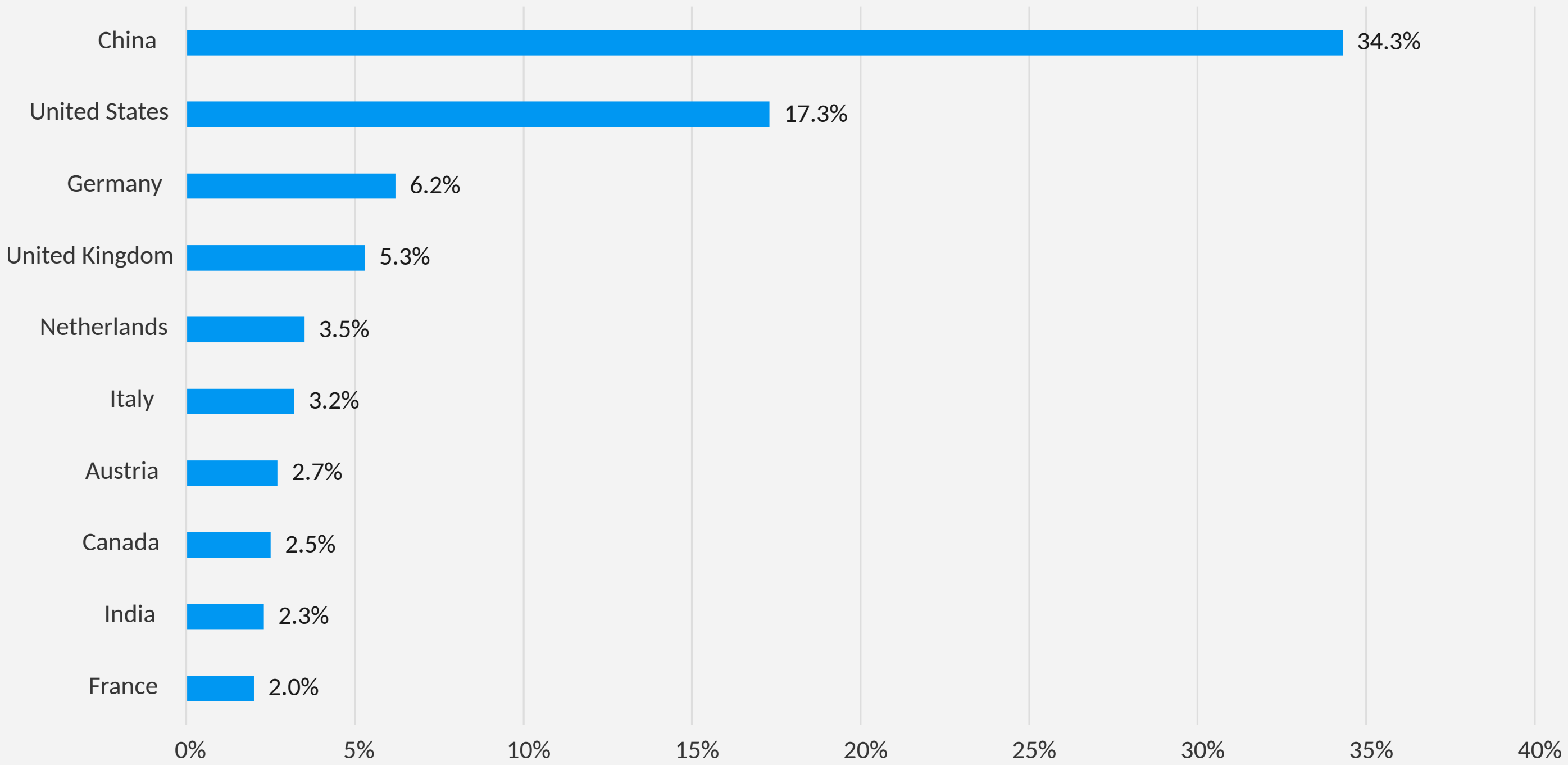

Among the top 10% of the most highly cited papers in quantum communications, China holds a commanding lead with 34.3% of all highly cited publications, nearly double the share of the United States, which ranks second at 17.3%.

Germany follows in third place with 6.2%, ahead of the United Kingdom at 5.3%, together forming a clear second tier well behind the two leaders but still ahead of the rest of the field. The Netherlands (3.5%) and Italy (3.2%) come next, closely matched, followed by Austria (2.7%) and Canada (2.5%).

India (2.3%) and France (2.0%) round out the top ten, with the gap between third and tenth place spanning just 4.2 points, a relatively compressed field once past the two dominant leaders.

Combined, the top ten countries account for 79.3% of all top-10%-cited papers in quantum communications, with China and the U.S. alone responsible for 51.6% of that total, signifying that just two countries produce more than half of the world's most influential quantum communications research.

China's dominance is consistent across all three metrics in this field: it leads in publication volume (35.8%), h-index (48), and now highly cited papers (34.3%), with each share landing in a similar range of around one-third of the global total. This is a notable contrast to quantum computing, where the U.S.'s citation-impact lead (31.2%) was proportionally much larger than its publication share (20.7%); in communications, China's citation-impact share (34.3%) tracks almost identically with its publication share (35.8%), suggesting its research influence in this subfield scales directly with its research volume.

## 2.2.4 | Quantum communications rankings overall

**Fig. 2.h** | Quantum communications research rankings overview, 2021–2025

MIT Quantum Index Report

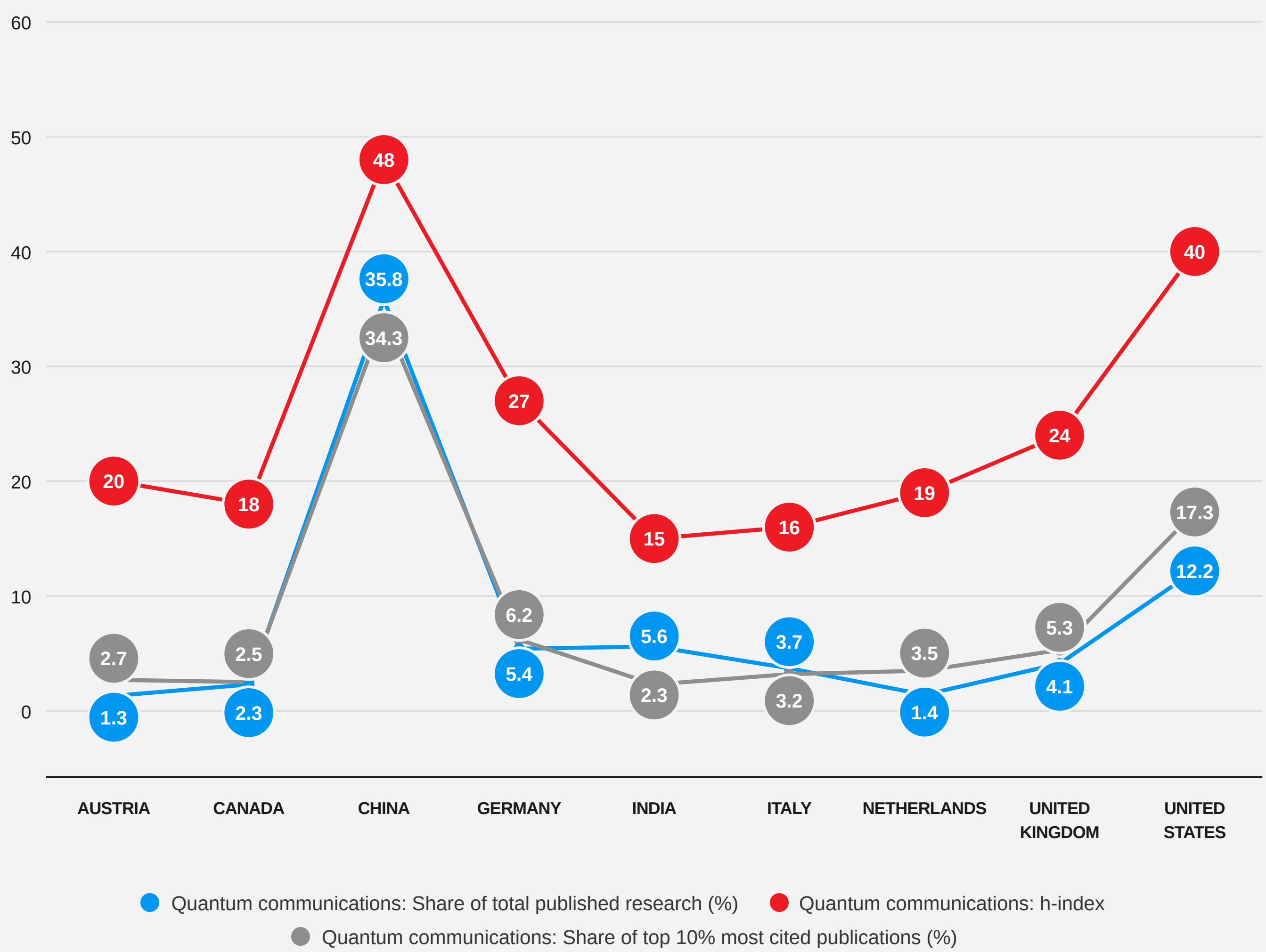

China dominates the quantum communications field across all three metrics, with 35.8% of total publications, significantly outpacing the next closest contributor, the United States, at 12.2% of global output. This nearly threefold gap in publication volume is reinforced by China's quality metrics, where it achieves an h-index of 48 and accounts for 34.3% of the top 10% most-cited works.

The United States holds a firm second place across the board, with 12.2% of publications, an h-index of 40, and 17.3% of highly cited papers. Notably, the U.S. converts its smaller publication share into a citation-impact share (17.3%) that is proportionally larger, roughly 1.4 times its output share, echoing the “small volume, outsized impact” pattern seen in quantum computing, though far less pronounced here given China’s significant lead in this field.

Germany emerges as the clear third-place research power in quantum communications. The United Kingdom follows a similarly stable pattern, placing fourth or fifth across all three metrics.

India presents a divergence: it ranks third in total publications (5.6%), ahead of Germany, yet falls to tenth in h-index (15) and ninth in highly cited papers (2.3%). This high-volume, low-impact pattern is even more pronounced in communications than it was in quantum computing.

Austria ranks in a lower tier in publications (1.3%) but climbs to fifth in h-index (20) and seventh in highly cited papers (2.7%), while the Netherlands shows an almost identical trajectory (1.4% publications, sixth in h-index at 19, fifth in highly cited papers at 3.5%). Both nations exemplify concentrated, high-efficiency research programs rather than broad-based output.

> Quantum communications is a field defined by China’s leadership across volume, quality, and impact alike, a more uniform and commanding leadership position than seen in quantum computing, where the U.S. retained the edge on research quality despite trailing in output. The U.S. remains the clear second place power, Germany and the UK provide steady mid-tier consistency, and countries like Austria and the Netherlands demonstrate that impact and volume don’t always move together, even within a field as concentrated as this one.

## 2.3 | Quantum computing vs. quantum communications research

**Fig. 2.i** | National share of published research, 2021–2025

MIT Quantum Index Report

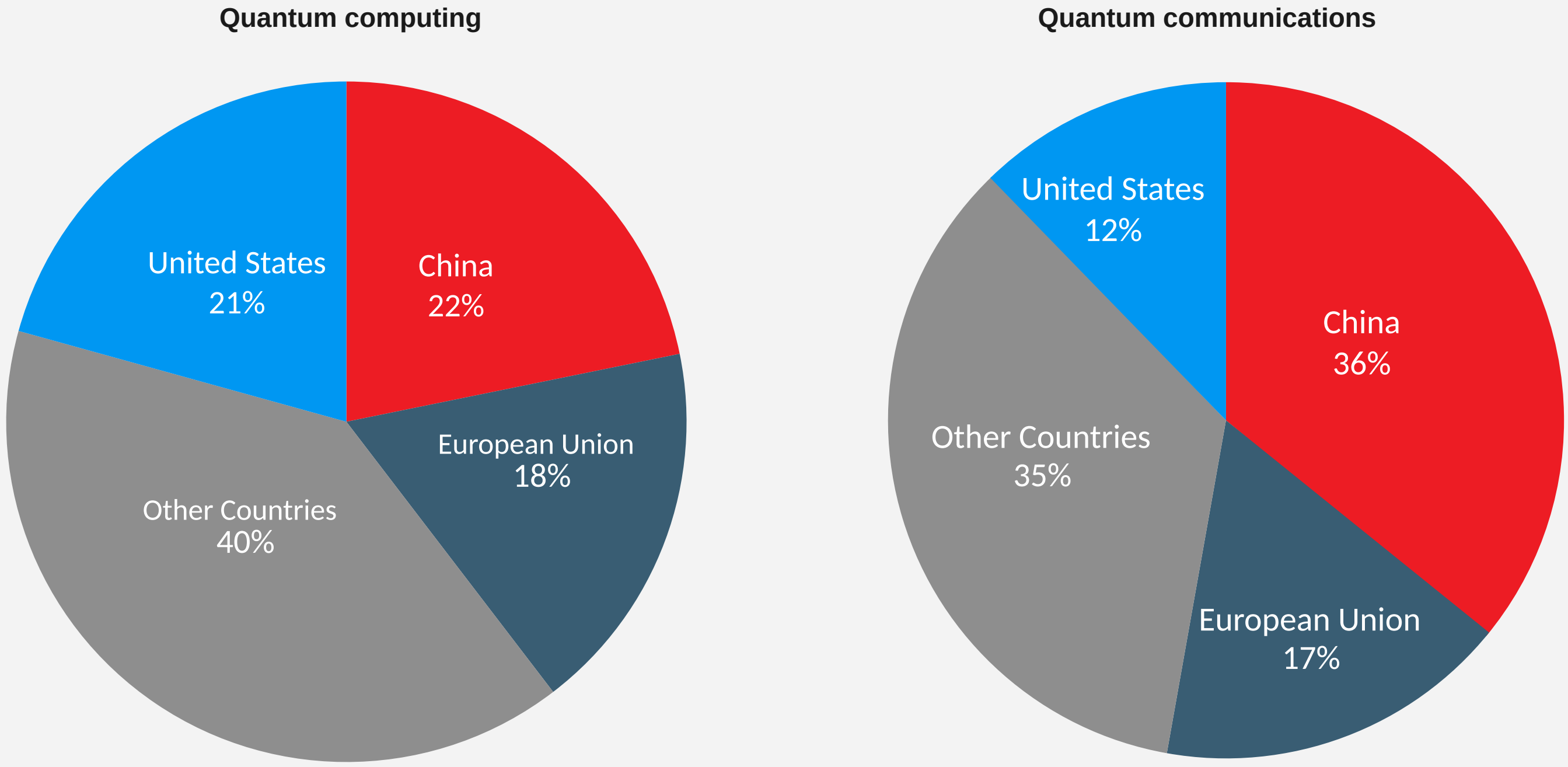

The regional comparison reveals a divergence in strategic focus between the world's two research superpowers: China's 14-point overweighting toward quantum communications contrasts with the U.S.'s 8.5-point overweighting toward quantum computing, while the European Union stands out as the most balanced major player, maintaining an almost identical share of global output across both quantum subfields.

China demonstrates clear strategic emphasis in quantum communications, holding 35.8% of publications compared to 21.8% in quantum computing, a difference of 14 percentage points. This gap might suggest a deliberate national focus on quantum communications, an area often tied to secure networking and satellite-based encryption infrastructure, where China has invested heavily and publicly for over a decade.

The United States shows the inverse pattern. It leads the “other” major players in quantum computing with 20.7% of publications, but drops to a distant second in quantum communications at just 12.2%, a difference of 8.5 percentage points in the opposite direction. This reversal might suggest that the U.S. research base is more heavily weighted toward computing hardware and algorithms than toward communications infrastructure, pointing to a divergent set of national priorities compared to China.[4]

The European Union maintains a remarkably stable position across both fields, holding 17.8% of quantum computing publications and 17.0% of quantum communications publications, a gap of just 0.8 percentage points. This consistency might suggest the EU’s quantum research strategy is broad-based rather than concentrated in one subfield, reflecting a more evenly distributed investment across quantum technologies as a whole.

## 2.4 | Future research

Looking ahead, we aim to expand our analysis by diving into specific subfields of quantum computing and quantum communications. By tracking how research output and quality differ across these specialized areas, we can better map the relationship between theoretical breakthroughs and practical applications. This granular focus would help uncover emerging trends and innovation opportunities that broader, field-level data often hides. We invite the quantum research community to contribute data and insights to future editions of this report.

You can reach us at contact@qir.mit.edu

## Footnotes

[1] Wong-Leung, J., Robin, S., & Cohen, L. (2025). ASPI's Critical Technology Tracker: 2025 updates and 10 new technologies; Gaida, J., Wong-Leung, J., & Robin, S. (2023). Critical technology tracker. Who Is Leading the Critical Technology Race. A Project by the Australian Strategic Policy Institute. Critical Technology Tracker, Australian Strategic Policy Institute (ASPI), (Accessed May 2026), https://techtracker.aspi.org.au.

[2] Ibid, h-index (Hirsch index) is an established performance metric used for analyzing the impact of scholarly output. It's a combined measure of quantity and impact. To calculate h-index, a set of papers (e.g. all those on quantum computing from a particular country over a certain time period) has an h-index of N if the relevant authors have published N papers that have N or more citations each. The h-index is based on Times Cited data from the Web of Science database.
It does not include citations from non-indexed resources.

[3] The top 10% of the most highly cited papers were analyzed to generate insights into which countries are publishing the greatest share of high-quality, innovative and high-impact research. Credit for each publication was divided among authors and their affiliations and not assigned only to the first author (for example, if there were two authors, they would each be assigned half the allocation). Fractional allocation of credit was used for all metrics.

[4] 'Vying for Quantum Supremacy: U.S.-China Competition in Quantum Technologies' (US-China Economic and Security Review Commission 2025).

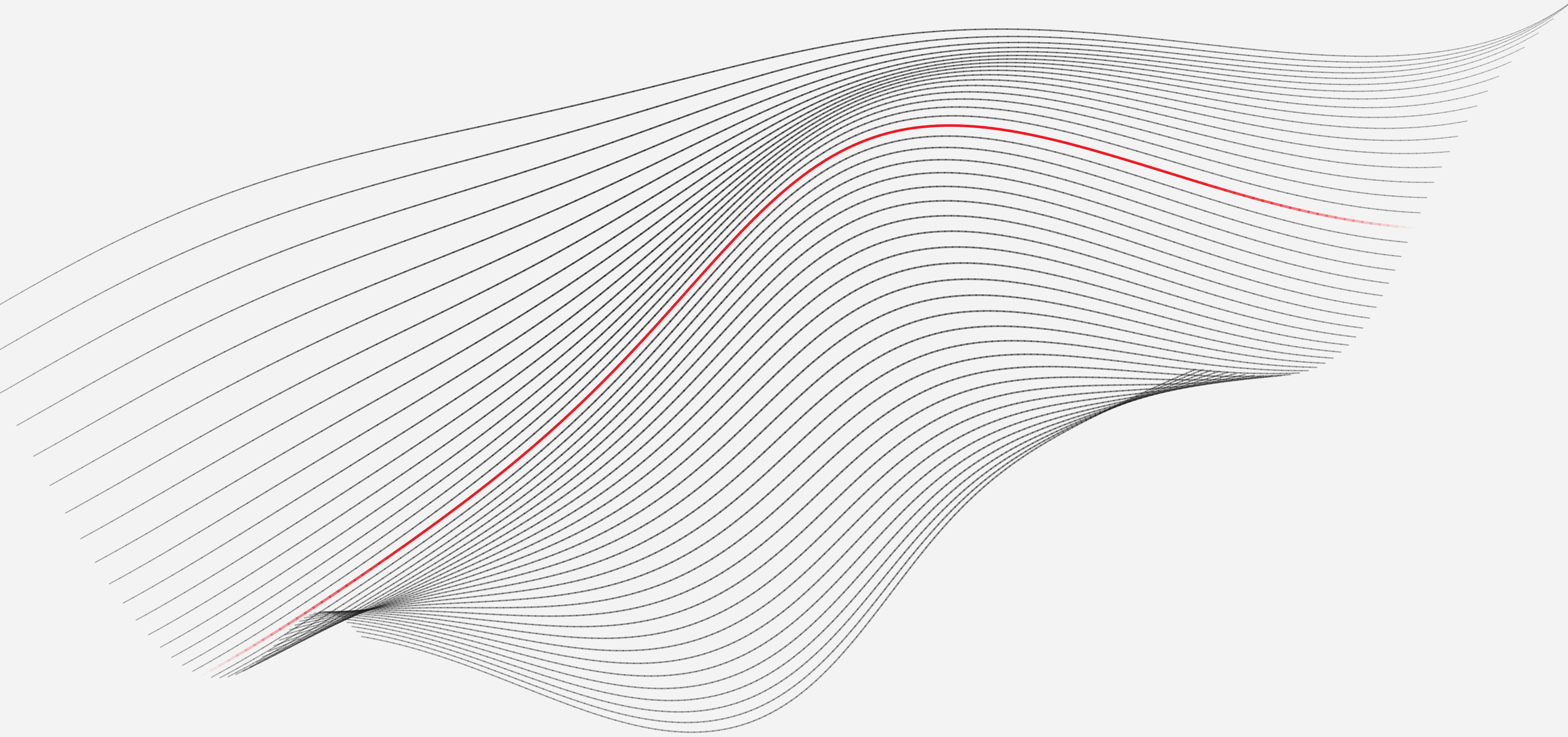

# 3 | Funding

Many early-stage quantum firms rely heavily on financial investment to advance their technology. Given that the field is technically complex and commercially nascent, private firms often source funds from specialized investors willing to take a patient approach.

> Quantum investment hits a record $10.9 billion in 2025, nearly quadrupling in a single year as the sector enters a new growth phase.

Our data, sourced from The Quantum Insider, draws mainly from public sources and press releases. However, gathering and classifying this information comes with limitations, as not all companies disclose funding details.

# **3.1** | Quantum technology funding landscape by financing stage

**Fig. 3.a** | Quantum technology funding landscape by financing stage, 2014–2025

MIT Quantum Index Report

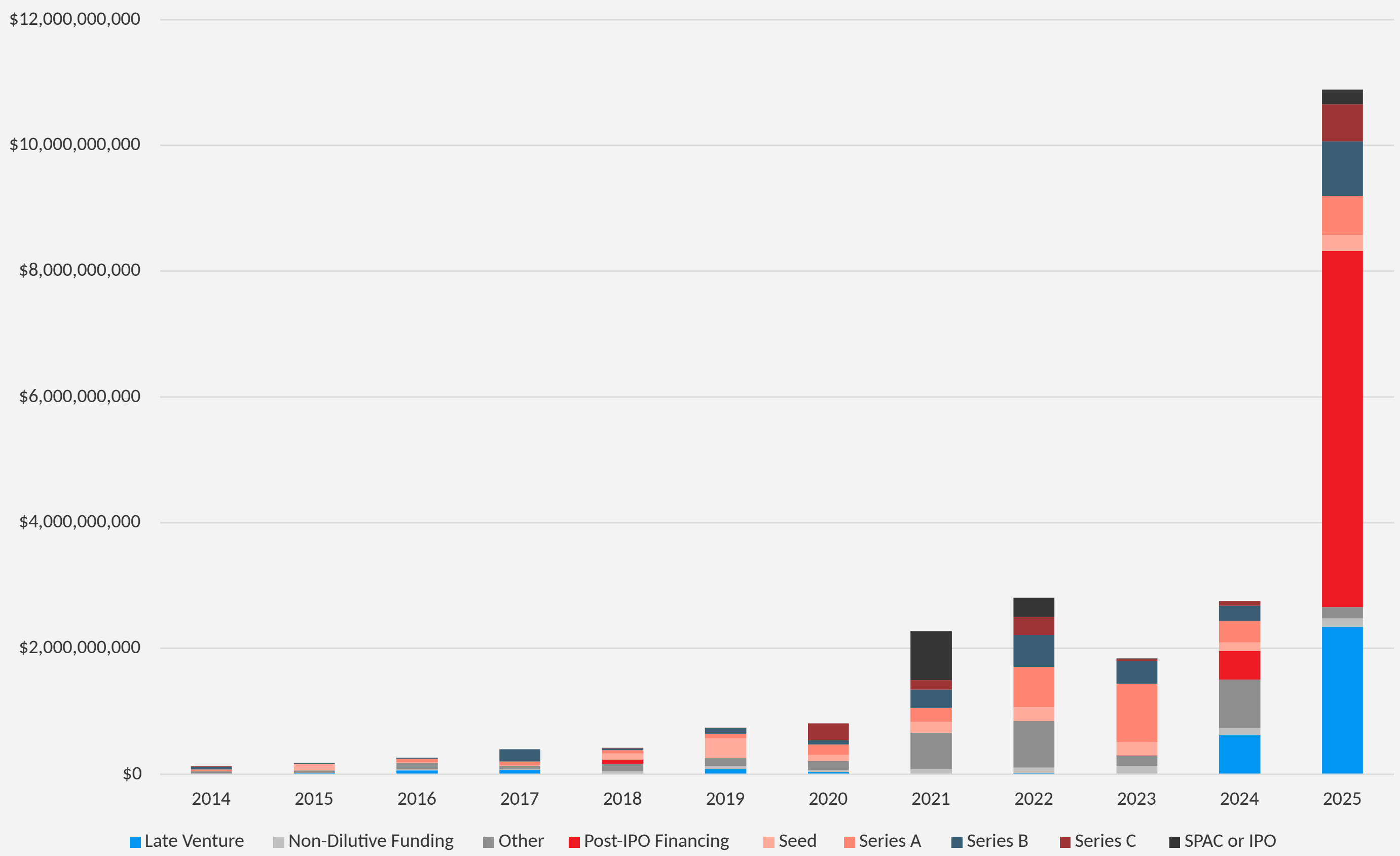

The data traces three distinct eras: a build-up phase (2014–2018) of steady but modest growth; a breakout phase (2019–2022) supercharged by SPAC listings; and a step-change phase (2023–2025), opening with the decade's only real correction before capital investment came back strongly. If 2021–2022 was quantum computing's public-market debut, 2024–2025 looks like its growth-equity coming-of-age.

The quantum technology funding landscape between 2014 and 2025 shows compounding momentum interrupted by a single sharp correction, followed by the most dramatic funding year on record.

Across all round types, the sector raised approximately $23.48 billion, with more than 46% of that total ($10.9 billion) landing in 2025 alone, a signal that quantum is becoming an increasingly attractive investment proposition.

**$10.9B**

2025 marked an unprecedented capital surge, with total funding hitting $10.9 billion, nearly four times 2024 levels, driven almost entirely by Post-IPO Financing ($5.7B) and Late Venture rounds ($2.3B), signaling that capital is now concentrating in relatively mature quantum players rather than new entrants.

The early years (2014–2017) show steady, compounding growth: funding roughly tripled from $126 million to $397 million as the technology moved from experimental to promising. 2018 stands out as a plateau year, with growth nearly flatlining (+4%) after several years of 40–50% expansion, before the sector reaccelerated sharply in 2019 (+78%).

The first real inflection came in 2021, when funding nearly tripled to $2.28 billion, a surge closely tied to the Special Purpose Acquisition Companies (SPAC) wave, which alone contributed $781 million that year as several quantum computing companies took the public-markets route rather than traditional late-stage venture rounds. Significant SPAC examples in this period included IonQ and Rigetti in 2021 and D-Wave in 2022.

2022 pushed totals to a then-record $2.81 billion, but this proved to be a local peak: 2023 delivered the decade's only real contraction, with funding falling 35% to $1.83 billion as the SPAC financing vehicle all but disappeared (dropping to just $500,000) and growth-stage capital broadly tightened.

What followed was a two-year rebound of historic scale. 2024 recovered 50% to $2.75 billion, and 2025 rose sharply to $10.9 billion, nearly four times the prior year and by far the largest single-year total.

Post-IPO Financing is the single largest category over the full period, despite appearing in only four of twelve years (2018, 2023, 2024, 2025), almost entirely on the back of a single staggering year: $5.67 billion in 2025, or 92% of the category's entire decade-long total. This is the clearest contributor to the 2025 surge: capital increasingly flowing to already-public quantum companies rather than early-stage bets.

Late Venture funding tells a similar story of concentration at the end of the timeline. After hovering in the $3–80 million range for most of the decade, it jumped to $618.5 million in 2024 and then to $2.34 billion in 2025, together accounting for over 90% of all late-venture capital ever raised. This, paired with the Post-IPO spike, suggests 2025 was defined less by new entrants and more by existing leaders raising unprecedented growth-stage capital.

Earlier-stage rounds show a steadier growth curve. Series A funding grew from $22.7 million in 2014 to a peak of $927.6 million in 2023; notably, Series A funding held up even as total market funding contracted that year, suggesting investor conviction in new-generation quantum startups persisted even as later-stage and SPAC capital retreated. Series B followed a similar late-decade acceleration, closing 2025 at $866.4 million, its highest year on record.

Seed funding has been comparatively volatile but structurally smaller, peaking early at $314.4 million in 2019, dipping to $95.9 million in 2020, and settling in the $138–254 million range through 2023–2025, a sign that seed-stage risk appetite has normalized below its 2019 highs even as the rest of the market ballooned.

Non-dilutive funding is the one category with an almost uninterrupted growth line, from $3.3 million in 2014 to $138.2 million in 2025, likely reflecting sustained support for quantum research from government grants and institutional contracts.

The SPAC/IPO channel, meanwhile, is a decisively 2021–2022 phenomenon: $781 million and $310.3 million respectively, just $500,000 in 2023, zero in 2024, and a modest $231.2 million reappearance in 2025. This confirms that the SPAC wave was an important liquidity event tied to a specific market window, and later-stage capital needs have since been met through late venture rounds and post-IPO financing instead.

Examples of investment growth in the 2024–2025 period include NVIDIA's venture arm investing[1] in QuEra ($230 million Series B), Quantinuum (Quantinuum's $600 million Series B), and PsiQuantum in 2025 (PsiQuantum's $1 billion Series E), IQM raising $300 million in Series B in 2025[2], and Quantum Machines raising $170 million in Series C funding in February 2025[3]. The large-scale financing also continued into 2026 with examples such as UK-based Oxford Quantum Circuits (OQC) raising US$350 million in Series C in June 2026[4].

Another important development in 2026 has been the emergence of public-market exit and growth-capital pathways for previously venture-backed quantum companies. IQM announced a public-market transaction in February 2026 and began trading on Nasdaq in July 2026, as the first European quantum computing company to be listed on a major U.S. stock exchange.[5] Quantinuum's shares rose 13.3% in their Nasdaq debut in 2026, with a valuation of $17.63 billion.[6] D-Wave announced it would move the listing of its common stock from the New York Stock Exchange to Nasdaq as of July 2026.[7] These developments were also preceded in 2026 by the Nasdaq listings of Xanadu as the first pure-play photonic quantum computing company to go public[8] and Horizon Quantum listing on Nasdaq following a SPAC merger.[9]

## 3.2 | Quantum technology funding by category

**Fig. 3.b** | Quantum technology funding landscape by category, 2014–2025

MIT Quantum Index Report

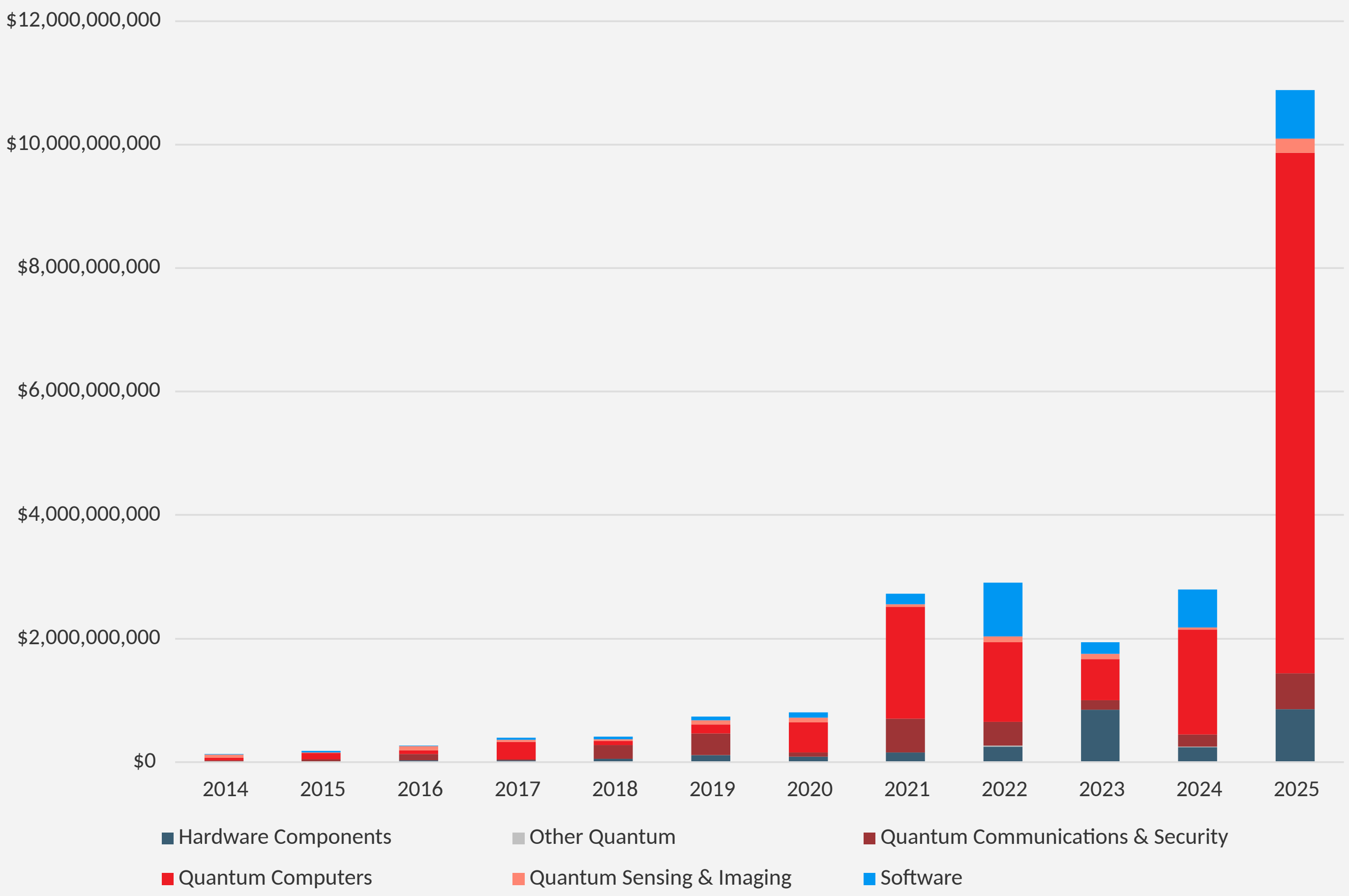

The quantum technology sector has experienced significant shifts in investment patterns over the past decade, and the category-level data makes clear where that capital has concentrated. Quantum Computers have generated the highest share of overall funding by a wide margin, pulling in $15.08 billion since 2014.

No other segment comes close: Quantum Communications & Security ($2.68B), Hardware Components ($2.67B), and Software ($2.90B) each represent roughly 11–12% of total funding, while Quantum Sensing & Imaging trails at $832 million (3.4%), and the miscellaneous “Other Quantum” bucket accounts for just $25.8 million.

**$15.1B**

▶ Quantum Computers have generated the highest share of overall funding by a wide margin, pulling in $15.1 billion since 2014, roughly 62% of all quantum investment across every category combined.

The data shows strong, sustained growth through 2024, particularly among Quantum Computers, which received $1.70 billion in investments across the year, more than 2.5 times their 2023 total ($670.5M) and enough on its own to drive the sector’s overall 44% year-over-year rebound. Software also showed remarkable strength in 2024, reaching $616.0 million, more than three times its 2023 level, while Hardware Components ($244.5M) and Sensing & Imaging ($34.7M) lagged behind, suggesting the 2024 recovery was led by both code-layer investment and Quantum Computers.

If 2024 was a recovery, 2025 was a structural break. Total funding across all categories hit $10.9 billion, and Quantum Computers alone accounted for $8.43 billion of it, nearly 77% of the year’s total and more than five times the category’s already-strong 2024 figure. This single category effectively defines the 2025 surge: strip it out, and the rest of the sector grew from roughly $1.1 billion in 2024 to $2.45 billion in 2025, still a healthy more-than-doubling, but nowhere near the scale of the compute-side sharp increase.

Communications & Security ($578.0M) and Hardware Components ($857.0M) posted their strongest years on record in 2025, while Sensing & Imaging ($229.8M) rebounded sharply from subdued 2024 levels, indicating that, even as Quantum Computers captured the largest share of capital, the broader ecosystem of adjacent technologies grew in tandem rather than being crowded out.

Quantum Computers have led every major growth cycle. Its funding jumped from $480M in 2020 to $1.8B in 2021, the single largest driver of that year's 238% overall surge, before dipping through 2022–2023 and then re-accelerating into the record-breaking 2024–2025 run. 2023 was the only broad-based correction of the decade, with total funding falling 33% as nearly every category retreated simultaneously (Quantum Computers -48%, Software -78%, Communications -61%), before all categories recovered together in 2024.

Software's rise has tracked, and increasingly rivaled Hardware. From near-negligible funding in 2016 ($224K), Software has grown into a $2.9 billion cumulative category, overtaking Hardware Components' decade-long total ($2.7B), a sign that investor confidence has broadened from physical qubit infrastructure toward the algorithms and tooling layer built on top of it.

# 3.3 | Quantum technology funding landscape by country

**Fig. 3.c** | Quantum technology funding landscape by top 10 countries, 2014–2025

MIT Quantum Index Report

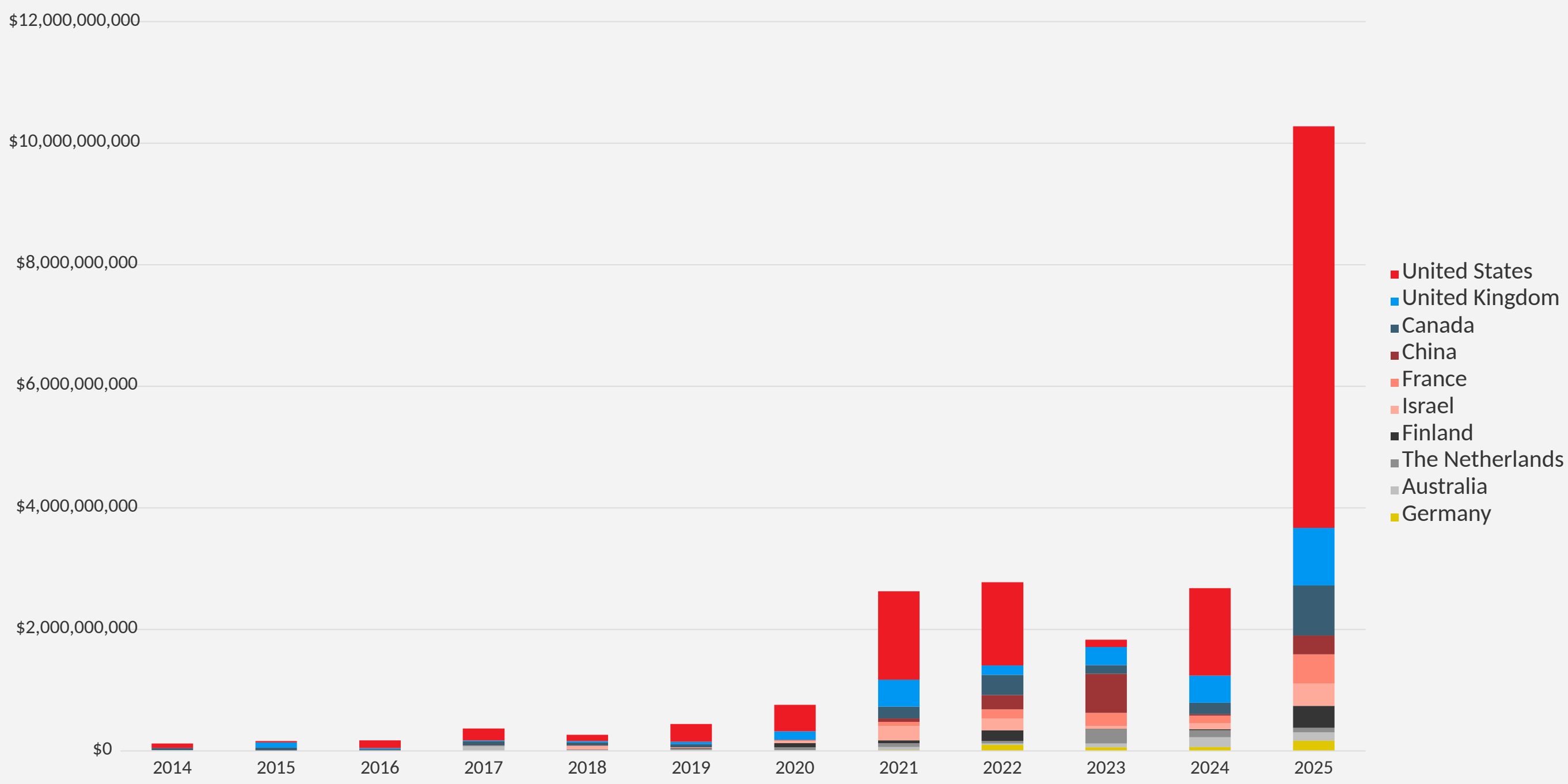

The United States ranks at the top by a wide margin in total equity funding across 2014–2025. U.S.-headquartered firms secured $12.23 billion, more than four times the second-placed nation (the UK, with $2.60 billion). Canada ranks third with $1.97 billion, rounding out a top tier that captures nearly 75% of all quantum investment across the top 10 countries.

At the next level down, China ($1.26B), France ($1.10B) and Israel ($1.03B) form a closely bunched middle tier, each having crossed the billion-dollar mark cumulatively despite starting from near-zero funding as recently as 2017–2018.

Finland ($692.6M), the Netherlands ($619.7M), Australia ($539.2M) and Germany ($443.1M) cluster just behind, having built their totals through steadier, less spike-driven growth.

The funding history over time shows that most countries maintained very modest, relatively volatile investment levels through 2016, after which the market entered a more consequential growth phase. 2017 marked the first meaningful acceleration, more than doubling the prior year's total on the strength of early Australian and U.S. activity, though this proved uneven: 2018 saw a 28% pullback, largely reflecting the absence of large U.S. rounds that year.

From 2019 onward, growth became more broad-based: nearly every country in the top 10 began contributing meaningfully, and by 2021 the market had entered its first major boom, with total funding nearly quadrupling to $2.63 billion. This was driven disproportionately by the U.S. ($1.46B) and a sudden emergence of Israeli funding ($235.5M, up from $2.5M in 2019), alongside the UK's breakout year ($437.4M). 2022 essentially held these gains (+6%), with China making its first major appearance ($233.8M) and Canada surging to $330.9M.

2023 stands out as the only true correction in the country-level data, with total funding falling 34% to $1.83 billion, driven almost entirely by a sharp U.S. pullback (from $1.37B to just $117.0M), even as several other countries posted their strongest years yet: China ($638.4M), France ($217.5M), and the Netherlands ($236.1M) all hit multi-year highs, suggesting capital was rotating toward Europe and Asia even as the U.S. market paused.

2024 brought a broad-based recovery (+47%), led by a rebound in U.S. funding to $1.44 billion, alongside record years for Australia ($160.5M) and steady gains across the UK, Canada and the Netherlands.

2025 marked a global surge: the U.S. ($6.61B), the UK ($946M), Canada ($826M), France ($480M), Israel ($367M), and Finland ($362M) all posted record-high years simultaneously, showing the funding boom broadened across geographies even as American capital scaled fastest.

**64%**

In 2025, top-10 country funding reached $10.28 billion, of which the U.S. contributed $6.61 billion, more than quadrupling its 2024 levels, and ended up accounting for 64% of the top-10 total.

## 3.4 | Quantum & AI data center colocation

The positioning of quantum systems alongside AI hardware in data centers suggests that quantum firms are beginning to access a new commercial market, with cloud providers and large enterprises becoming potential buyers of quantum hardware and services.

Across 2025–2026, quantum computing has started to take on a new role in AI data centers as an additional form of specialized processing. The physical colocation of quantum processing units (QPUs) inside commercial data center environments has accelerated in this time frame.

In August 2026, it was announced that Quantinuum's Helios quantum computer will be deployed inside a U.S.-based Oracle Cloud Infrastructure (OCI) AI data center, where it will operate alongside OCI's existing GPU and high-performance computing infrastructure. The stated objective was to make hybrid quantum-AI workloads available as an OCI service, allowing classical and quantum processors to contribute to different stages of the same computational problem. Quantinuum subsequently described the deployment as an "industry-first," making it an important example of quantum hardware becoming part of the physical and computational architecture of an AI-focused data center.[10]

Other deployments suggest that this development is becoming a broader infrastructure trend. In June 2026, Oxford Quantum Circuits (OQC), JPMorganChase and AMD announced a dedicated Quantum-AI Data Centre in London that combines OQC quantum hardware with AI and high-performance classical computing resources supplied in part by AMD. JPMorganChase researchers are using the environment to investigate hybrid quantum-classical algorithms, portfolio optimization, quantum machine learning, and AI models designed to improve quantum-circuit performance.[11]

A different model can be seen at Aramco, where Pasqal has installed a 200-qubit neutral-atom quantum processor directly in the company's data center in Dhahran. The system entered active operation through a commercial Quantum Computing as a Service platform in May 2026 and is being explored for optimization, simulation, energy and industrial applications. [12]

The interest in using quantum hardware in data centers is not entirely novel in itself, as in 2023 Quandela announced the delivery of its MosaiQ photonic quantum computer to an OVHcloud data center, providing an earlier, smaller-scale example of a cloud and data-center operator installing quantum hardware within its own facilities.[13][14]

The increased frequency of these collaborations points toward the emergence of a hybrid computing architecture in AI data centers. In such a system, CPUs, GPUs, HPC systems, and quantum processing units perform complementary tasks, with workloads moving between them according to the type of computation required. This approach is also being supported by dedicated interconnection technologies.

NVIDIA's NVQLink architecture, for example, is designed to connect quantum processors directly with GPU-accelerated supercomputing systems through low-latency, high-throughput links, allowing GPUs to support functions such as quantum control, calibration, and error correction while also enabling hybrid quantum-classical applications.[15]

HPE and Quantinuum have similarly begun developing frameworks for integrating quantum computers with existing HPC and AI environments.[16] These developments reinforce the idea that the QPU may eventually become another specialized accelerator within the data center, comparable not in function but in architectural role, similar to how GPUs are used alongside CPUs today.

This convergence could become even more significant for AI data-center design, as facilities that host quantum processors would need to support specialized control systems and communication links required to reliably connect quantum and classical hardware. For instance, BTQ Technologies and data-center developer TIDAL PWR announced in July 2026 that they are developing reference architectures for facilities intended to support AI, HPC, and future quantum workloads, including requirements for networking, security, procurement, and trusted supply chains.[17]

## 3.5 | Future research

With relatively few quantum venture deals occurring globally each year, reporting gaps naturally exist, offering a great opportunity for collective data improvement. If you possess data that could strengthen future Quantum Index Reports, we welcome your input.

Looking ahead, our analysis points to several high-priority research areas. Key topics include comparative cross-country studies on how research investment and talent pipelines fuel startup growth, impact assessments correlating funding with technical milestones, and a deeper look at the strategic role of corporate investors.

A related and currently underexplored area is the question of realized returns.

As the earliest quantum investments begin to mature, tracking exits, valuations and follow-on financing will be pivotal in understanding what financial performance from quantum investment actually looks like.

You can reach us at contact@qir.mit.edu

---

## Footnotes

[1] Matt Swayne, ‘How Is NVIDIA “Betting” on Quantum? A Look at NVentures Quantum Investment Playbook From Skepticism to Strategic Stakes’ (*The Quantum Insider*, 13 September 2025) <https://thequantuminsider.com/2025/09/13/how-is-nvidia-betting-on-quantum-a-look-at-nventures-quantum-investment-playbook-from-skepticism-to-strategic-stakes/> accessed 6 July 2026.

[2] ‘IQM Quantum Computers Raises over $300 million in Series B Funding Round’ (*HPCwire*) <https://www.hpcwire.com/off-the-wire/iqm-quantum-computers-raises-over-300m-in-series-b-funding-round/> accessed 20 June 2026.

[3] ‘Quantum Machines Raises $170 Million in Series C Funding’ (*Quantum Machines*) <https://www.quantum-machines.co/press-release/quantum-machines-raises-170-million-in-series-c-funding/> accessed 16 July 2026.

[4] ‘OQC Raises £260m in Europe’s Largest Ever Private Quantum Computing Funding Round – OQC’ (17 June 2026) <https://oqc.tech/company/newsroom/series-c> accessed 16 July 2026.

[5] ‘IQM Joins Nasdaq as Public Quantum Computing Company’ (*The Quantum Insider*, 3 July 2026) <https://thequantuminsider.com/2026/07/03/iqm-begins-trading-on-nasdaq/> accessed 29 July 2026.

[6] ‘Honeywell’s Quantinuum Valued at $17.6 Billion as Shares Rise in Nasdaq Debut’ *Reuters* (4 June 2026) <https://www.reuters.com/business/honeywells-quantinuum-heads-nasdaq-debut-quantum-computing-gains-momentum-2026-06-04/> accessed 16 August 2026.

[7] Matt Swayne, ‘D-Wave Announces Transfer of Stock Exchange Listing to Nasdaq’ (*The Quantum Insider*, 15 July 2026) <https://thequantuminsider.com/2026/07/15/d-wave-announces-transfer-of-stock-exchange-listing-to-nasdaq/> accessed 6 August 2026.

[8] ‘Xanadu | Xanadu Becomes First Pure-Play Photonic Quantum Computing Company to Go Public’ <https://www.xanadu.ai/press/xanadu-becomes-first-pure-play-photonic-quantum-computing-company-to-go-public> accessed 11 June 2026.

[9] Matt Swayne, ‘Horizon Quantum Goes Public via SPAC, Raises $120 Million’ (*The Quantum Insider*, 19 March 2026) <https://thequantuminsider.com/2026/03/19/horizon-quantum-goes-public-via-spac-raises-120-million/> accessed 6 August 2026.

[10] Quantinuum’s most advanced quantum computer and others, ‘Quantinuum and Oracle Partner to Accelerate Hybrid Quantum Compute Adoption on Oracle Cloud Infrastructure’ <https://www.oracle.com/news/announcement/quantinuum-and-oracle-partner-to-accelerate-hybrid-quantum-compute-adoption-on-oracle-cloud-infrastructure-2026-08-11/> accessed 20 August 2026.

[11] ‘OQC, JPMorganChase and AMD Commence Research Collaboration to Develop New Quantum-AI Platform in London’ <https://www.jpmorganchase.com/about/technology/blog/oqc-jpmc-amd-quantum-ai-platform> accessed 19 August 2026.

[12] ‘Aramco and Pasqal Launch Commercial Quantum Computing as a Service Platform’ (18 May 2026) <https://www.aramco.com/en/news-media/news/2026/aramco-and-pasqal-launch-commercial-quantum-computing-as-a-service-platform> accessed 20 August 2026.

[13] ‘First Quandela Quantum Computer Delivered and Installed in OVHcloud Datacenter’ (*Quandela*) <https://www.quandela.com/about-us/newsroom/first-quandela-quantum-computer-delivered-and-installed-in-ovhcloud-datacenter/> accessed 20 August 2026.

[14] ‘OVHcloud Purchases Its First Quandela Quantum Computer to Spur Innovation in the Quantum Ecosystem’ (16 March 2023) <https://corporate.ovhcloud.com/en/newsroom/news/ovhcloud-quantum-quandela-mosaiq/> accessed 21 August 2026.

[15] ‘NVIDIA Introduces NVQLink — Connecting Quantum and GPU Computing for 17 Quantum Builders and Nine Scientific Labs’ (*NVIDIA Newsroom*) <https://nvidianews.nvidia.com/news/nvidia-nvqlink-quantum-gpu-computing> accessed 21 August 2026.

[16] ‘HPE Advances Quantum Computing at Scale with Expanded Industry Collaborations | HPE’ <https://www.hpe.com/us/en/newsroom/press-release/2026/06/hpe-advances-quantum-computing-at-scale-with-expanded-industry-collaborations.html> accessed 20 August 2026.

[17] BTQ Technologies Corp, ‘BTQ Technologies Announces Strategic Collaboration with TIDAL PWR to Advance Trusted Quantum Data Center Architecture’ <https://www.newswire.ca/news-releases/btq-technologies-announces-strategic-collaboration-with-tidal-pwr-to-advance-trusted-quantum-data-center-architecture-870426888.html> accessed 21 August 2026.

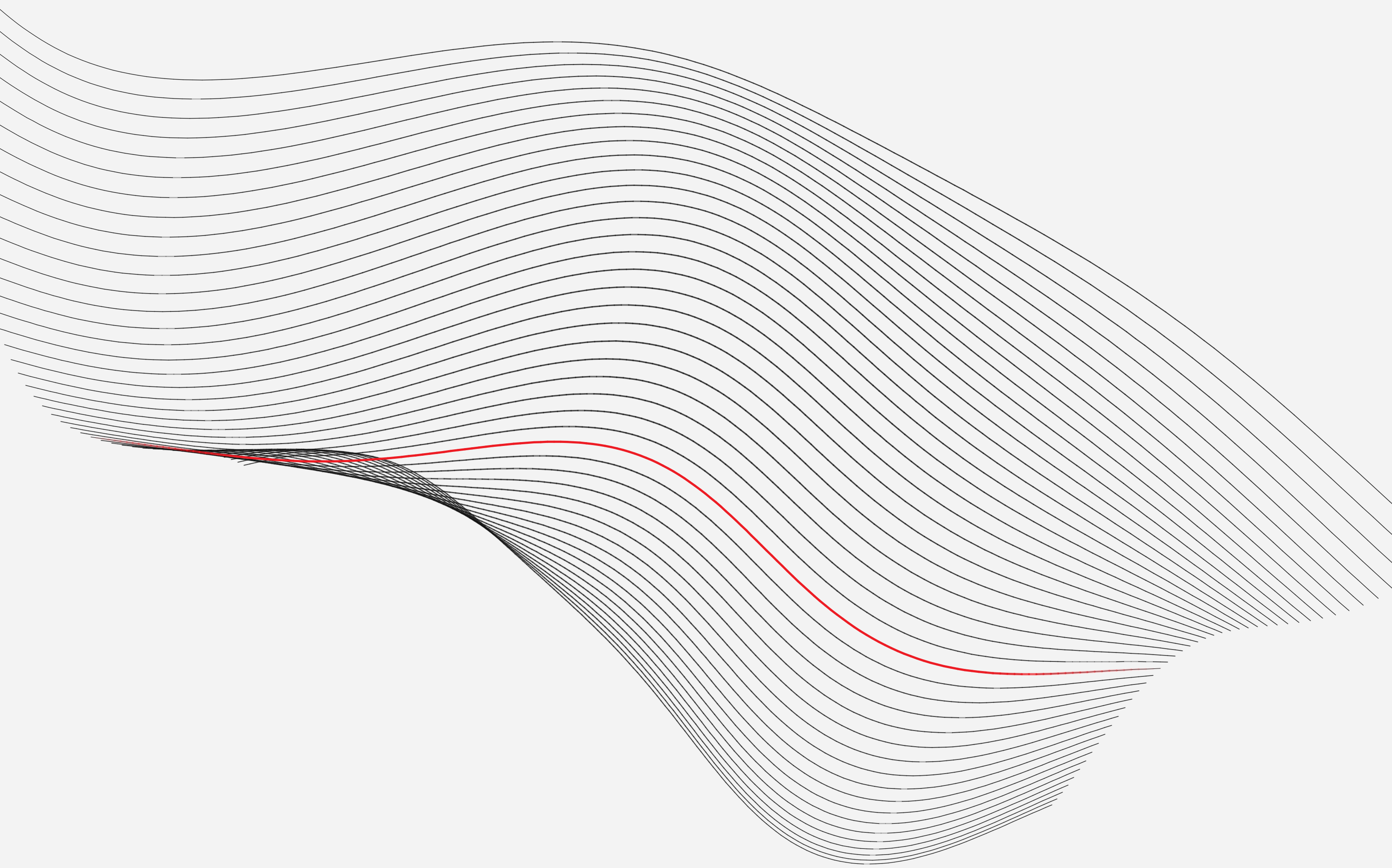

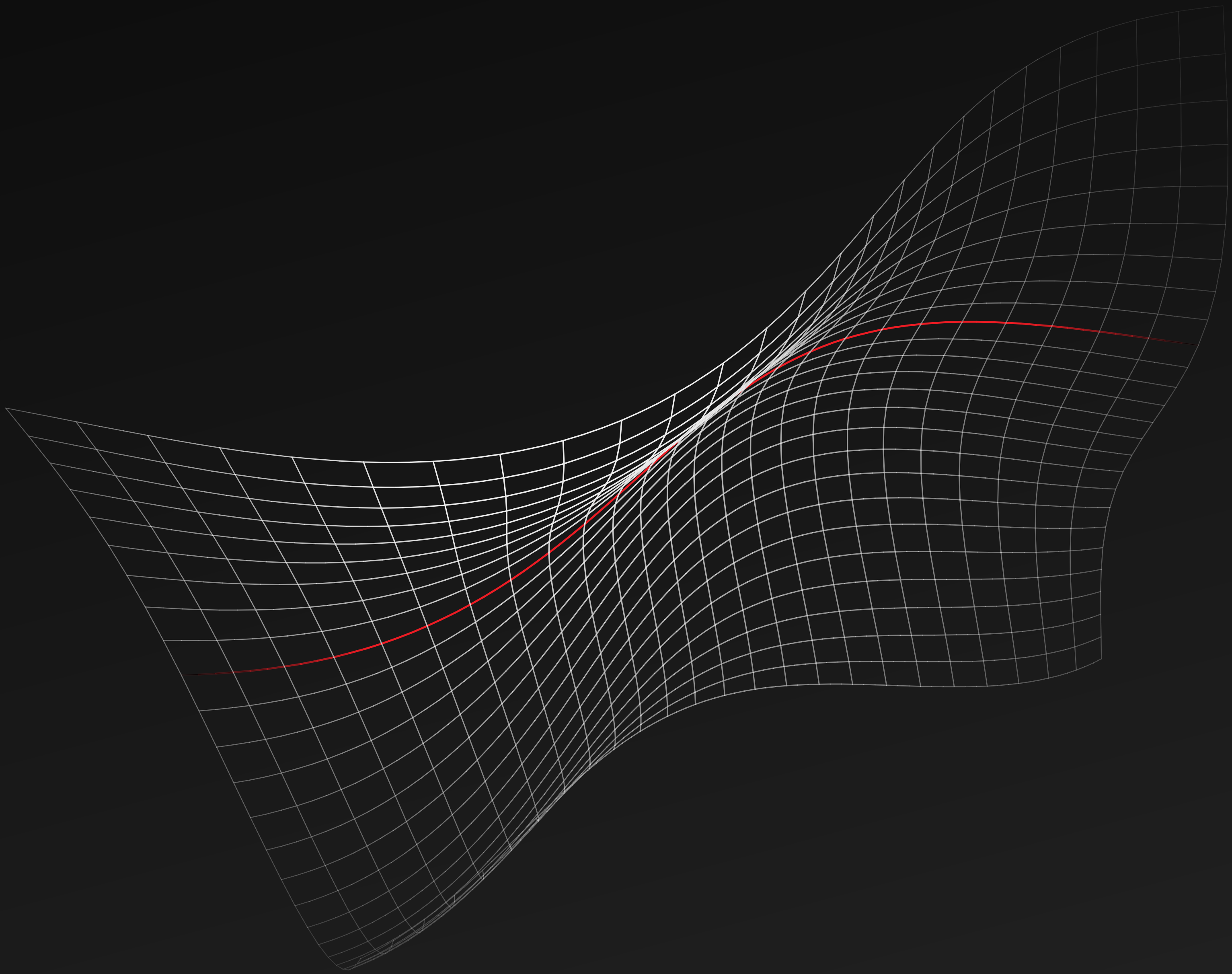

# 4 | Quantum in Corporate Communications

Mentions of quantum computing in corporate communications have risen sharply over the past two years, with news documents and company documents exhibiting a noticeable increase. This upward trajectory likely points to expanding commercial interest in quantum technology. The data for this section was compiled in collaboration with Accenture.

# 4.1 | Mentions in corporate communications

**Fig. 4.a** | Total documents mentioning "quantum computing" by month, April 2024–March 2026

MIT Quantum Index Report

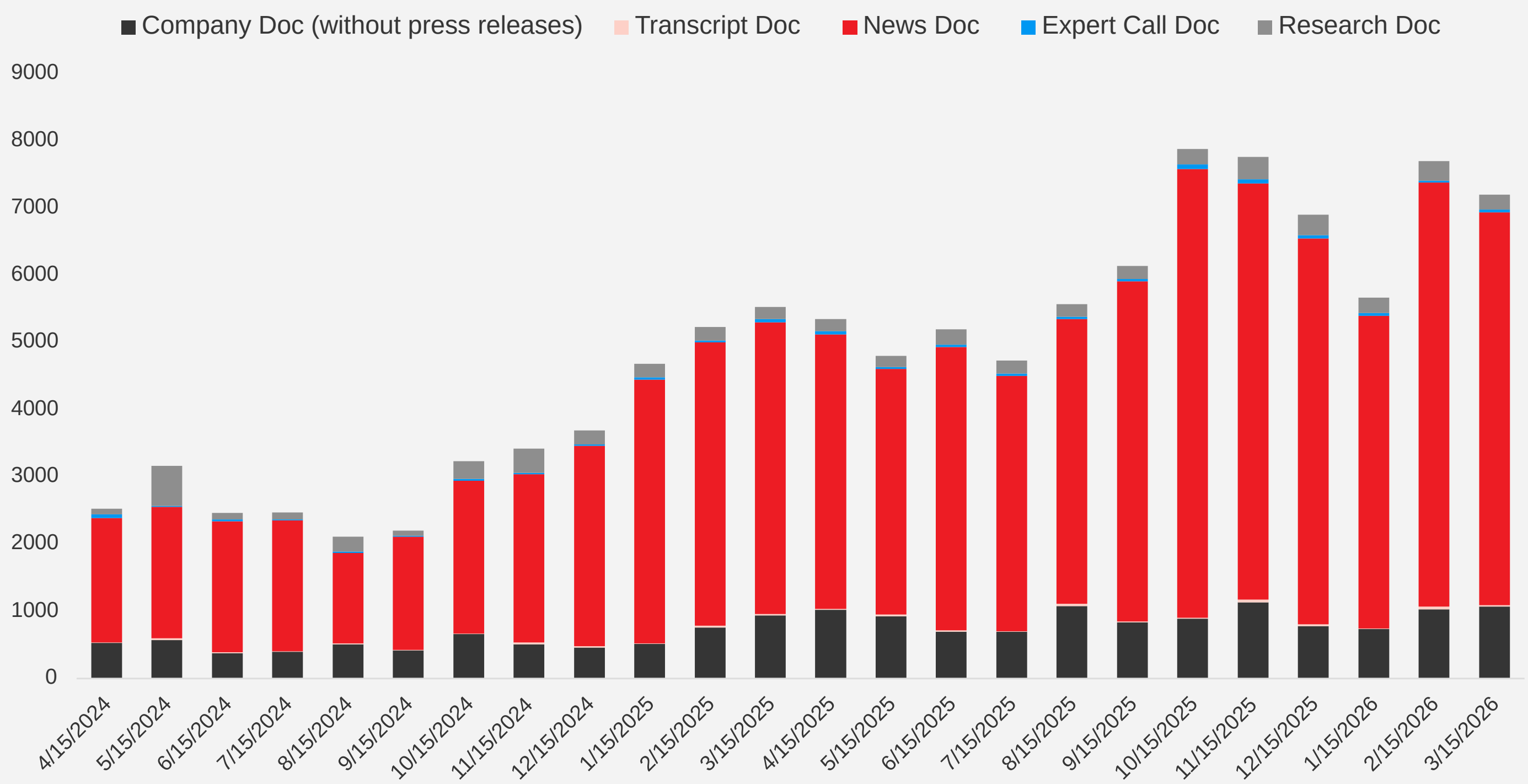

Considering this dataset, covering 24 months of document activity (April 2024 through March 2026) across five document types (Company Docs, Transcripts, News, Expert Calls, and Research) and totaling 115,534 documents, the frequency of references has shown consistent upward momentum, reinforcing the broader pattern of expanding integration into mainstream business and media discourse. The trajectory reveals distinct phases of acceleration, brief plateaus, and recurring seasonal surges that together show how this topic has diffused across different communication channels.

Total monthly document counts nearly tripled over the observation window, rising from a six-month average of 2,486 documents (April–September 2024) to 7,181 documents (October 2025–March 2026).

This growth was not always steady, as a modest dip appears in mid-2025 (April–July), following the strong first-quarter 2025 run-up, before activity reaccelerated into a clear peak cluster in October–November 2025 (7,877 and 7,757 total documents respectively) and a second surge in February–March 2026 (7,695 and 7,198). This pattern of repeated peaks on top of a rising baseline suggests the topic is no longer just steadily diffusing into discourse on its own momentum; it is increasingly reactive to specific news cycles, product announcements, or triggering events, generating more pronounced spikes layered over sustained underlying growth.

News Docs account for the overwhelming majority of total volume, 79.1% of all documents (91,408 of 115,534), and also posted by far the steepest growth of any category, climbing from a 1,788-document monthly average in the first six months to 5,901 in the last six, a 230% increase. This is a notable pattern in the data, given that media coverage, far more than any internal or expert-driven channel, is responsible for the topic's growing visibility, and public/press attention has outpaced internal corporate discussion in both absolute and relative terms. This suggests that the narrative around the topic is currently being shaped primarily by external media framing rather than by companies' own disclosures, which has implications for how quickly and in what form the topic reaches the broader public versus specialist audiences.

Company Docs (excluding press releases), the second-largest category at 15.1% of total volume, grew more moderately but still substantially, up 102% from a 464-document monthly average early on to 937 in the most recent period, with a particularly sharp acceleration in early 2025 (crossing 900 documents by March 2025 for the first time, and peaking at 1,129 in November 2025). This might suggest that the topic is moving beyond isolated technical mentions and into recurring business communications, internal reporting, and strategic disclosures. This is likely evidence that companies themselves are treating the topic as increasingly material to their own narratives, even if their pace of documentation still trails the media's.

Expert Call Docs, while a small share of total volume (0.7%), nearly doubled in average monthly frequency (25.0 to 47.8, +91%), suggesting growing appetite among analysts and investors for direct expert insight, which could be seen as a signal of rising institutional and investment interest running alongside, rather than strictly ahead of, broader market awareness. Research Docs grew more modestly (+38%, from 197 to 271 monthly average), and the series includes a notable outlier in May 2024 (594 documents, roughly triple the surrounding months), which likely reflects a one-off publication cycle or major report release rather than a sustained shift in research output. Transcript Docs remain the smallest category by volume (0.4% of total) but showed comparable relative growth to Expert Calls (+89%, from 13.5 to 25.5 monthly average), indicating gradual but real uptake in earnings calls and other recorded discussions.

## 4.1.1 | Mentions in earnings calls

**Fig. 4.b** | Total earnings calls mentioning "quantum computing", 2012–2025

MIT Quantum Index Report

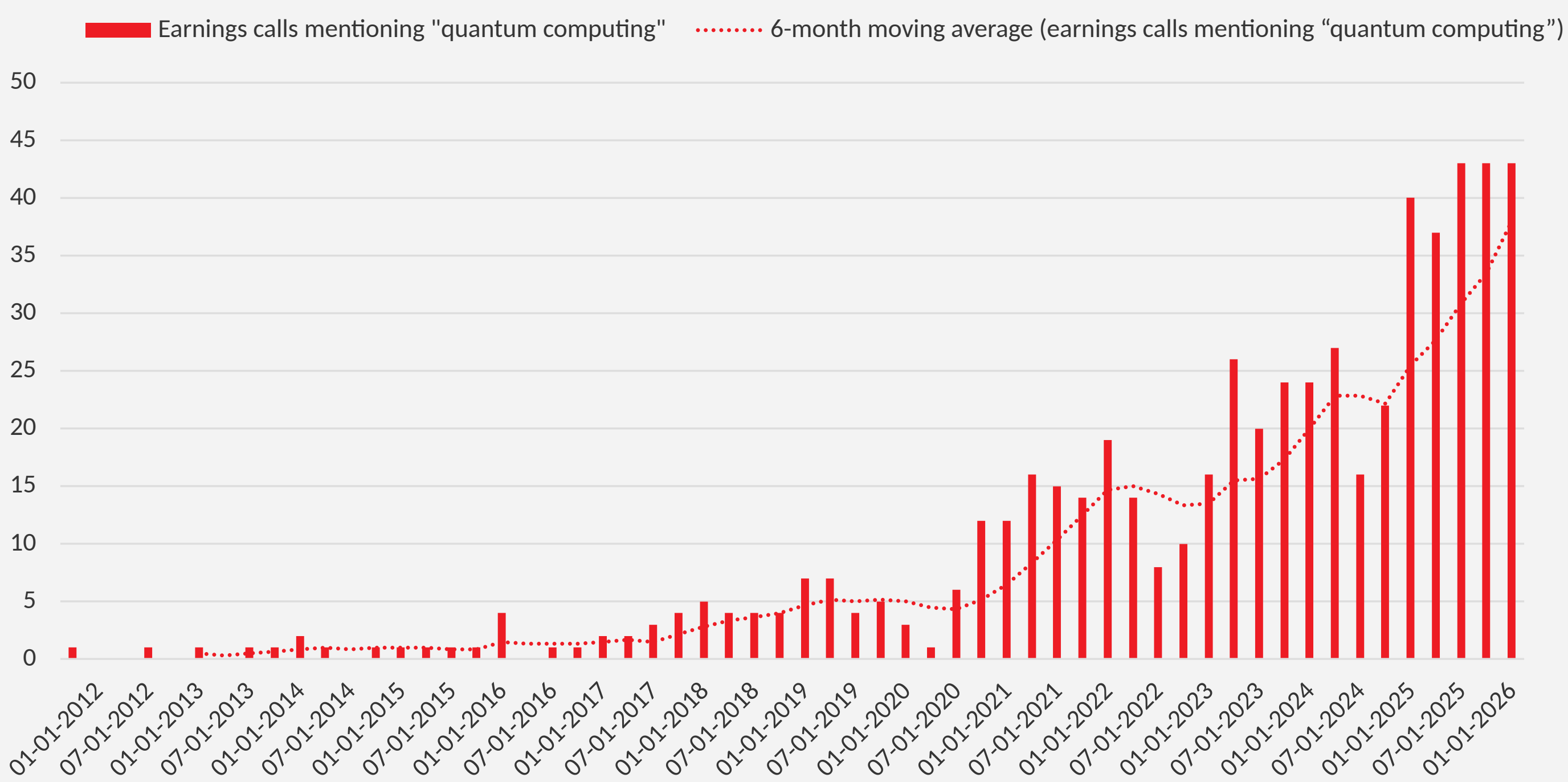

The growth in quantum computing mentions during earnings calls represents a significant trend in corporate communications within the period 2011–2026. After a negligible number of mentions in the earliest years, averaging fewer than one per quarter between Q4 2011 and Q4 2015, the topic began gaining traction starting from 4 mentions in Q1 2016, growing substantially to reach 24 mentions in Q1 2024, an approximately sixfold increase over this eight-year period. This upward trend has continued even further since then, with quarterly mentions climbing to 40 by early 2025 and holding steady at 43 mentions per quarter through the most recent data (Q3 2025–Q1 2026), representing roughly a tenfold increase from the Q1 2016 starting point.

The trend reveals three distinct phases. During the dormant phase (2011–2015), mentions rarely exceeded one or two per quarter, reflecting quantum computing's status as a niche academic or research topic largely absent from mainstream corporate discourse. A steady growth phase followed from 2016 through 2020, with annual totals climbing from 6 mentions in 2016 to 22 in 2020, briefly interrupted by a pandemic-related dip in Q2 2020 (just 1 mention, down from 3 in the prior quarter) as earnings calls shifted focus to COVID-19 disruptions. From 2021 onward, an acceleration phase took hold: annual mentions jumped from 22 in 2020 to 57 in 2021, a 159% single-year increase, before continuing to climb to 89 in 2024 and 163 in 2025, nearly triple the 2021 total in just four years.

Most recently, the data suggests the trend may be entering a plateau at a substantially higher baseline: mentions have held in the low-to-mid 40s per quarter since Q3 2025 (43, 43, and 43 through mid-March 2026), compared to sharper quarter-over-quarter swings seen earlier in the series.

This suggests quantum computing may be transitioning into a stable, recurring fixture of corporate earnings discourse. The strengthening of the quantum computing market is also supported by findings in the Quantum Economic Development Consortium's 2026 Report which estimates a 30% annual growth rate to drive the global quantum computing market to reach $3 billion in 2028.[1] This has been accompanied by supporting developments, for example, in its debut quarterly earnings report as a public company, Quantinuum announced a second-quarter revenue of $8 million, up 279% from the same period last year, and stated expected annual 2026 revenue to be in the $28 million to $32 million range.[2]

**10×**

Earnings call mentions have grown approximately 10 times since 2016.

## 4.1.2 | Quantum computing mentions by company

**Fig. 4.c** | Total document count by company (top 50)

MIT Quantum Index Report

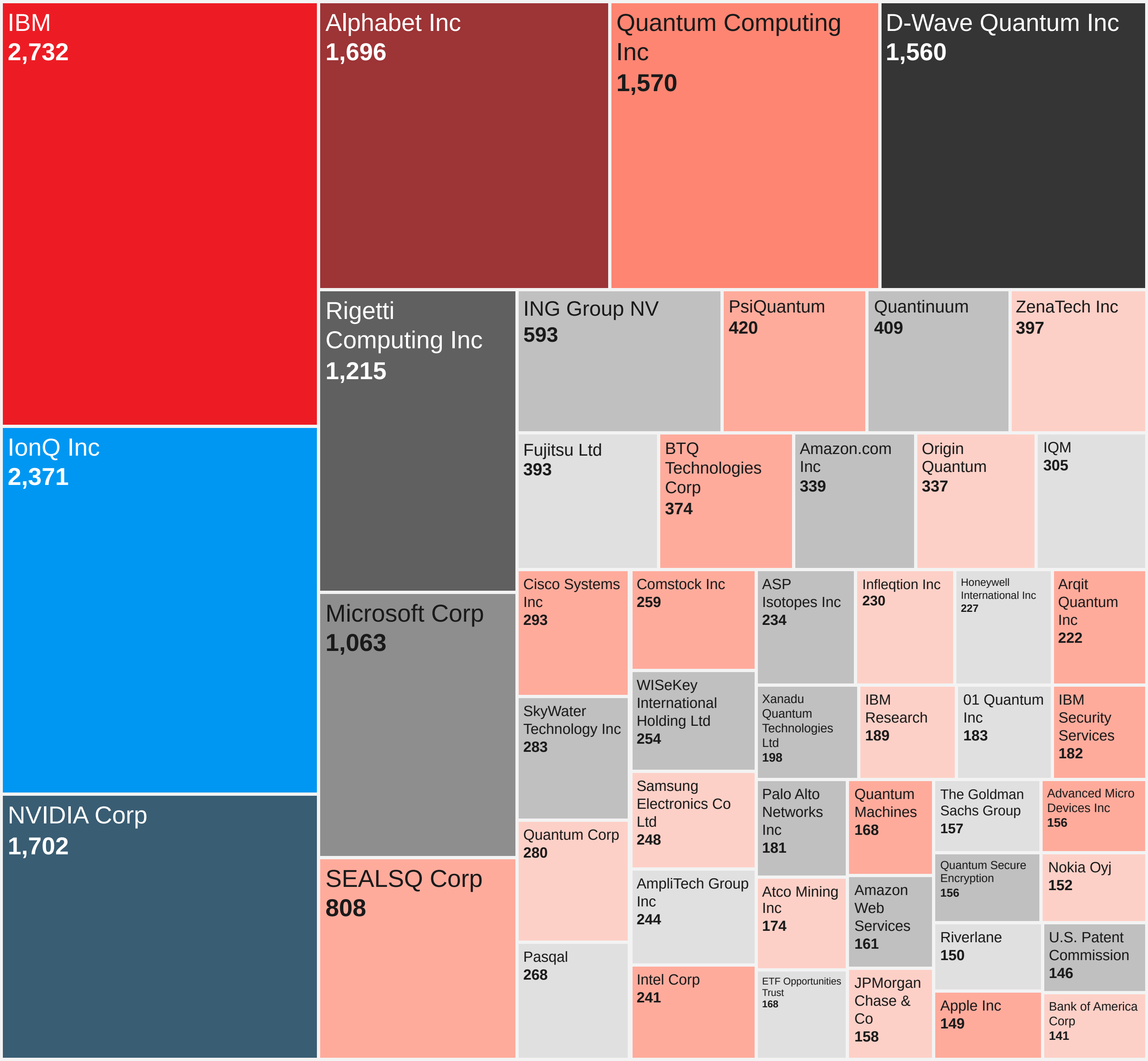

IBM and IonQ emerged as the leading firms in terms of producing external communications, as measured by number of documents (e.g. press releases or earnings calls) mentioning quantum computing. They are followed by NVIDIA in third place, Alphabet in fourth place, and Quantum Computing Inc. in fifth place, completing the top five. D-Wave, Rigetti, Microsoft, SEALSQ and ING Groep complete the top ten companies by document count.

**Fig. 4.d** | Total document count by company (top 10), 2025

MIT Quantum Index Report

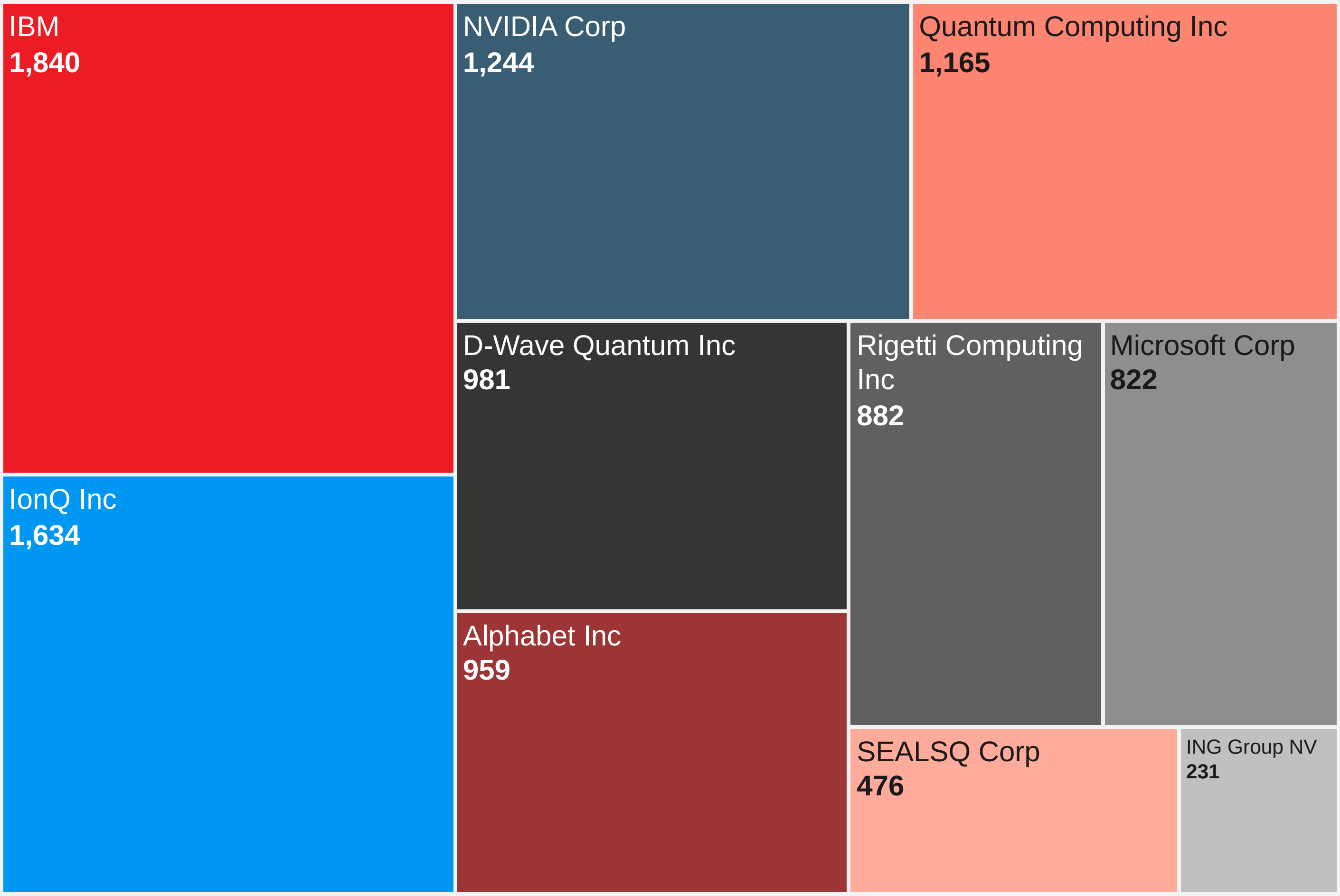

Considering the top 10 companies by 2025 document volume, a clear and broad-based intensification of quantum computing-related disclosure across the sector is observed, though with different growth rates depending on company type.

Combined document output across these 10 companies rose from 4,032 in 2024 to 10,234 in 2025, meaning total volume among the leading producers increased roughly two-and-a-half-fold over just one year.

**2.5×**

Combined document output across the top 10 companies rose from 4,032 in 2024 to 10,234 in 2025, roughly two-and-a-half times in a single year.

**Fig. 4.e** | Total document count by company (top 10), 2025 vs. 2024

MIT Quantum Index Report

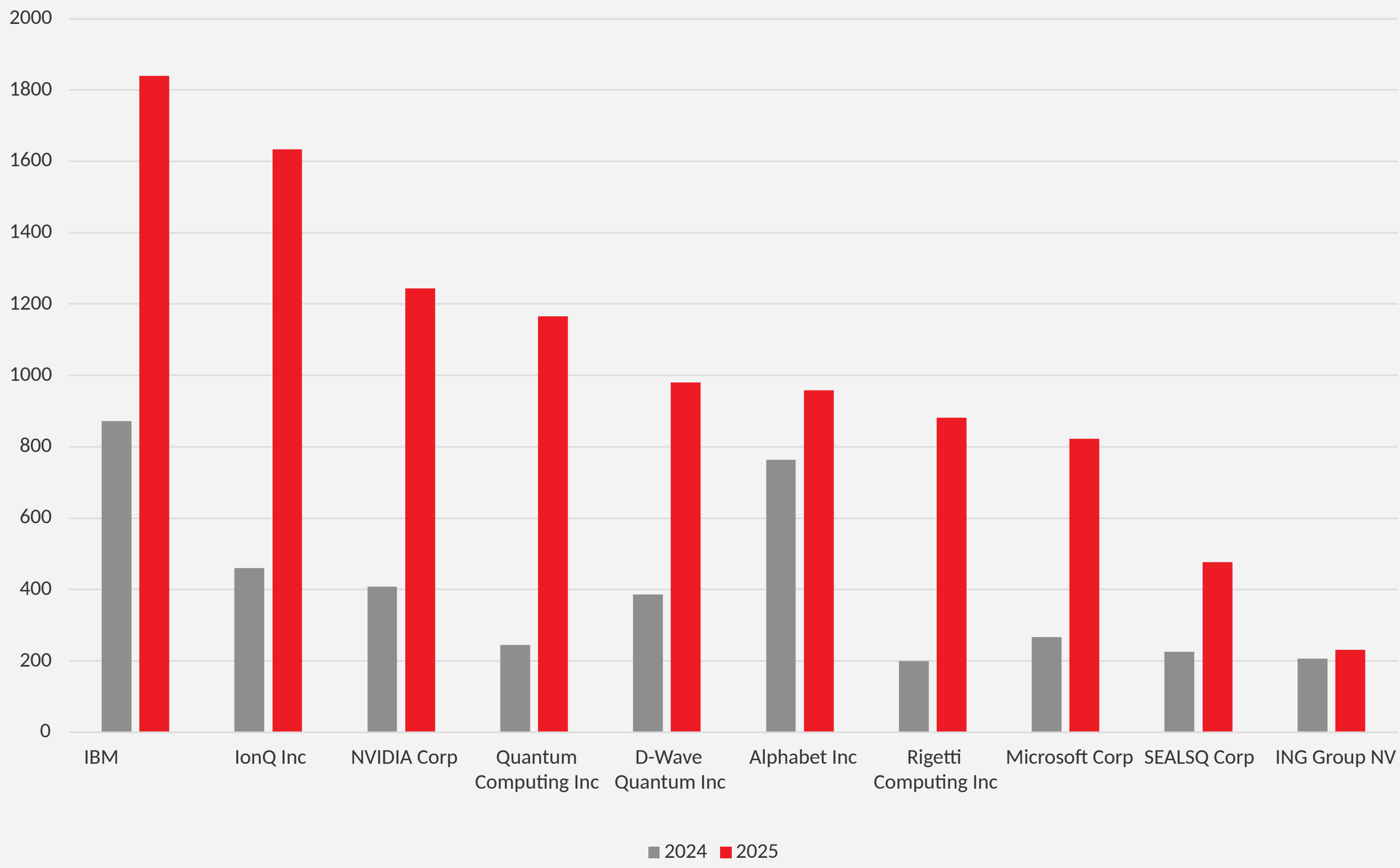

IBM remains the largest single producer in both years, more than doubling in number from 872 to 1,840 documents and maintaining its position at the top of the list, consistent with IBM's long-standing positioning as a quantum computing leader. Another notable pattern is in the relative growth rates among smaller, quantum-native companies, several of which grew far faster than IBM and effectively closed the gap: IonQ Inc more than tripled its output (460 to 1,634, +255%), overtaking NVIDIA Corp to become the second-largest producer in 2025 despite trailing it in 2024. Rigetti Computing Inc (+341%) and Quantum Computing Inc (+376%) posted the two highest growth rates in the entire group, with Quantum Computing Inc's count rising nearly fivefold from just 245 to 1,165 documents.

## 4.2 | Future research

We plan to expand this research in future editions of this report. Areas of particular interest include analyzing industry-specific patterns in quantum technology adoption messaging, examining cross-industry collaboration trends in quantum-related communications, and exploring additional trends that may be reflected in how companies communicate publicly.

You can reach us at contact@qir.mit.edu

---

### Footnotes

[1] 'State of the Global Quantum Industry 2026' (QED-C 2026).

[2] 'Quantinuum Reports Second Quarter 2026 Results' (Quantinuum 2026).

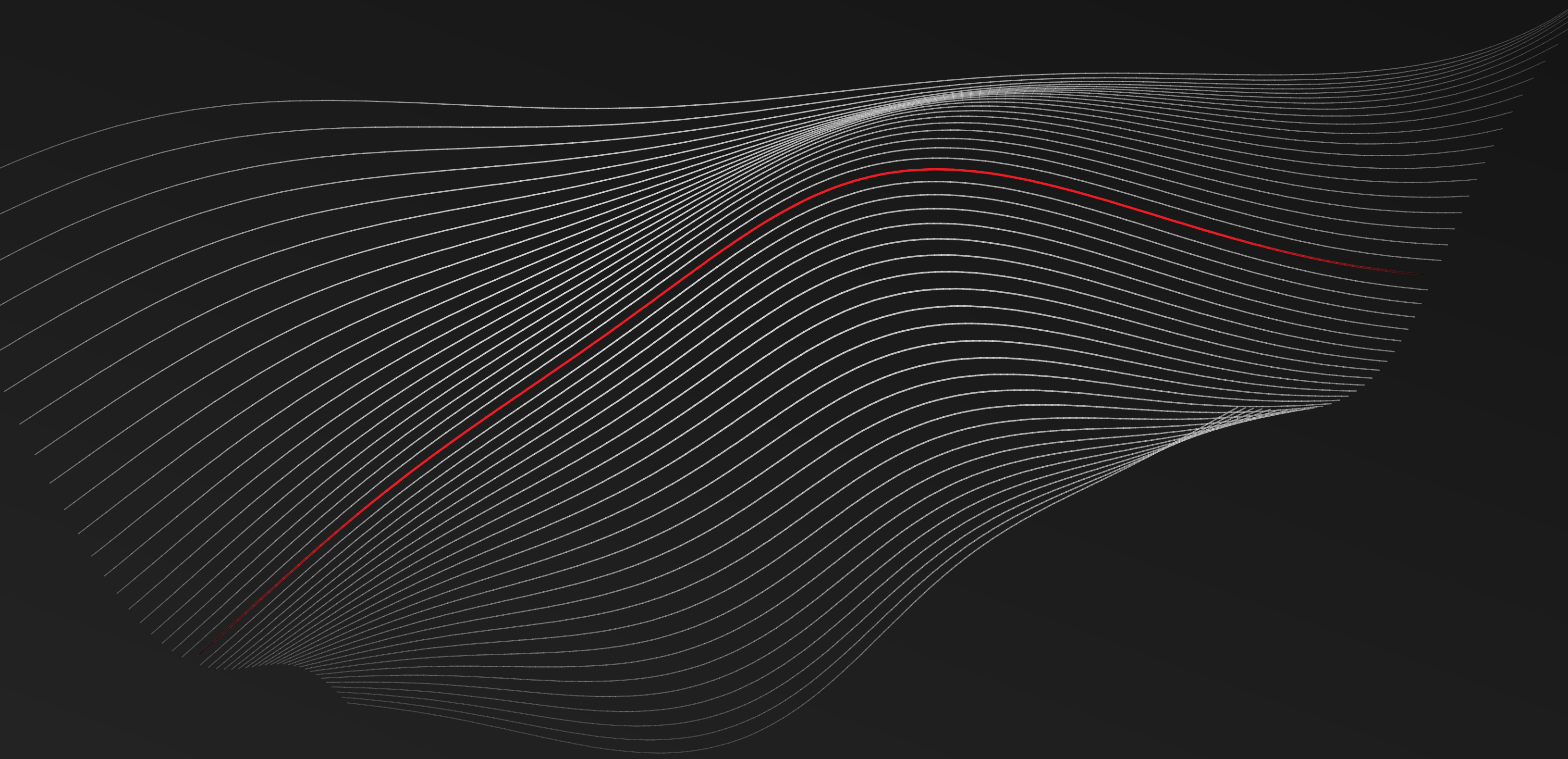

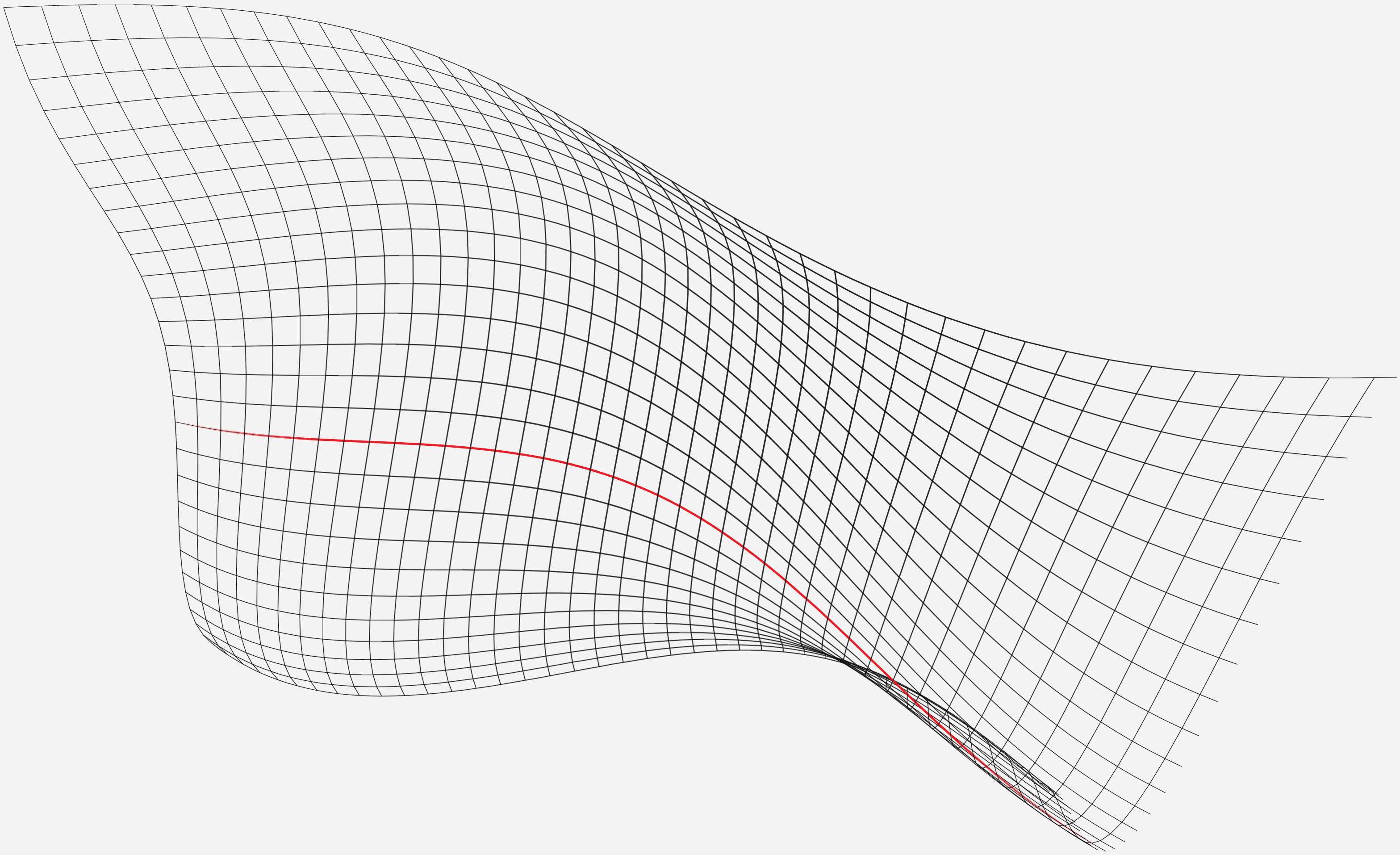

# 5 | Policy

Our 2025 report examined the broader policy landscape across a selection of countries. This year's report focuses more specifically on policy developments that have emerged in the last two years. This enables a more systematic assessment of recent developments, and complements our previous analysis by examining changes in policy direction, newly introduced measures, and emerging trends that may have implications for future policy direction.

## 5.1 | Overview of national quantum policy efforts, 2024–2026

| | |
|---|---|
| MAY 2024<br>🇸🇬<br>**Singapore** | Singapore launched its national quantum strategy, which commits nearly S$300 million to quantum research, engineering, commercialization and talent development. Measures include quantum processors, sensing, communications/security, scholarships and strengthening the national quantum ecosystem.[1] |
| AUG 2024<br>🇷🇴<br>**Romania** | Romania launched its national quantum strategy, which establishes a national framework and quantum ecosystem connecting research institutions, government and industry, with emphasis on research, technological applications, skills and participation in European quantum initiatives.[2] |
| APR 2025<br>🇪🇸<br>**Spain** | Spain launched its national quantum strategy backed by €808 million in public investment. It focuses on quantum computing, communications and sensing; R&D and technology transfer; development of quantum companies; talent; infrastructure; and preparation for post-quantum cybersecurity.[3] |
| APR 2025<br>🇫🇮<br>**Finland** | Finland launched its national quantum strategy containing eight measures covering R&D, education and talent, quantum-computing infrastructure, quantum-secure communications, commercialization, private investment, international cooperation and integration with AI/HPC.[4] |
| JUL 2025<br>🇮🇹<br>**Italy** | Italy announced its national quantum strategy for establishing a national framework for research, industrial applications, skills and national security. It seeks to create an integrated quantum ecosystem, strengthen technological sovereignty and competitiveness, attract capital and talent, and protect critical infrastructure.[5] |

DEC 2025

**Czech Republic**

Czech Republic's national quantum strategy aims to create a national quantum ecosystem covering computing and simulation, communications, sensing, metrology and materials. Strategic pillars include R&D, education, competitiveness, international cooperation, security/defense and quantum infrastructure.[6]

DEC 2025

**Chile**

Chile announced its national quantum strategy including a ten-year roadmap covering quantum computing, advanced sensing, secure communications, materials simulation and cybersecurity. Priorities include human capital, scientific infrastructure, industrial development, governance and technological sovereignty.[7]

MAR 2026

**United Kingdom**

The UK announced a major expansion of the 2023 National Quantum Strategy with new support worth up to £2 billion spanning research, skills, procurement and deployment. A central element is a world-first procurement commitment of up to £1 billion for commercial-scale quantum computers, with an ambition to make and deploy quantum computers at scale in the early 2030s.[8]

JUN 2026

**United States**

The Executive Order (EO) 'Ushering in the Next Frontier of Quantum Innovation' brought a significant update to the U.S. National Quantum Strategy. The new strategy gives greater weight to commercialization, deployment, industry partnerships and quantum-enabling technologies. The EO also creates the QC-ADDS initiative for a discovery-scale quantum computer, requires plans for sensing, networking and supply chains, strengthens quantum security and international cooperation, and establishes a government-wide quantum workforce strategy.[9]

## 5.2 | United States

The United States continues to develop its quantum technology policy through the National Quantum Initiative (NQI), established in 2018 as a coordinated federal program.[10] In 2025 and 2026, U.S. policy moved toward deployment, commercialization, manufacturing and sovereign technological capability.

This shift was formalized in June 2026 with the Executive Order ’Ushering in the Next Frontier of Quantum Innovation’ which underlined an updated National Quantum Strategy focused on commercial deployment, quantum-enabling technologies and stronger partnerships with U.S. industry.[11]

Federal investment in the domestic quantum ecosystem has simultaneously expanded. In November 2025, the Department of Energy (DOE) announced $625 million to renew its five National Quantum Information Science Research Centers for up to five years. These centers support quantum computing, sensing, networking, materials, workforce development and partnerships with industry, while seeking to accelerate technologies toward practical applications.[12] In 2026, the DOE also advanced domestic production capabilities, in line with the Genesis Mission’s goals, for high-purity silicon and germanium materials important to quantum devices, reflecting greater policy attention to quantum supply-chain resilience.[13]

**$625M**

In November 2025, the Department of Energy announced $625 million to renew its five National Quantum Information Science Research Centers for up to five years.

Commercialization and government adoption became particularly prominent under the 2026 Executive Order. It established the Quantum Computer for Application Development and Discovery Science effort to pursue a quantum computer capable of economically significant scientific applications, including through private-sector partnerships and potentially advance market commitments. In June 2026, NIST announced a $20 million investment in establishing the Quantum Manufacturing Engineering Center (QMEC) to accelerate manufacturing of scalable, high-performance quantum components and systems to drive significant growth in the U.S. quantum industry.[14]

International cooperation has also broadened. In 2025, the United States concluded Technology Prosperity Deals with the United Kingdom[15], Japan and South Korea that include cooperation on quantum technologies, research and innovation.[16] The 2026 Executive Order further calls for trusted international supply chains, improved market access for U.S. quantum companies and closer coordination with allies on technology protection.

## 5.3 | Australia

Australia continues to implement its National Quantum Strategy, released in May 2023, which aims to position the country as a global leader in quantum technologies by 2030. The strategy remains organized around five themes covering research and development, infrastructure and materials, skills, standards and regulatory frameworks, and a trusted and inclusive quantum ecosystem.[17] In 2025 and 2026, Australian policy moved toward commercialization, domestic scale-up and deployment.

The Department of Industry's 2025–29 Corporate Plan confirms continued implementation through government investments in PsiQuantum, Silicon Quantum Computing and the establishment of Quantum Australia, the national center for quantum industry growth.[18] In August 2026, Quantum Australia announced that its national programs have created $83.1M in economic value and that it has played an active role in establishing 15 new quantum companies.[19]

Commercialization has become a particularly important priority. The Critical Technologies Challenge Program is designed to connect quantum researchers, companies and end-users, demonstrate Australian capabilities and reduce barriers to adoption. In February 2026, the program provided $12.7 million for quantum projects in areas including mining, healthcare and data-center optimization[20], as part of the broader $35.9 million program.[21]

Quantum technologies are also becoming more closely connected with defense innovation. Under AUKUS Pillar II, four Australian-developed quantum optical atomic clocks were successfully trialed in the United States in 2025, demonstrating progress toward integrating Australian quantum technologies into allied defense capabilities.[22]

Regarding further cooperation with international partners, in July 2026, Australia and Japan signed a dedicated quantum cooperation memorandum covering research, commercialization, investment, workforce development and supply-chain resilience.[23]

## 5.4 | Canada

Canada continues to implement its National Quantum Strategy (NQS), launched in 2023 with $360 million in dedicated funding over seven years. The strategy remains organized around three interrelated pillars: research, talent and commercialization.[24] In 2025 and 2026, Canadian policy showed more focus on scaling domestic companies, accelerating technology adoption and linking quantum capabilities with industrial and defense policy.

The most significant development came through Budget 2025, which committed an additional $334.3 million through the Defence Industrial Strategy to strengthen Canada's quantum ecosystem.[25] Building on this initiative, in December 2025, the government launched Phase 1 of the Canadian Quantum Champions Program, providing up to $92 million to Anyon Systems, Nord Quantique, Photonic and Xanadu Quantum Technologies, with up to $23 million available to each company, aiming to anchor leading Canadian quantum companies and talent domestically and to accelerate development of fault-tolerant quantum computers that demonstrate industrial application.[26]

Quantum technologies also became more closely integrated with defense-industrial policy. Canada's 2026 Defence Industrial Strategy, which establishes mechanisms to accelerate domestic R&D, procurement and commercialization, identifies quantum as a frontier and sovereign-capability technology.[27] In August 2026, Canada launched its first Quantum Defence Innovation Secure Hub in Calgary, backed by $20.3 million over two years, to develop and test technologies including quantum sensing, navigation, algorithms and hardware assurance for defense applications.[28]

International cooperation has simultaneously expanded.

During Canada's 2025 G7 Presidency, G7 leaders adopted the Kananaskis Common Vision for the Future of Quantum Technologies, strengthening cooperation on quantum innovation.[29] Canada subsequently agreed with Germany[30] to support joint commercialization projects in quantum computing and sensing, and in May 2026 signed a quantum science and technology cooperation statement with France covering R&D, talent, market development and resilient supply chains.[31]

## 5.5 | United Kingdom

The United Kingdom continues to implement its National Quantum Strategy, published in March 2023, which sets out a ten-year vision for the UK to become a leading quantum-enabled economy, backed by £2.5 billion of public investment.[32] In 2025 and 2026, UK quantum policy moved toward a focus on commercialization, scale-up, procurement and deployment.

**£2B**

In March 2026, the UK announced a major expansion of its National Quantum Strategy with new support worth up to £2 billion.

Commercialization became a major priority through the UK's Modern Industrial Strategy[33]. In June 2025, the government committed £670 million to accelerate quantum computing applications and provided the National Quantum Computing Centre with a long-term funding settlement.[34] This was followed in March 2026 by a package worth up to £2 billion for quantum technology, including more than £500 million for quantum computing, over £400 million for sensing and navigation, additional infrastructure and skills funding, and approximately £1 billion for procurement of large-scale quantum computers.[35]

International cooperation has simultaneously expanded.

The 2025 UK–U.S. Technology Prosperity Deal includes quantum technologies among its strategic areas of cooperation.[36] The UK also expanded bilateral quantum cooperation with Germany in 2025 and hosted the Quantum Development Group in 2026, where members prioritized investment, supply-chain resilience, standards and commercialization.[37][38]

## 5.6 | The Netherlands

The Netherlands continues to develop its quantum technology ecosystem through Quantum Delta NL (QDNL), a National Growth Fund program focused on establishing the country as an internationally leading center for quantum research, technology development and commercialization. The program is structured around quantum computing and simulation, quantum networking and quantum sensing, supported by activities in research and innovation, ecosystem development, human capital and societal impact. In March 2025, the Dutch government approved a revised programming of Quantum Delta NL, concentrating resources on core activities and strengthening its governance structure.[39] In recent years, commercialization and company scale-up have become important priorities for the Netherlands.

In 2025, the Invest-NL Deep Tech Fund invested €5 million in QuiX Quantum as part of a €15 million financing round to support development of a full-stack photonic quantum computer.[40] In May 2026, Invest-NL provided a further €19.5 million to QuantWare within a €152 million funding round to expand quantum-processor production in the Netherlands.[41] This was reinforced in June 2026 when the Deep Tech Fund was expanded to €610 million, with quantum technology explicitly identified among its target strategic technologies.[42]

Sovereign technological capability also gained greater prominence in 2026. The new coalition agreement commits the Netherlands to public-private investment in quantum and other key technologies, with an emphasis on scalability, productivity and adoption in critical sectors, and explicitly states the ambition to build a quantum computer in the Netherlands.[43]

International cooperation has also expanded.

In May 2025, Quantum Delta NL and Japan's AIST agreed to deepen long-term cooperation in quantum technology and develop an international quantum innovation ecosystem.[44] In 2026, the Netherlands and India also included quantum technologies within their broader Strategic Partnership Roadmap.[45]

## 5.7 | China

China continues to pursue quantum technology through a strongly state-directed science and industrial policy framework. The 14th Five-Year Plan identified quantum information as a frontier field for national laboratory development and called for accelerated progress in quantum computing and quantum communications.[46] In 2025, quantum technology received greater industrial-policy prominence when the Government Work Report formally identified it as an "industry of the future" and called for a mechanism to increase funding for future industries alongside demonstration and large-scale application initiatives.[47]

This approach was reinforced as China entered its 15th Five-Year Plan period. The 2026 Government Work Report again lists quantum technology among future industries and calls for funding and risk-sharing mechanisms, greater use of government investment funds as "patient capital," and support for technology-focused SMEs and unicorn companies.[48]

Commercialization and company scale-up are also receiving more attention. In December 2025, China formally launched its National Venture Capital Guidance Fund, designed as a long-term financing mechanism for strategically important technologies, including quantum technology.[49]

Sovereign technological capability remains central to the strategy. In March 2025, Chinese researchers unveiled the Zuchongzhi 3.0 superconducting quantum-computing prototype[50], while Hefei has continued developing a concentrated industrial cluster encompassing quantum computing, communications and measurement companies.[51]

Internationally, China combines technological self-reliance with selective openness. In 2026, Premier Li Qiang highlighted that Chinese large-scale scientific facilities in areas including quantum technology remain open to international users.[52]

## 5.8 | Ireland

Ireland continues to implement Quantum 2030: A National Quantum Technologies Strategy for Ireland, launched in 2023, to establish the country as an internationally competitive hub for quantum technologies by 2030. The strategy is organized around five pillars covering research, talent, national and international collaboration, innovation and entrepreneurship, and public awareness.[53] In 2025 and 2026, Irish policy moved toward building research infrastructure, supporting commercial applications and creating a more concentrated national quantum ecosystem.

Commercialization became more prominent through the Disruptive Technologies Innovation Fund (DTIF). In 2025, the QUBIC project received €13.7 million to bring together Irish companies, multinationals and research organizations to develop quantum algorithms, hardware and software for applications in energy, climate modeling, advanced materials and pharmaceuticals.[54] The project illustrates Ireland's increasing emphasis on translating quantum research into commercially relevant applications and strengthening industry–research collaboration.

Domestic research capability also expanded substantially in 2026. The government announced €460 million for seven new Research Ireland centers, including Rinn Quantum, a dedicated national center for quantum information science and technology intended to bridge fundamental research and real-world applications, develop industry partnerships and strengthen the quantum workforce.[55] This follows the 2026 Digital and AI Strategy commitment to establish a Quantum Centre of Excellence.[56] In 2026, the government also approved a €100 million expansion of the Tyndall National Institute to strengthen capabilities in semiconductors, photonics, quantum technologies and next-generation computing.[57]

Regarding international cooperation, Ireland signed the European Quantum Pact in June 2025, committing to closer EU cooperation on research, infrastructure, investment, skills and the development of a competitive European quantum ecosystem.[58]

## 5.9 | India

India continues to implement its National Quantum Mission (NQM), approved in 2023 with an outlay of ₹6,003.65 crore through 2030–31. The mission aims to build a domestic quantum ecosystem spanning computing, communications, sensing and metrology, and quantum materials, including the development of intermediate-scale quantum computers.[59] In 2025 and 2026, Indian policy moved toward infrastructure construction, technology translation, startup development and domestic hardware capability.

The mission's institutional structure became fully operational during 2025. Four Thematic Hubs were established at IISc Bengaluru for quantum computing, IIT Madras with C-DOT for quantum communications, IIT Bombay for quantum sensing and metrology, and IIT Delhi for quantum materials and devices. Together, these activities involve researchers across dozens of Indian institutions and combine R&D with entrepreneurship, workforce development and industry collaboration.[60]

Commercialization and domestic company development have simultaneously become more prominent. NQM introduced dedicated startup-support guidelines providing funding, infrastructure access and mentorship, with eight quantum startups supported by early 2026.[61] In April 2025, NQM-supported QpiAI[62] launched QpiAI-Indus, a 25-superconducting-qubit full-stack quantum-computing system; the government subsequently reported further development of a 64-qubit quantum processor.

International cooperation has become more structured. In April 2025, India released its first International Technology Engagement Strategy for Quantum, aimed at guiding partnerships, commercialization, supply-chain engagement and participation in international standards.[63] In 2026, India deepened cooperation with Japan on quantum hardware, research and industrial deployment[64] and launched a U.S.–India funding call[65] supporting joint quantum technology projects with commercial potential.

## 5.10 | Germany

Germany continues to develop its quantum technology ecosystem through the Federal Government's Action Concept for Quantum Technologies, which coordinates federal activities for 2023–2026 and mobilizes approximately €3 billion from the government and research organizations. The framework focuses on quantum computing, sensing, communications, enabling technologies, technology transfer and the development of a competitive domestic ecosystem.[66] In 2025 and 2026, German policy moved toward commercialization, technological sovereignty and translating its strong research base into industrial capabilities.

This was reinforced through the High-Tech Agenda Germany, adopted in July 2025, which identifies quantum technologies as one of six strategic technologies.[67] The agenda emphasizes faster transfer from research into industry, public-private partnerships, improved conditions for startups and a stronger role for the state as an anchor customer. In May 2026, the government presented dedicated roadmaps for the six technologies, further institutionalizing quantum as a long-term industrial-policy priority.[68]

At the same time, domestic quantum-computing capability has expanded. The DLR Quantum Computing Initiative encompasses 17 hardware projects, more than 60 use cases and over 100 partners, providing companies and researchers with access to German quantum systems and supporting industrial applications.[69] In March 2026, the government-backed SQale project was launched with €18.4 million in federal funding to develop a modular ion-trap quantum computer with more than 200 interconnected qubits, with a longer-term objective of establishing internationally competitive quantum-computing capacity in Germany.[70]

International cooperation has also broadened. Germany and the United Kingdom announced a £6 million joint quantum R&D program[71] in December 2025, accompanied by cooperation on quantum standards. In July 2026, Germany and France agreed to strengthen bilateral quantum cooperation and signed common positions regarding the planned EU Quantum Act.[72]

## 5.11 | France

France continues to implement its National Quantum Strategy, launched in 2021 to establish the country among the leading global powers in quantum technologies. The original strategy combined more than €1 billion in public funding with an overall public-private commitment of approximately €1.8 billion, supporting quantum computing, sensing, communications, enabling technologies, research and talent development.[73] In 2025 and 2026, French policy moved toward industrialization, domestic scale-up, sovereign computing capability and the creation of European quantum champions.

In May 2026, the government announced additional funding through the France 2030 plan for the 2026–2030 period. The updated strategy brings the public funding commitment to approximately €1.7 billion and the overall public-private commitment to more than €3 billion by 2030.[74]

Commercial scale-up has also received direct state support; in January 2025, the Defense Innovation Fund participated in a €100 million financing round for Alice & Bob, supporting development of its quantum-computing technology.[75]

Achieving sovereign computing capability has also been prioritized. France's PROQCIMA program uses long-term public procurement to support five French quantum-computing companies: Alice & Bob, C12, Pasqal, Quandela and Quobly.[76] In 2026, the government doubled planned defense and security procurement funding for the program's 2026–2030 phases and raised its 2032 objective from 128 to 1,024 logical qubits. Public procurement and equity investment are intended to accelerate industrialization and mobilize additional private capital.[77]

Regarding international cooperation, in May 2026, France and Canada signed a dedicated quantum cooperation statement covering joint R&D, talent, knowledge exchange, market development and resilient supply chains.[78] France is also advocating together with Germany for European pooled procurement, aligned technology roadmaps and quantum-chip pilot lines under the forthcoming EU Quantum Act.[79]

## 5.12 | South Korea

South Korea continues to develop its quantum ecosystem through the Quantum Science and Technology Strategy and the Quantum Technology and Industry Act, which established a statutory framework for coordinated government support, commercialization, research infrastructure, talent and international cooperation.[80] In January 2026, this framework was substantially strengthened through Korea's first Comprehensive Plan for Fostering Quantum Science and Technology and the Quantum Industry, which sets objectives to become a world-leading quantum-chip manufacturing country, cultivate 10,000 quantum professionals and develop 2,000 quantum-related companies by 2035.[81]

Commercialization became a further priority in 2025. The Ministry of Science and ICT (MSIT) allocated KRW 198 billion to quantum science and technology, a 54% increase from 2024, supporting 24 projects covering computing, components, measurement, international research and industrial infrastructure.[82] A new Quantum Computing Service and Utilization Infrastructure program was also established to install verified quantum computers domestically and provide access to research organizations and companies. The 2026 budget further increased the proposed Quantum Flagship quantum-computing program from KRW 9.8 billion in 2025 to KRW 50 billion, reinforcing Korea's move toward large-scale mission-oriented development.[83]

Sovereign technological capability is also a focus of the new policy. The 2026 Comprehensive Plan calls for a domestically developed full-stack "K-Quantum Computer", hybrid infrastructure combining quantum computing, HPC and AI, and quantum clusters specializing in computing, sensors, algorithms and enabling components. The government also launched a Quantum Technology Council involving Samsung Electronics, LG Electronics, SKT, KT, Hanwha, LIG Nex1 and other major companies to accelerate industrial adoption.[84] By mid-2026, the government was targeting a domestically developed 50-qubit computer during 2026 and a 100-qubit error-correcting system by 2029.[85]

International cooperation has expanded. The 2025 U.S.–Korea Technology Prosperity Deal provides for cooperation on quantum institutions, standards and supply-chain resilience.[86] Korea also participated in the Quantum Development Group in 2026.[87]

## 5.13 | Japan

Japan continues to develop its quantum ecosystem through a layered policy framework comprising the Quantum Technology and Innovation Strategy, the Vision of Quantum Future Society, and the Strategy of Quantum Future Industry Development. The latter established objectives to be achieved by 2030, including 10 million quantum-technology users, quantum-enabled production worth ¥50 trillion and the creation of quantum unicorn companies.[88]

Commercialization and industrial scale-up received substantially greater support in late 2025 and 2026. The government's economic package called for accelerated development of domestic quantum computers, expansion of Quantum Technology Innovation Hubs, use-case creation and stronger testbed infrastructure at AIST's G-QuAT, RIKEN and other national institutes.[89] In 2026, the Ministry of Economy, Trade and Industry confirmed that approximately ¥100 billion had been secured in the supplementary budget to accelerate next-generation quantum-computer development and establish a globally competitive Japanese industry.[90]

Sovereign technological capability has advanced. In April 2025, RIKEN and Fujitsu unveiled a 256-qubit superconducting quantum computer, significantly expanding Japan's domestic computing capability.[91] In 2026, AIST expanded its ABCI-Q quantum–classical infrastructure and installed MoQuren, a Japan-developed optical quantum computer intended for future commercialization.[92]

International cooperation has also broadened.

Japan and the EU signed a quantum cooperation arrangement in May 2025 covering joint research, infrastructure, education and industrialization.[93] This produced the first joint Japan–EU quantum project, Q-NEKO, launched in 2026.[94] Japan also concluded new cooperation arrangements with Singapore and India in 2026, including collaboration on research, industrial applications, commercialization and talent.[95]

Cooperation with the United States is a well-resourced strand of Japan's external quantum strategy. The U.S.–Japan Technology Prosperity Deal[96], signed in October 2025, makes quantum information science a pillar of the bilateral technology relationship and commits both sides to align on securing technology and supply chains. Funding followed in June 2026 with a five-year, $1 billion partnership[97] between the Department of Energy, MEXT and METI (subject to appropriations), making Japan the first international partner in the DOE's Genesis Mission. Eleven joint teams link twelve DOE national laboratories with twelve Japanese institutions, including RIKEN, the University of Tokyo and KEK, across six domains spanning quantum information science, fusion and advanced materials.

## 5.14 | International cooperation and bilateral agreements

The bilateral agreements on quantum between 2021 and 2026 mark the emergence of quantum technology as a distinct axis of international diplomacy. The network diagram below visualizes the recent state of bilateral agreements on or referring to quantum technologies in order to better showcase emerging diplomatic links.

The United States and United Kingdom sit at the center as the most prominent players, followed by the next tier of Japan and India, while most countries such as New Zealand, Slovakia, Czech Republic, Italy, China, and Thailand have much smaller numbers of bilateral agreements in this domain. This might suggest that a small number of Western-aligned technology powers, each running parallel bilateral campaigns and frequently reusing the same structure, such as the U.S.'s “Joint Statement on Cooperation in QIST” and later “Technology Prosperity Deal” account for a large share of the agreements on their own.

The strongest single link in the network, U.S.–Japan, reflects a recurring bilateral channel whereas China, despite being among the leading quantum research powers, is not a prominent player in the diplomatic network of bilateral agreements between 2021 and 2026.

**Fig. 5.a** | Bilateral cooperation in quantum, 2021–2026

MIT Quantum Index Report

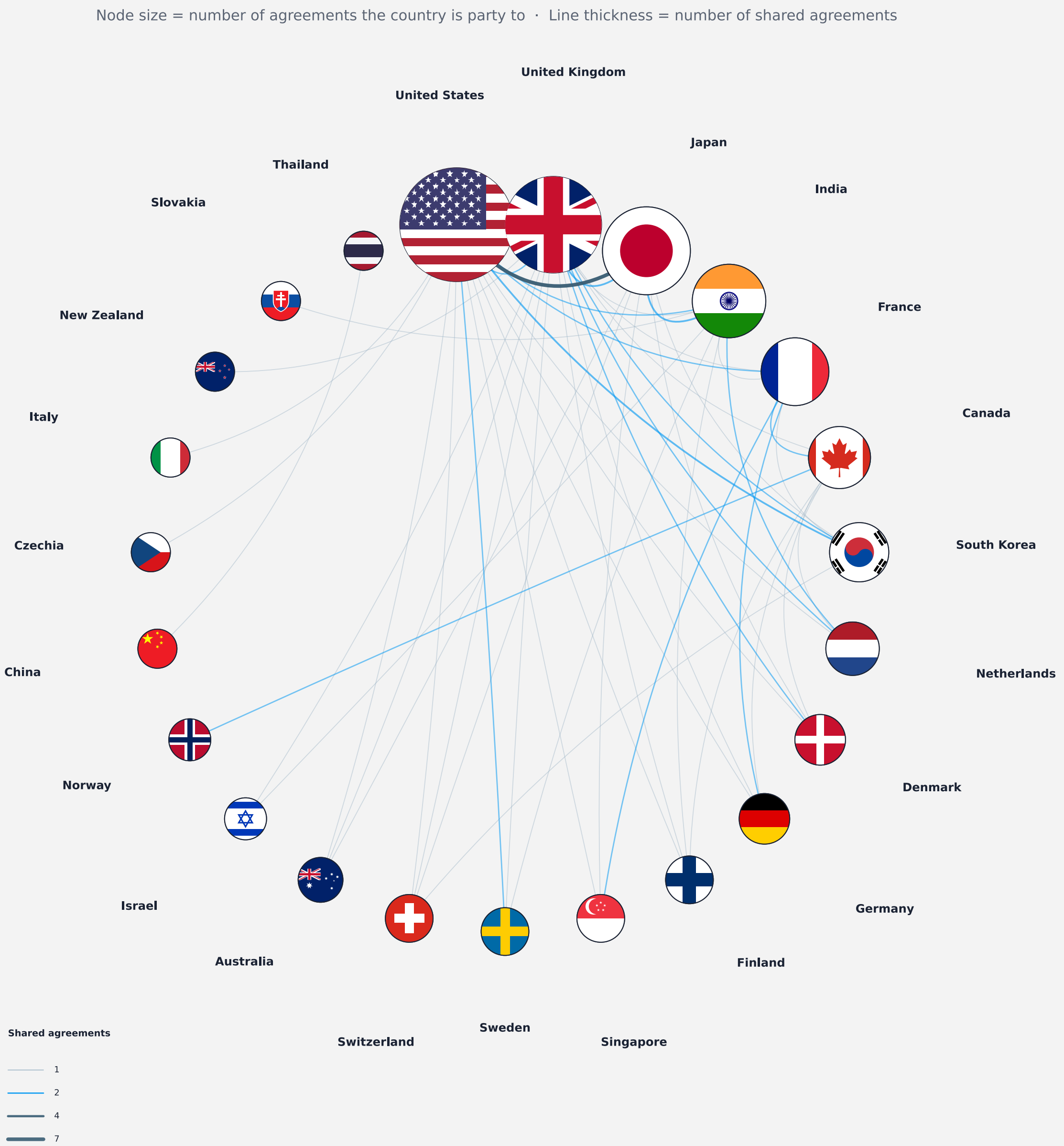

Looking at the year-over-year and specific type of the bilateral agreements in this time period, it is observed that international cooperation became much more frequent in 2023–2026.

Stand-alone bilateral quantum cooperation agreements are outnumbered by bilateral general technology-alliance diplomacy (AI, semiconductors, critical-tech pacts) where quantum technologies were explicitly included.

---

## Footnotes

[1] ‘DPM Heng Swee Keat at the Asia Tech x Singapore 2024 Opening Ceremony’ (*Prime Minister’s Office Singapore*) <https://www.pmo.gov.sg/newsroom/dpm-heng-swee-keat-at-the-asia-tech-x-singapore-2024-opening-ceremony/> accessed 18 August 2026.

[2] ‘STRATEGIE 21/08/2024’ <https://legislatie.just.ro/Public/FormaPrintabila/00000G2L6213FLICHXE2MGZYDAS2MVVR?> accessed 18 August 2026.

[3] ‘El Gobierno lanza la primera Estrategia de Tecnologías Cuánticas de España, con una inversión de 800 millones de euros’ <https://www.lamoncloa.gob.es/serviciosdeprensa/notasprensa/transformacion-digital-y-funcion-publica/paginas/2025/240425-lopez-estrategia-cuantica.aspx> accessed 18 August 2026.

[4] ‘Strategy: Quantum Technology Is Finland’s New Driver of Growth and Builder of a Sustainable Future’ (*Finnish Government*, 24 April 2025) <https://valtioneuvosto.fi/en/-/1410877/strategy-quantum-technology-is-finland-s-new-driver-of-growth-and-builder-of-a-sustainable-future> accessed 18 August 2026.

[5] ‘Tecnologie quantistiche: una Strategia per l’Italia’ (*mimit.gov.it*, 10 July 2025) <https://www.mimit.gov.it/it/notizie-stampa/tecnologie-quantistiche-una-strategia-per-litalia> accessed 18 August 2026.

[6] ‘Národní strategie pro kvantové technologie 2026+’ <https://vlada.gov.cz/cz/vlada/clenove-vlady/pri-uradu-vlady/marek_zenisek/aktualne/narodni-strategie-pro-kvantove-technologie-2026-224029/> accessed 18 August 2026.

[7] MinCiencia, ‘Chile Presenta Sus Estrategias Nacionales De Biotecnología Y Tecnologías Cuánticas’ (*MinCiencia*) <https://www.minciencia.gob.cl/noticias/chile-presenta-sus-estrategias-nacionales-de-biotecnologia-y-tecnologias-cuanticas/> accessed 18 August 2026.

[8] ‘UK’s “Quantum Leap” to Help Beat Disease, Deliver High-Paid Jobs, and Strengthen National Security, as First Country in the World to Roll out Quantum Computers at Scale’ (*GOV.UK*) <https://www.gov.uk/government/news/uks-quantum-leap-tohelp-beat-diseasedeliver-high-paid-jobs-and-strengthen-national-security-as-first-country-in-the-world-to-roll-out-quantum> accessed 18 August 2026.

[9] Executive Order No 14413, 91 Fed Reg 38487, ‘Ushering in the Next Frontier of Quantum Innovation’ (22 June 2026).

[10] 115th Congress, ‘H.R.6227 - National Quantum Initiative Act’ (21 December 2018).

[11] Executive Order No 14413, 91 Fed Reg 38487 (22 June 2026).

[12] ‘Energy Department Announces $625 Million to Advance the Next Phase of National Quantum Information Science Research Centers’ (*Energy.gov*, 4 November 2025) <https://www.energy.gov/articles/energy-department-announces-625-million-advance-next-phase-national-quantum-information> accessed 17 August 2026.

[13] ‘U.S. Department of Energy Advances Domestic Capabilities for Producing Quantum Materials’ (*Energy.gov*, 4 March 2026) <https://www.energy.gov/science/articles/us-department-energy-advances-domestic-capabilities-producing-quantum-materials> accessed 17 August 2026.

[14] ‘NIST Launches Center to Drive the Manufacture of Quantum Technologies’ (NIST, 2026) <https://www.nist.gov/news-events/news/2026/06/nist-launches-center-drive-manufacture-quantum-technologies> accessed 20 August 2026.

[15] ‘Memorandum of Understanding Between the Government of The United States of America and the Government of The United Kingdom of Great Britain and Northern Ireland Regarding the Technology Prosperity Deal’ (*The White House*, 18 September 2025) <https://www.whitehouse.gov/presidential-actions/2025/09/memorandum-of-understanding-between-the-government-of-the-united-states-of-america-and-the-government-of-the-united-kingdom-of-great-britain-and-northern-ireland-regarding-the-technology-prosperity-de/> accessed 17 August 2026.

[16] The White House, ‘The United States Signs Technology Prosperity Deals with Japan and Korea’ (*The White House*, 29 October 2025) <https://www.whitehouse.gov/releases/2025/10/the-united-states-signs-technology-prosperity-deals-with-japan-and-korea/> accessed 17 August 2026.

[17] Department of Industry, Science and Resources, ‘National Quantum Strategy’ (2023) <https://www.industry.gov.au/publications/national-quantum-strategy> accessed 9 June 2026.

[18] Department of Industry, Science and Resources, ‘Corporate Plan 2025–29 | Department of Industry, Science and Resources’ (*https://www.industry.gov.au/node/95084*, 5 September 2025) <https://www.industry.gov.au/publications/corporate-plan-2025-29> accessed 18 August 2026.

[19] ‘Quantum Australia Reveals $83.1M Economic Impact and 15 New Quantum Companies’ (*Quantum Australia*) <https://www.quantum-australia.com/news/quantum-australia-reveals-831m-economic-impact> accessed 18 August 2026.

[20] Department of Industry, Science and Resources, ‘$12.7 Million to Support Innovative Quantum Technology Projects | Department of Industry, Science and Resources’ (*https://www.industry.gov.au/node/96216*, 27 July 2026) <https://www.industry.gov.au/news/127-million-support-innovative-quantum-technology-projects> accessed 18 August 2026.

[21] Department of Industry, Science and Resources, ‘$5.74 Million to Fast-Track Aussie Quantum Solutions | Department of Industry, Science and Resources’ (*https://www.industry.gov.au/node/94910*, 22 July 2026) <https://www.industry.gov.au/news/574-million-fast-track-aussie-quantum-solutions> accessed 18 August 2026.

[22] ‘World-Leading Australian Quantum Clocks Successfully Trialled under AUKUS Pillar II’ (4 November 2025) <https://www.minister.defence.gov.au/media-releases/2025-11-04/world-leading-australian-quantum-clocks-successfully-trialled-under-aukus-pillar-ii> accessed 18 August 2026.

[23] Department of Industry, Science and Resources, ‘Australia and Japan Strengthen Quantum Partnership | Department of Industry, Science and Resources’ (*https://www.industry.gov.au/node/96830*, 31 July 2026) <https://www.industry.gov.au/news/australia-and-japan-strengthen-quantum-partnership> accessed 18 August 2026.

[24] Government of Canada, ‘Canada’s National Quantum Strategy’ (Innovation, Science and Economic Development Canada 2023).

[25] ‘Budget 2025’ <https://budget.canada.ca/2025/home-accueil-en.html> accessed 18 August 2026.

[26] Innovation, Science and Economic Development Canada, ‘Minister Solomon Announces Major New Quantum Initiative’ (15 December 2025) <https://www.canada.ca/en/innovation-science-economic-development/news/2025/12/minister-solomon-announces-major-new-quantum-initiative.html> accessed 18 August 2026.

[27] National Defence, ‘Canada’s Defence Industrial Strategy’ (17 February 2026) <https://www.canada.ca/en/department-national-defence/corporate/reports-publications/industrial-strategy/security-sovereignty-prosperity.html> accessed 18 August 2026.

[28] National Defence, ‘Minister McGuinty Launches Canada’s First Quantum Defence Innovation Secure Hub to Strengthen Defence Capabilities’ (6 August 2026) <https://www.canada.ca/en/department-national-defence/news/2026/08/minister-mcguinty-launches-canadas-first-quantum-defence-innovation-secure-hub-to-strengthen-defence-capabilities.html> accessed 18 August 2026.

[29] Innovation, Science and Economic Development Canada, ‘Canada to Host G7 Industry, Digital and Technology Ministers’ Meeting in Montréal’ (31 October 2025) <https://www.canada.ca/en/innovation-science-economic-development/news/2025/10/canada-to-host-g7-industry-digital-and-technology-ministers-meeting-in-montreal.html> accessed 18 August 2026.

[30] Innovation, Science and Economic Development Canada, 'Canada and Germany Deepen Their Collaboration on Advanced Technologies' (8 December 2025) <https://www.canada.ca/en/innovation-science-economic-development/news/2025/12/canada-and-germany-deepen-their-collaboration-on-advanced-technologies.html> accessed 18 August 2026.

[31] Innovation, Science and Economic Development Canada, 'Government of Canada and Government of the French Republic Sign Joint Statement on Quantum Science and Technologies Cooperation' (29 May 2026) <https://www.canada.ca/en/innovation-science-economic-development/news/2026/05/government-of-canada-and-government-of-the-french-republic-sign-joint-statement-on-quantum-science-and-technologies-cooperation.html> accessed 18 August 2026.

[32] UK Department for Science, Innovation and Technology, 'National Quantum Strategy' (March 2023).

[33] 'The UK's Modern Industrial Strategy 2025' (*GOV.UK*, 9 June 2026) <https://www.gov.uk/government/collections/the-uks-modern-industrial-strategy-2025> accessed 18 August 2026.

[34] 'Tech Innovators Backed to Set up and Scale up in Britain through Industrial Strategy' (*GOV.UK*) <https://www.gov.uk/government/news/tech-innovators-backed-to-set-up-and-scale-up-in-britain-through-industrial-strategy> accessed 18 August 2026.

[35] 'UK's "Quantum Leap" to Help Beat Disease, Deliver High-Paid Jobs, and Strengthen National Security, as First Country in the World to Roll out Quantum Computers at Scale' (n 8).

[36] 'Memorandum of Understanding Between the Government of The United States of America and the Government of The United Kingdom of Great Britain and Northern Ireland Regarding the Technology Prosperity Deal' (*The White House*, 18 September 2025).

[37] 'UK and Germany Deepen Science and Tech Ties with £14 Million to Unlock Quantum's Vast Potential' (*GOV.UK*) <https://www.gov.uk/government/news/uk-and-germany-deepen-science-and-tech-ties-with-14-million-to-unlock-quantums-vast-potential> accessed 18 August 2026.

[38] 'Leading Quantum Nations to Boost Cooperation Following UK Summit' (*GOV.UK*) <https://www.gov.uk/government/news/leading-quantum-nations-to-boost-cooperation-following-uk-summit> accessed 18 August 2026.

[39] Ministerie van Economische Zaken, 'Kabinet kent definitief € 182,2 miljoen toe aan 3 projecten uit Nationaal Groeifonds | Rijksoverheid.nl' (18 March 2025) <https://www.rijksoverheid.nl/actueel/nieuws/2025/03/18/kabinet-kent-definitief-1822-miljoen-euro-toe-aan-drie-projecten-uit-nationaal-groeifonds> accessed 18 August 2026.

[40] 'With an investment in QuiX Quantum, Europe's first quantum computer is accelerating at light speed' (*Invest-NL*) <https://www.invest-nl.nl/en/news/with-investment-in-quix-quantum-the-deep-tech-fund-accelerates-europes-first-quantum-computer-at-light-speed> accessed 18 August 2026.

[41] 'Invest-NL Makes Follow-on Investment in QuantWare to Scale up Quantum Hardware' (*Invest-NL*) <https://invest-nl.nl> accessed 18 August 2026.

[42] 'Deep Tech Funds grow to € 610 million for Dutch key technology companies' (*Invest-NL*) <https://www.invest-nl.nl/en/news/deep-tech-fund-grows-to-610-million-for-dutch-key-technology-companies> accessed 18 August 2026.

[43] Ministerie van Algemene Zaken, 'Coalition Agreement "Let's Get to Work. Building a Better Netherlands" | Government.Nl' (23 February 2026) <https://www.government.nl/government/coalition-agreement> accessed 18 August 2026.

[44] Ministerie van Economische Zaken, 'Nederlandse hightechsector sluit strategische deals in Japan: nieuwe samenwerkingen rond chips, kwantum en 6G | Rijksoverheid.nl' (22 May 2025) <https://www.rijksoverheid.nl/actueel/nieuws/2025/05/22/nederlandse-hightechsector-sluit-strategische-deals-in-japan-nieuwe-samenwerkingen-rond-chips-kwantum-en-6g> accessed 18 August 2026.

[45] Ministerie van Buitenlandse Zaken, 'Joint Statement by India and the Netherlands | Government.Nl' (16 May 2026) <https://www.government.nl/documents/2026/05/16/joint-statement-by-india-and-the-netherlands> accessed 18 August 2026.

[46] ‘Outline of the 14th Five-Year Plan (2021–2025) for National Economic and Social Development and Vision 2035 of the People’s Republic of China New Initiatives Foreign Affairs Office of the Fujian Provincial People’s Government’ <https://wb.fujian.gov.cn/English/usefulinformation/NewInitiatives/202108/t20210809_5666236.htm> accessed 18 August 2026.

[47] ‘Report on the Work of the Government New Initiatives Foreign Affairs Office of the Fujian Provincial People’s Government’ <https://wb.fujian.gov.cn/English/usefulinformation/NewInitiatives/202503/t20250314_6779252.htm> accessed 18 August 2026.

[48] ‘Report on the Work of the Government News Fujian Provincial People’s Government’, Delivered at the Fourth Session of the 14th National People’s Congress of the People’s Republic of China on March 5, 2026 <https://www.fujian.gov.cn/english/news/202603/t20260319_7113602.htm> accessed 18 August 2026.

[49] ‘China Unveils National Venture Capital Guidance Fund to Boost Innovation’ <https://english.www.gov.cn/news/202512/26/content_WS694e4e56c6d00ca5f9a08486.html> accessed 18 August 2026.

[50] ‘China Hits New Landmark in Global Quantum Computing Race’ <https://english.www.gov.cn/news/202503/04/content_WS67c656dbc6d0868f4e8f0489.html> accessed 18 August 2026.

[51] ‘How Hefei’s ‘Quantum Avenue’ became a thriving innovation corridor’ <https://ydyl.gansu.gov.cn/enggsydyl/news/topnews/202605/t20260521_36581.html> accessed 18 August 2026.

[52] ‘Address by H.E. Li Qiang Premier of the State Council of the People’s Republic of China at the Opening Plenary of the Annual Meeting of the New Champions 2026’ <https://sydney.china-consulate.gov.cn/eng/zgxw/202606/t20260625_11952034.htm> accessed 18 August 2026.

[53] Government of Ireland, ‘Quantum 2030: A National Quantum Technologies Strategy for Ireland’ (2023).

[54] ‘€159 Million Awarded under Disruptive Technologies Fund’ <https://enterprise.gov.ie/en/news-and-events/department-news/2025/december/20251208.html> accessed 18 August 2026.

[55] ‘Government Announces €460 Million Investment for Seven New “Rinn” Research Ireland Centres | Research Ireland’ (10 June 2026) <https://www.researchireland.ie/news/rinn-network/> accessed 18 August 2026.

[56] ‘Government Publishes New Digital and AI Strategy’ (*gov.ie*, 18 February 2026) <https://www.gov.ie/en/department-of-the-taoiseach/press-releases/government-publishes-new-digital-ai-strategy-90-actions-to-strengthen-irelands-position-as-a-digital-leader-and-ai-hub/> accessed 18 August 2026.

[57] ‘Government Approves Major Expansion of Tyndall National Institute’ (*gov.ie*, 13 January 2026) <https://www.gov.ie/en/department-of-further-and-higher-education-research-innovation-and-science/press-releases/government-approves-major-expansion-of-tyndall-national-institute/> accessed 18 August 2026.

[58] ‘Minister Lawless Signs the European Quantum Pact’ (*gov.ie*, 19 June 2025) <https://www.gov.ie/en/department-of-further-and-higher-education-research-innovation-and-science/press-releases/minister-lawless-signs-the-european-quantum-pact/> accessed 18 August 2026.

[59] ‘National Quantum Mission’ <https://dst.gov.in/national-quantum-mission-nqm> accessed 18 August 2026.

[60] ‘Parliament Question: Achievements under NQM’ <https://www.pib.gov.in/Pressreleaseshare.aspx?PRID=2227070> accessed 18 August 2026.

[61] Ibid.

[62] ‘Startup selected under NQM launches one of India’s most powerful quantum computers’ <https://dst.gov.in/startup-selected-under-nqm-launches-one-indias-most-powerful-quantum-computers> accessed 18 August 2026.

[63] ‘Office of Principal Scientific Adviser to GoI Releases First Edition of India’s International Technology Engagement Strategy (ITES) for Quantum’ <https://www.pib.gov.in/PressReleasePage.aspx?PRID=2121527&reg=48&lang=2> accessed 18 August 2026.

[64] ‘Fact Sheet 2.0: India-Japan Economic Security Cooperation’ <https://www.mea.gov.in/press-releases?dtl/41396> accessed 18 August 2026.

[65] 'Call for Proposals on "Building U.S.-India Partnerships to Drive Innovation in Critical Minerals and Quantum Technology"' <https://dst.gov.in/callforproposals/call-proposals-building-us-india-partnerships-drive-innovation-critical-minerals> accessed 18 August 2026.

[66] German Federal Ministry of Education and Research, 'Quantum Technologies Conceptual Framework Programme' (2023).

[67] 'Hightech Agenda Deutschland beschlossen | Bundesregierung' (*Die Bundesregierung informiert | Startseite*) <https://www.bundesregierung.de/breg-de/aktuelles/hightech-agenda-deutschland-2366912> accessed 18 August 2026.

[68] 'Quantentechnologien' (2026) <https://hightech-agenda-deutschland.de/roadmaps/quantentechnologien/> accessed 18 August 2026.

[69] 'DLR Quantum Computing Initiative – We Enable Quantum' <https://qci.dlr.de/en/> accessed 18 August 2026.

[70] 'Project SQale' (*Fraunhofer Institute for Silicon Technology*) <https://www.isit.fraunhofer.de/en/news/project-sqale.html> accessed 18 August 2026.

[71] 'UK and Germany Deepen Science and Tech Ties with £14 Million to Unlock Quantum's Vast Potential' (*GOV.UK*) <https://www.gov.uk/government/news/uk-and-germany-deepen-science-and-tech-ties-with-14-million-to-unlock-quantums-vast-potential> accessed 18 August 2026.

[72] 'Conclusion of the 26th Franco-German Ministerial Council' (2026).

[73] 'La Stratégie Nationale Quantique - Portail de La Stratégie Nationale Quantique' <https://quantique.france2030.gouv.fr/> accessed 18 August 2026.

[74] 'Souveraineté technologique : la France investit 1 milliard d'euros supplémentaire dans la stratégie nationale quantique' (*info.gouv.fr*) <https://www.info.gouv.fr/actualite/souverainete-technologique-la-france-investit-1-milliard-d-euros-supplementaire-dans-la-strategie-nationale-quantique> accessed 19 August 2026.

[75] 'Quantique : le Fonds Innovation Défense participe à une levée de fonds de 100 millions d'euros en faveur du développement de la société Alice & Bob | Ministère des Armées et des Anciens combattants' (31 January 2025) <http://www.defense.gouv.fr/dga/actualites/quantique-fonds-innovation-defense-participe-levee-fonds-100-millions-deuros-faveur-du-developpement-societe-alice> accessed 19 August 2026.

[76] 'Programme PROQCIMA - Portail de La Stratégie Nationale Quantique' <https://quantique.france2030.gouv.fr/acces-aux-marches/programme-proqcima/> accessed 19 August 2026.

[77] 'Souveraineté technologique : la France investit 1 milliard d'euros supplémentaire dans la stratégie nationale quantique' (*info.gouv.fr*) <https://www.info.gouv.fr/actualite/souverainete-technologique-la-france-investit-1-milliard-d-euros-supplementaire-dans-la-strategie-nationale-quantique> accessed 19 August 2026.

[78] Innovation, 'Government of Canada and Government of the French Republic Sign Joint Statement on Quantum Science and Technologies Cooperation' (n 31).

[79] 2026, 'France and Germany: Joint Position Paper on the European Quantum Act' <https://www.bmftr.bund.de/SharedDocs/Downloads/DE/2026/joint-position-paper-eqa.pdf?__blob=publicationFile&v=1>.

[80]Ministry of Science and ICT Press Release - 'Act on Promotion of Quantum Science, Technology, and Industry Takes Effect' <https://www.msit.go.kr/eng/bbs/view.do?bbsSeqNo=42&mId=4&mPid=2&nttSeqNo=1053&sCode=eng&searchOpt=ALL> accessed 19 August 2026.

[81]Ministry of Science and ICT Press Release - 'Korea Unveils First Quantum Master Plan for the Next-AI Era' <https://www.msit.go.kr/eng/bbs/view.do?bbsSeqNo=42&mId=4&mPid=2&nttSeqNo=1222> accessed 19 August 2026.

[82]Ministry of Science and ICT Press Release - 'MSIT to Invest KRW 198 Billion in Quantum Science and Technology for 2025 >' <https://www.msit.go.kr/eng/bbs/view.do?bbsSeqNo=42&mId=4&nttSeqNo=1079&sCode=eng> accessed 19 August 2026.

[83]Ministry of Science and ICT Press Release - 'Driving Future Growth with the Twin Engines of AI and Science & Technology' <https://www.msit.go.kr/eng/bbs/view.do?bbsSeqNo=42&mId=4&mPid=2&nttSeqNo=1163> accessed 19 August 2026.

[84]Ministry of Science and ICT Press Release - ‘Korea Unveils First Quantum Master Plan for the Next-AI Era’ <https://www.msit.go.kr/eng/bbs/view.do?bbsSeqNo=42&mId=4&mPid=2&nttSeqNo=1222> accessed 19 August 2026.

[85]Ministry of Science and ICT Press Release - ‘An Unrivaled Korea, Happy Together Through AI and Science and Technology’ <https://www.msit.go.kr/eng/bbs/view.do?bbsSeqNo=42&mId=4&mPid=2&nttSeqNo=1284&sCode=eng> accessed 19 August 2026.

[86] ‘U.S.–Korea Technology Prosperity Deal’ (*The White House*, 29 October 2025) <https://www.whitehouse.gov/releases/2025/10/u-s-korea-technology-prosperity-deal/> accessed 19 August 2026.

[87] ‘Leading Quantum Nations to Boost Cooperation Following UK Summit’ (*GOV.UK*) <https://www.gov.uk/government/news/leading-quantum-nations-to-boost-cooperation-following-uk-summit> accessed 19 August 2026.

[88] Secretariat of Science, Technology and Innovation Policy, Cabinet Office, ‘Strategy of Quantum Future Industry Development’ (2023).

[89] Cabinet Decision, ‘Comprehensive Economic Measures to Build a “Strong Japanese Economy”’ (2025).

[90] ‘New Year Greetings 2026 - Minister of Economy, Trade and Industry (METI) / METI Ministry of Economy, Trade and Industry’ <https://www.meti.go.jp/english/speeches/2026newyeargreetings.html> accessed 19 August 2026.

[91] ‘RIKEN and Fujitsu Unveil World-Leading 256-Qubit Quantum Computer’ <https://www.riken.jp/en/news_pubs/news/2025/20250422_1/index.html> accessed 19 August 2026.

[92]The National Institute of Advanced Industrial Science and Technology, ‘OptQC and AIST Commence Initial Operations of “MoQuren”’ <https://www.aist.go.jp/aist_e/news/au20260722.html> accessed 19 August 2026.

[93] ‘Signing of the Letter of Intent on Japan-EU Cooperation in Quantum Science and Technology - Science, Technology and Innovation’ (*Cabinet Office Home Page*) <https://www8.cao.go.jp/cstp/stmain/20250513ryoshi_en.html> accessed 19 August 2026.

[94] ‘Q-NEKO Project Kick off Meeting Held’ <https://www8.cao.go.jp/cstp/stmain/20260413ryoshi_en.html> accessed 19 August 2026.

[95] ‘Signing of the Letter of Intent on Cooperation between Japan and India in Quantum Science and Technology’ <https://www8.cao.go.jp/cstp/stmain/20260504ryoshi_en.html> accessed 19 August 2026.

[96] ‘U.S.–Japan Technology Prosperity Deal’ (*The White House*, 28 October 2025) <https://www.whitehouse.gov/releases/2025/10/u-s-japan-technology-prosperity-deal/> accessed 26 August 2026.

[97] ‘United States and Japan Announce Historic $1 Billion Partnership Under President Trump’s Genesis Mission’ (*Energy.gov*, 4 June 2026) <https://www.energy.gov/articles/united-states-and-japan-announce-historic-1-billion-partnership-under-president-trumps> accessed 26 August 2026.

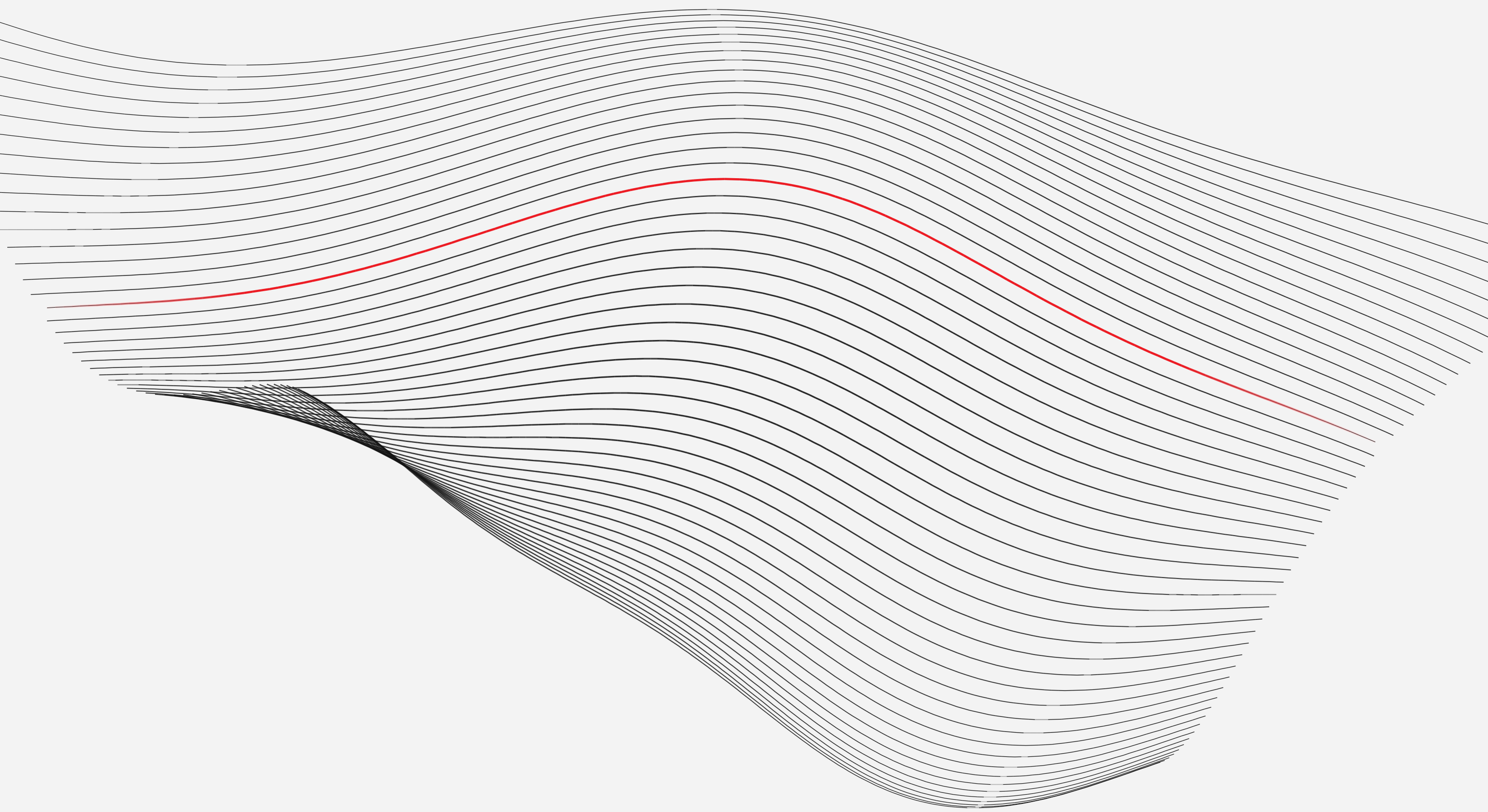

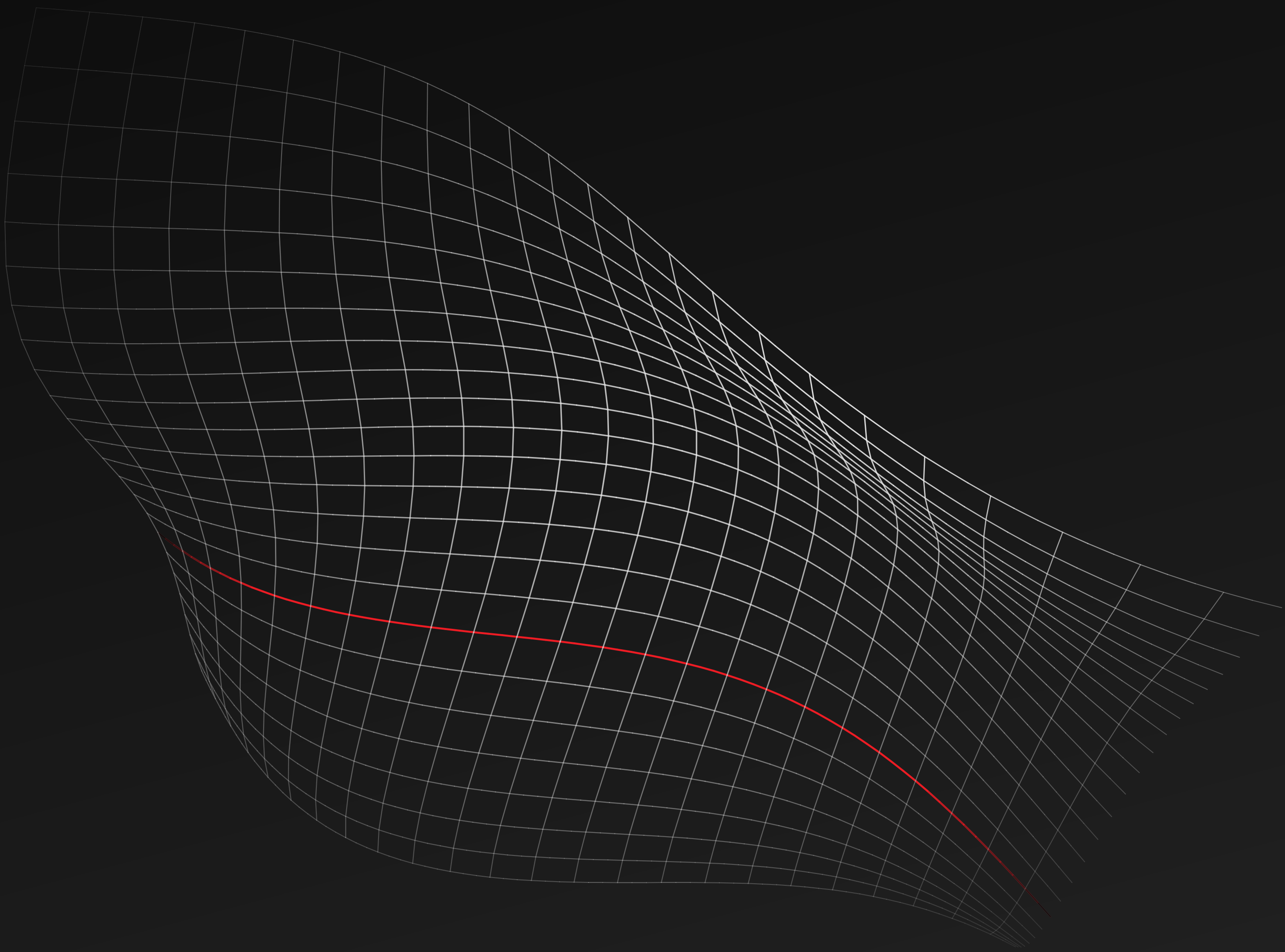

# 6 | Workforce

Between 2024 and 2026, the U.S., UK, Australia and the European Union each moved workforce development into named programs with dedicated budgets. The key driver is the belief that the workers that the quantum sector needs are engineers and technicians rather than just physicists. This also suggests there is increasing belief that talent demand is arriving fast enough to justify building supply now.

In the U.S., the 2026 Executive Order on quantum innovation treats workforce development as essential to maintaining U.S. leadership in quantum information science and technology (QIST).[1] It guides the National Science Foundation (NSF) to begin creating a network of National QIST Workforce Development Institutes to expand training opportunities and coordinate efforts across federal, state and local governments. In addition, it directs the Office of Personnel Management to create a government-wide strategy for recruiting and retaining quantum talent, potentially using special pay and stronger hiring incentives. This follows previous NSF efforts, such as ExpandQISE, launched in 2024 and backed by $39 million, with the aim of strengthening the quantum workforce and investing in the scientific and technological advances that will be the foundation for a quantum-enabled future.[2]

The United Kingdom published its Quantum Skills Taskforce report in June 2025, emphasizing the need for engineers, technicians and professionals in quantum-adjacent disciplines to work with emerging systems and recommending greater use of apprenticeships, continuing professional development, industry placements and quantum modules embedded in engineering degrees.[3] The government's Digital and Technologies Sector Plan included increased funding in 2026 for quantum apprenticeships and a competition to establish a Centre for Quantum Commercialisation Skills.[4]

Australia offers another model in which workforce development is positioned as part of broader ecosystem formation. Its National Quantum Strategy[5] identifies a "skilled and growing quantum workforce" as one of five central themes and recognizes that future demand will encompass engineers, software developers, educators and managers as well as physicists. Government projections suggest that Australia's quantum industry could support approximately 8,700 jobs by 2030 and 19,400 by 2045. In 2026, the Australian Research Council (ARC) launched its Training Centre in Current and Emergent Quantum Technologies (QuTech) aiming to prepare master's and PhD students to work across research, industry and applied environments.[6]

At the European Union level, the Quantum Europe Strategy of 2025 has as one of its five themes "building a diverse, world-class workforce through coordinated education, training, and talent mobility across the EU".[7] The strategy also announces support for a pilot Quantum Apprenticeship Programme to prepare a pipeline of quantum specialists trained on real-world projects. In 2026, Europe also launched the European Quantum Academy, a pan-European ecosystem for quantum education and training.[8]

## 6.1 | Quantum skills in job postings

**Fig. 6.a** | U.S. job postings requiring "quantum" skills as a share of total job postings, 2011 to March 2026

MIT Quantum Index Report

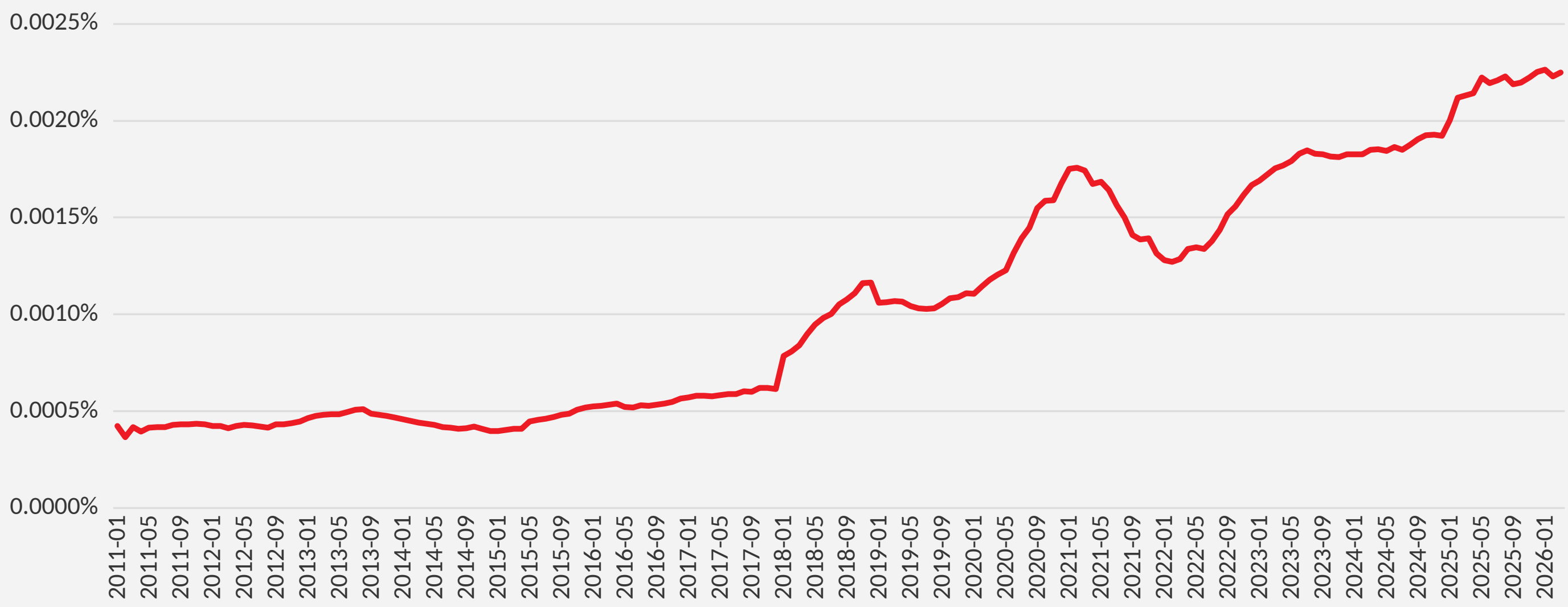

+16%

▶ Growth nearly stalled in 2024 (+4%) right after a strong 2023 (+26%), which could have been read as the market maturing or cooling. Instead, 2025 came back at +16%, and Q1 2026 is already at the series' all-time high.

Based on the Lightcast data on U.S. job postings requiring "quantum" skills as a share of total job postings from 2011 to March 2026, the share has grown more than fivefold from the start of 2011 to the first quarter of 2026. It should be noted that the measure quoted is a share rather than a count, and so moves with total U.S. posting volume as well as with quantum hiring.

The first phase, 2011 to 2017, was one of limited and uneven growth. The yearly average barely moved, even dipping in 2014, before drifting up in 2017, for cumulative growth of only about 42% across seven years, with year-over-year swings as large as +18% (2016) and -12% (2014).

The second phase, 2018 to 2020, marks a sharp acceleration. The yearly average jumped 66% in a single year in 2018, then kept climbing through the period, with annual growth of 8% in 2019 and 29% in 2020.

In the third phase, mid-2021 through 2022, after peaking in the first months of 2021, the monthly share fell for roughly a year and a half, dragging the 2022 annual average down, marking the first full-year decline since 2014, a contraction of about 9.5%.

The fourth phase, from 2023 onward, is a renewed and increasingly steady climb. 2023 rebounded sharply (+26% on 2022), 2024 growth cooled to a plateau (+4%, with monthly values essentially flat for most of the year), and then the trend re-accelerated: 2025 posted +16% growth, and the first three months of 2026 are already running at the highest level in the series.

In short, what looked like a stabilizing, moderate upward trend as of mid-2024 has since resolved into a second acceleration, following the classic pattern of a demand surge, a partial pullback, and then renewed, broader-based growth as quantum skills become more embedded in the job market.

> Job postings requiring quantum skills have increased more than fivefold since 2011, reaching their highest recorded share of total job openings in early 2026.

# 6.2 | Quantum mentions in job postings

**Fig. 6.b** | U.S. job postings with mentions of "quantum" as a share of total job postings, 2012 to March 2026

MIT Quantum Index Report

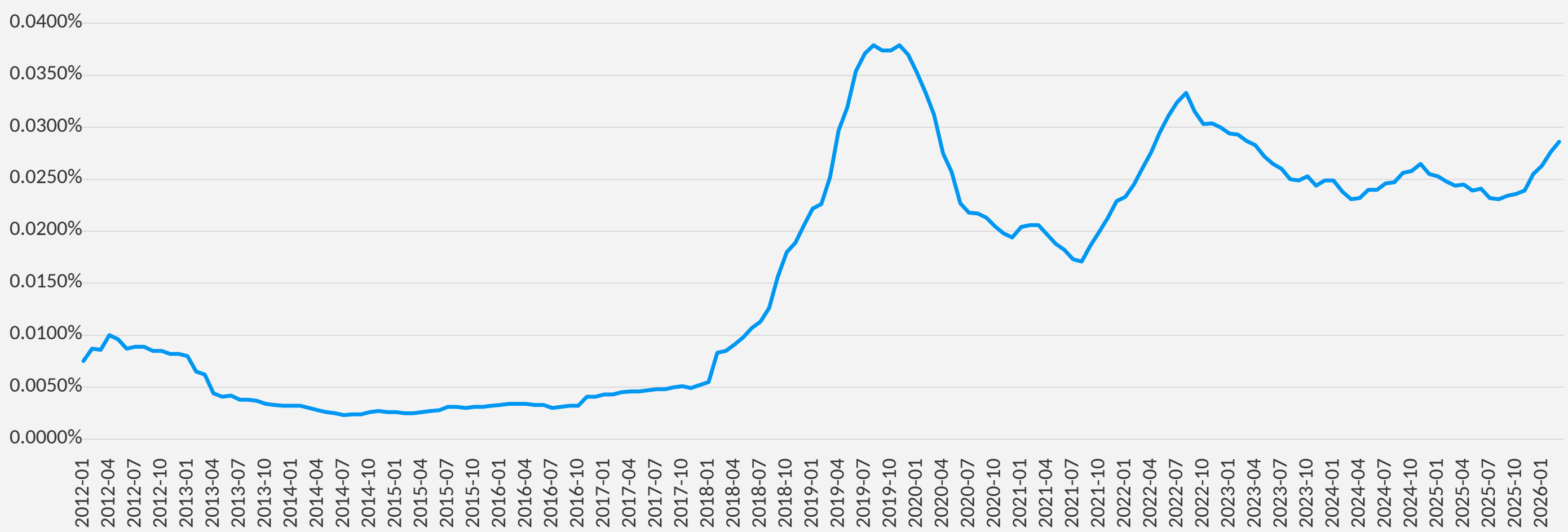

Based on the Lightcast data on U.S. job postings mentioning "quantum" as a share of total job postings from 2012 to March 2026, this series tells a different story from the required-quantum-skills data: rather than a steady climb, it is a boom-bust cycle. Overall, the share ends about 3.2 times higher than it started, but that net gain masks two full cycles rather than one continuous trend.

The first phase, 2012 to 2014, is actually a decline. The second phase, 2015 to 2017, is a slow, unremarkable recovery. The yearly average showed roughly 65% growth over three years but off a very low base.

The third phase, 2018 to 2019, shows a sharp surge with a more than sixfold increase in just two years, and the single highest point in the entire 14-year series.

The fourth phase, 2020 to 2021, is an equally sharp downward correction. The average fell by 23% in 2020 and a further 22% in 2021, bottoming out around August 2021, nearly halving from the 2019 peak in two years. This decline is steeper and starts two years earlier than the dip seen in the skills series, suggesting the "mentions" of quantum cooled off faster than actual quantum-skill demand.

The fifth phase is a second, smaller wave: a rebound through 2022 (+49%), peaking mid-year in August, nearly matching the 2019 high, before easing off again.

Since then, 2023 through 2025 has been a plateau/decline, roughly -3% to -8% per year, essentially flat compared to the volatility earlier in the series. The most recent signal is a fresh uptick starting in late 2025 and accelerating into Q1 2026, the highest level since the 2022 rebound, mirroring the renewed growth seen in the skills data over the same window, but with much more amplitude.

Unlike the “required quantum skills” data, which shows a fairly orderly, mostly upward climb, “mentions” of quantum are more cyclical, with two distinct hype-and-correction waves (2018–2021 and a smaller 2021–2023 one) sitting on top of a more modest long-run growth trend.

**Fig. 6.c** | U.S. job postings share mentioning “quantum” by year, 2011–2025

MIT Quantum Index Report

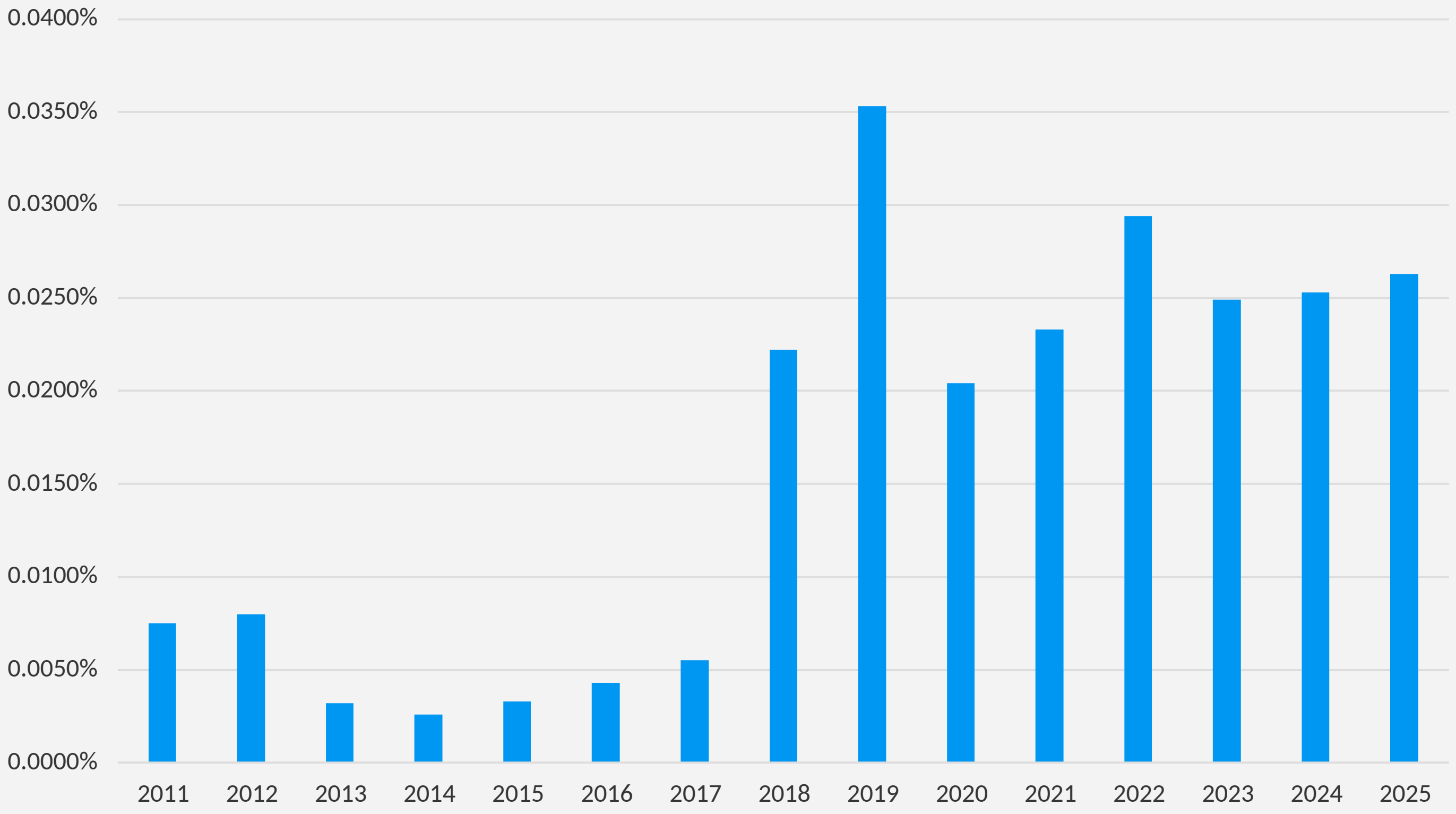

> The share of postings mentioning “quantum” declined in 2023 following its 2022 peak and has since remained broadly flat through 2025, whereas the share of postings requiring quantum-related skills has continued to increase over the same period. This divergence might indicate a shift from general interest toward more concrete skill requirements.

## 6.3 | Future research

We plan to keep tracking this data to provide the community with clearer insights into evolving workforce trends. Our findings from the Lightcast dataset highlight the need for regular updates to the quantum skills library, reflecting a dynamic market where job roles and technical demands change rapidly. If you are interested in collaborating on skills taxonomy, suggesting new resources to track, or helping build international datasets, we welcome you to reach out.

You can reach us at contact@qir.mit.edu

---

### Footnotes

[1] Executive Order No 14413, 91 Fed Reg 38487 (22 June 2026).

[2] 'Quantum Science and Engineering Expands across the Nation with $39M from NSF | NSF - U.S. National Science Foundation' (29 August 2024) <https://www.nsf.gov/news/quantum-science-engineering-expands-across-nation-39m-nsf> accessed 6 June 2026.

[3] 'UK Quantum Skills Taskforce Report' (*GOV.UK*) <https://www.gov.uk/government/publications/uk-quantum-skills-taskforce-report/uk-quantum-skills-taskforce-report> accessed 16 July 2026.

[4] 'Centre for Quantum Commercialisation Skills' (6 May 2026) <https://www.ukri.org/opportunity/centre-for-quantum-commercialisation-skills/> accessed 6 August 2026.

[5] Department of Industry, Science and Resources, 'National Quantum Strategy' (2023).

[6] 'Building Australia's Quantum Workforce | Australian Research Council' (22 July 2026) <https://www.arc.gov.au/news-and-publications/media/building-australias-quantum-workforce> accessed 6 August 2026.

[7] 'Quantum Europe Strategy: Quantum Europe in a Changing World' (European Commission 2025).

[8] 'European Quantum Academy' <https://ec.europa.eu/info/funding-tenders/opportunities/portal/screen/opportunities/projects-details/43152860/101298535> accessed 8 August 2026.

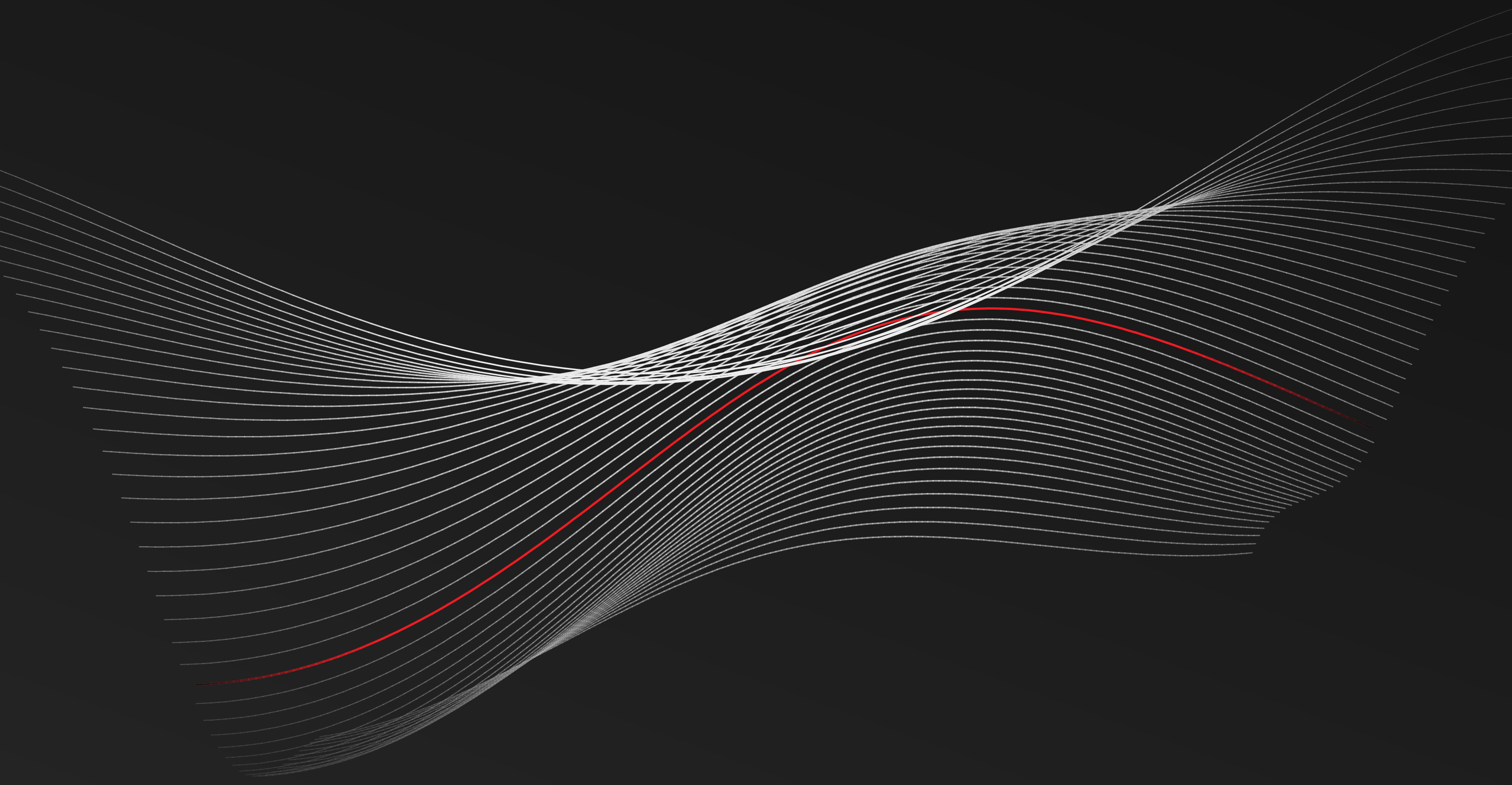

# 7 | Education

The creation of a mature quantum ecosystem depends not only on scientific breakthroughs and the identification of commercially valuable use cases, but also on the development of a multidisciplinary workforce capable of building, operating and governing quantum technologies. This remains a particular challenge because quantum education has traditionally been concentrated at the postgraduate level and associated with advanced mathematics, engineering and physics. Between 2024 and 2026, education policies widened globally with a focus on distinct areas such as training teachers, introducing quantum concepts earlier, creating undergraduate pathways, expanding practical access to hardware and laboratories, and supporting professionals who may need quantum literacy without becoming quantum physicists. This shift is taking place against a continuing shortage of qualified quantum talent, particularly people able to bridge quantum science with engineering and software skills.

In the United States, the National Q-12 Education Partnership[1] remains one of the clearest examples of an attempt to build a pre-university quantum pipeline. Launched in 2020 by the National Science Foundation (NSF) and the White House Office of Science and Technology Policy, the partnership provides K-12 educators and students with learning materials, career resources and links to internships and mentorship opportunities. In 2025, NSF presented Q-12 as part of a broader workforce portfolio that also included teacher-development initiatives such as Quantum for All and Quantum Education for Students and Teachers, alongside the ExLENT program[2], which supports experiential learning from high-school level through professional upskilling.

Europe has moved even further toward system-level coordination. Earlier Quantum Flagship projects[3], such as QTIndu (Quantum Technology Courses for Industry) and DigiQ (Digitally Enhanced Quantum Technology Master), developed educational modules, degree pathways, industry training and the European Competence Framework for Quantum Technologies.[4] The framework was updated to version 3.0 in April 2025, providing a common reference for describing quantum competences, proficiency levels and qualifications. A larger step followed in May 2026 with the launch of the European Quantum Academy (EQA).[5] Backed by €19.8 million and involving more than 70 partner institutions and over 100 affiliated organizations, the EQA is intended to coordinate education from school outreach and harmonized bachelor's and master's programs through doctoral training and professional upskilling. Its stated targets include training at least 600 advanced quantum professionals and reaching 5,000 learners, with six Regional Quantum Academies intended to reduce uneven access across Europe.

The United Kingdom has similarly placed skills at the center of national quantum policy. Its Quantum Skills Taskforce report, published in June 2025, argues that the sector will need engineers, technicians and professionals in adjacent disciplines who possess sufficient quantum knowledge to work with emerging systems. The Taskforce recommends greater use of apprenticeships, continuing professional development, industry placements and quantum modules embedded in engineering degrees, as well as continued school outreach and teacher training. It also links these pathways to the UK National Quantum Strategy commitment to support more than 1,000 PhD studentships between now and 2034.[6]

Recent approaches suggest that a scalable quantum economy will not rely exclusively on PhD-level physicists, and that quantum education is becoming broader, more structured and more coordinated, offering differentiated pathways for school students, teachers, technicians, engineers and industry users.

## 7.1 | Postgraduate education

**Fig. 7.a** | Master's degrees with a specific reference to "quantum" in the degree name, 2025 vs. 2026

MIT Quantum Index Report

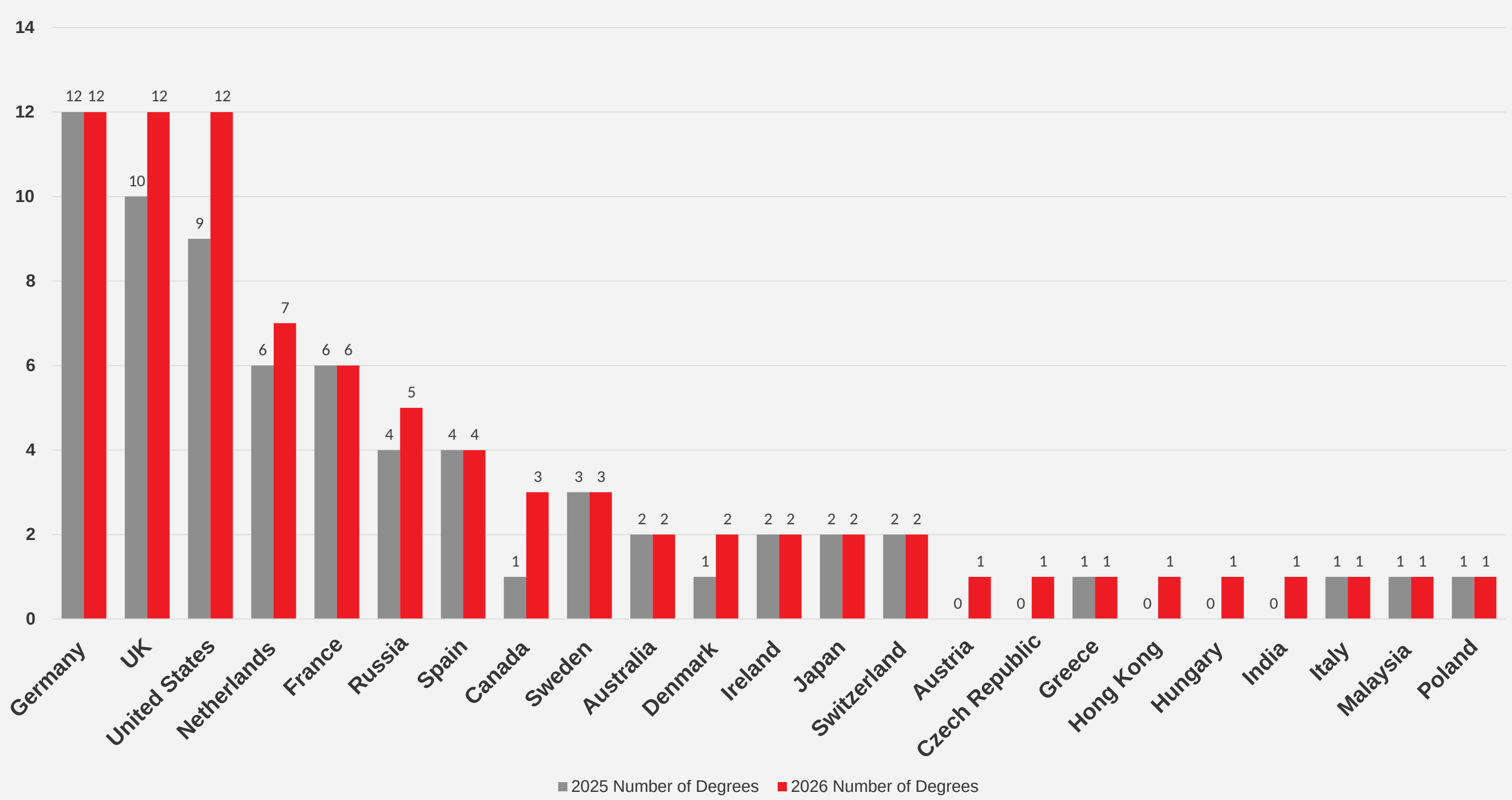

In 2026, the master's degree landscape[7] specifically referencing "quantum" in the postgraduate degree title is led by three countries in a tie at the top: Germany, the UK, and the United States each offer 12 programs. This marks a shift from a single clear leader in 2025 to a three-way tie at the top. The Netherlands sits just behind in fourth place with 7 programs, while France holds fifth with 6.

Together, these top five countries account for 49 of the 83 total quantum-titled degrees on offer globally, roughly 59% of all programs tracked. Russia follows with 5 programs, and Spain, Canada, and Sweden form a next tier with 3-4 programs each. A broader cluster of countries, Australia, Denmark, Ireland, Japan, and Switzerland, each offer 2 programs, while a long tail of Austria, Czech Republic, Greece, Hong Kong, Hungary, India, Italy, Malaysia and Poland each offer a single program.

> Austria, the Czech Republic, Hong Kong, Hungary and India each launched their first master's degree referencing "quantum" in the degree title in 2026, extending country-level representation in the dataset to 23 countries.

Comparing the 2025 vs 2026 data reveals a market that has grown substantially rather than simply reshuffled. Total quantum-titled master's degrees worldwide climbed from 68 to 83, a 22% increase, suggesting the field is scaling up broadly rather than merely consolidating around early leaders.

12

In 2025, Germany offered the most degrees (12); in 2026, the UK and the United States caught up to match it exactly at 12 programs each in 2026, creating a three-way tie at the top.

## 7.2 | Enrollment numbers

Historically, the United States has successfully retained global talent, with roughly 70% of international STEM PhD graduates electing to stay in the country as of 2017.[8] However, cultivating new quantum talent is a long-term endeavor, typically demanding around ten years of higher education and specialized training. To meet expanding labor demands, the U.S. must adopt a two-pronged approach: strengthening domestic educational pathways while continuing to attract and retain top international talent. According to the 2021 U.S. Report “The Role of International Talent in Quantum Information Science”, physics, electrical engineering, and computer science represent the primary degree fields for quantum information science and technology (QIST). These disciplines were identified using a dual methodology: filtering online job listings for key industry terms and evaluating doctoral dissertation titles, abstracts and keywords.

To better understand emerging enrollment trends for physics, electrical engineering, and computer science courses (broad proxies for quantum) we analyzed data from the National Student Clearinghouse (NSC) Research Center (June 2026 update).

**Fig. 7.b** | Undergraduate degree enrollments in the U.S., 2019–2026

MIT Quantum Index Report

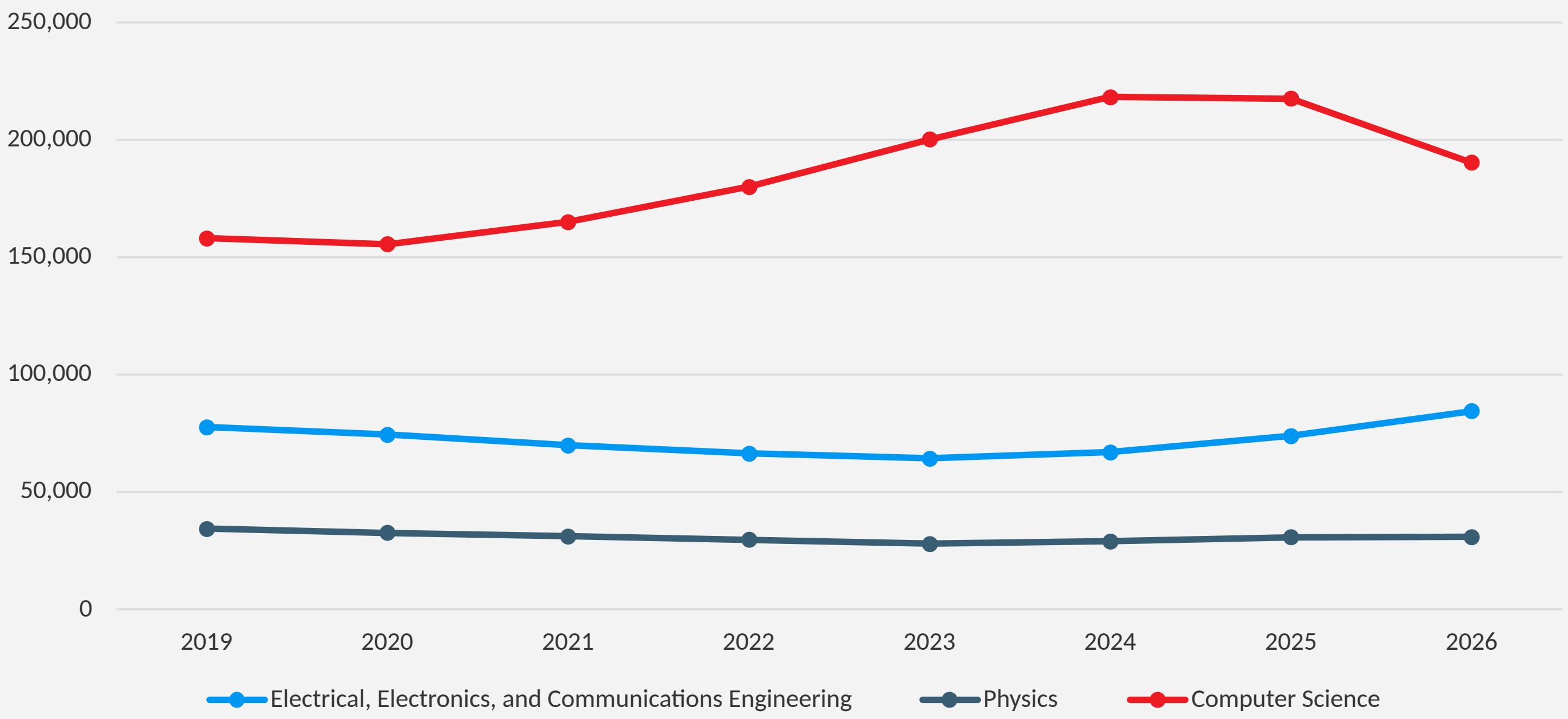

Based on the enrollment data from 2019 to 2026, computer science exhibits by far the highest and most dynamic enrollment trajectory among the three fields. Numbers climbed steadily from 158,122 in 2019 to a peak of 218,190 in 2024, a gain of roughly 60,000 students, or 38%, before pulling back over the following two years to 190,285 in 2026.

Electrical, Electronics, and Communications Engineering follows a different pattern: enrollment contracts from 77,527 in 2019 down to a low of 64,234 in 2023, before rebounding to 84,415 by 2026, its highest point in the entire period, and a swing of over 20,000 students between trough and peak.

Physics, by contrast, remains the smallest field throughout and shows the most contained movement, narrowing from 34,256 in 2019 to a low of 27,952 in 2023 before a modest recovery to 30,832 in 2026, a range of roughly 6,300 students between its lowest and highest years, reinforcing physics as the most stable, if slowly contracting, of the three fields.

As in our 2025 report, categorization presents some challenges: related degrees such as “Computer and Information Science, General” and “Astronomy and Astrophysics” fall outside these subject-level figures, meaning the field-level counts likely understate the true scale of activity in each area. To provide a fuller picture, the corresponding field-family totals—Engineering, Physical Sciences, and Computer and Information Sciences and Support Services, tell a broadly consistent story.

**Fig. 7.c** | Undergraduate major field family enrollments in the U.S., 2019–2026

MIT Quantum Index Report

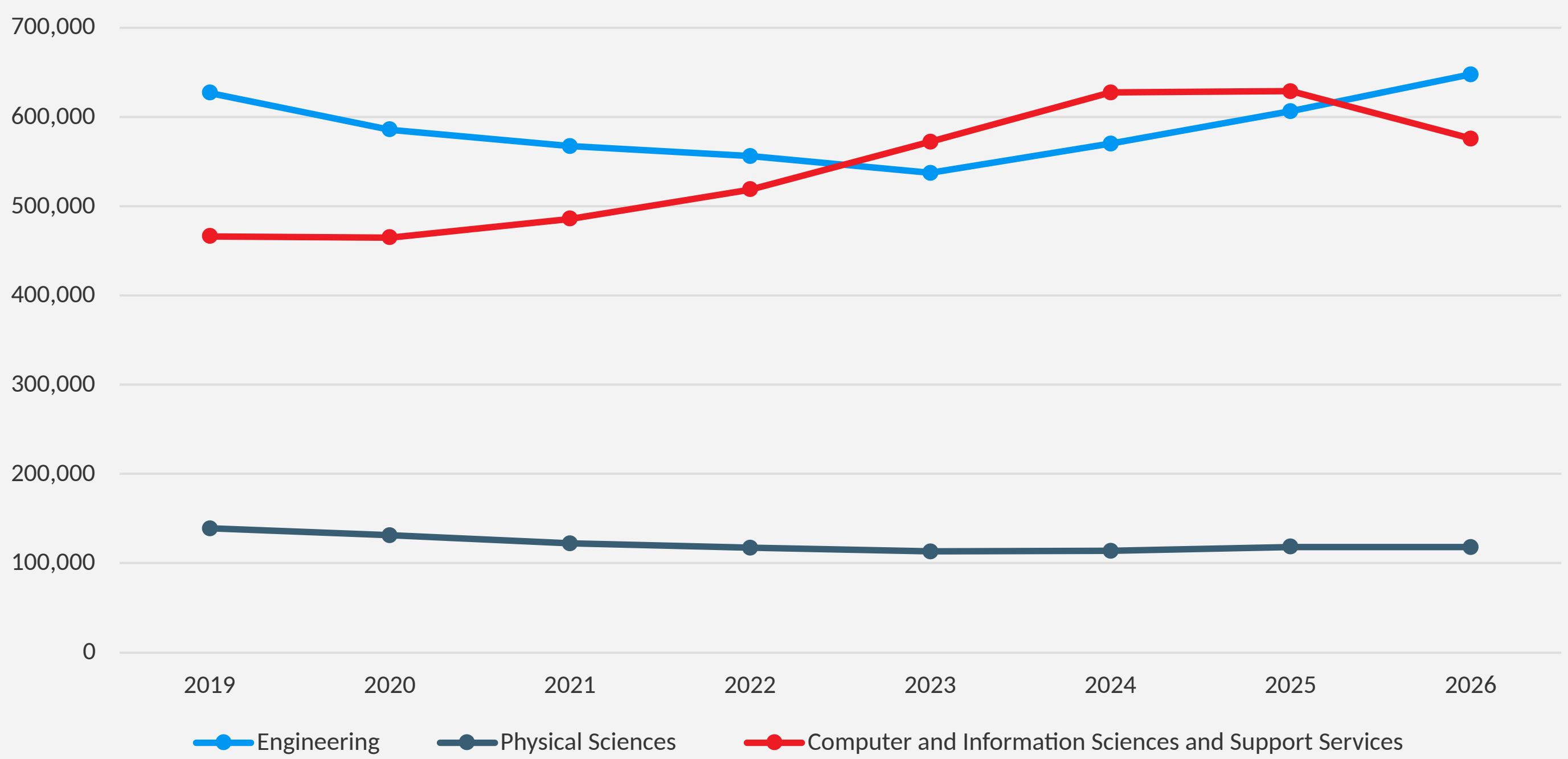

Computer and Information Sciences and Support Services shows the most substantial growth, rising from 466,306 in 2019 to a peak of 628,770 in 2025, an increase of nearly 35%, before easing back to 575,747 in 2026, mirroring the pullback seen at the subject level, though the decline arrives one year later at the family level.

Engineering follows a dip-then-recover pattern similar to its subject-level counterpart: enrollment falls from 627,221 in 2019 to a low of 537,294 in 2023, then climbs steadily to a period-high 647,877 in 2026, surpassing its 2019 starting point.

Physical Sciences again proves the most stable of the three, moving within a comparatively narrow band from a 2019 high of 138,947 down to a 2023 low of 113,075, before completing a mild recovery to 118,027 in 2026, a pattern that closely tracks physics at the subject level, though at roughly four times the enrollment scale. Taken together, the family-level data confirms the subject-level trends while smoothing out some of their sharper year-to-year swings, and helps contextualize the subject-specific figures within their broader disciplinary categories.

## 7.3 | Future research

Although the quantum education space is expanding quickly, it remains fragmented and inconsistently tracked. While informal feedback suggests growing interest and rising enrollments in quantum programs, solid data on student demographics, institutional funding, and career trajectories is urgently needed. Gathering these metrics is vital for pinpointing best practices, addressing equity and access gaps, and guiding data-driven policymaking.

To deepen and broaden these insights in future editions of this report, we welcome contributions from the quantum education community. We encourage researchers and educators to reach out if they can share enrollment figures, curriculum models, or updates on emerging programs. Through this community-driven approach, we hope to build a fuller global picture of quantum education efforts and support better learning strategies across the field.

You can reach us at contact@qir.mit.edu

## Footnotes

[1] 'National Q-12 Education Partnership' <https://q12education.org/> accessed 16 August 2026.

[2] 'Experiential Learning for Emerging and Novel Technologies (ExLENT) | NSF - U.S. National Science Foundation' <https://www.nsf.gov/funding/opportunities/exlent-experiential-learning-emerging-novel-technologies> accessed 16 August 2026.

[3] 'Quantum Flagship' <https://qt.eu/ecosystem/education-and-training> accessed 16 August 2026.

[4] European Commission: Directorate-General for Communications Networks, Content and Technology, Quantum Flagship, Greinert, F. and Müller, R., 'European Competence Framework for Quantum Technologies (CFQT)' (Publications Office of the European Union 2025).

[5] 'The Quantum Academy Launches to Train Europe's Next Generation of Talent' (*Quantum Flagship*) <https://qt.eu/news/2026/2026-05-06_The_Quantum_Academy_launches_to_train_Europes_next-generation_of_talent> accessed 16 July 2026.

[6] 'UK Quantum Skills Taskforce Report' (*GOV.UK*) <https://www.gov.uk/government/publications/uk-quantum-skills-taskforce-report/uk-quantum-skills-taskforce-report> accessed 16 July 2026.

[7] As sourced from StudyPortals database, following the country coverage therein.

[8] Subcommittee on Economic and Security Implications of Quantum Science, Committee on Homeland and National Security of the National Science & Technology Council, 'The Role of International Talent in Quantum Information Science'.

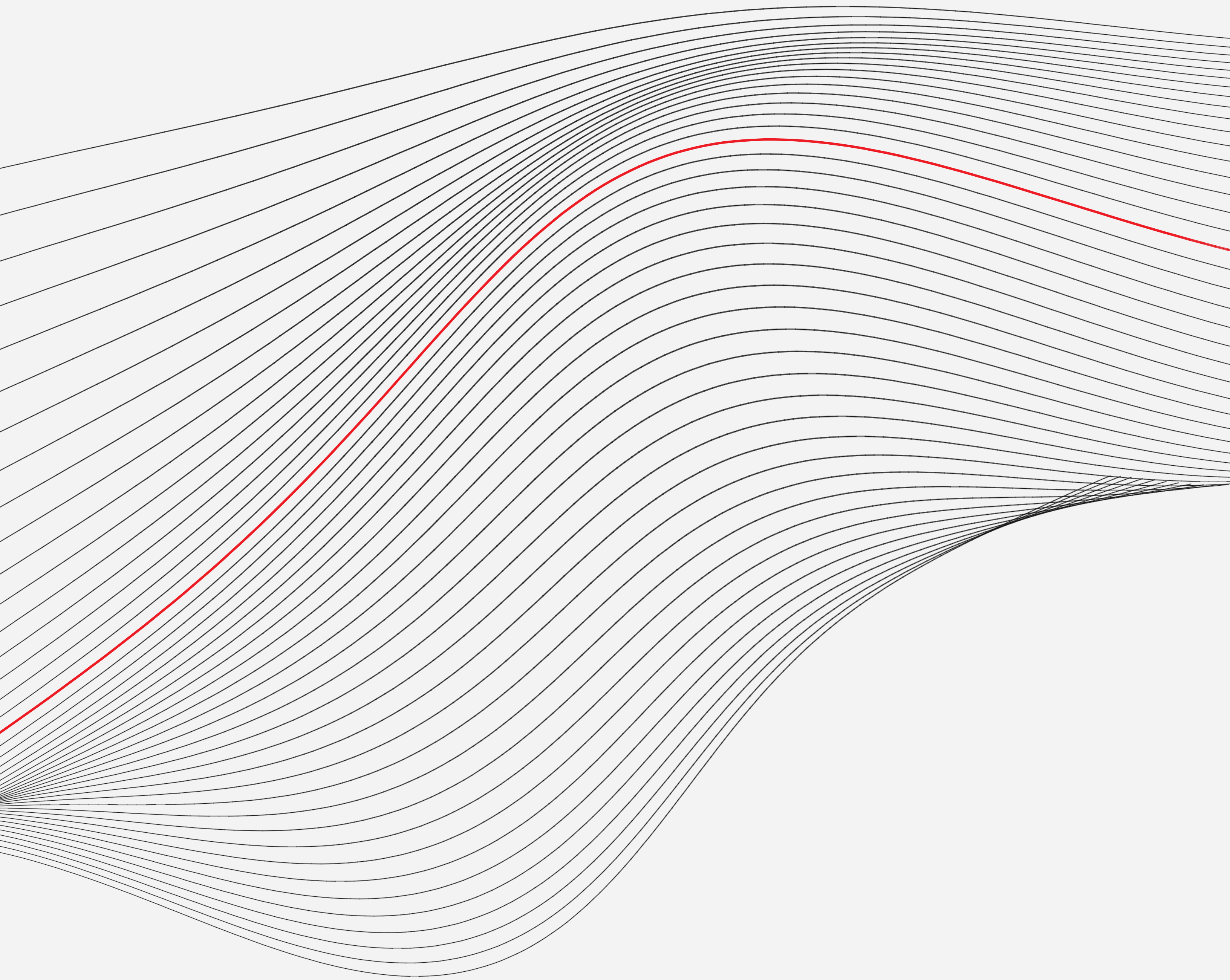

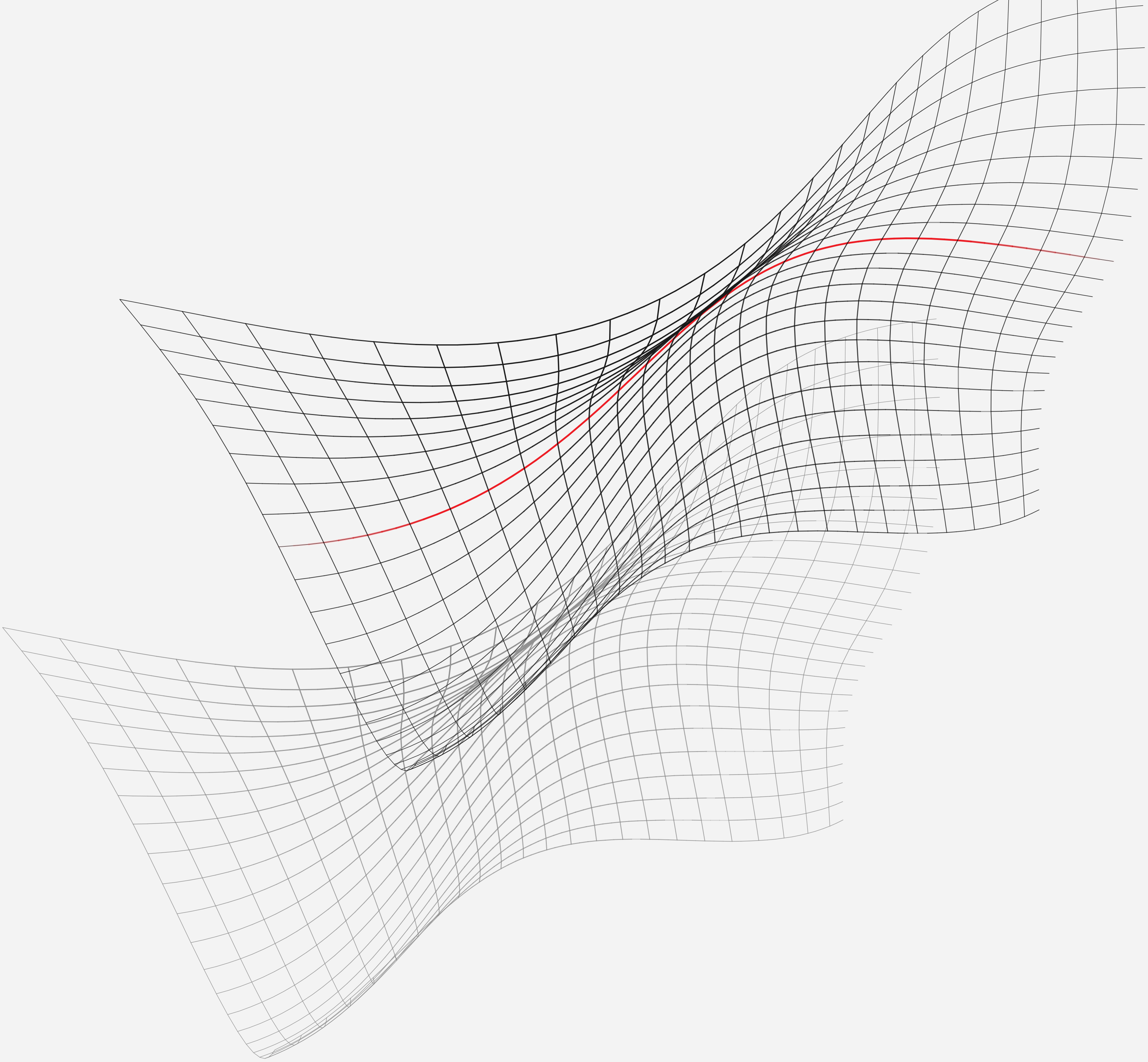

# 8 | Public Opinion

In December 2025, we surveyed 1,309 U.S. residents to assess public knowledge, expectations and attitudes toward quantum technologies. The sample was aligned with U.S. Census Bureau age and gender benchmarks to maintain representative demographic coverage. With quantum innovations increasingly shifting toward commercialization, cultivating public trust and engagement will play a decisive role in driving adoption and shaping future policy.

# 8.1 | Quantum computing

## 8.1.1 | Familiarity

**Fig. 8.a** | How familiar are you with quantum computing?

MIT Quantum Index Report

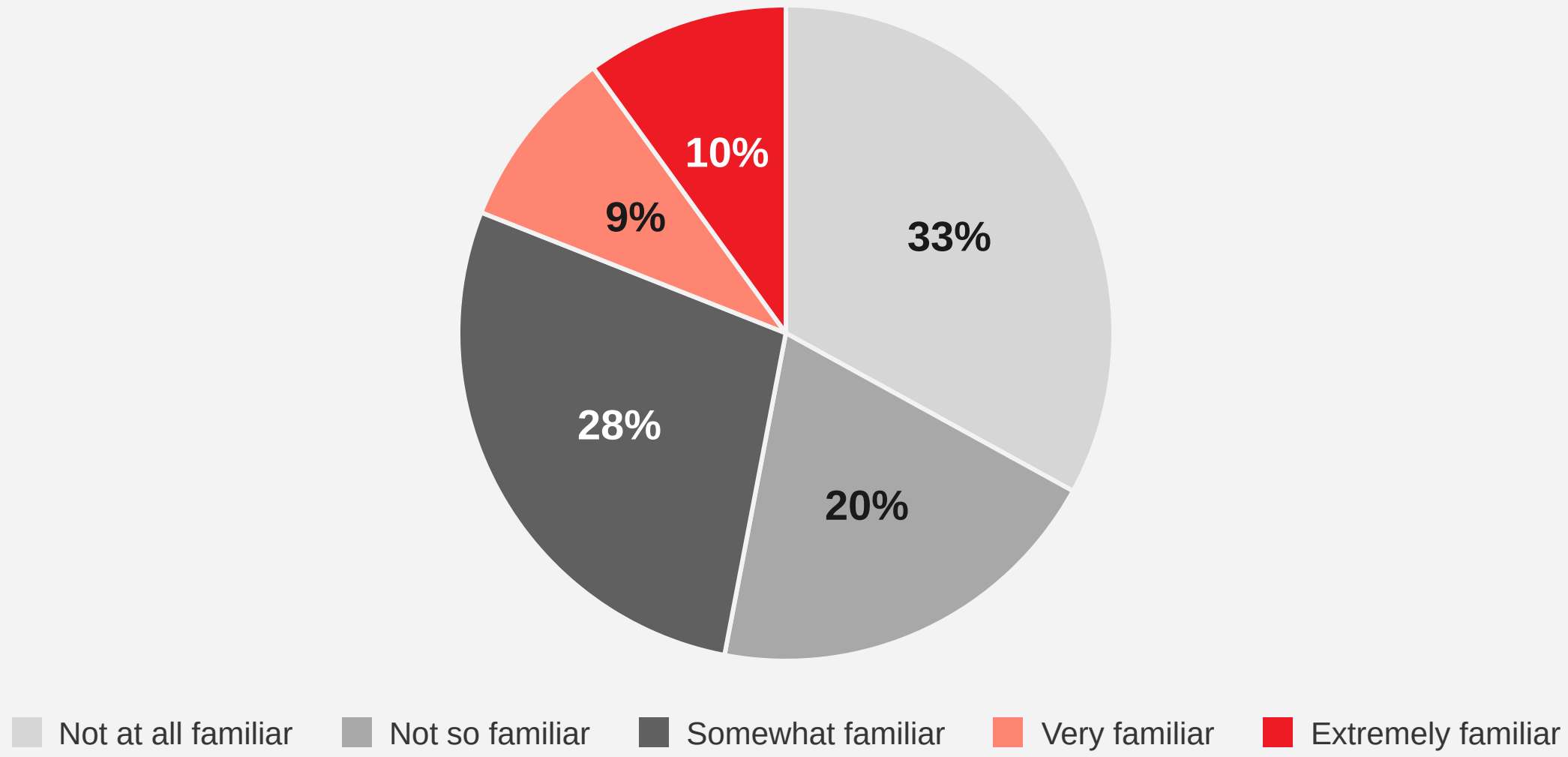

When asked about how familiar they were with quantum computing, from the 1,309 survey respondents, 33% responded they were "not at all familiar," 20% said "not so familiar," 28% indicated they were "somewhat familiar," 9% claimed to be "very familiar," and 10% reported being "extremely familiar."

Taken together, the data points to a general lack of familiarity: more than half of respondents (53%) placed themselves in the "not at all" or "not so familiar" categories, while only about one in five described themselves as "very" or "extremely" familiar. The largest single group was "not at all familiar" at 33%, suggesting this remains an unfamiliar topic for a substantial share of the population. "Somewhat familiar" was the second most common response at 28%, indicating a sizable middle group with at least a passing understanding, while the highest levels of familiarity were comparatively rare, with "very familiar" the smallest category overall at just 9%.

## 8.1.2 | Emotional responses

**Fig. 8.b** | Quantum computing can be potentially used for different areas. To what extent do you feel nervous or excited regarding the potential uses listed below?

MIT Quantum Index Report

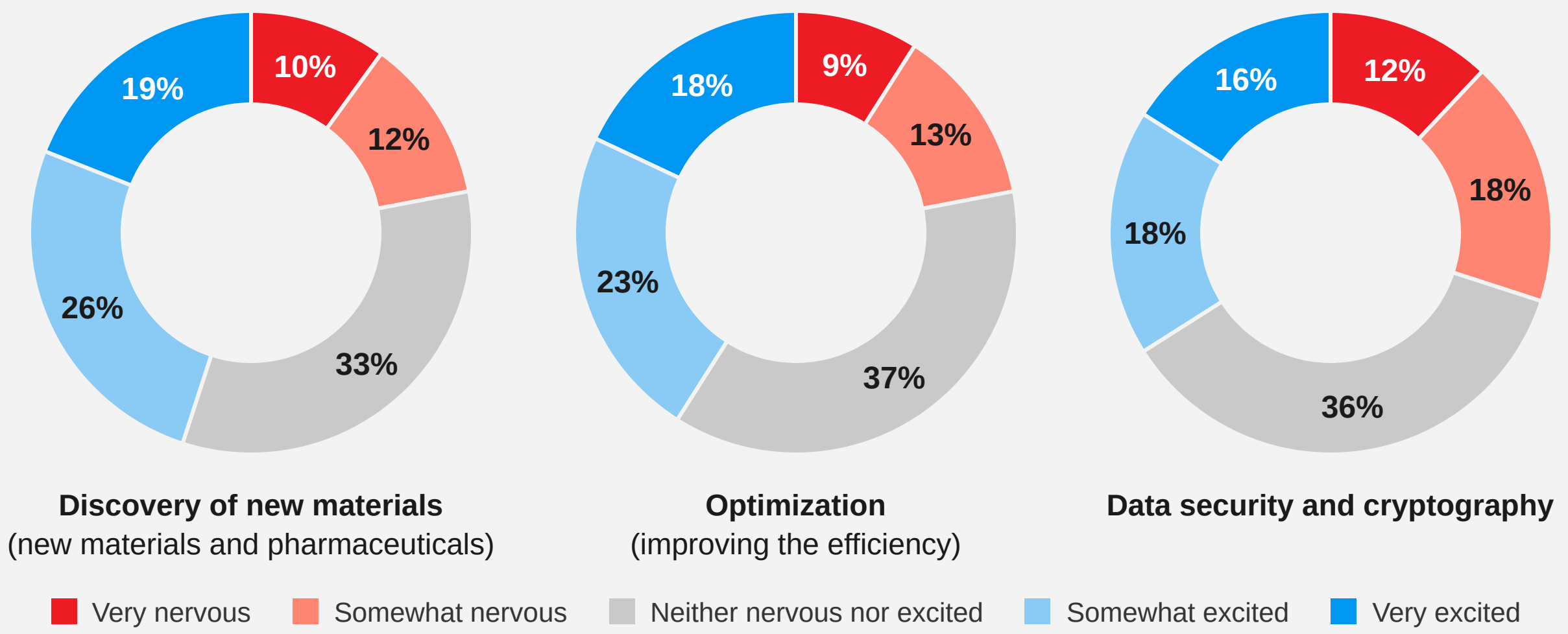

Among the 1,309 survey respondents, emotional responses to quantum computing applications in 2025 continued to vary across use cases. When asked to report on the extent to which they feel nervous or excited regarding potential uses of quantum computing, for materials and pharmaceuticals development, 10% felt very nervous, 12% somewhat nervous, 33% neutral, 26% somewhat excited, and 19% very excited. Regarding optimization, 9% reported to feel very nervous, 13% somewhat nervous, 37% neutral, 23% somewhat excited, and 18% very excited. For data security and cryptography, responses showed 12% very nervous, 18% somewhat nervous, 36% neutral, 18% somewhat excited, and 16% very excited.

Across all three applications, “neutral” was the single largest response category in 2025, ranging from 33% (discovery of new materials) to 37% (optimization). Using quantum computing for the discovery of new materials and pharmaceuticals development remained the application generating the most positive sentiment, with 45% of respondents reporting some degree of excitement (26% somewhat, 19% very) compared to just 22% expressing any nervousness.

Optimization followed a similar but slightly more muted pattern, with 41% expressing excitement versus 22% nervousness.

Data security and cryptography stood out as the application generating the most ambivalence and unease: it had the lowest share of excited respondents (34% combined) and the highest share of nervous respondents (30% combined) among the three categories, continuing to reflect the perceived risks associated with quantum computing's potential to disrupt encryption and security systems.

The graphs below compare and contrast the number of positive and negative sentiments (neutral answers are not represented in these graphs).

**Fig. 8.c** | Quantum computing uses: nervous vs. excited (neutral responses excluded)

MIT Quantum Index Report

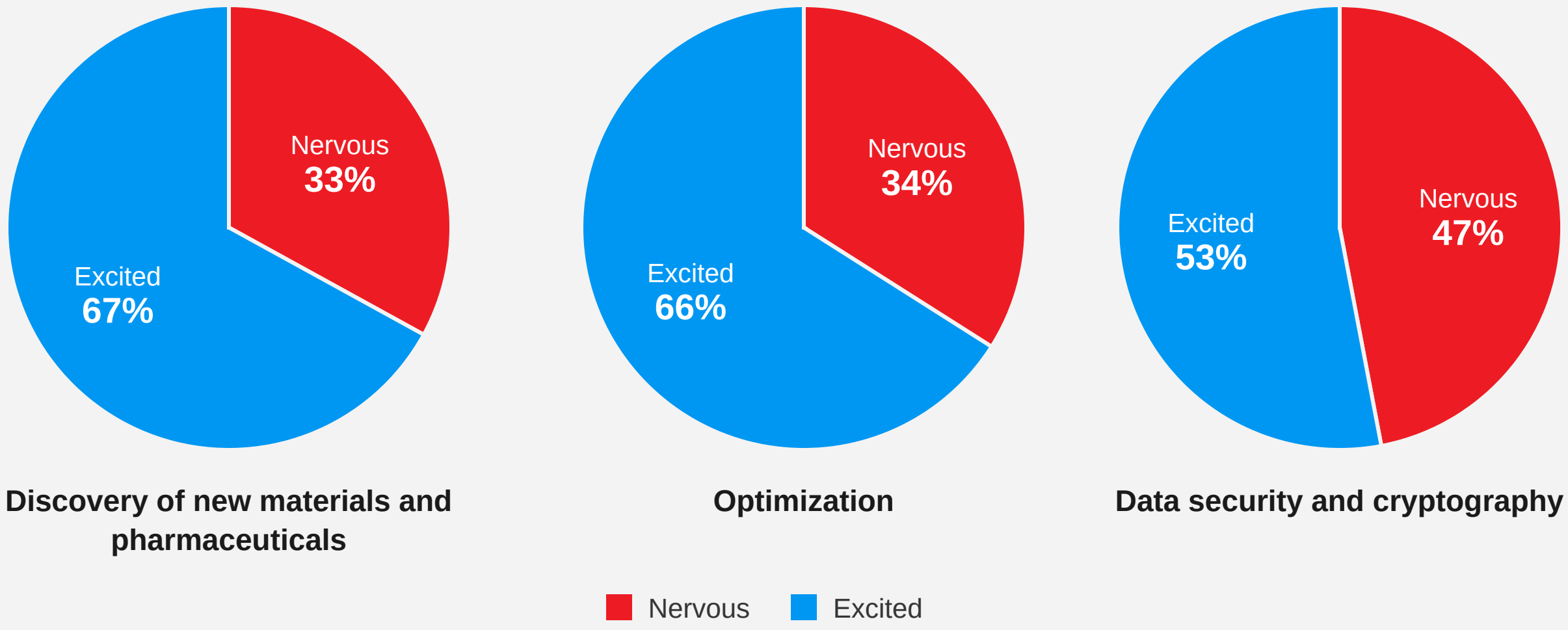

For materials and pharmaceuticals development, among the respondents who expressed a directional sentiment, 33% felt nervous while 67% felt excited, more than a two-to-one margin favoring excitement. Optimization showed a nearly identical split: of the respondents with a directional response, 34% felt nervous compared to 66% excited. Data security and cryptography told a different story: among the respondents who leaned one way or the other, 47% felt nervous versus 53% excited, a far narrower margin, with sentiment nearly evenly split.

This comparison sharpens the contrast between applications considerably. When neutral responses are set aside, discovery of new materials/pharmaceuticals and optimization both show excitement outweighing nervousness by roughly two to one, suggesting that among respondents who hold an opinion, enthusiasm clearly dominates for these use cases. Data security and cryptography, by contrast, stands apart as the only application where sentiment approaches an even split: just 6 percentage points separate nervous from excited, underscoring that this remains the most emotionally contested application of the three.

> Discovery of new materials and pharmaceuticals and optimization both show excitement outweighing nervousness by roughly two to one among respondents with a directional opinion.

- Comparing the 2025 directional sentiment data (neutral responses excluded) to our 2024 survey[1] reveals a consistent trend: excitement has declined and nervousness has risen across all three applications, with the shift most pronounced for data security and cryptography.

# 8.2 | Quantum networking

## 8.2.1 | Familiarity

**Fig. 8.d** | How familiar are you with quantum networking?

MIT Quantum Index Report

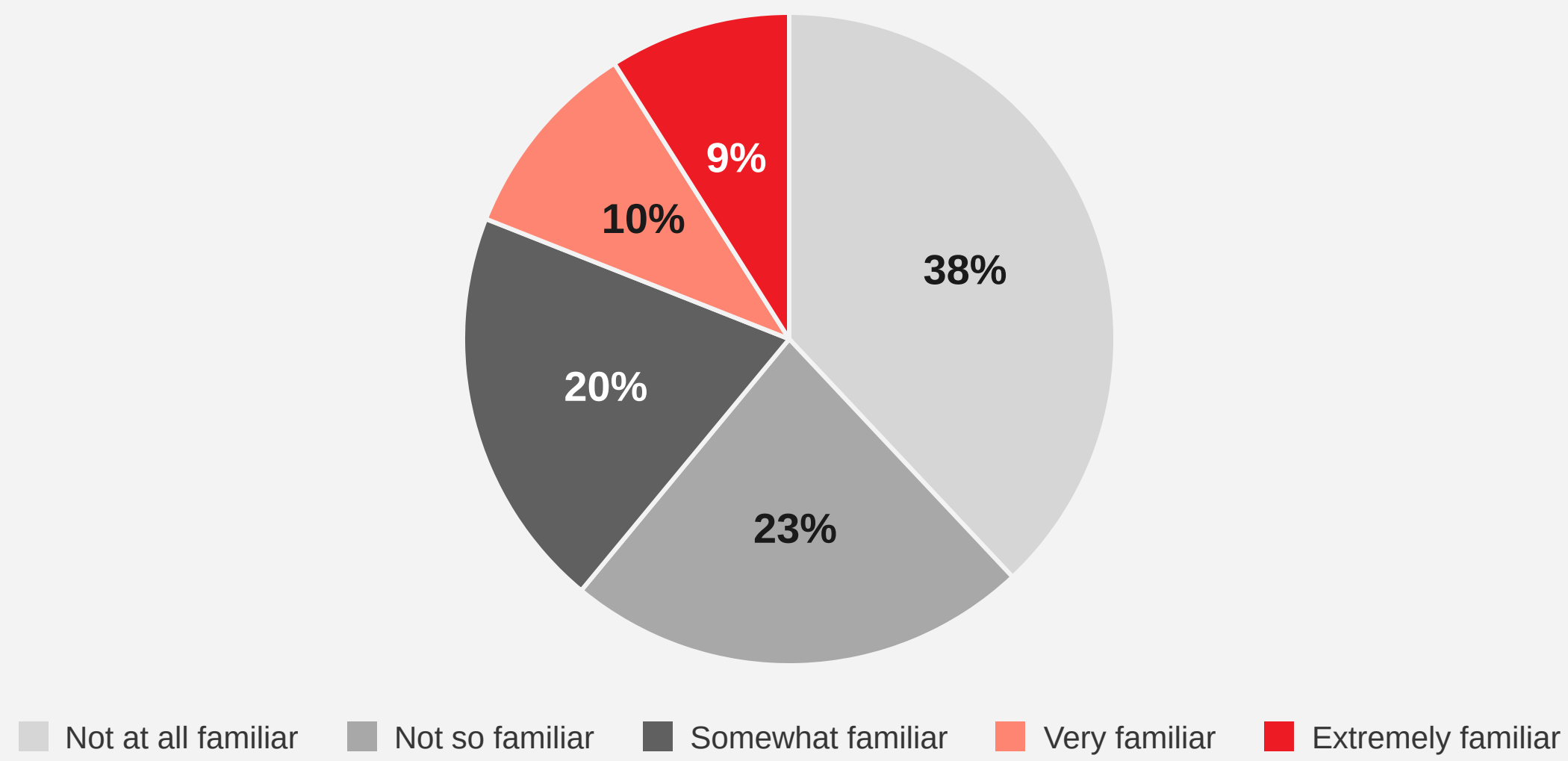

When asked about how familiar they are with quantum networking, from the 1,309 survey respondents, 38% reported being "not at all familiar" with quantum networking, 23% said they were "not so familiar," 20% indicated they were "somewhat familiar," 10% claimed to be "very familiar," and 9% reported being "extremely familiar" with the technology.

Taken together, the data points to a pronounced lack of familiarity with quantum networking: 61% of respondents placed themselves in the "not at all" or "not so familiar" categories, while only 19% described themselves as "very" or "extremely" familiar. "Not at all familiar" was by far the largest single response, claimed by more than a third of respondents, underscoring how unfamiliar quantum networking remains for the general population. "Somewhat familiar" made up a fifth of responses, representing a moderate middle group with at least some passing awareness, while the two highest familiarity categories were comparatively rare and nearly identical in size, together accounting for less than one in five respondents.

38%

In 2025, “not at all familiar” was the single largest response category for both topics, but was notably higher for quantum networking (38%) than for quantum computing (33%).

## 8.2.2 | Emotional responses

**Fig. 8.e** | Quantum networking can be potentially used for different areas. To what extent do you feel nervous or excited regarding the potential uses listed below?

MIT Quantum Index Report

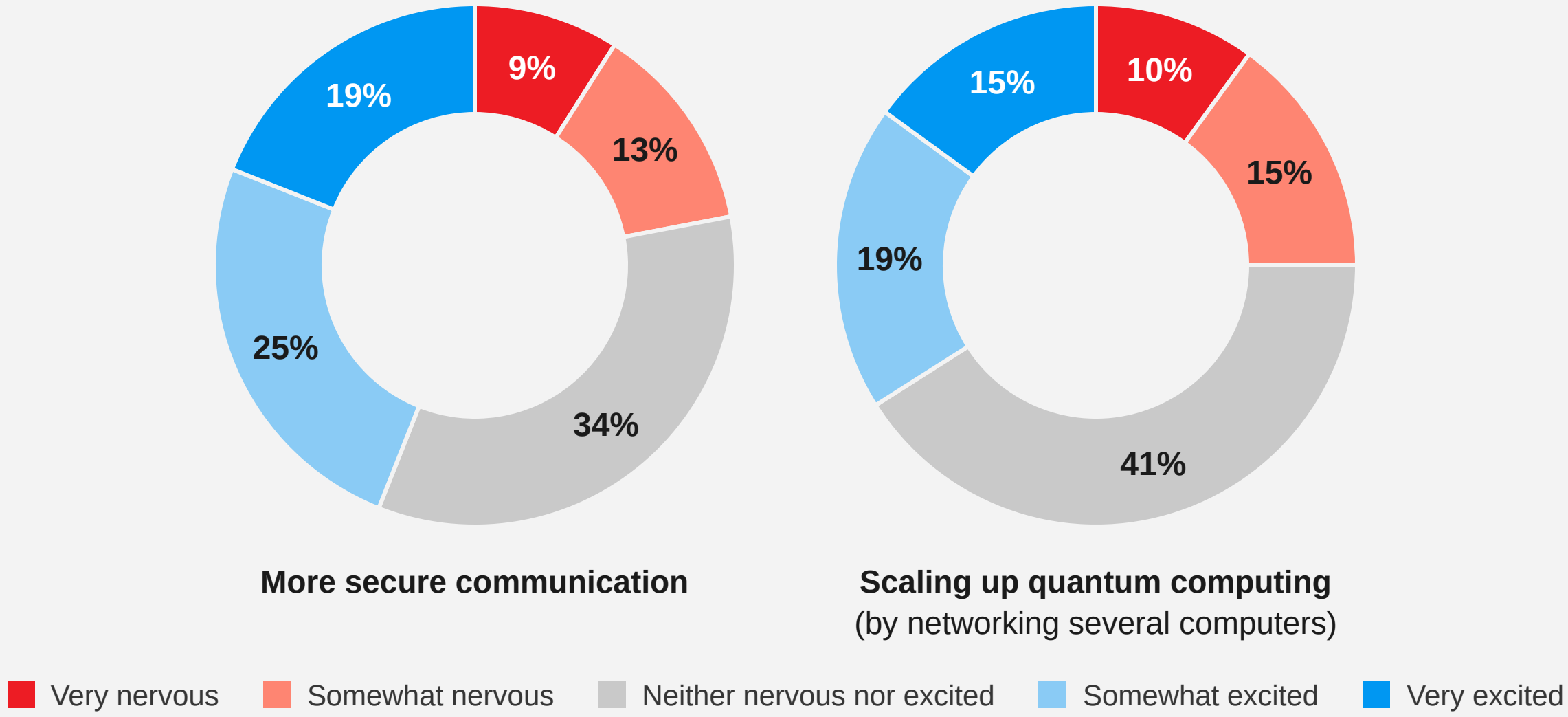

Among the 1,309 survey respondents, the emotional responses varied across different potential uses of quantum networking. When asked to report to what extent they feel nervous or excited regarding potential uses of quantum networking for its relevance to secure communication: 9% of the respondents reported very nervous, 13% somewhat nervous, 34% neither nervous nor excited, 25% somewhat excited, and 19% very excited.

For its relevance to scaling up quantum computing by networking: 10% of the respondents reported being very nervous, 15% somewhat nervous, 41% neither nervous nor excited, 19% somewhat excited, and 15% very excited.

Across both applications, “neither nervous nor excited” was the most common response, and notably more so for scaling up quantum computing (41%) than for secure communication (34%), suggesting respondents feel more ambivalent about networking’s role in computing infrastructure than about its communication use case.

Secure communication generated the more favorable emotional profile of the two, with 44% of respondents expressing some degree of excitement (25% somewhat, 19% very) compared to 22% expressing any nervousness, a two-to-one margin favoring excitement. Scaling up quantum computing showed a more muted response: only 34% reported excitement (19% somewhat, 15% very) against 25% nervousness, a narrower gap that points to greater uncertainty around this particular application. Notably, “very excited” responses were higher for secure communication (19%) than for scaling up quantum computing (15%), while nervousness levels were comparable across both applications (22% vs. 25%), together indicating that respondents view quantum networking’s communication benefits somewhat more favorably than its role in expanding quantum computing capacity, even as neutrality dominates sentiment on both fronts.

The graphs below compare and contrast the number of positive and negative sentiments (neutral answers are not represented in these graphs).

**Fig. 8.f** | Quantum networking uses: nervous vs. excited (neutral responses excluded)

MIT Quantum Index Report

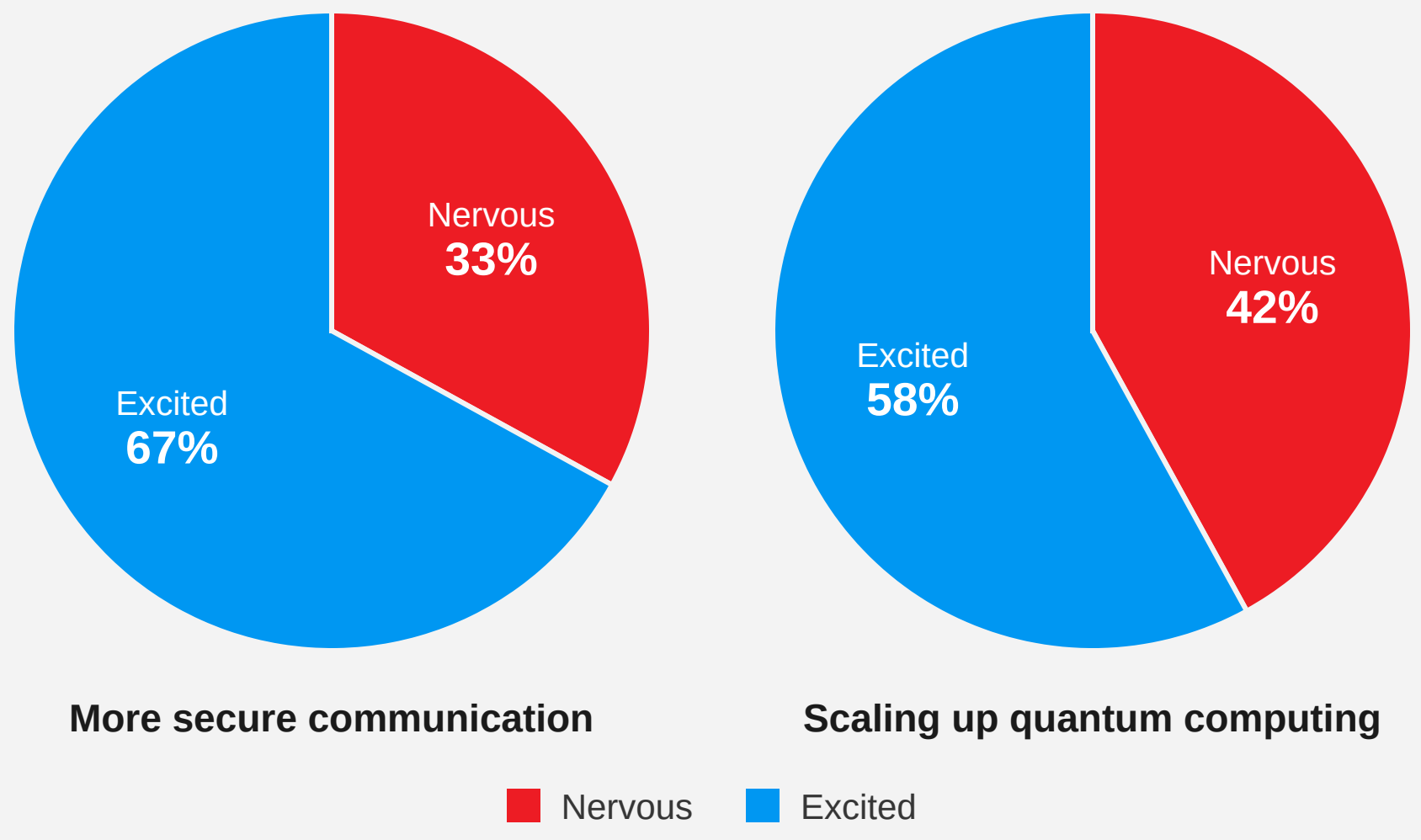

For secure communication, among the respondents who expressed a directional sentiment, 33% felt nervous while 67% felt excited, a two-to-one margin favoring excitement. Scaling up quantum computing showed a narrower split: of the respondents with a directional response, 42% felt nervous compared to 58% excited. This comparison sharpens the contrast between the two applications.

> Secure communication elicits a higher excitement share and lower nervousness among respondents in comparison to scaling up quantum computing.

Scaling up quantum computing, by contrast, shows a meaningfully tighter margin: just 16 percentage points separate nervous from excited, compared to a 34-point gap for secure communication, indicating that among respondents who hold an opinion, sentiment toward using quantum networking to scale up computing infrastructure is more divided.

Comparing the 2025 directional sentiment data to 2024 shows a decrease of excitement and rise in nervousness for both quantum networking applications, with the shift more pronounced for scaling up quantum computing.

The survey results allow the interpretation that regardless of whether the technology is quantum computing or quantum networking, applications tied to expanding or scaling the underlying infrastructure (optimization, scaling up) or touching sensitive systems (data security) are consistently viewed with more ambivalence than applications with clearer, more relatable benefits (pharma, secure communication), and that ambivalence has intensified for both technologies over the past year.

## 8.3 | Governance

We asked the same 1,309 participants to report to what extent they agree or disagree with the following three statements: (1) the state can be trusted to exert effective control over organizations and companies using quantum technologies, (2) the U.S. government should fund the research and development of quantum technologies, and (3) U.S. companies should fund the research and development of quantum technologies.

**Fig. 8.g** | To what extent do you agree or disagree with the following statements?

MIT Quantum Index Report

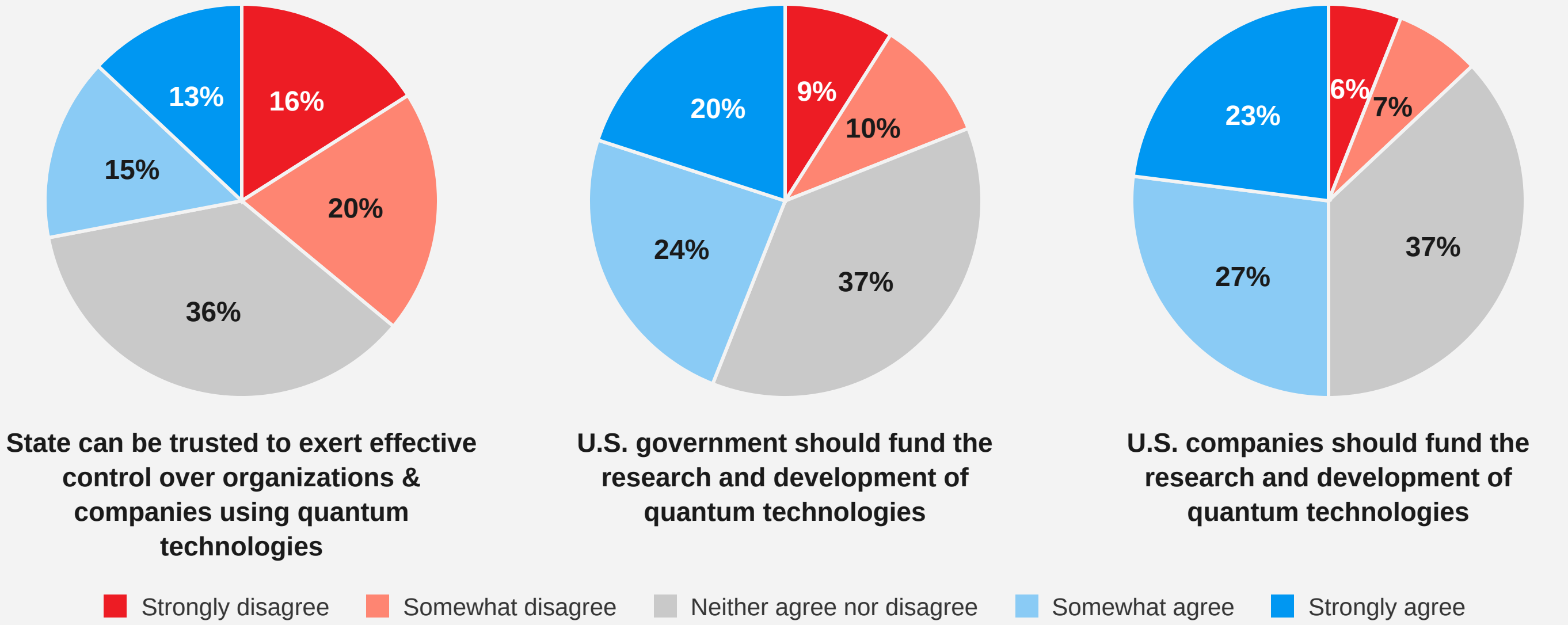

For the statement about state control over quantum technologies: 16% strongly disagree, 20% somewhat disagree, 36% neither agree nor disagree, 15% somewhat agree, and 13% strongly agree. For government funding of quantum technologies: 9% strongly disagree, 10% somewhat disagree, 37% neither agree nor disagree, 24% somewhat agree, and 20% strongly agree. For company funding of quantum technologies: 6% strongly disagree, 7% somewhat disagree, 37% neither agree nor disagree, 27% somewhat agree, and 23% strongly agree.

Across all three statements, “neither agree nor disagree” was the most common response, ranging from 36% to 37%, indicating that a substantial share of respondents remain undecided on questions of quantum governance and funding.

The statement on whether the state can be trusted to exert effective control over organizations and companies using quantum technologies stands out as the only one of the three statements where skepticism exceeded support.

By contrast, both funding statements drew clearly favorable responses: government funding saw 44% agree against just 19% disagree, while company funding drew the strongest support of the three, with 50% agreeing compared to only 13% disagreeing. Notably, company funding edged out government funding as the more broadly supported statement, suggesting respondents are somewhat more comfortable with private-sector investment in quantum research than with public funding, and considerably more comfortable with either form of funding than with the idea of state control over the technology itself.

**Fig. 8.h** | Governance statements: disagree vs. agree, 2024 vs. 2025 survey (neutral responses excluded)

MIT Quantum Index Report

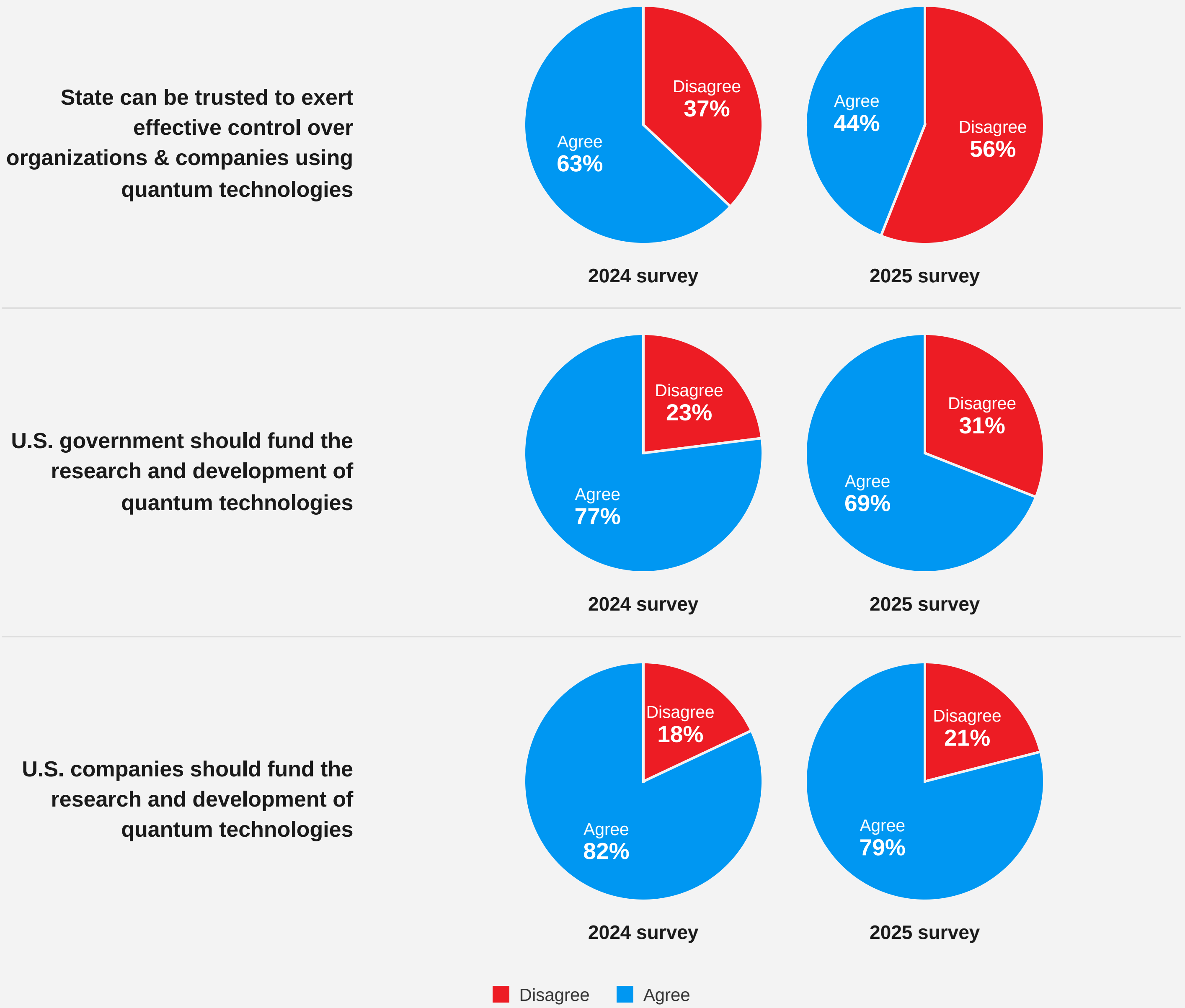

Comparing the 2025 directional sentiment data (neutral responses excluded) to 2024 reveals a dramatic reversal on state control and more modest softening on funding support.

For state control over quantum technologies, sentiment flipped entirely: in 2024, among directional respondents, 63% agreed that the state can be trusted to exert effective control, against 37% who disagreed, a clear majority in favor. By 2025, that had reversed to 56% disagreeing and just 44% agreeing, a swing of roughly 19 points away from trust in state control.

**79%**

Support for funding quantum R&D remained comparatively resilient, with company funding still commanding strong majority backing (79%, down just 3 points) even as enthusiasm cooled slightly across the board.

Government funding also saw a decline in support, though far less dramatic: agreement fell from 77% in 2024 to 69% in 2025, while disagreement rose from 23% to 31%. Company funding showed the smallest movement of the three, with agreement slipping only modestly from 82% to 79% and disagreement rising from 18% to 21%; support remained strong and comparatively stable.

The results suggest that respondents' confidence in funding quantum research, whether public or private, has remained relatively resilient, while confidence in government's ability to effectively regulate or control the technology has deteriorated over the past year.

## 8.4 | Future research

Running this study on an ongoing basis will let us map long-term shifts in public perceptions and knowledge of quantum technologies. Gathering longitudinal data will help highlight emerging concerns, clear up misconceptions, and pinpoint where focused outreach and education are needed most. It will also provide meaningful insight into societal readiness and institutional trust. We invite readers across the public, academic, industry and policy spheres to share their feedback and propose themes or questions for upcoming surveys.

Your input will ensure our research stays relevant, inclusive, and attuned to the needs of the broader community.

You can reach us at contact@qir.mit.edu

---

## Footnotes

[1] Ruane, J. et al., "The Quantum Index Report 2025", MIT Initiative on the Digital Economy, Massachusetts Institute of Technology, Cambridge, MA, May 2025.

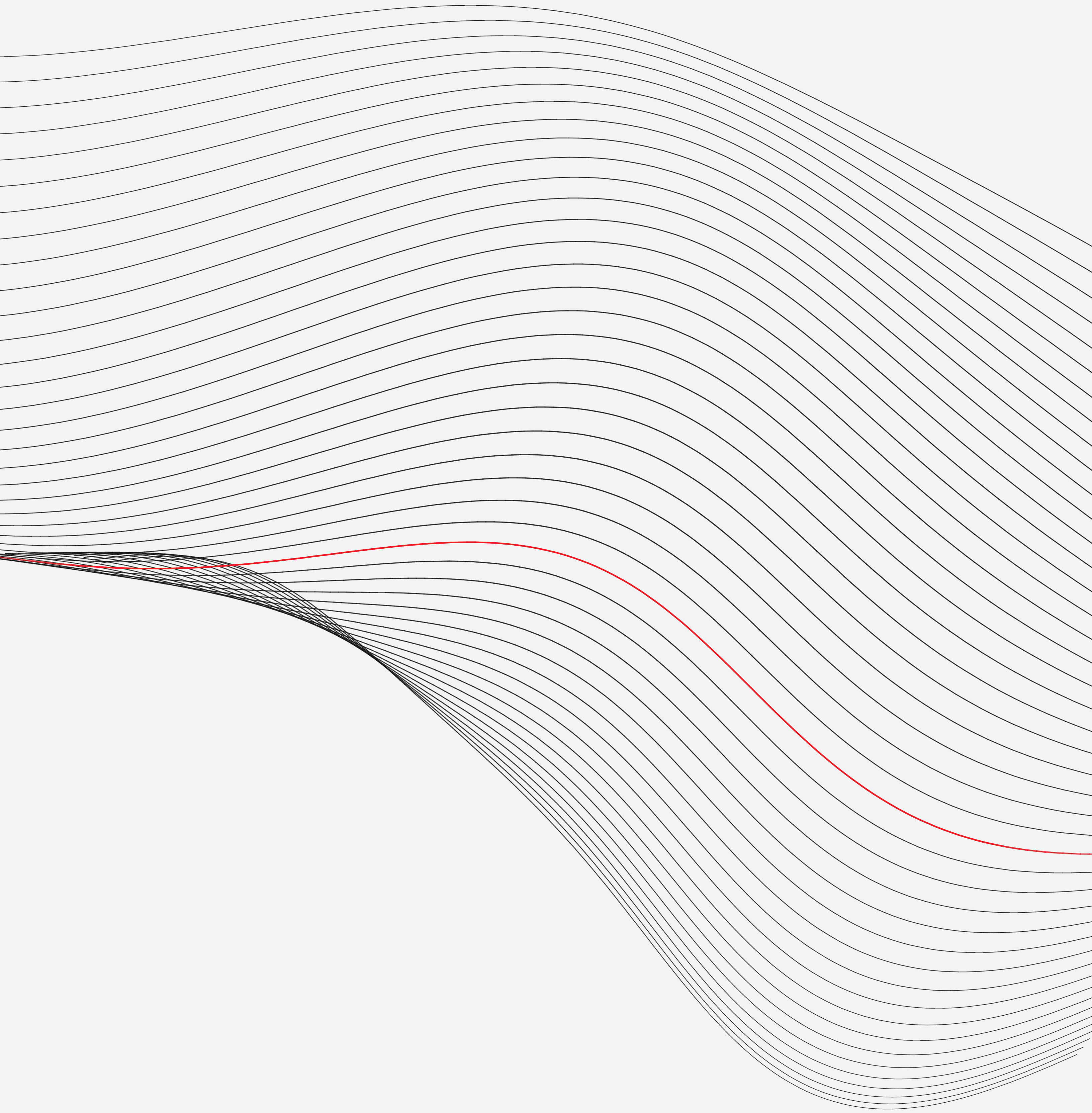

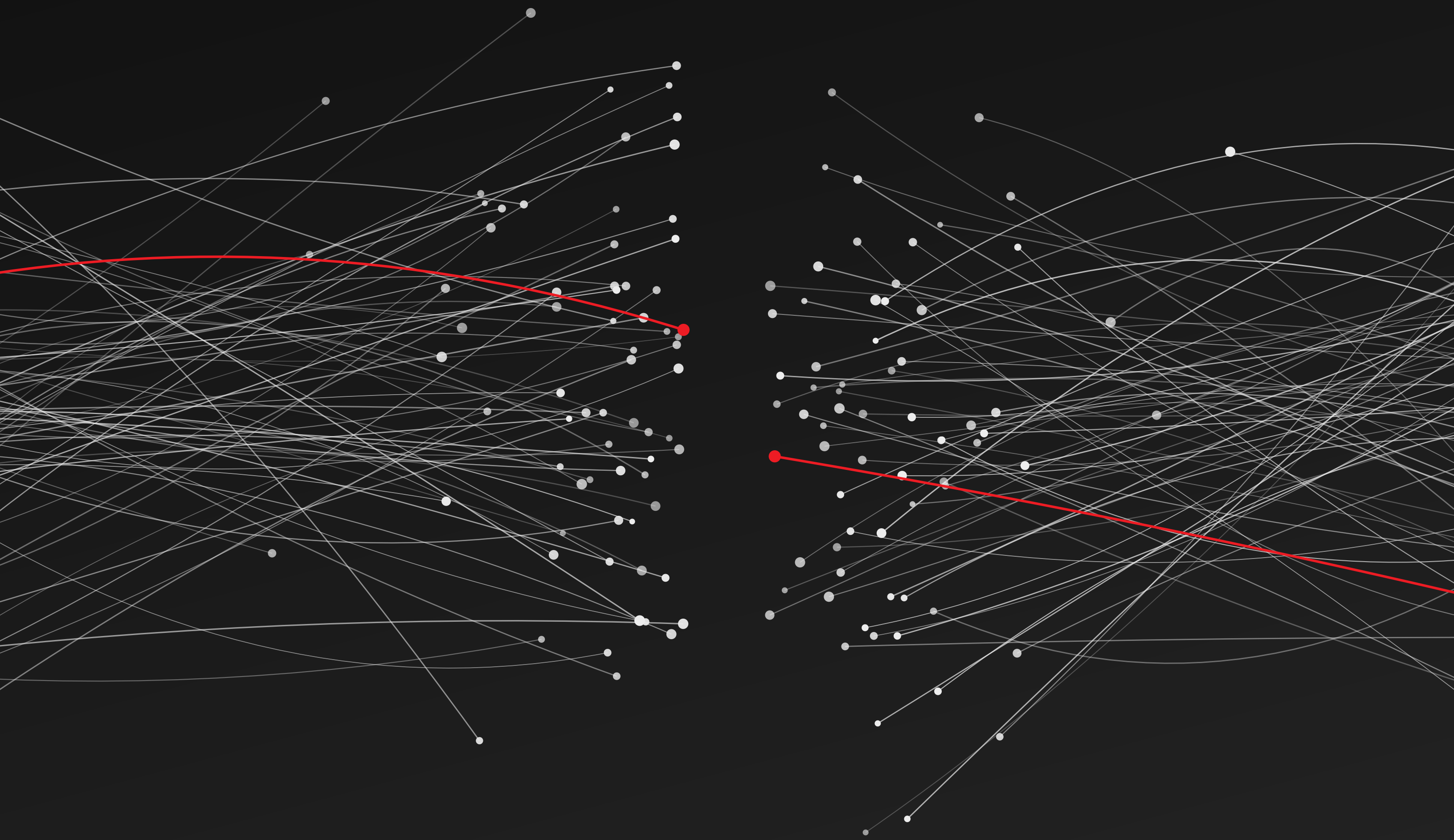

# 9 | Quantum Networking

## 9.1 | Quantum networks

Quantum networks are emerging communication systems that use quantum-mechanical properties to transmit quantum states and distribute entanglement between quantum information systems located at different points.[1] Unlike conventional networks, which primarily transmit classical bits, quantum networks are intended to distribute quantum states or entanglement as an additional communications resource.

This capability could eventually allow geographically separated quantum devices to work together for applications such as distributed quantum computing and distributed quantum sensing.

The U.S. National Quantum Coordination Office stated that quantum networking has significant implications for national security, scientific discovery, and economic competitiveness.[2]

Potential applications most frequently discussed today include quantum key distribution, distributed quantum computing, and distributed quantum sensing. For distributed computing, entanglement could allow multiple quantum processors to participate in joint quantum computations and may provide a way of overcoming architectural limits on the number of qubits that can be integrated into a single processor.[3] For distributed sensing, entanglement shared among separate sensors could potentially improve measurement sensitivity or precision beyond what can be achieved with individual classical or quantum sensors.[4] Given the strategic importance of quantum-networking capabilities, the National Quantum Initiative Advisory Committee recommended continued government investment in fundamental quantum-networking research, enabling technologies, performance metrics, network architectures, testbeds, industry participation, international cooperation and workforce development.

The European Commission's Quantum Europe Strategy, adopted in July 2025, calls for support in 2026 for a pilot facility for the European Quantum Internet to test quantum-safe components, early use cases, secure quantum-cloud services, distributed computing, and validation environments bridging research and deployment, with a longer-term objective of a fully operational quantum-safe communications network by 2030 as a first step toward a federated quantum internet.[5]

## 9.2 | Quantum networking testbeds

Quantum networking testbeds are a cornerstone of building a future quantum internet. The National Quantum Initiative Advisory Committee (NQIAC) therefore treats testbeds as distinct from prototypes, demonstrations, and general user facilities and defines them as “a platform or facility that is accessible to multiple users to conduct replicable and rigorous testing of component technologies, protocols, and systems integration”.[6] They allow quantum technologies to be tested under conditions that more closely resemble real-world networks. Researchers can use these environments to determine whether different components work reliably together, how well they perform as part of a larger system, and whether they can be scaled beyond small laboratory experiments. The NQIAC has emphasized that well-selected testbeds, introduced at the appropriate stage of technological development, can help advance not only the hardware needed for quantum networks but also the underlying theory, security approaches, and potential applications.[7] [8]

Testbeds can also reduce the financial and technical risks of moving too quickly toward large-scale deployment. NQIAC recommends investing in “right-sized” testbeds.[9] In practice, this means designing each testbed around a clearly defined research goal and limiting its scale and cost accordingly. This approach allows researchers and policymakers to evaluate technologies at an appropriate level before committing greater resources to larger networks, helping to ensure that only sufficiently mature and promising technologies are expanded further.[10]

### 9.2.1 | Overview of newly announced testbeds, 2025–2026

In August 2025, the Air Force Research Laboratory (AFRL) announced three reconfigurable Quantum Local Area Networks (QLANs) operating at telecommunications-band frequencies to support its mission of building heterogeneous quantum networks.[11]

In September 2025, the National Science Foundation advanced SCY-QNet, led by Stony Brook University with Columbia University, Yale University and Brookhaven National Laboratory, from its initial National Quantum Virtual Laboratory (NQVL) pilot into the design phase, awarding the team $4 million over two years to develop a long-distance, ten-node quantum network intended to connect atomic quantum processors and investigate quantum communication and distributed processing.[12] [13]

In September 2025, New York State committed $300 million to establish a 150,000-square-foot Quantum Research and Innovation Hub at Stony Brook University, with a particular emphasis on quantum communication and networking and an anticipated opening around 2029.[14] The investment also supports the expansion of the Long Island quantum-internet testbed, which is ultimately planned to extend for more than 1,000 miles across New York State.[15]

In November 2025, New Mexico's first quantum network, ABQ-Net, was announced by Qunnect and was subsequently brought online in February 2026 with two active nodes connected over commercial fiber.[16]

In Europe, in January 2025, Germany's first quantum internet node was announced by the Fraunhofer Institute for Laser Technology ILT in Aachen, in cooperation with Dutch TNO.[17] In April 2026, the Czech Quantum Communication Infrastructure announced the launch of a completed quantum backbone linking Prague, Brno and Ostrava, following pilot operation of the network as part of the broader European EuroQCI program.[18] In June 2026, the Netherlands added the Quantum Communication Fieldlab Rotterdam, launched as a real-world environment for testing, validating and operationalizing quantum-secure communications for critical and public infrastructure.[19]

**Fig. 9.a** | Quantum network testbeds — United States

MIT Quantum Index Report

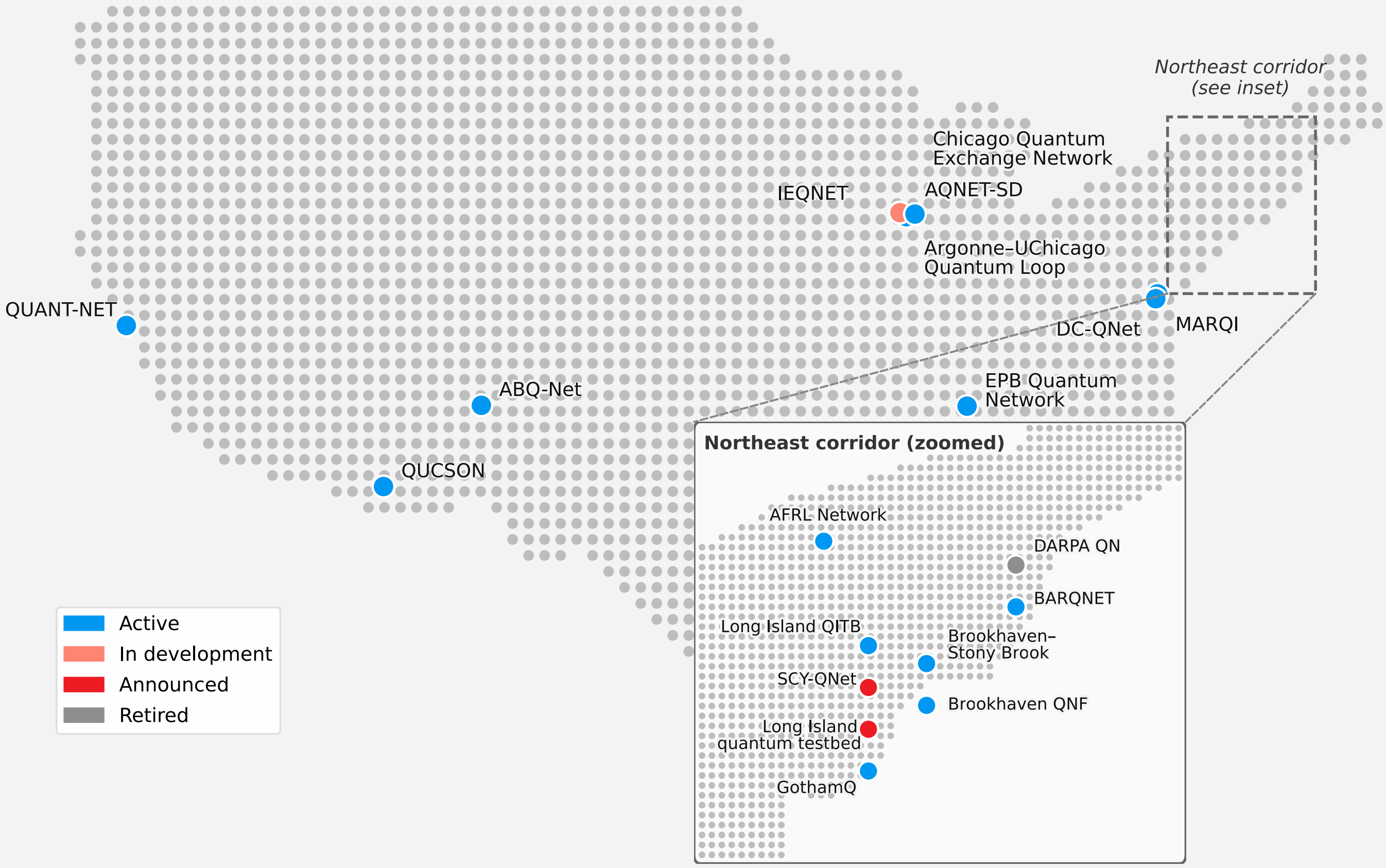

**Fig. 9.b** | Quantum network testbeds — Europe

MIT Quantum Index Report

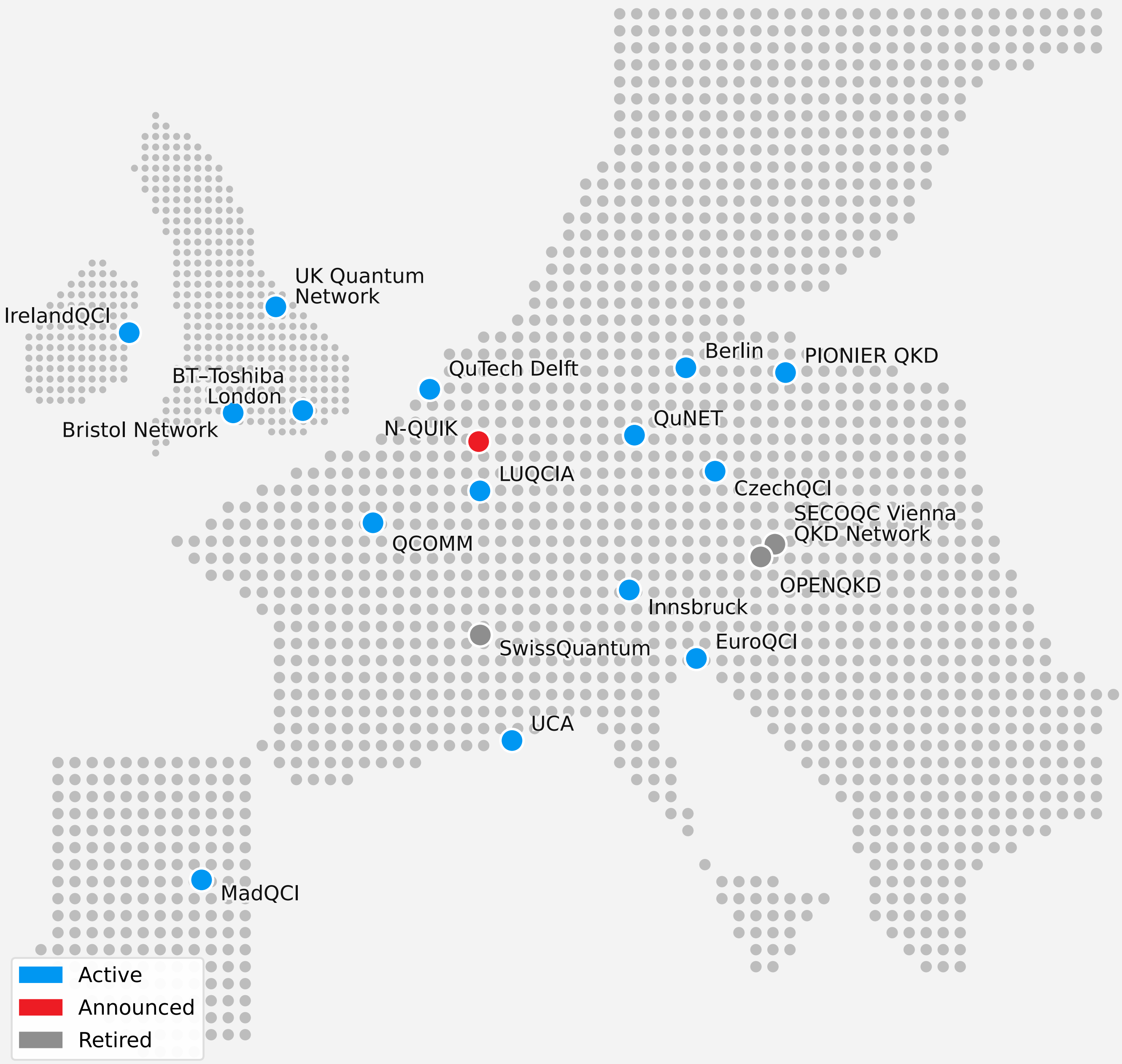

**Fig. 9.c** | Quantum network testbeds — Asia

MIT Quantum Index Report

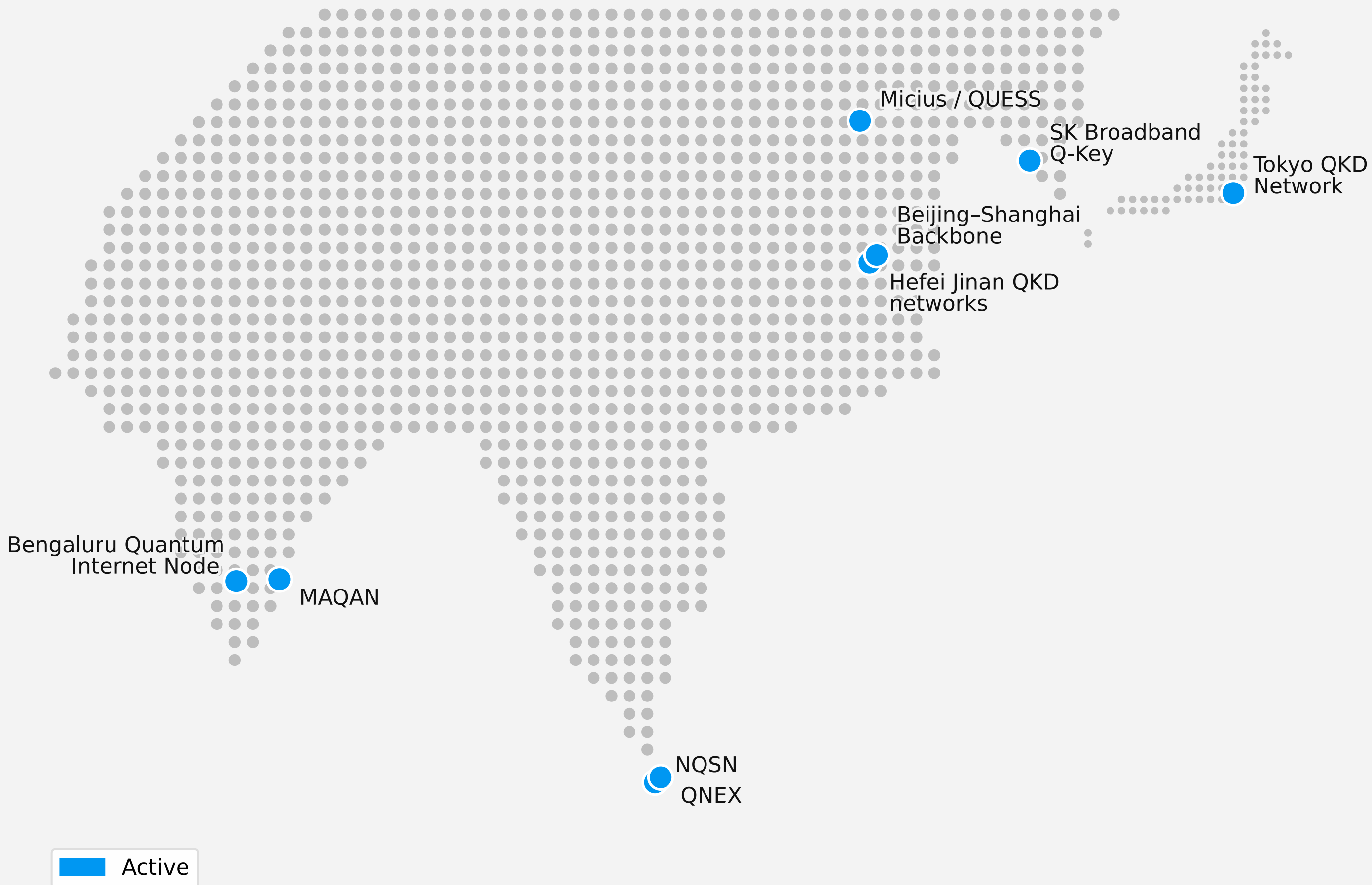

## 9.3 | Quantum Networking Applications Roadmap 2026

In April 2026, the Center for Quantum Networks (CQN) published the Quantum Networking Applications Roadmap, developed in collaboration with the Quantum Economic Development Consortium (QED-C). The Roadmap takes an application-driven approach to assessing how quantum networks may develop from today's experimental systems into commercially useful infrastructure.[20]

The Roadmap evaluates the network capabilities required for specific applications and compares those requirements with what current quantum-networking technologies can achieve. The report defines seven principal network-performance metrics: fidelity, synchronization, qubit transmission rate, transmission distance, number of nodes, node size, and connectivity, and identifies eight core enabling technologies: quantum memories, quantum light sources, light-matter interfaces, transduction or frequency converters, single-photon detectors, quantum repeaters, quantum satellite infrastructure, and quantum optical network switches. Its analysis is based on contributions from more than 50 experts from industry, government, academia, national laboratories and the defense sector, including a July 2025 workshop and subsequent consultations; however, the final quantitative readiness and commercialization estimates rely on detailed survey responses from seven experts and therefore have limitations.

**Fig. 9.d** | Readiness vs. time to commercialization, by category

MIT Quantum Index Report

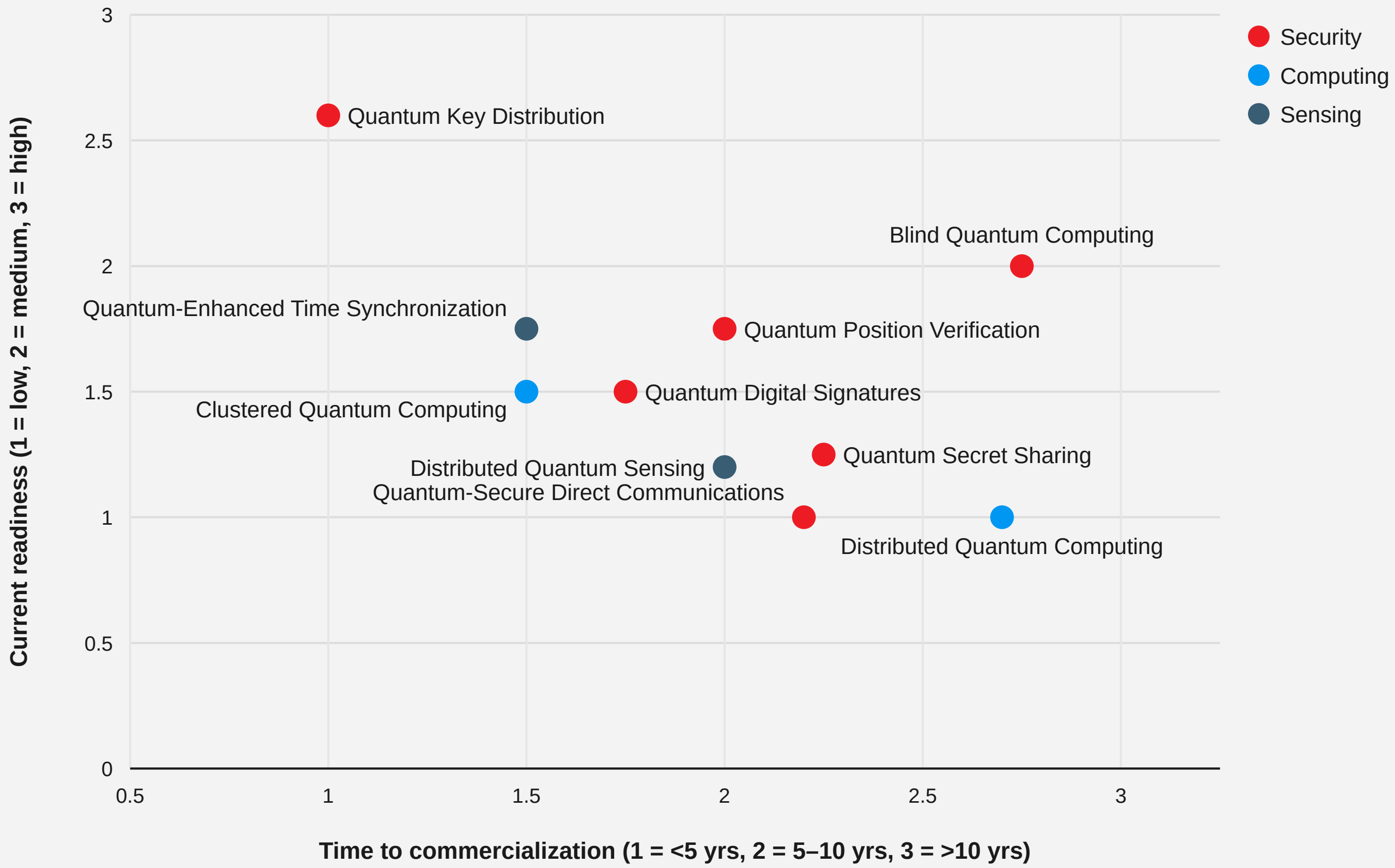

The Roadmap identifies 10 potentially high-impact quantum-network applications, divided into three broad categories: network security, networked quantum computing, and networked quantum sensing.

Network security applications include quantum key distribution (QKD), blind quantum computing (BQC), quantum-secure direct communications (QSDC), quantum position verification (QPV), quantum secret sharing (QSS), and quantum digital signatures (QDS); networked quantum computing applications comprise clustered quantum computing (CQC) within a data center and distributed quantum computing (DQC) between geographically separated data centers; and networked quantum sensor applications comprise quantum-enhanced time synchronization (QETS) and distributed quantum sensing (DQS).

The report's central finding is that these applications are not progressing toward commercialization at the same rate: short-distance QKD and quantum digital signatures have comparatively small gaps between present and required capabilities, while long-distance distributed quantum computing and blind quantum computing face much larger technical gaps and are estimated to be roughly a decade from commercialization. Clustered quantum computing and quantum-enhanced time synchronization occupy an intermediate position and may become commercially relevant on approximately a five-year timescale.

> Short-distance QKD and quantum digital signatures have comparatively small gaps between present and required capabilities, while long-distance distributed quantum computing and blind quantum computing face much larger technical gaps and are estimated to be roughly a decade from commercialization.

The readiness-versus-commercialization chart shows this uneven development: QKD sits at the highest current readiness and shortest commercialization horizon, whereas distributed quantum computing combines low readiness with one of the longest expected commercialization timelines.

Current network performance is already sufficient to support QKD and some forms of distributed quantum sensing over short-distance, point-to-point links, although broader functionality remains constrained by network scale and, in the case of distributed sensing, by the maturity of the sensors themselves.

QKD is the clearest near-term example: commercial point-to-point systems already operate on several networks, but today's long-distance deployments generally depend on trusted intermediate nodes, while future untrusted long-distance networks would require entanglement swapping, quantum repeaters and potentially satellite links.

The Roadmap therefore distinguishes between applications that can extract value from relatively simple networks today and applications that require scalable entanglement-distribution infrastructure. This distinction is especially apparent for distributed quantum computing, which the report identifies as the most technologically demanding application because it requires high-performance networking between geographically separated quantum processors.

**Fig. 9.e** | Component technologies: achieved today vs. remaining headroom (scale L1–L4)

MIT Quantum Index Report

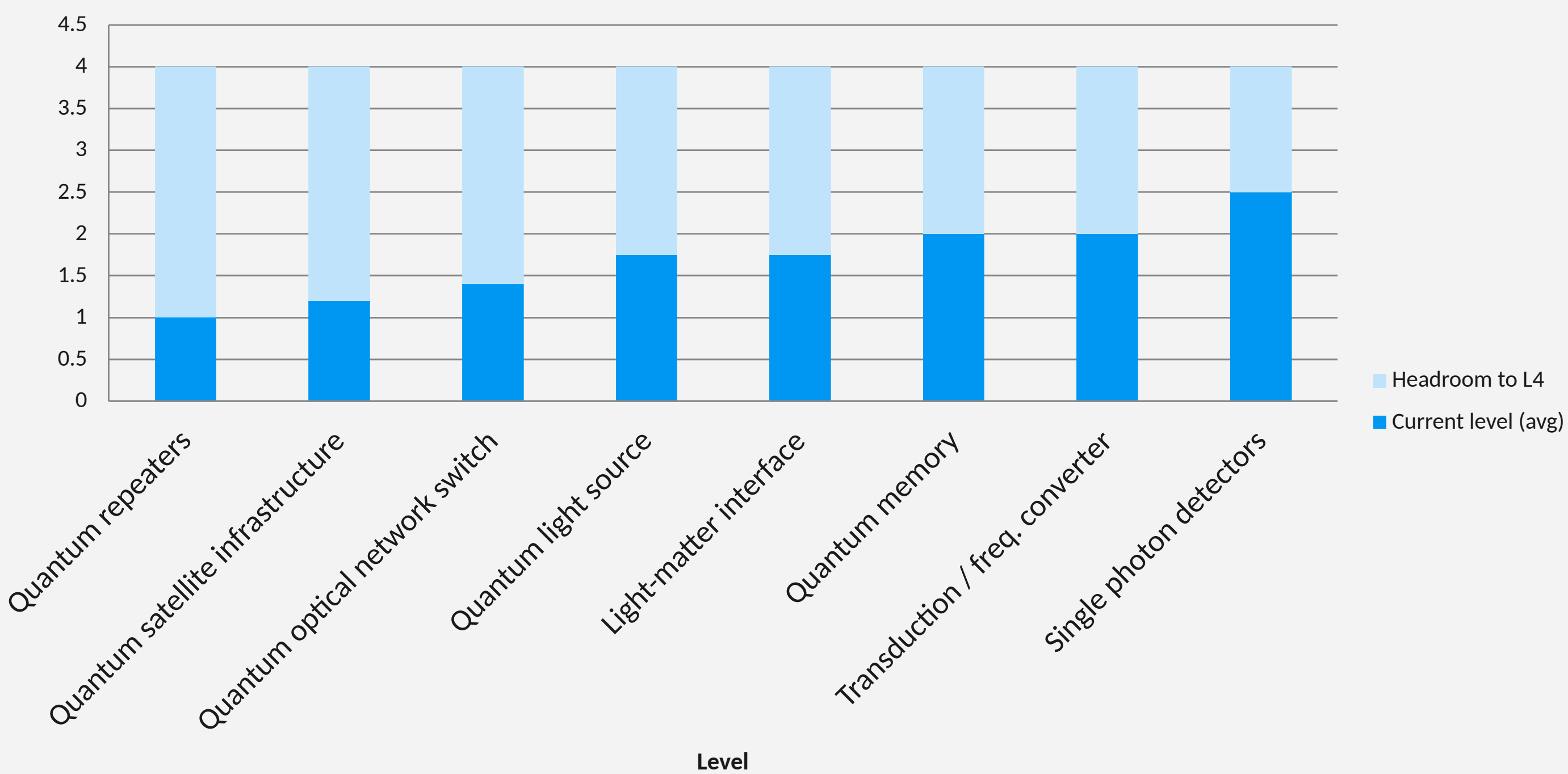

The report highlights which underlying technologies might be most attractive for investors as some are crucial for the improvement of several different applications. More specifically, advancements in quantum optical network switches, quantum repeaters and quantum satellite infrastructure would impact nine of the ten applications studied, while quantum light sources and light-matter interfaces are relevant for eight of them.

**9 of 10**

▶ Advancements in quantum optical network switches, quantum repeaters and quantum satellite infrastructure would impact nine of the ten applications studied.

**Fig. 9.f** | Network metrics: achieved today vs. remaining headroom (scale L1–L4)

MIT Quantum Index Report

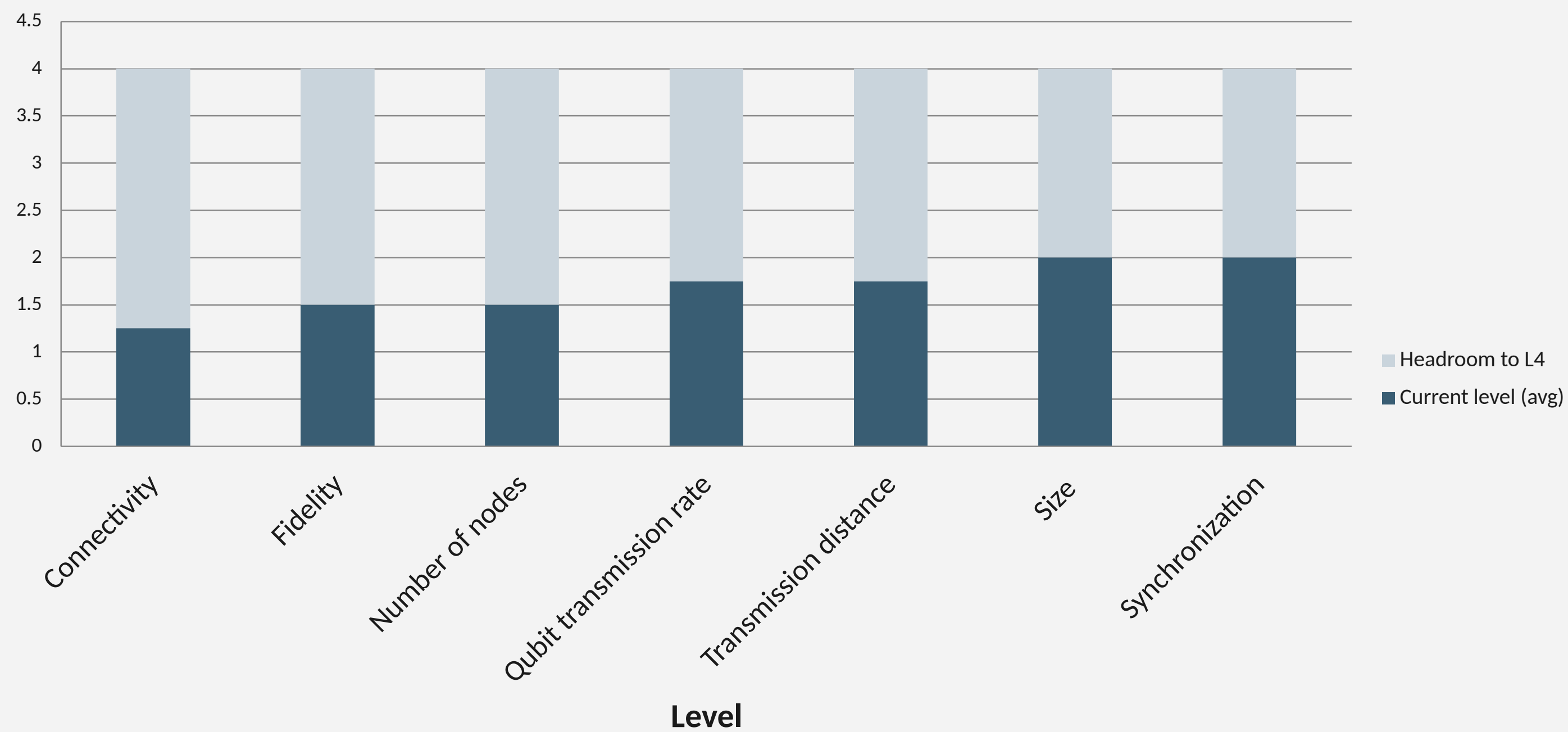

The network metrics most commonly requiring improvement are qubit transmission rate and transmission distance, affecting eight applications each, followed by fidelity and synchronization, which would be required for advancing seven applications.

In particular, extending networks beyond metropolitan or regional scales will depend heavily on three capabilities: quantum repeaters for extending the range of the quantum network on an optical fiber network, satellite infrastructure for very long distances, and quantum-compatible optical switches.

The Roadmap also stresses that existing telecommunications infrastructure will remain important: much installed single-mode metropolitan and long-haul fiber may potentially support quantum signals, but conventional network equipment such as some ROADMs and last-mile GPON infrastructure is generally not quantum compatible, creating an additional systems-integration challenge.

The report presents quantum networking as an incremental technology-development and commercialization process rather than a single transition to a fully realized quantum internet. Its analysis suggests that early commercialization may first emerge from relatively constrained security, sensing, timing and intra-data-center applications, while long-distance distributed computing and more sophisticated entanglement-based services will depend on substantial improvements across the networking technology stack.

**Fig. 9.g** | Sum of technology gaps vs. time to commercialization

MIT Quantum Index Report

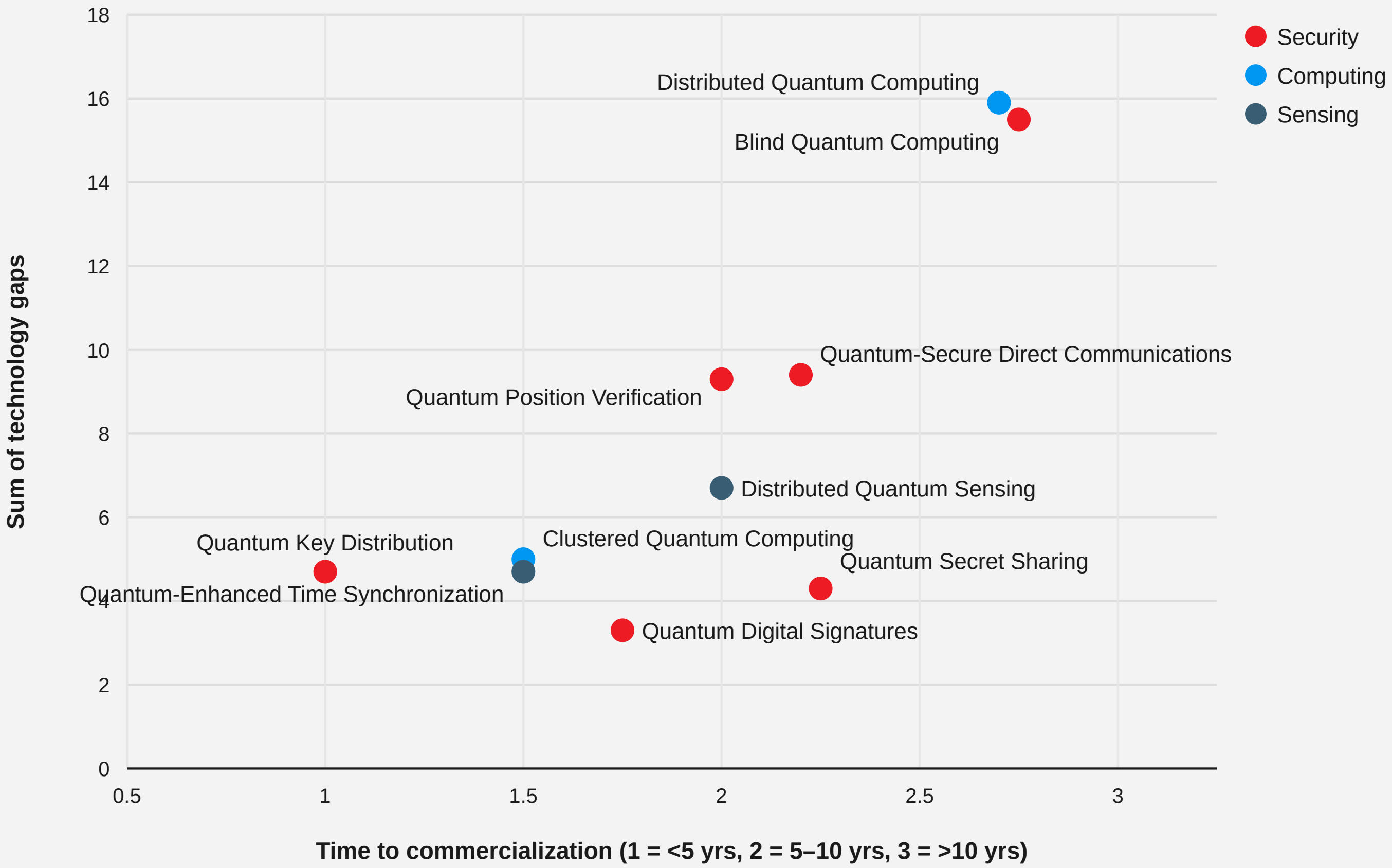

The Roadmap cautions that technical readiness alone will not determine commercial adoption: systems must also be manufacturable and deployable at costs that markets are willing to support; therefore, the Roadmap serves primarily as a tool for identifying technical bottlenecks and helping governments, researchers and private investors prioritize resources.

## Footnotes

[1] 'Quantum Networking: Findings and Recommendations for Growing American Leadership' [2024] National Quantum Initiative Advisory Committee.

[2] The White House National Quantum Coordination Office, 'A Strategic Vision for America's Quantum Networks' (2020).

[3] ‘Quantum Networking: Findings and Recommendations for Growing American Leadership’ (n 1).

[4] Ibid.

[5] ‘Quantum Europe Strategy | Shaping Europe’s Digital Future’ <https://digital-strategy.ec.europa.eu/en/library/quantum-europe-strategy> accessed 20 August 2026.

[6] ‘Quantum Networking: Findings and Recommendations for Growing American Leadership’ (n 1).

[7] Ibid.

[8] ‘Network Function’ [2022] NIST <https://www.nist.gov/pml/productsservices/quantum-networks-nist/technologies-quantum-networks/network-function> accessed 20 August 2026.

[9] Ibid.

[10] Ibid.

[11] Erin Sheridan and others, ‘Telecommunications Fiber-Optic and Free-Space Quantum Local Area Networks at the Air Force Research Laboratory’ (arXiv, 4 September 2025) <https://doi.org/10.48550/arXiv.2508.01030> accessed 20 August 2026.

[12] ‘NSF National Quantum Virtual Laboratory Speeds into the Design Phase | NSF - U.S. National Science Foundation’ (4 September 2025) <https://www.nsf.gov/news/nsf-national-quantum-virtual-laboratory-speeds-design-phase> accessed 20 August 2026.

[13] Michael Gasparino, ‘SBU-Led Team Receives $4M NSF Grant to Develop 10-Node Quantum Network - SBU News’ (4 September 2025) <https://news.stonybrook.edu/university/sbu-led-team-receives-4m-nsf-grant-to-develop-10-node-quantum-network/> accessed 21 August 2026.

[14] ‘Governor Hochul Announces $300 Million Investment in SUNY at Stony Brook’s Quantum Research and Innovation Hub | Governor Kathy Hochul | New York State’ <https://www.governor.ny.gov/news/governor-hochul-announces-300-million-investment-suny-stony-brooks-quantum-research-and> accessed 20 August 2026.

[15] ‘Stony Brook Announcement March 2026’ <https://www.stonybrook.edu/provost/about/2026.03-quantum.html> accessed 20 August 2026.

[16] ‘ABQ-Net Goes Live in Albuquerque as the First Open-Access, Entanglement-Based Quantum Network in the United States’ <https://www.qunnect.inc/press-releases/2026-02-24> accessed 20 August 2026.

[17] ‘First Node for the Quantum Internet of the Future - Fraunhofer ILT’ (*Fraunhofer Institute for Laser Technology ILT*) <https://www.ilt.fraunhofer.de/en/press/press-releases/2025/1-23-quantum-internet-node.html> accessed 20 August 2026.

[18] ‘The Czech Republic Is Launching the Infrastructure for the Future of Communication’ (*IT4Innovations*, 15 April 2026) <https://www.it4i.cz/en/about/infoservice/press-releases/the-czech-republic-is-launching-the-infrastructure-for-the-future-of-communication> accessed 20 August 2026.

[19] ‘Quantum Communication Fieldlab Rotterdam Moves Quantum-Secure Communication from Pilot to Operational Use’ (*Eurofiber*) <https://www.eurofiber.com/press/quantum-communication-fieldlab-rotterdam-moves-quantum-secure-communication-from-pilot-to-operational-use/> accessed 20 August 2026.

[20] ‘Quantum Networking Applications Roadmap’ (CQN – QED-C, 2026).

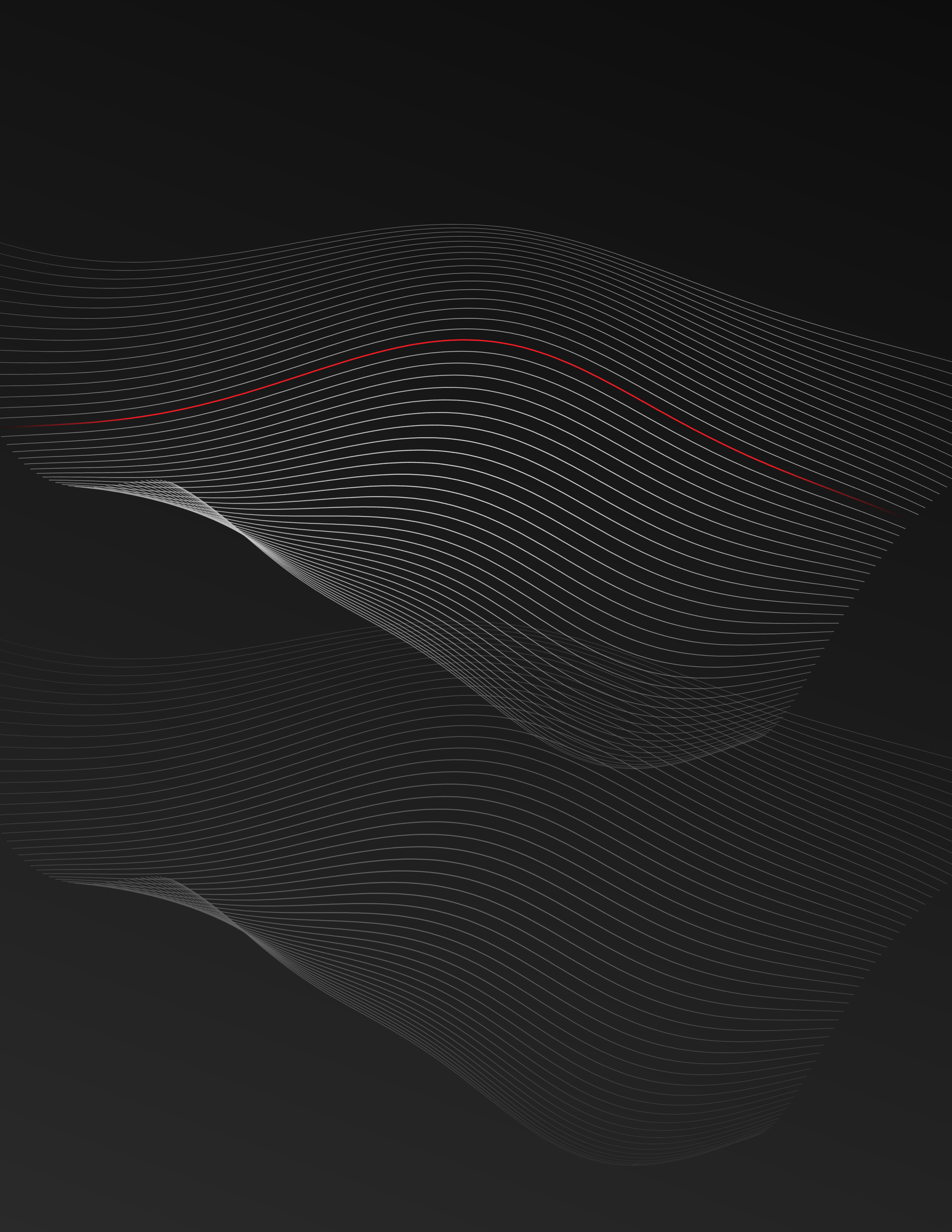

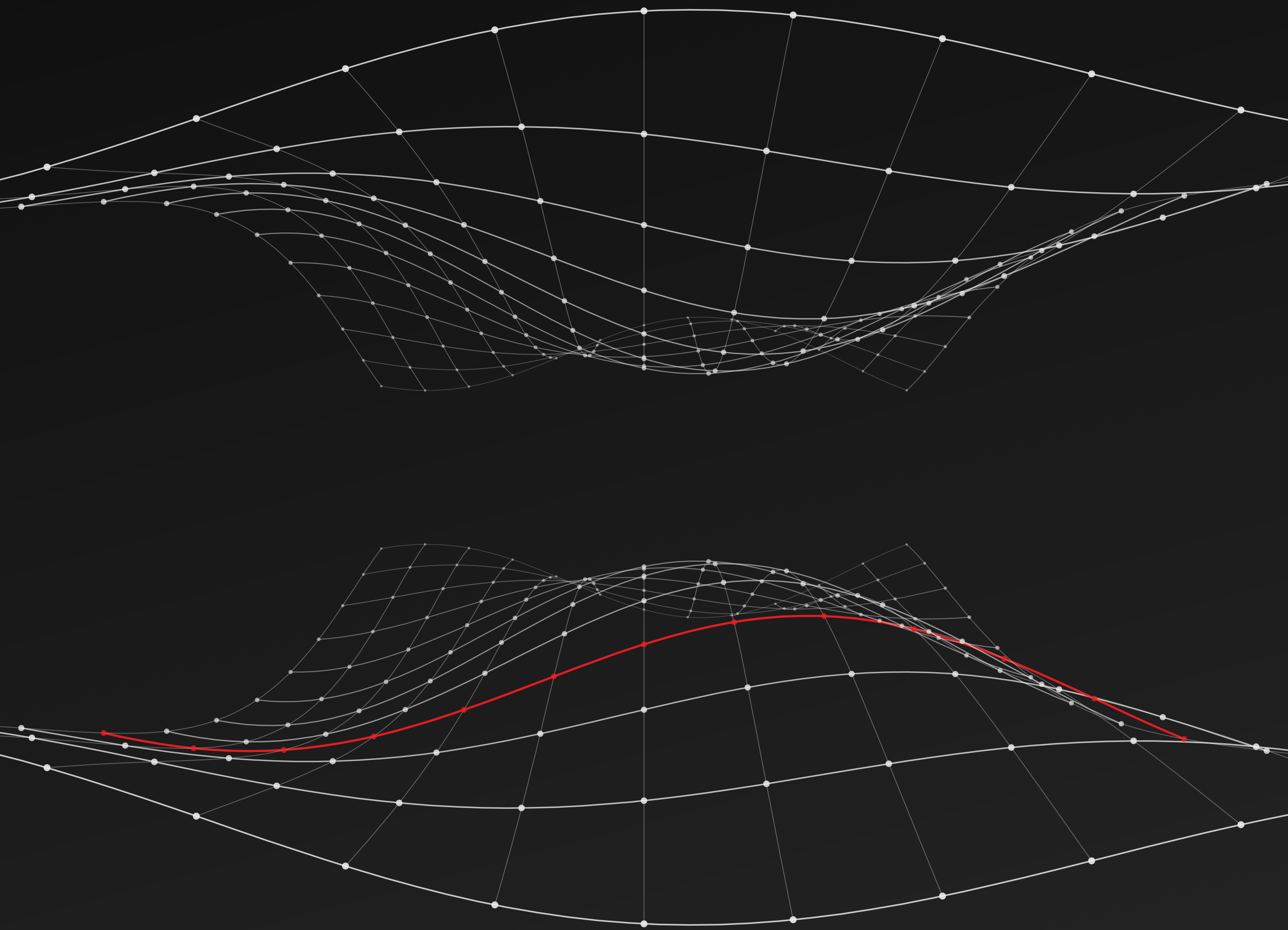

# 10 | Quantum Processor Benchmarking

It remains difficult to assess the performance of quantum computing today, especially for non-experts, which poses a challenge for making informed decisions about investment, commercial deployment, use case testing, and overall strategy. The technology is still nascent, several competing modalities exist, there is no independently verified universal performance metric, and the link between a hardware device and a given quantum algorithm is highly context-dependent.

## Benchmarks

To give better insight into where quantum computing performance actually stands, we indexed and analyzed published data on nearly 300 Quantum Processing Units (QPUs) from 17 countries, spanning retired, prototype, currently available, and announced QPUs. As of August 2026, we observed 56 commercially available QPUs from 24 manufacturers.[1]

It is worth stating up front that most experts agree that quantum computers remain below the capabilities of classical computers for all useful tasks today. A helpful way to think about QPU benchmarking is to consider racecars. A racecar benefits from high horsepower and torque, light construction, independent suspension, and good aerodynamics, but lap time on a specific circuit is the best single measure of how those elements work together. QPUs have comparable construction, design and performance idiosyncrasies. Three main categories of benchmarks are useful to consider:

Physical benchmarks: e.g., qubit count and the fidelity of qubit gates.

These are the core metrics of a QPU, akin to a racecar's weight and torque. They are objective, but in isolation they reveal only part of the likely overall performance. Manufacturers are most willing to disclose physical benchmarks, and they are the focus of the analysis below.

Aggregated benchmarks: e.g., Quantum Volume, CLOPS, and Logical Qubits.

These combine physical benchmarks, much as power-to-weight ratios do for cars. They are more informative than any single measure but still fall short of capturing a QPU's full performance.

Application-level benchmarks: e.g., Q-Score and RACBEM.

These measure how QPUs perform on specific problems, in the way LINPACK is used to rank classical supercomputers. They allow QPUs to be compared with one another and with classical devices, and are analogous to a racecar's lap time at a given track in defined conditions. As with lap times, different benchmarks reward different algorithmic strengths: a Formula 1 car set up for Monaco would be ineffective on a NASCAR oval.

Manufacturers still rarely publish application-level benchmarks, as today's QPUs cannot run sizable applications. We expect to track them in future reports as devices become more capable.

### Benchmarks

Our QPU dataset was assembled primarily from manufacturer announcements, publicly available datasets, online searches, and earlier direct queries to providers. It captures a range of benchmarks, of which the most important are:

Qubit Counts: Qubit count can mislead, because error rates tend to rise alongside it. This has at times prompted manufacturers to reduce qubit counts to improve overall performance. IBM followed its 1,121-qubit Condor in 2023 with the higher-performing 133-qubit Heron, and continued in that direction with the 120-qubit Nighthawk in 2025, a device built for circuit quality rather than scale. In short, qubit count alone is not a definitive measure of capability and must be read alongside other benchmarks.

Coherence: This refers to how long a qubit holds its quantum state. Through interaction with their environment, qubits inevitably lose their quantum information, a process known as decoherence, characterized by two timescales: $T_1$ (energy relaxation) and $T_2$ (dephasing). Both matter because they determine the execution time for a calculation. Trapped-ion and, to a slightly lesser degree, neutral-atom modalities show $T_2$ times several orders of magnitude longer than superconducting qubits, an inherent advantage for applications needing longer coherence.

Fidelity: Quantum computers still experience error rates orders of magnitude higher than classical systems. These errors originate in imperfect control pulses, inter-qubit couplings, and qubit state measurement, and reflect the engineering limits of current hardware. Several benchmarks characterize quantum error rates: single-qubit gate fidelity, two-qubit gate fidelity, readout fidelity, state preparation and measurement (SPAM) error, decoherence-related errors, and crosstalk error. Manufacturers describe these differently (mid-circuit, median, or average), which hinders direct comparison, and the problem has grown rather than receded as more vendors publish. Each of these benchmarks captures inaccuracies in a specific quantum operation essential to running a quantum circuit.

For instance, two-qubit gate fidelity remains among the most critical metrics, as it is usually the bottleneck in large circuits, where two-qubit gates are both more prevalent and more error-prone than single-qubit gates.

Quantum Volume (QV): Introduced by IBM in 2017, QV reflects several physical-level benchmarks at once, including gate fidelity, qubit count, and connectivity. Unlike the volume of a cube, it is not a simple multiplication but the result of a set of statistical tests. QV identifies the largest square-shaped random circuits, where qubit count equals circuit depth, that a device can implement with high fidelity. It has been criticized by scholars and by IBM itself as unhelpful for larger devices and for relying on square circuits that poorly represent real applications. Despite this, many hardware providers have adopted it, including for spec sheets and marketing.

Gate speed: This is the time taken to perform a single-qubit or two-qubit gate operation. It connects directly to decoherence, since together they determine how many operations a system can complete before the qubit becomes ineffective. Systems with fast gate speeds, such as superconducting qubits and electron spins, generally have shorter coherence times than trapped ions or neutral atoms, which have slower gates but far longer coherence times. Superconducting quantum computers operate at raw gate speeds in the 1–100 MHz range, and 1–10 kHz when fully burdened with error correction and overhead, significantly slower than classical CPUs at 2–5 GHz, though the comparison is only partly useful given the fundamentally different computational approaches.

### Execution time

Gate speed is critical but under-reported, and many hardware vendors do not disclose it at all. It nonetheless sets a hard limit on circuit runtime. Consider molecular simulation using a Quantum Phase Estimation algorithm, which can require circuits exceeding $10^{13}$ logical gates. On a trapped-ion processor with gate speeds around 10 microseconds, running such a circuit once would take several days. Since a single molecule may need thousands of full executions for statistical confidence, and since error correction significantly increases circuit depth and gate count, total runtime could stretch into years, well beyond practical limits for most applications. Businesses evaluating quantum computing should estimate execution time from circuit size, hardware gate speed, and error-correction overhead. Gate speed imposes a fundamental limit, but total runtime can be reduced by optimizing algorithms for parallelism, reducing circuit depth, and improving qubit fidelity to lower the cost of error correction.

### Error correction

Error correction is fundamental to quantum computing. Methods such as surface codes require growing numbers of physical qubits to form a single logical one: thousands, or tens of thousands for very large circuits. In 2024, Google announced a relevant breakthrough, demonstrating a system operating below the fault-tolerance threshold, meaning that adding qubits and correction cycles produces a net decrease in logical error rates.

Composing a detailed QPU list remains challenging, made harder by vendors highlighting only their most favorable benchmarks. Trapped-ion gates are generally orders of magnitude slower than the fastest superconducting gates. IBM is still among the most transparent firms on QPU benchmarking, publishing most relevant metrics including data for individual QPU instances and even individual qubits.

## Modalities

Physical quantum computing systems are built and operated using several distinct technical architectures, referred to as modalities. Although each modality utilizes a different method for encoding, processing and reading out quantum information, they achieve comparable functionality (such as gate-based or measurement-based models). Each comes with distinct strengths and limitations, measurable across metrics like qubit count, speed and fidelity. Below, the table outlines the leading commercial or prototype devices for each modality, where a clear frontrunner has yet to be established.

| | Trapped Ions | Superconducting | Neutral Atoms | Photonics | Electron Spin / NV Center |
|---|---|---|---|---|---|
| **Qubit Count** | 98 | 1,121 | 1,225 | 216 | 424 |
| **2Q Gate Fidelity** | 99.92% | 99.70% | 99.35% | 93.80% | 99.20% |
| **1Q Gate Fidelity** | 99.998% | 99.98% | 99.90% | 99.60% | 99.995% |
| **2Q Gate Speed (LogHz)** | N/A | 7.32 | 6.18 | N/A | N/A |
| **1Q Gate Speed (LogHz)** | 4.00 | 7.70 | 5.39 | N/A | N/A |
| **Coherence – T1 (μs)** | 10,000,000 | 305 | 1,000,000 | N/A | N/A |

Best ▭ ▭ ▭ ▭ Worst ▭ N/A

*Vendors report different statistics under different conditions, these figures mark the frontier of each modality, not a ranking of individual devices.

A quantum circuit is a sequence of operations, or quantum gates, that a QPU follows to solve a problem. It is the foundation of quantum algorithms, which use these circuits to process information.

Superconducting QPUs are electronic circuits made with the lithography techniques used in classical computing fabrication. They are cooled to millikelvin temperatures, which suppresses thermal noise and permits coherent quantum behavior. They excel in gate speed and qubit count with reasonable fidelities, but require extensive cooling. They remain the most common modality, accounting for half of commercially available QPUs.

Trapped-ion QPUs implement gate-based quantum computing using individual ions held by radiofrequency traps, with gates performed by lasers or microwaves that manipulate the ions' internal quantum states. Trapped ions offer high fidelity, coherence, and qubit connectivity, but have slower gate speeds. They have not yet scaled to large qubit counts: the leading trapped-ion device reached 98 qubits in 2025 without sacrificing its fidelity lead.

Photonic QPUs use photons as qubits, propagating through photonic integrated circuits containing linear optical elements such as beam splitters and phase shifters, which can implement certain quantum gates.

Neutral Atom QPUs use atoms, typically alkali or alkaline-earth metals, that are laser-cooled and confined in vacuum chambers by optical and magnetic trapping. The atoms are ultracold while the hardware operates near room temperature. Relative to other modalities, they show promise in high qubit counts but have slower gate speeds and lower fidelity, though development is rapid.

Electron Spin QPUs use the quantum state of single electrons as qubits, offering relatively long coherence times and the potential for high-fidelity control, with the added benefit of compatibility with existing semiconductor fabrication techniques. Scalable control and readout of large arrays remain active research areas.

Nitrogen-Vacancy (NV) centers in diamond are a promising solid-state platform, where a nitrogen impurity adjacent to a lattice vacancy hosts a localized electron spin used as a qubit. NV-center QPUs can operate at room temperature, unlike most other quantum computer hardware.

Nuclear Magnetic Resonance (NMR) QPUs use the nuclear spins of atoms within molecules as qubits, manipulated by radiofrequency pulses inside a strong magnetic field. Because they operate at room temperature and are comparatively simple to build, NMR systems produced some of the earliest experimental demonstrations of quantum algorithms. They are difficult to scale, however, since measurements are taken across a large ensemble of molecules and the usable signal weakens sharply as qubits are added, which has kept devices to a handful of qubits. Today they are used mainly for teaching and research.

### Cats, Rails and Flux

Superconducting quantum processors are among the most promising modalities, and several sub-architectures are being pursued. Amazon and Alice & Bob are working on "cat qubits," which are harder to fabricate but improve qubit fidelity because only one type of error occurs: the phase flip (no bit flip error). Quantum Circuits Inc. takes a different approach with Dual Rail QPUs, adding physical redundancy to transmon qubits with a second cavity / rail to protect from errors. Atlantic Quantum uses Fluxonium qubits rather than traditional transmons; Fluxonium offers longer coherence and error-correction possibilities, again at the cost of fabrication complexity.

### Annealers

A distinct class of quantum computer is the adiabatic quantum computer, or annealer, named after the metallurgical process. Quantum annealing rests on the adiabatic theorem, which holds that a quantum system remains in its lowest-energy state if its parameters change slowly enough and noise stays low. An optimization problem can be mapped onto an energy landscape of possible solutions, with the lowest energy the best solution; by annealing (adjusting the system parameters), the system is guided toward the lowest-energy state, which, if reached, yields the optimum. D-Wave produced the first commercial annealer in 2011 with 128 qubits. In this report, annealers are treated separately, as their architecture is not directly comparable to gate-based quantum computers. They reach much larger qubit counts but do not implement universal gate-based control, which restricts them to a narrower class of problems compared to gate-based QPUs.

## Majorana Qubits

Another type of superconducting qubit is the Majorana qubit, which drew significant attention in early 2025. Microsoft has invested in the approach for over a decade and remains the primary industry player pursuing it. The design uses superconducting nanowires hosting Majorana zero modes, exotic quasiparticles predicted to appear at the ends of the wire under specific conditions. Microsoft's Majorana 1 QPU represents a milestone toward a Majorana-based processor[2], though the February 2025 announcement met skepticism from parts of the scientific community, as conclusive evidence for the topological nature of the zero modes remains under debate.

## Logical Qubits

Most quantum algorithms assume a qubit is "perfect," behaving flawlessly throughout the algorithm. In reality, qubits are error-prone and short-lived, so many physical qubits are combined using error correction techniques such as surface codes to form a more stable unit known as a logical qubit. The term can mislead, since practical utility depends heavily on the size and complexity of the circuit the logical qubit can reliably support. To be viable for applications such as simulating complex molecules, a logical qubit would need to support circuits millions to billions of gates long, several orders of magnitude beyond current capability. When presented with a number of logical qubits for a QPU, the key follow-up question is what circuit depth it can support, which is essential for meaningful interpretation.

### Quantum emulators

Quantum circuits can be simulated on classical computers, allowing quantum algorithms to be developed and tested without access to a quantum processing unit. Emulators do not physically use quantum effects such as entanglement or superposition. They are particularly useful for testing, debugging, and benchmarking quantum algorithms. Classical supercomputers can typically simulate general quantum circuits containing around 50 qubits, although the practical limit depends heavily on the circuit's structure and the simulation method used. For classical computers, each additional qubit is exponentially harder, whereas for QPUs it requires only incremental logical qubits. Today's best quantum computers remain orders of magnitude slower and more expensive to run than equivalent CPUs.

> The number of commercially available QPUs globally now stands at 56, from 24 manufacturers. Countries are in a strategic race to achieve high-performance QPUs. The U.S. continues to lead this race in 2026.

# 10.1 | QPUs per country and modality

**Fig. 10.a** | Commercially available QPU models per country

MIT Quantum Index Report

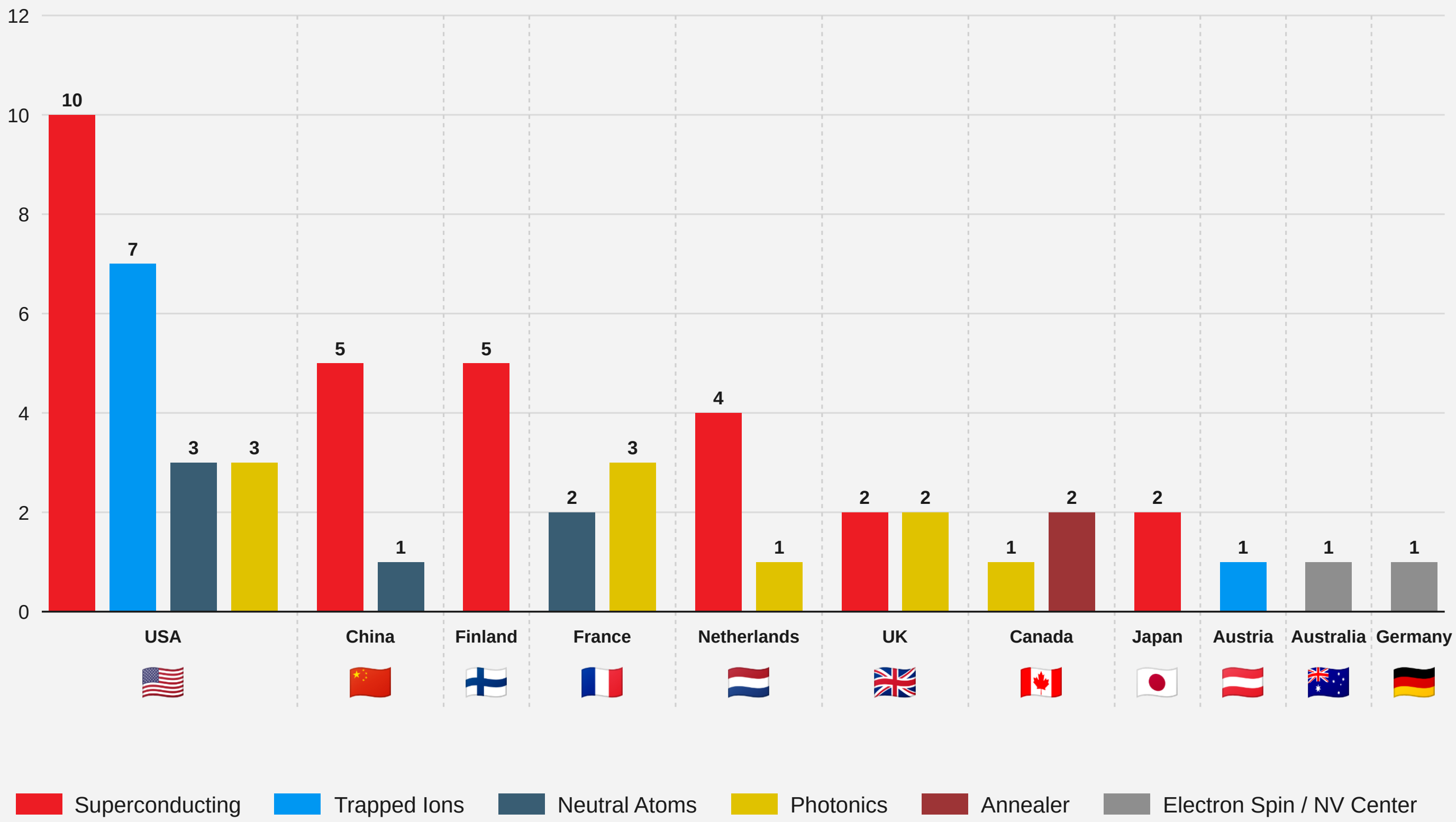

Countries remain in a strategic race to field high-performance QPUs.

The number of commercially available QPUs globally is 56 from 24 manufacturers across eleven countries. The total number of commercially available QPUs is up from 40 in our 2025 report.

The United States retains a clear lead with 23 and is the only country present in four modalities, its superconducting fleet having grown from six devices to ten. China holds second place with 6 and is the only country to have added new manufacturers rather than new devices from existing ones.

Below them, Finland, France and the Netherlands hold five each, the UK four, Canada three, Japan two, and Australia, Austria and Germany remain at one.

Data considerations: we list QPUs that are commercially available per country and per quantum computing modality. Country data is allocated based on the location of the manufacturer's headquarters as described in their official materials and website. QPUs are classified as commercially available if there is public access to the QPU either via on-premise or cloud. This also includes QPUs that may not be available on public clouds, but where access is provided to specific partner companies for commercial use. However, the device must be intended as a useful quantum computer for commercial use and not solely for experimental purposes. The number of QPUs is determined as uniquely differentiated products actively provided and marketed by the provider, e.g., IBM Eagle and IBM Heron are two distinct QPUs, but Eagle r3-Brussels and Eagle r3-Sherbrooke are considered as one QPU. On the same basis, a trapped-ion system newly listed on a public cloud in 2025 is recorded against the existing product rather than as a new one. The number of QPUs is not necessarily an indication of the progress of each country in quantum computing, as some manufacturers have made several very small QPUs available for basic academic research and teaching, while others have retired smaller but powerful QPUs from their offerings (e.g., IBM).

# 10.2 | QPUs per modality

**Fig. 10.b** | Commercially available QPUs per modality

MIT Quantum Index Report

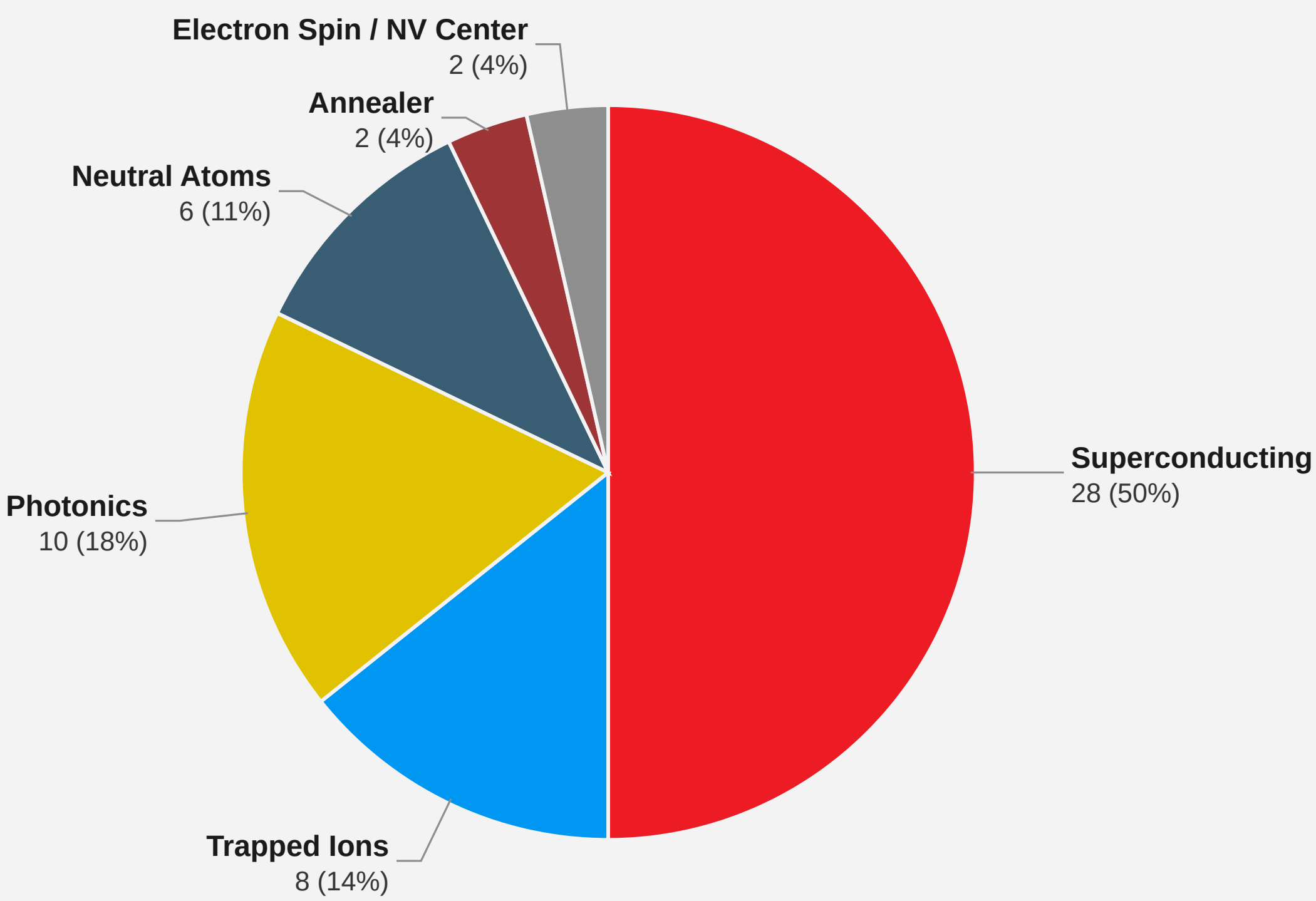

Superconducting remains dominant at 28 of 56 devices, exactly 50%, a share essentially unchanged (48% in 2025) despite the sample growing by 40%, which might be driven by the inherent manufacturing benefits and a historic head start in R&D. Photonics is next at 18%, followed by trapped ions at 14%. Neutral atoms hold 11%, annealers and electron spin / NV centers 3.6% each.

## 10.3.1 | Qubit count per modality

**Fig. 10.c** | Largest QPU released, prototyped, or planned per year

MIT Quantum Index Report

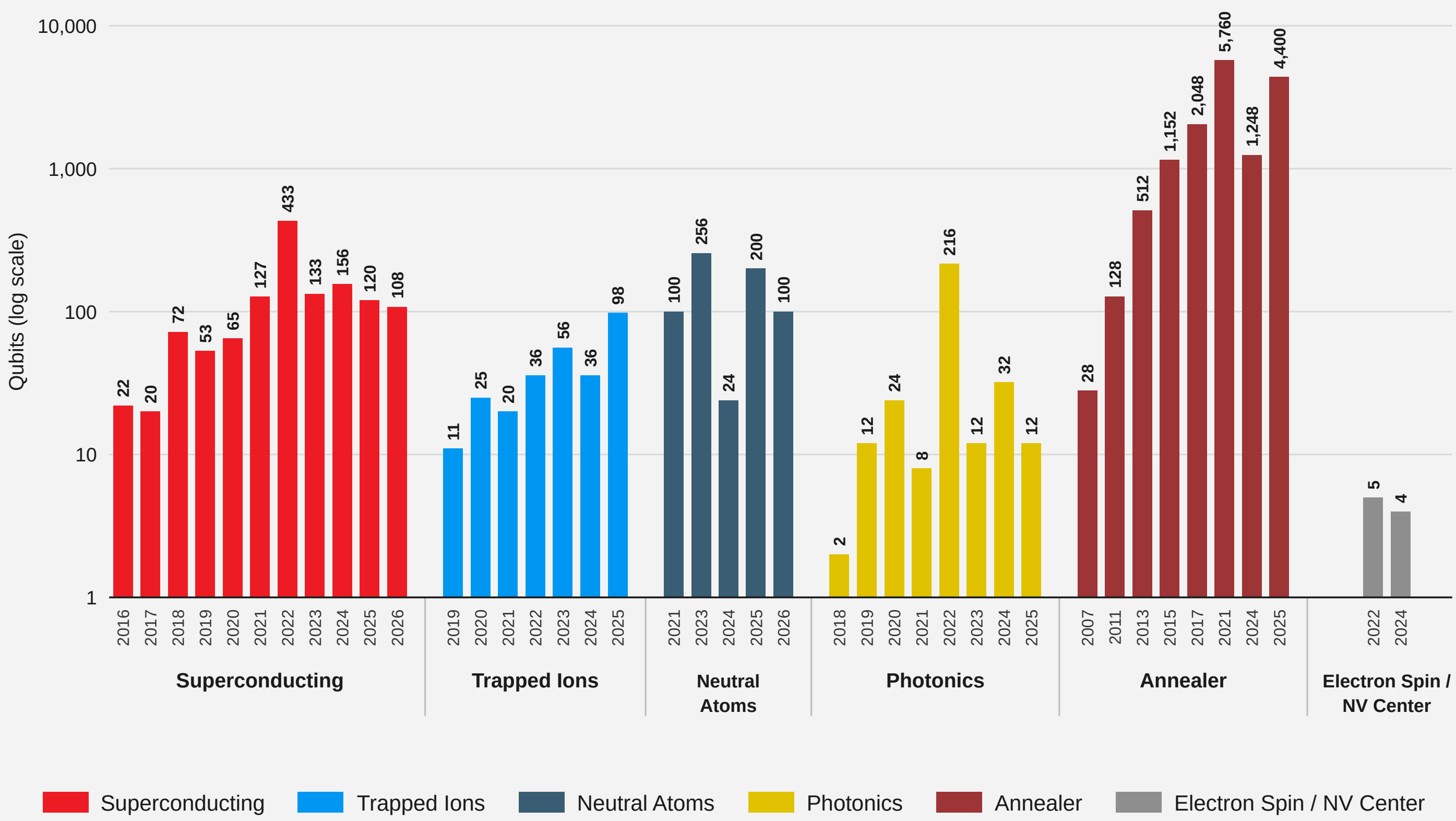

Superconducting counts peaked at 433 in 2022 and have generally declined since (133, 156, 120, 108), which might reflect a sustained shift toward error correction and fidelity rather than scale.

Trapped ions tell the opposite story, growing more consistently to 98 qubits in 2025, and remain on an upward trend. Neutral atoms reached 200 qubits in 2025 and 100 in 2026. Annealers sit at 4,400 for 2025, below the 5,760-qubit device of 2021. Photonics remains the least regular series, moving between 2 and 216 across eight years.

Data considerations: this graph shows the progression of the number of qubits in our dataset over time, considering only the largest QPU announced per modality in that year. For a QPU to be considered, it needed to be officially announced by a manufacturer and made commercially available in the given year (or expanded, e.g., Quantinuum H1 to H1-1). Retired QPUs are included, as they were available in the year shown. The data does not always show a steady increase, as some calendar years only contained new QPUs that were smaller than previously available.

## 10.3.2 | Fidelity per modality over time

**Fig. 10.d** | Best 2Q-gate error rate over time (commercially available QPUs)

MIT Quantum Index Report

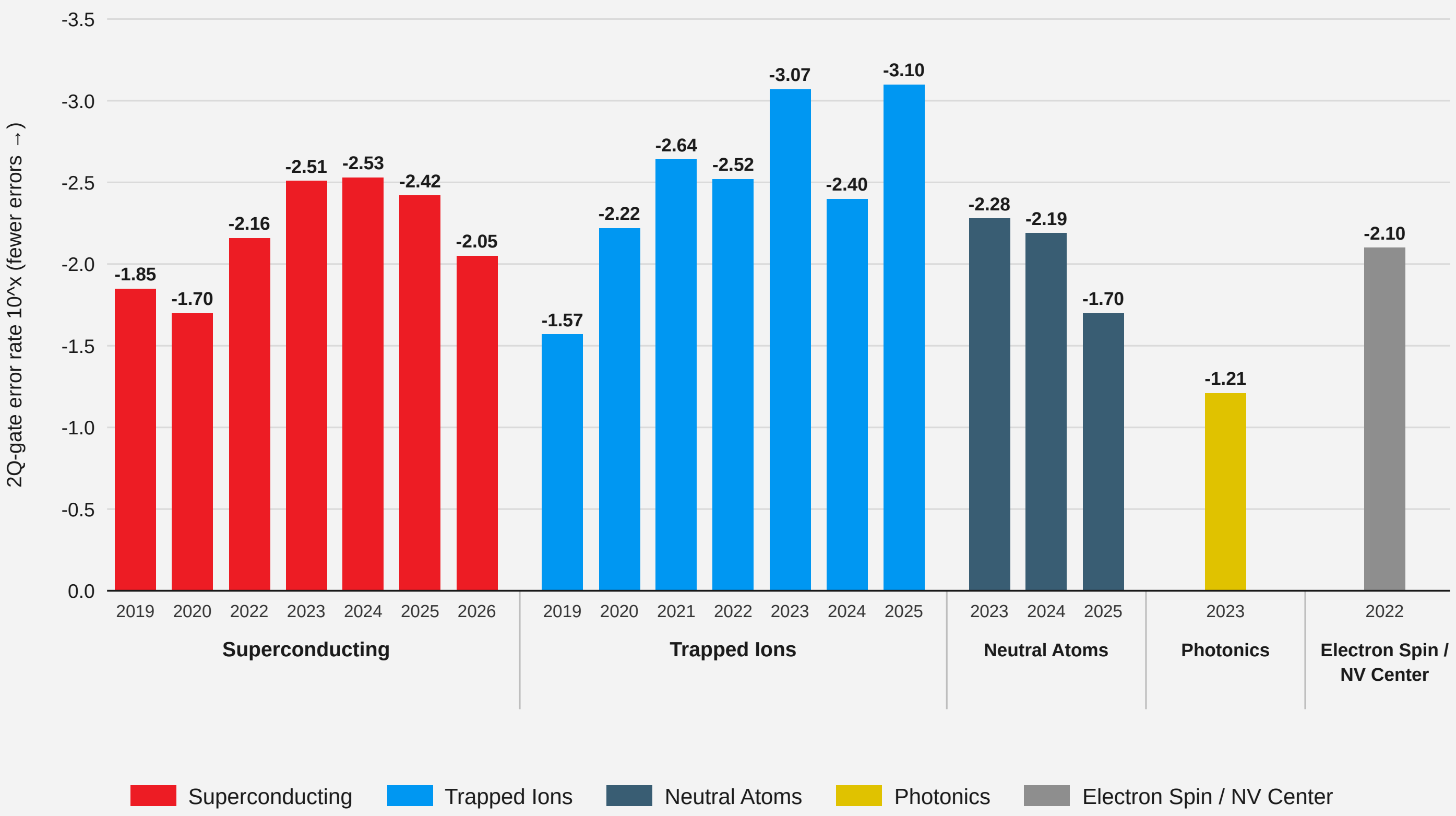

Fidelity for 2-Qubit is a key metric of performance improvement. Trapped ions have shown consistent growth and the highest overall fidelity.

Data considerations: this graph shows the progression of fidelity levels in new, commercially available QPUs in our dataset over time. We only considered the highest fidelity announced per modality per year, so a year's bar reflects that year's releases rather than the best result achieved to date. Error rates are given in a log10 scale, i.e. -3 translates to a 0.001 error rate which corresponds to a 99.9% or 0.999 fidelity. The noted error rates should be treated with caution as there are significant differences in the way they are measured for each QPU, e.g., mid-circuit vs first-gate-measurement, average vs median of several measurements across qubits, different gates (CZ, SWAP, etc.), and different iterations of the same QPU that give different values.

# 10.4 | Gate time versus gate fidelity

Fig. 10.e | Gate time versus gate fidelity (2-qubit)

MIT Quantum Index Report

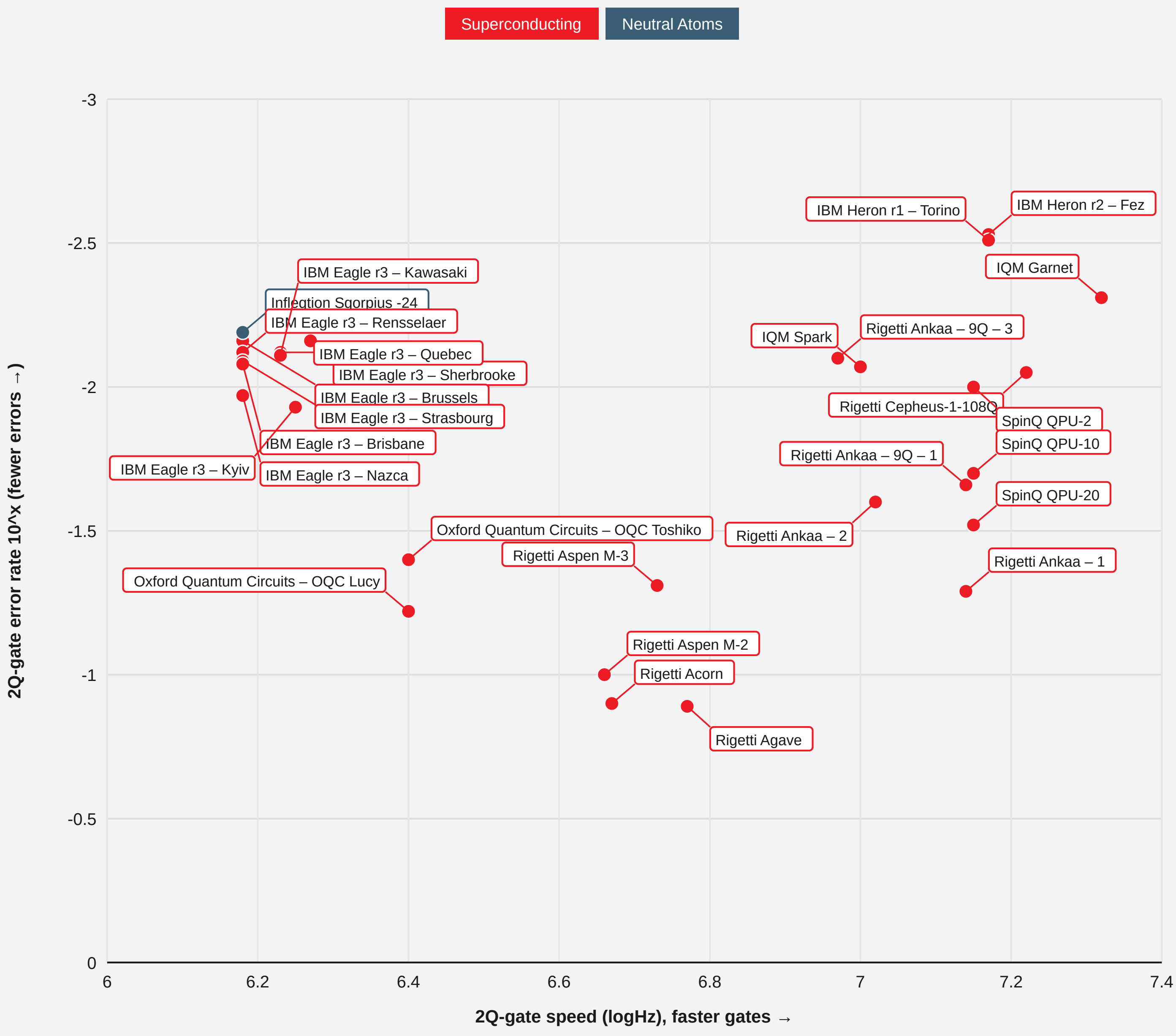

To determine the maximum length of a circuit for a QPU, an important metric is the comparison of the speed of executing a single gate to how accurate this gate is (fidelity). For real-life scenarios such as Shor's algorithm for decrypting information utilizing RSA-2048, more than $10^{13}$ logical gates are required. Slow gate speeds at that size lead to calculation times of days or even months with some modalities. Speed alone is not the answer either: a fast gate that makes mistakes forces the machine to repeat work, so the useful QPUs are those that sit toward the right and the top of this chart at once.

Devices are shown for both one- and two-qubit gates; the two-qubit case matters more because such gates are more common in large circuits for most algorithms. The superconducting IQM Garnet is the fastest device to publish a figure, at 48 nanoseconds per two-qubit gate, with Rigetti's newly released Cepheus-1-108Q at 60 nanoseconds and IBM's Heron at 68. IBM Heron pairs its slightly slower gate with the best accuracy on the chart at 99.70%, so no single device now leads on both counts. Below them sits a group of processors running at comparable speeds but with fidelities between 94.9% and 99%, which illustrates the point directly: gate speed by itself says little about whether a machine can finish a long calculation correctly.

Notably, trapped ions are absent entirely and neutral atoms nearly so, as manufacturers tend not to disclose exact gate speeds, which are expected to be orders of magnitude slower than superconducting QPUs (as can be seen in the 1-Qubit graph).

Of the 36 trapped-ion QPUs in our dataset, only three publish any gate speed at all, and none publish a two-qubit figure. The 1-Qubit graph shows what the gap looks like where the numbers do exist. Superconducting devices there run their fastest single-qubit gates in 20 to 32 nanoseconds, while the trapped-ion systems that publish a figure take around 100 microseconds – several thousand times longer, though those same systems hold their quantum states for seconds rather than fractions of a millisecond, which is the compensating advantage.

**3 of 36**

Of the 36 trapped-ion QPUs in our dataset, only three publish any gate speed at all, and none publish a two-qubit figure.

Data considerations: we chose 2-Qubit-Gates like CZ, CNOT, and SWAP as they are used for most quantum algorithms and make up the majority of gates for these circuits. The 2Q Gate time is the time required for the execution of a 2-Qubit gate and is given in Hertz (Hz) in a logarithmic scale, where higher values mean faster gate speeds, i.e., 7 logHz corresponds to a 100ns gate speed and 3 logHz to 1,000,000ns. The error rates are given in a log10 scale where higher is better, i.e., $10^{-3}$ translates to a 0.001 error rate, which is 99.9% fidelity. The data points with missing labels in the graphs are closely related QPUs (e.g., different instances of IBM Eagle). Roadmap devices are excluded, as a projected gate speed paired with a projected fidelity describes an intention rather than a measurement. To illustrate the performance comparison, we included the 1-Qubit Gate graph, which shows that these would likely land in the top-left quadrant, trading high fidelity for low gate speeds. Notably absent are ion traps and neutral atom QPUs, as manufacturers tend not to disclose exact gate speeds.

**Fig. 10.f** | Gate time versus gate fidelity (1-qubit)

MIT Quantum Index Report

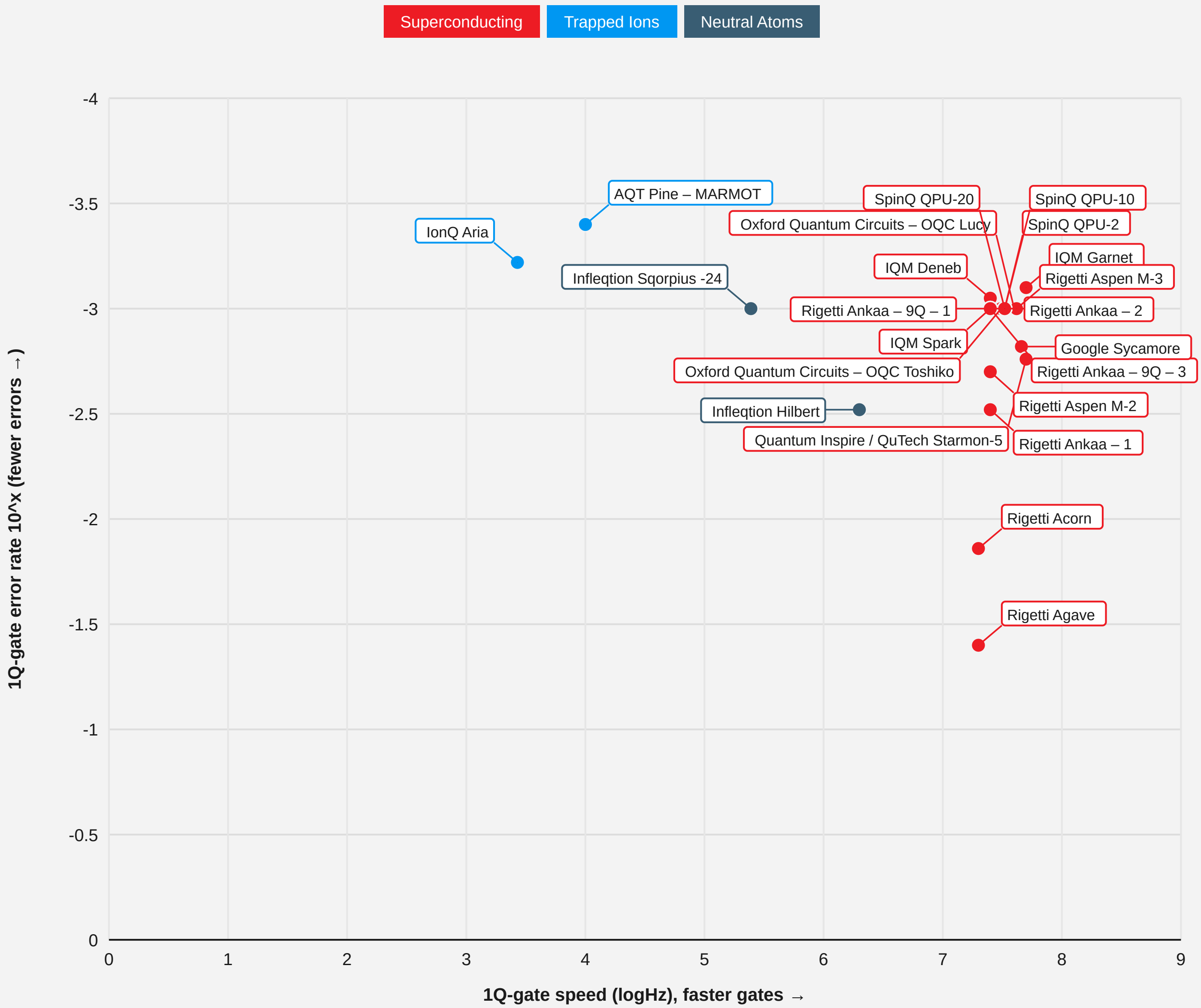

The companion 1-Qubit graph now shows the missing modalities directly, with trapped ions at 3.4 to 4.0 logHz achieving the best fidelities on the chart, three to four orders of magnitude slower than superconducting.

This is the direct expression of the relationship between gate speed and decoherence: fast gates and long coherence remain alternatives rather than a combination any modality has delivered.

Data considerations: we chose 2-Qubit-Gates like CZ, CNOT, and SWAP as they are used for most quantum algorithms and make up the majority of gates for these circuits. The 2Q Gate time is the time required for the execution of a 2-Qubit gate and is given in Hertz (Hz) in a logarithmic scale, where higher values mean faster gate speeds, i.e., 7 logHz corresponds to a 100ns gate speed and 3 logHz to 1,000,000ns. The error rates are given in a log10 scale where higher is better, i.e., $10^{-3}$ translates to a 0.001 error rate, which is 99.9% fidelity. The data points with missing labels in the graphs are closely related QPUs (e.g., different instances of IBM Eagle). To illustrate the performance comparison, we included the 1-Qubit Gate graph, which shows that these would likely land in the top-left quadrant, trading high fidelity for low gate speeds.

## 10.5 | Quantum Volume

**Fig. 10.g** | Max QPU Quantum Volume (log2) over time

MIT Quantum Index Report

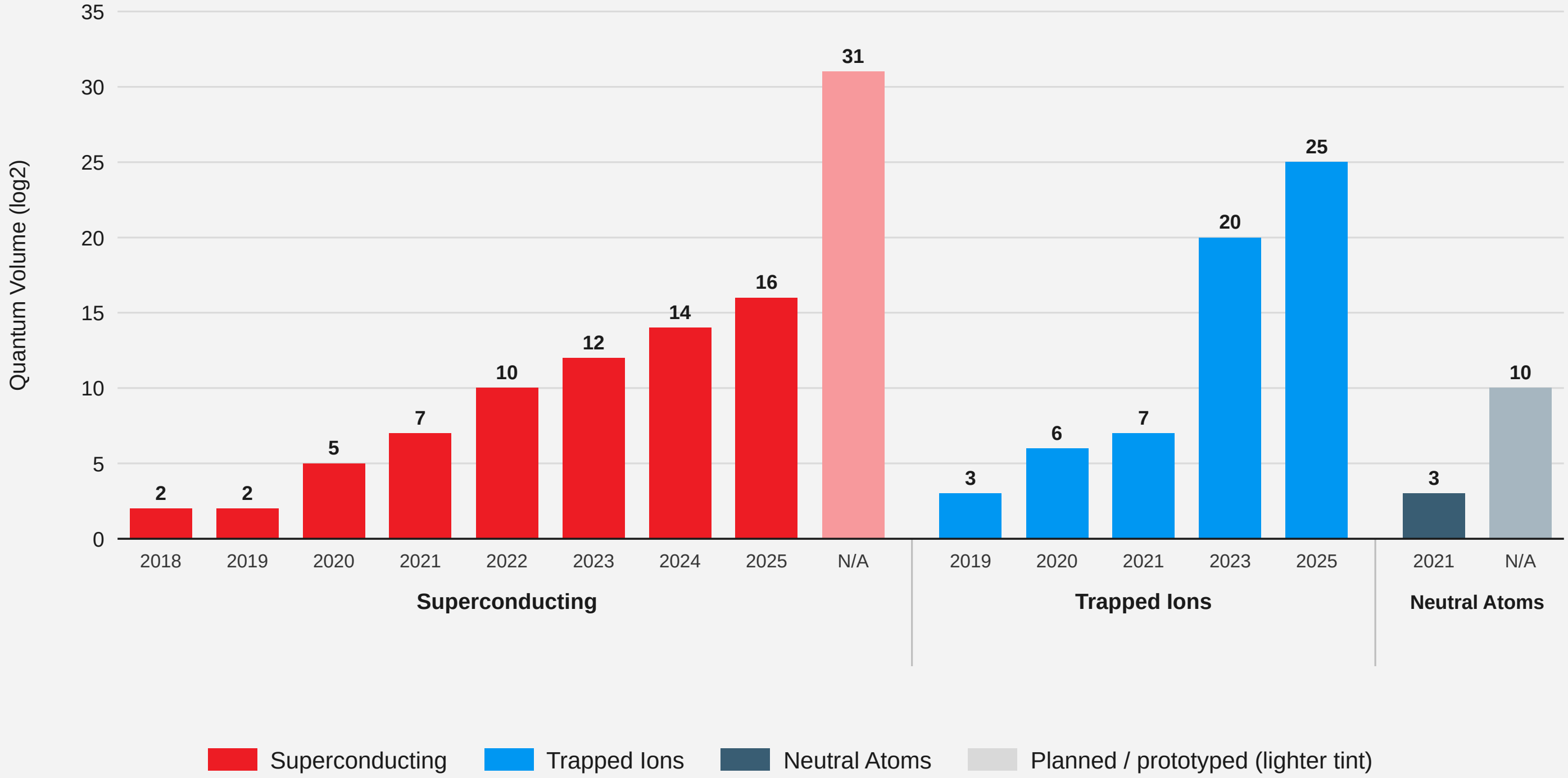

As described in an earlier section, Quantum Volume (QV) is an aggregated metric designed to reflect a more holistic view of overall performance of a QPU.

Although its usefulness is heavily debated, manufacturers have been reporting this metric. There is a substantial and consistent increase in QV performance of the top-performing QPU across modalities. However, the stated QV values are hard to validate, and in some cases the only data has been manufacturer claims.

Data considerations: The Quantum Volume (in log2 basis) is listed for each QPU at the given year. QV is itself a debated benchmark (even by IBM itself), which has been used less in the last few years. Despite all these caveats, we chose to include QV, as out of all aggregated metrics (like RACBEM, Algorithmic Qubit, CLOPS, etc.) it is the one that has published values for a sizable number of QPUs, and as such a progression on quantum computing capabilities can be roughly traced over time.

## 10.6.1 | A look into the future: QPUs per country and modality

**Fig. 10.h** | Planned, prototyped, and available QPUs per country

MIT Quantum Index Report

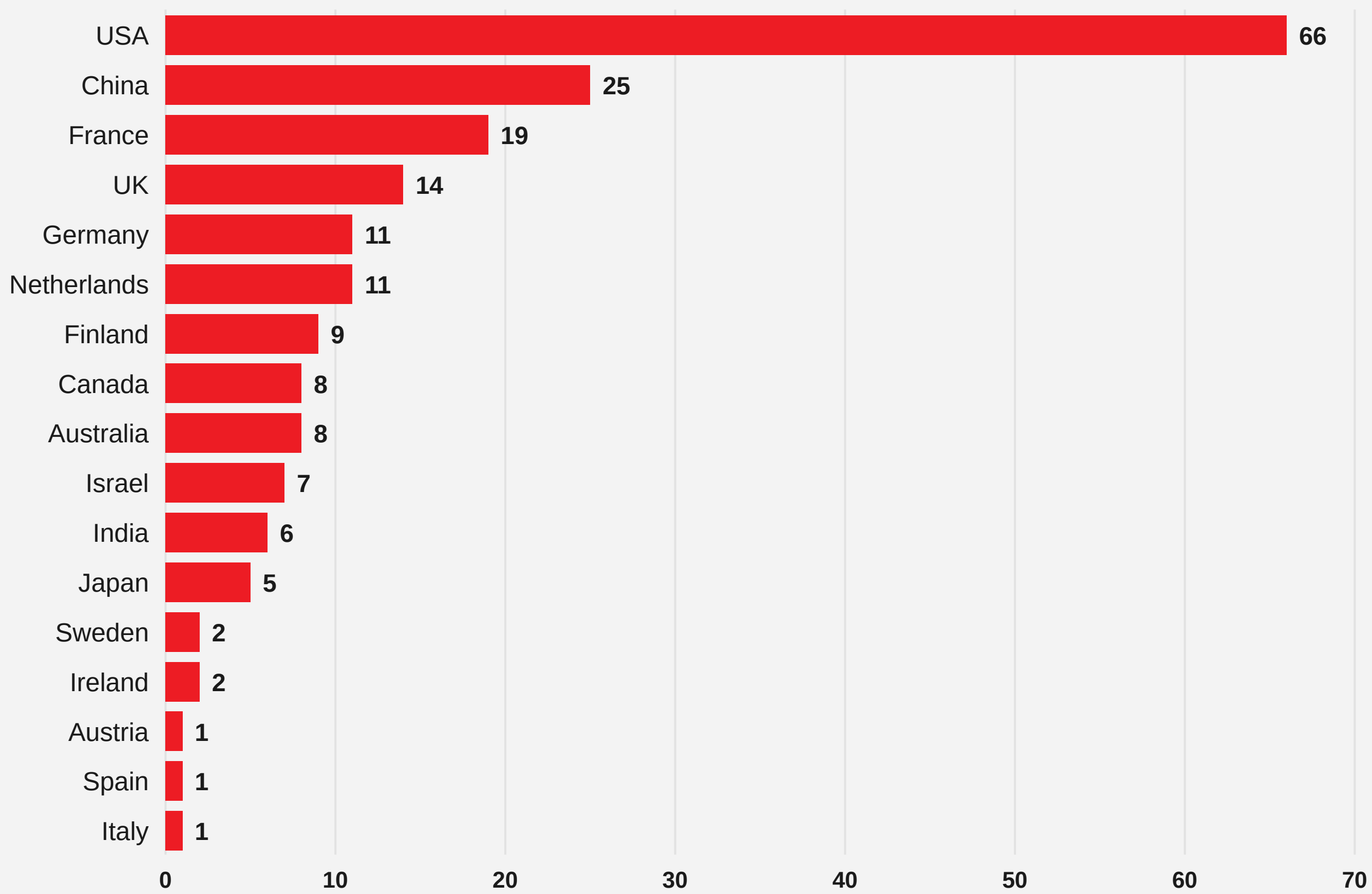

Quantum computing remains in the early stages of technical development. Numerous startups and established companies are actively developing prototypes and announcing roadmaps for future QPUs. A comprehensive overview of these efforts provides valuable insight into the evolving technological landscape and strategic directions within the field. The U.S. is the clear leader in the number and diversity of QPUs announced, with 66; China is a distant second with 25, followed by France with 19 and the UK with 14. Germany and the Netherlands have 11 each, Finland 9, and Canada and Australia 8 each. This data gives an approximate measure of country activity but is not an indication of QPU quality. It also reflects planned QPU announcements and does not fully reflect the probability of their successful release.

**Fig. 10.i** | Planned, prototyped, and available QPUs per modality

MIT Quantum Index Report

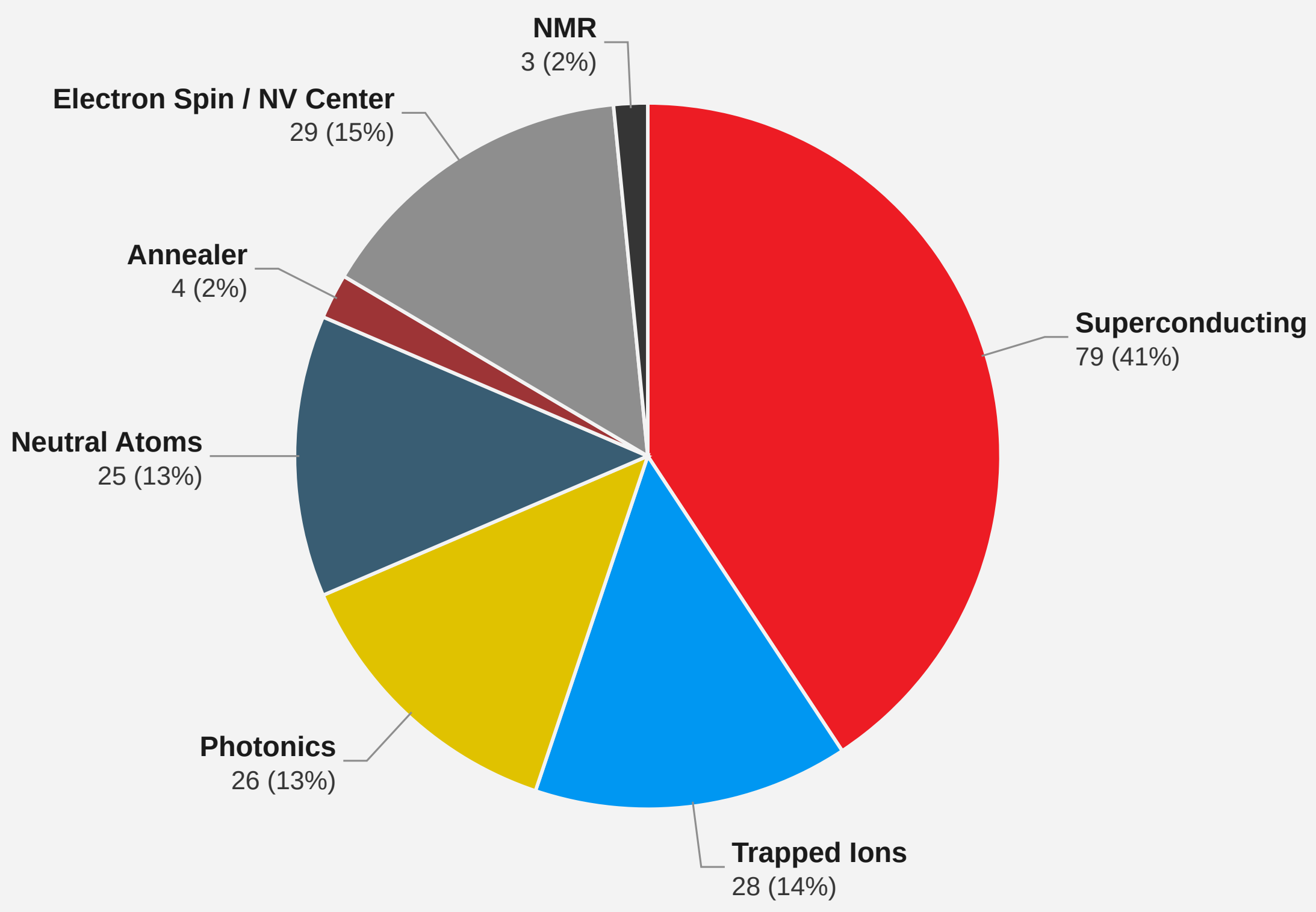

Looking at the future for different modalities, electron spin, NV Centers and neutral atoms are planned to become increasingly prevalent while NMRs and annealers are more stagnant and may be phased out. Photonics, superconducting, and trapped ion QPUs may have lower overall shares in the future due to the higher growth levels of other modalities. Superconducting retains the largest pipeline but a lower share of announced devices than of available ones, consistent with the expectation that other modalities will grow faster.

Data considerations: the data in these graphs includes prototype devices, which are not intended for commercial usage and are not available to the wider community of researchers but are used by a manufacturer for research to develop a new product. It also includes Future Planned QPUs, which are announced in a manufacturer roadmap or interview. Due to the dynamic evolution of startups in the space, recent announcements and changes in QPU roadmaps may not be fully captured in our dataset. The number of QPUs is not necessarily an indication of the progress of each country in quantum computing, as some manufacturers have made several very small QPUs available for basic academic research and teaching, while others have retired smaller but powerful QPUs from their offerings.

## 10.6.2 | A look into the future: qubit count and fidelity

**Fig. 10.j** | A look into the future: qubit count and fidelity

MIT Quantum Index Report

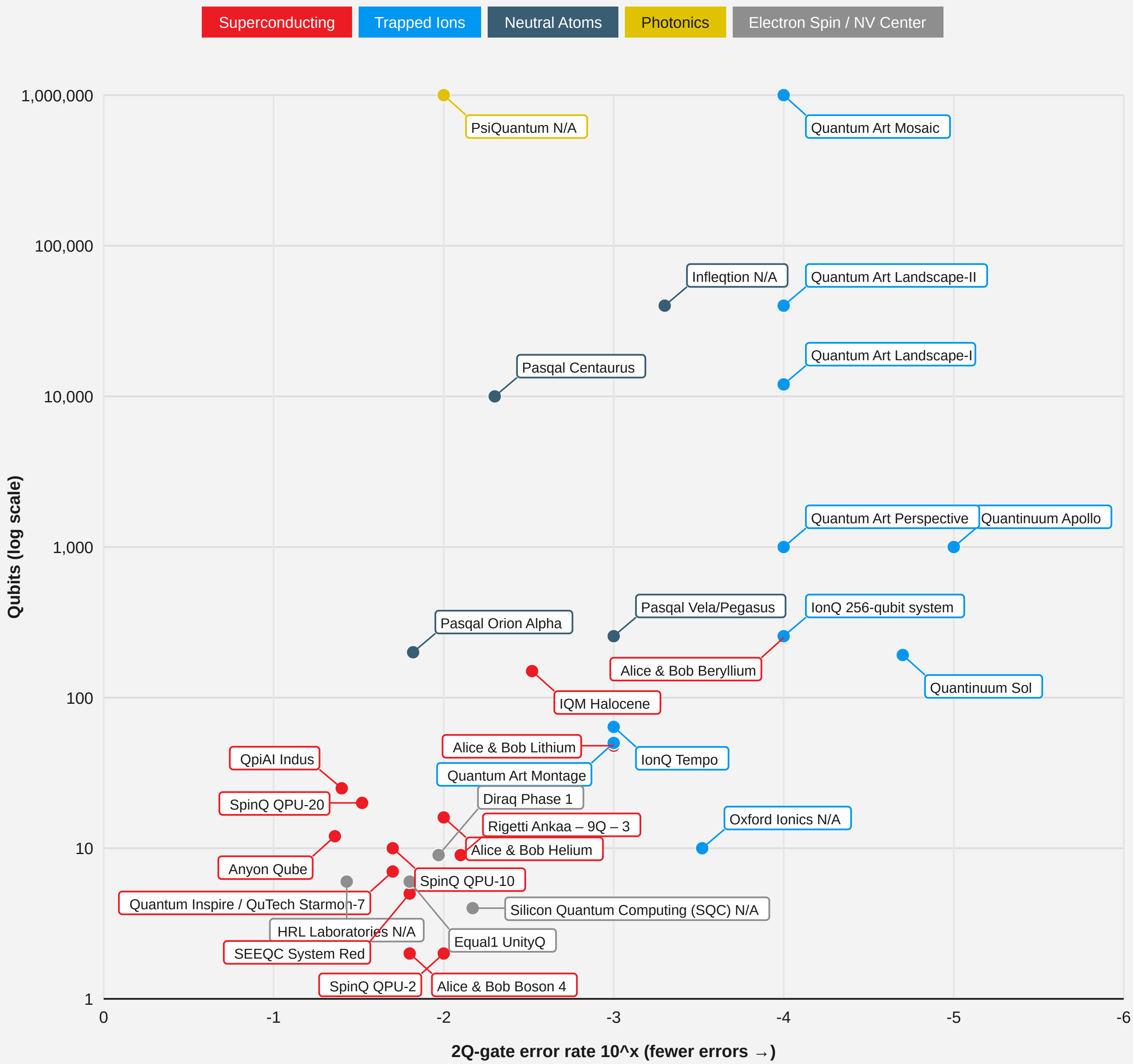

QPU manufacturers are beginning to share more forward-looking roadmaps, offering insights into how their systems might compare across benchmarks. PsiQuantum leads in projected qubit count, Quantinuum excels in error rates, while Quantum Art and Infleqtion position themselves as balanced performers across both metrics.

**Fig. 10.k** | Largest QPU released, prototyped, or planned per year (incl. future plans)

MIT Quantum Index Report

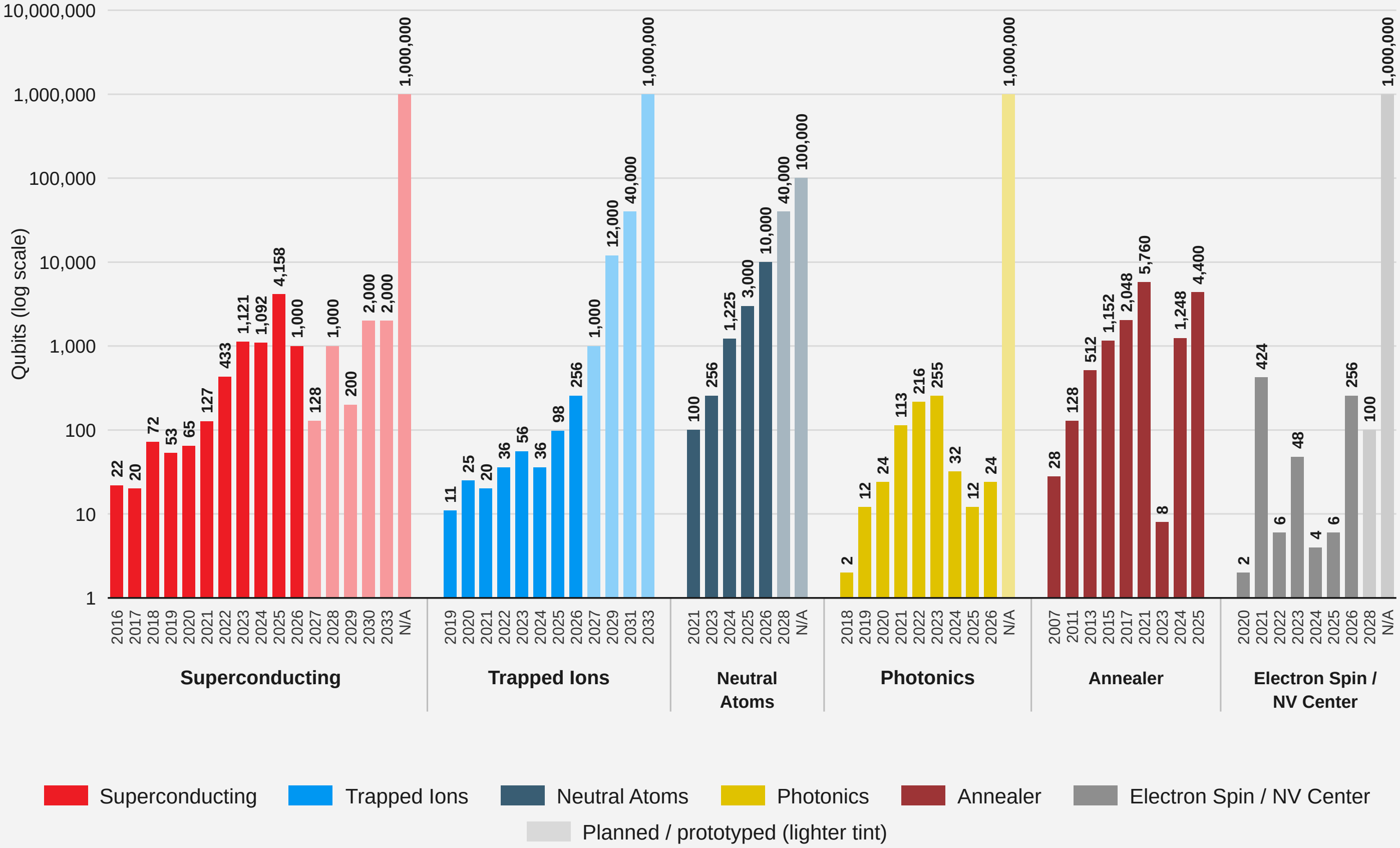

While qubit counts are expected to continue rising, the pace of growth is moderating. This reflects a shift in focus toward improving performance through better error correction and higher qubit fidelity rather than simply scaling up qubit numbers.

**Fig. 10.l** | Best 2Q-gate error rate over time (available, planned, and prototyped QPUs)

MIT Quantum Index Report

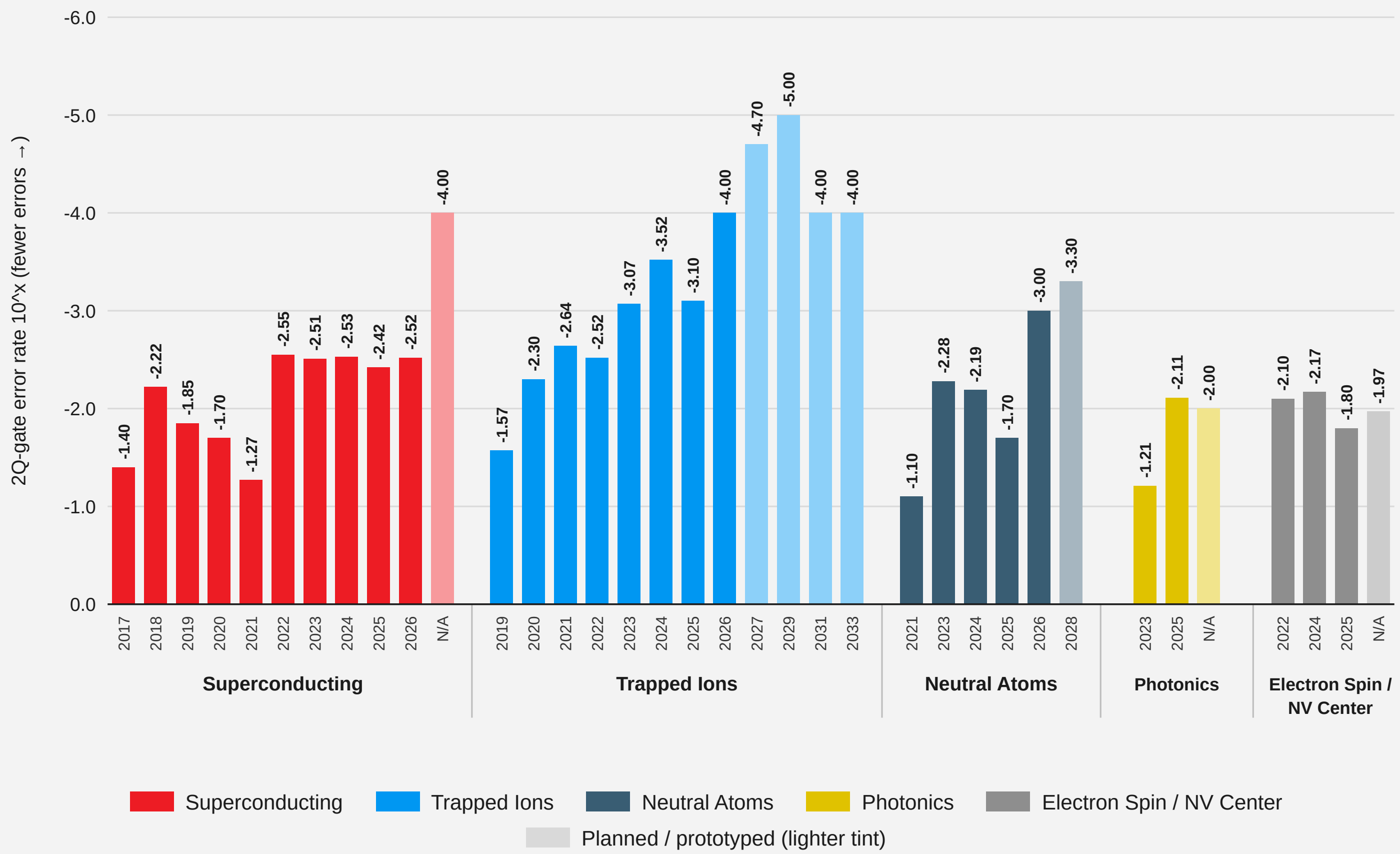

## Trapped ion systems aim for exponential gains in fidelity and are on track to continue outperforming other modalities in that benchmark. Neutral atom platforms also show strong ambitions in this area, while other technologies appear more conservative in their likely fidelity trajectories.

Data considerations: these graphs show the progression of the fidelity and qubit counts for published QPUs over time, including any future plans, considering only the largest QPU announced per modality in that year. The data does not demonstrate a constantly increasing trend, as some calendar years saw smaller QPUs than previously available. Although approximately 60 manufacturers have announced approximately 95 future QPU models, only 19 QPUs have been provided with both target qubit count and fidelities, which is why there are fewer QPUs in the first graph, QPU vs 2-Qubit Fidelity. Devices announced without a target year appear in the “N/A” column at the end of each modality block; these hold the largest qubit counts in the dataset, so these would reflect the boldest projections that do not yet carry a delivery date.

## 10.7 | Future research

Benchmarking plays a vital role in advancing quantum technology, guiding strategic decisions, and establishing shared industry standards. This public overview is designed to increase transparency and spark meaningful dialogue across the community regarding how different systems perform. Because standardized data is still limited in this fast-moving field, we actively encourage industry stakeholders to share their metrics over time. Keeping this repository up to date will close existing data gaps while fostering greater collaboration and progress across the ecosystem.

Qubit count versus two-qubit gate fidelity graph is not included in this edition because vendors report this benchmark using different statistics, best pair, median, or mean, under different operating conditions, which leads to comparability issues. For future editions we look forward to working with the industry to present this data in a more detailed table that makes each vendor's reported statistic explicit.

You can reach us at contact@qir.mit.edu

---

### Footnotes

[1] The gap between the 288 QPUs indexed here and the 56 counted as commercially available breaks down as follows. Of 288 records, 53 are separate instances of a QPU counted only once as a product (IBM Eagle r3-Brussels and Eagle r3-Sherbrooke are one QPU, while IBM Eagle and IBM Heron are two), leaving 235 uniquely differentiated products. Of those, 95 are planned but not yet released, 39 have been retired, 25 are prototypes not intended for commercial use, and 20 are available for research or teaching only. The remaining 56 are commercially available. These 56 come from 24 manufacturers headquartered in 11 countries, out of 86 manufacturers and 17 countries represented in the dataset as a whole.

[2] In June 2026, Microsoft followed with Majorana 2, a four-qubit array on a new InAs/Pb tetron material stack reporting parity lifetimes of around twenty seconds, roughly a thousandfold improvement on Majorana 1, and revised its fault-tolerance target from 2033 to 2029.

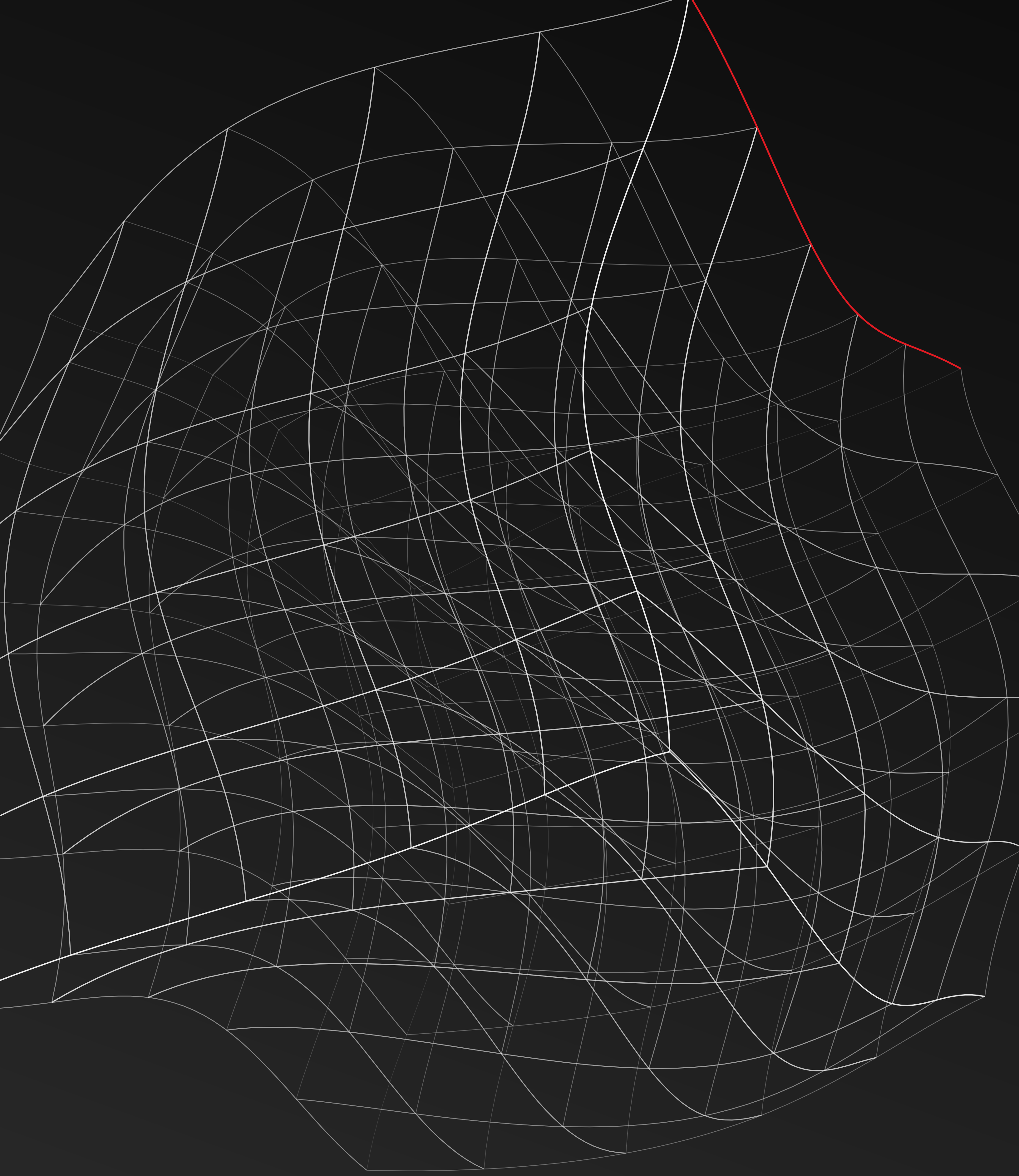

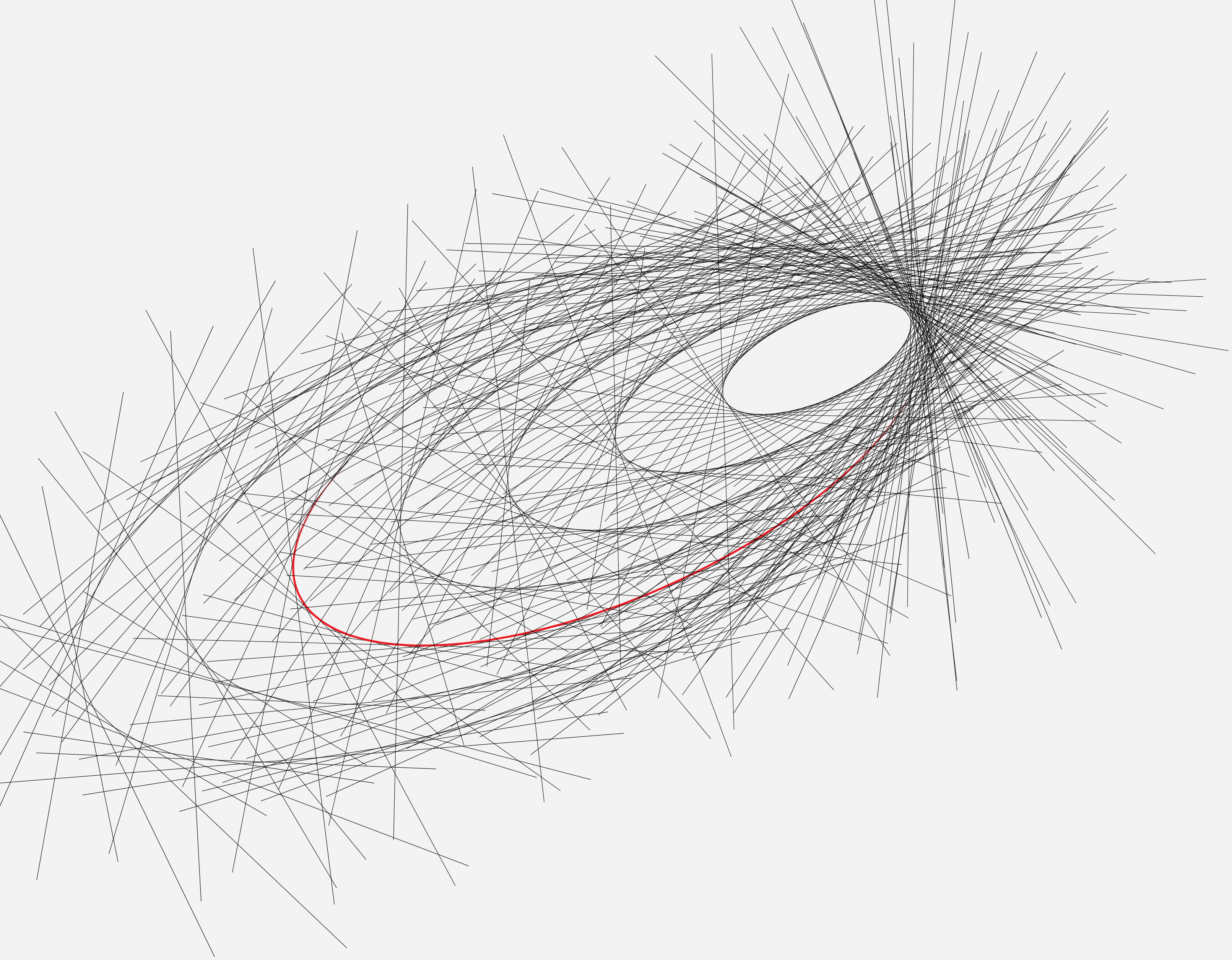

# 11 | Quantum in Defense

The role of quantum technologies in defense and national-security policies became more prominent between 2023 and 2026. Although quantum computing attracts much of the strategic attention, advanced military applications are also emerging in quantum sensing, positioning, navigation and timing (PNT), and communications. The advantages of advanced quantum capabilities in defense would span across enhanced precision in operations, ultra-secure data transfer, faster decision-making and improved radar and imaging.[1] Governments are putting more implementation plans in place that focus on field trials, procurement, dedicated defense programs, and closer integration between military users, national laboratories and commercial quantum companies.

Quantum technology is increasingly being treated as a strategic defense capability given its military significance across several distinct domains, which have different levels of maturity and different operational implications. NATO reflects this broader strategic interpretation, describing quantum technologies as an area of technological competition capable of providing major advantages in computing, communications and situational awareness, while simultaneously creating new vulnerabilities if competitors achieve those capabilities first.[2]

In November 2025, the Pentagon reorganized its critical-technology priorities and established Quantum and Battlefield Information Dominance (Q-BID) as one of six Critical Technology Areas.[3] This was further formalized by the June 2026 Executive Order, which requires dedicated activities to prepare for national-security applications of quantum computing, potentially including a specialized center, while the Department of Defense is directed to participate in government-wide assessment of commercial quantum-computing systems and their supply chains.[4]

The U.S. National Security Science and Technology Strategy, released in August 2026, is the first White House publication under that title since 1995,[5] and it classifies quantum information technologies among “potentially transformative emerging technologies”.[6] The Strategy states that future breakthroughs in quantum technologies could have profound implications for sensing, communications, computing, and information security, and therefore argues that U.S. resilience depends in part on maintaining leadership in the field.

Quantum computing’s potential defense applications include optimization of complex logistics and operational planning, simulation of advanced materials and chemical processes, analysis of large and complicated systems, and potentially elements of intelligence and decision support. Large fault-tolerant quantum computers would also have implications for existing public-key cryptography, making their development strategically important even before direct battlefield applications emerge.

The Defense Advanced Research Projects Agency's (DARPA) Quantum Benchmarking Initiative (QBI) focuses on the verification of whether particular systems can deliver measurable utility, and it is designed to determine whether any technical approach can produce an industrially useful quantum computer by 2033, defined as one whose computational value exceeds its cost.[7] By March 2026, DARPA reported that QBI had evaluated approaches from 20 commercial companies; eleven had reached Stage B[8], while Microsoft and PsiQuantum, originating from the predecessor US2QC program, had advanced to the equivalent of Stage C for system-level verification and validation. Rather than selecting a preferred qubit architecture prematurely, DARPA is evaluating superconducting, trapped-ion, neutral-atom, photonic and silicon-based approaches in parallel.[9]

DARPA is also creating a national testing infrastructure around QBI. Its agreement with Maryland provides for potential matching contributions of up to $100 million each from DARPA and Maryland over four years for the Capital Quantum Benchmarking Hub.[10] Similar arrangements support the Illinois Quantum Proving Ground, where DARPA and Illinois may each contribute up to $140 million over five years, and the New Mexico Quantum Frontier Project, where matching contributions may reach $60 million each over four years.[11]

Effective use of these capabilities would require integration with conventional high-performance computing, AI, secure cloud infrastructure and military data systems, and it would also depend on specialized supply chains involving cryogenics, photonics, control electronics, precision manufacturing and metrology. Therefore the enabling ecosystem is also treated as part of quantum readiness, noting that quantum computers depend on specialized manufacturing capabilities and highly precise supporting technologies.[12]

Quantum communications and quantum networking are developing in parallel. An advanced quantum network would distribute entanglement between remote nodes, potentially connecting quantum computers, memories and sensors. Such networks could eventually support distributed quantum computation, networked quantum sensing and advanced forms of information exchange.

For defense capabilities, the immediate attraction is the possibility of making communications infrastructure more resilient and harder to compromise, particularly when combined with existing networks. The practical challenge is therefore to integrate quantum technologies into existing communications architectures without creating isolated systems that are difficult to operate in military environments. In the United States, DARPA's Quantum-Augmented Network (QuANET) program is explicitly exploring how quantum and classical networking can be combined. In 2025, program participants demonstrated a functioning network in which classical and quantum links operated together, reflecting an emphasis on incremental integration.[13] This hybrid model is particularly relevant to defense because existing command, control and communications infrastructure might remain predominantly classical for the near future.

The European Defence Fund has also supported quantum-related projects aimed at secure communications and networked sensing, via its 2026 program calling for quantum-based network architectures for secure tactical networks, in addition to raising technology readiness and eventually integrating quantum components into European military systems.[14]

The potential military significance of quantum networking extends beyond secure point-to-point communications. Future entanglement-based networks could link geographically separated quantum sensors, improve synchronization and permit distributed quantum-computing resources to operate collaboratively.[15] For armed forces increasingly dependent on distributed sensors, autonomous platforms, satellites and data-intensive command systems, such capabilities could eventually become strategically important. Nevertheless, major technical obstacles remain, including limited quantum-memory performance, the difficulty of developing practical quantum repeaters, and the need to operate hardware under several constraints.[16] The European Commission therefore stresses that defense quantum technologies must function in high-risk environments and meet stringent constraints such as size, weight, power consumption, reliability and integration requirements. [17]

Allied Foreign Ministers approved NATO's first Quantum Technologies Strategy in November 2023[18], establishing the ambition of becoming a "quantum-ready Alliance." The strategy calls for identifying promising military applications, conducting experiments, developing interoperable hardware and software standards, coordinating Allied technology development and strengthening the quantum industrial ecosystem. NATO explicitly advocates a "learn-by-doing" approach in which quantum considerations are incorporated into experimentation, defense planning, capability development and standardization rather than waiting until the technologies are fully mature.[19] A central mechanism for achieving this is NATO's Transatlantic Quantum Community (TQC) bringing together governments, universities, research organizations, funding bodies and industry, including assessing new defense implications of quantum technologies, accelerating adoption and integration, and strengthening pathways between quantum companies and applications.[20]

Parallel international frameworks have also emphasized defense applications in recent years. Under AUKUS Pillar II, Australia, the United Kingdom and the United States have made quantum technologies one component of their advanced-capabilities cooperation and are attempting to reduce barriers to technology sharing and defense-industrial integration. The partnership has already used joint experimentation to accelerate positioning, navigation and timing (PNT) technologies and has broader programs aimed at interoperable communications.[21] In 2025, Australian-developed quantum clocks were subjected to trials in the United States to evaluate their performance for military communications and navigation. Australia views the technology as a means of detecting GPS spoofing and jamming and maintaining PNT in contested environments.[22]

In Canada, the 2026 Defence Industrial Strategy identifies quantum as a sovereign defense capability, and the government announced a call in February 2026 for the establishment of Defence Innovation Secure Hubs (DISHs) for quantum technology for secure environments where companies, researchers and military users can test and integrate defense-relevant technologies.[23] In the UK, in January 2025, the Defence Science and Technology Laboratory announced the first testing outside a laboratory of a UK-built optical atomic clock, with an objective of deploying such technology on military operations within five years.[24] In 2026, France organized its first Quantum Defence Forum to connect military requirements with industrial and scientific capabilities.[25]

## 11.1 | Department of Defense funding for quantum awards & grants

**Fig. 11.a** | U.S. Department of Defense awards for award descriptions including “quantum”, 2008–2025

MIT Quantum Index Report

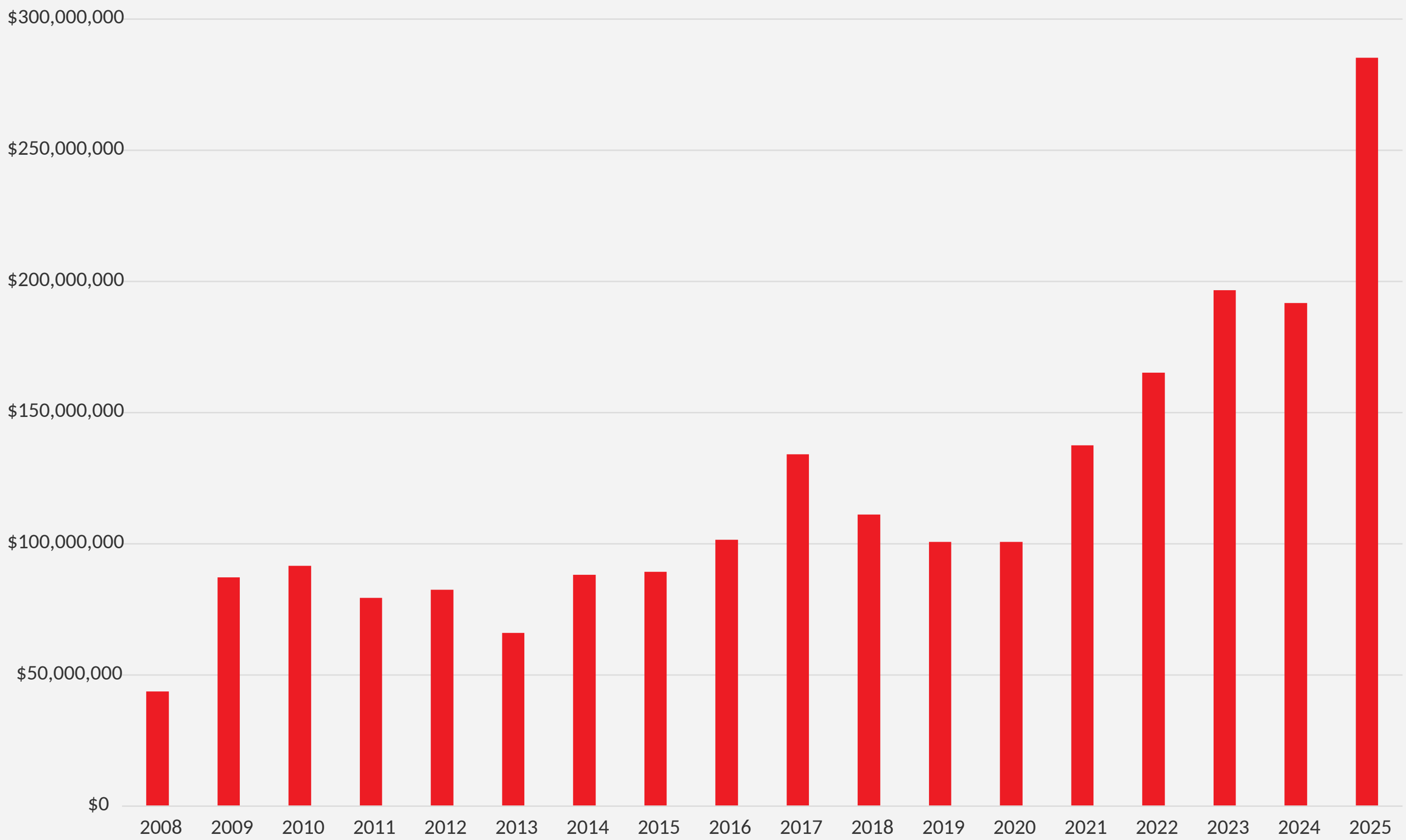

U.S. Department of Defense (DoD) award spending on quantum-related descriptions totaled roughly $2.15 billion across fiscal years 2008–2025.

Funding started small, at just $43.7 million in 2008, and has climbed dramatically since then, reaching a peak of $285.3 million in 2025, about six and a half times the starting amount. While spending has generally trended upward, there have been periods of steady, modest growth followed by stretches of much faster increases.

From 2008 to 2013, awards were volatile but contained, nearly doubling in 2009 and then drifting back to $66 million by 2013, with no year exceeding $92 million, as the DoD's quantum efforts were still getting off the ground. Then, from 2014 to 2020, funding leveled off, and the real turning point came after 2021, when spending began increasing more significantly, more than doubling from $137.5 million to $285.3 million by 2025. In fact, 2025 alone makes up more than 13% of all the money spent across the entire eighteen-year span, and it marks a roughly 49% jump over the previous year, the largest single-year increase in dollar terms in the time period. The only real speed bump in this recent surge was a small dip in 2024, when funding slipped slightly by about 2.5% before rebounding strongly.

> The last five years alone account for nearly half of all the money spent since 2008, underscoring how much the DoD's investment in quantum technology has accelerated recently.

**6.5×**

▶ Overall spending grew 6.5 times from 2008 to 2025.

**$285.3M**

Spending more than doubled in just four years, rising from $137.5 million in 2021 to $285.3 million in 2025.

**Fig. 11.b** | DoD quantum awards by place of performance, 2008–2025

MIT Quantum Index Report

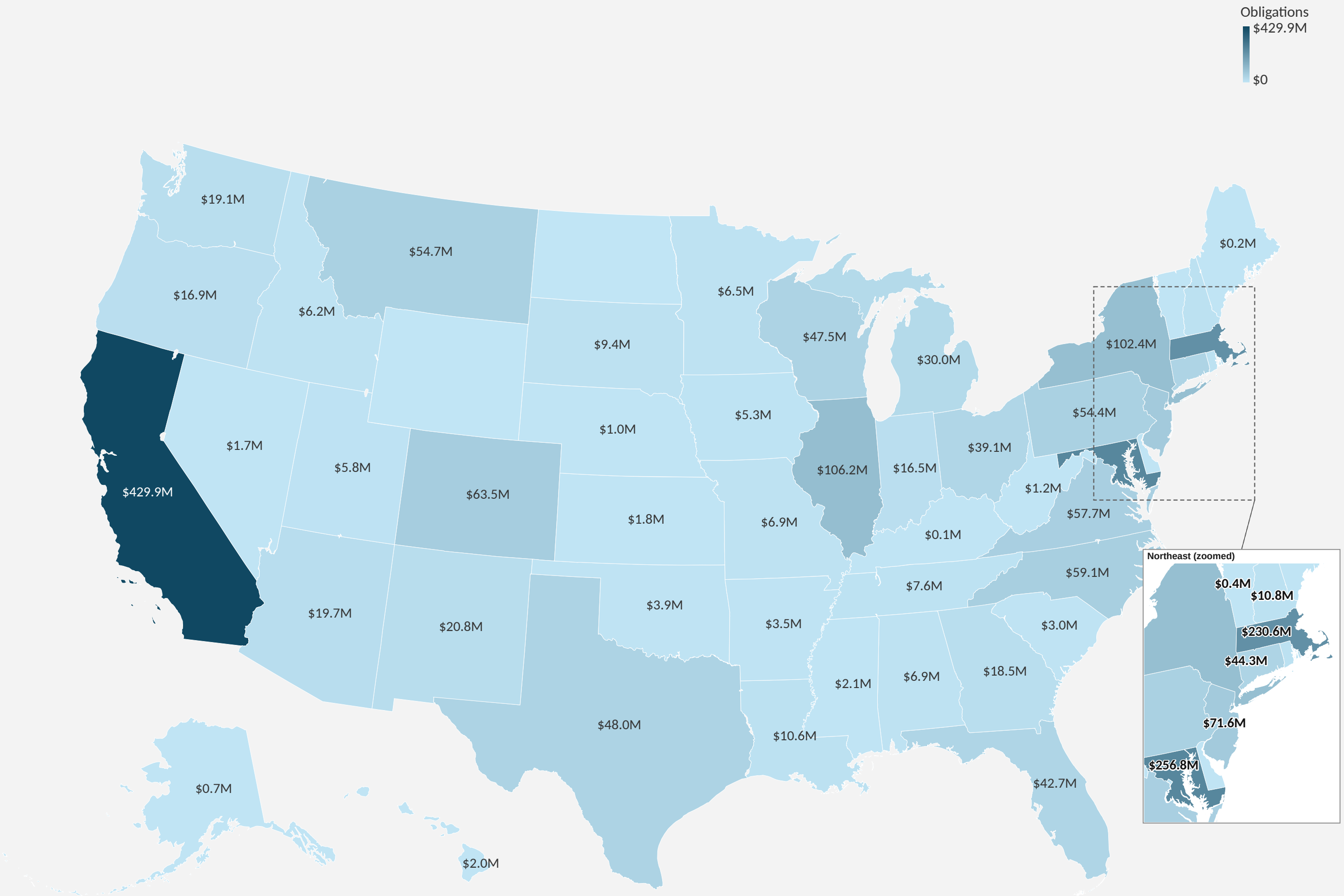

DoD quantum technology spending by place of performance is heavily concentrated in a small handful of states, led decisively by California at nearly $430 million, followed by Maryland at $256.8 million and Massachusetts at $230.6 million. Together these top three states account for over $917 million, more than 40% of total obligations across all locations.

A second tier of states, including Illinois ($106.2 million), New York ($102.4 million), and New Jersey ($71.6 million), along with Colorado, North Carolina, Virginia, and Montana each pulling in roughly $54–63 million, reflects established hubs of federal research activity, defense contracting and national laboratories. Beyond these leaders, the distribution drops off: mid-tier states like Pennsylvania, Texas, Wisconsin, and Connecticut each received $44–54 million, while the majority of states, more than 30 of them, received under $20 million, and many, including Kentucky, Maine, Vermont, and Alaska, saw only small amounts under $1 million.

**Fig. 11.c** | DoD quantum award recipients top 10, 2008–2025

MIT Quantum Index Report

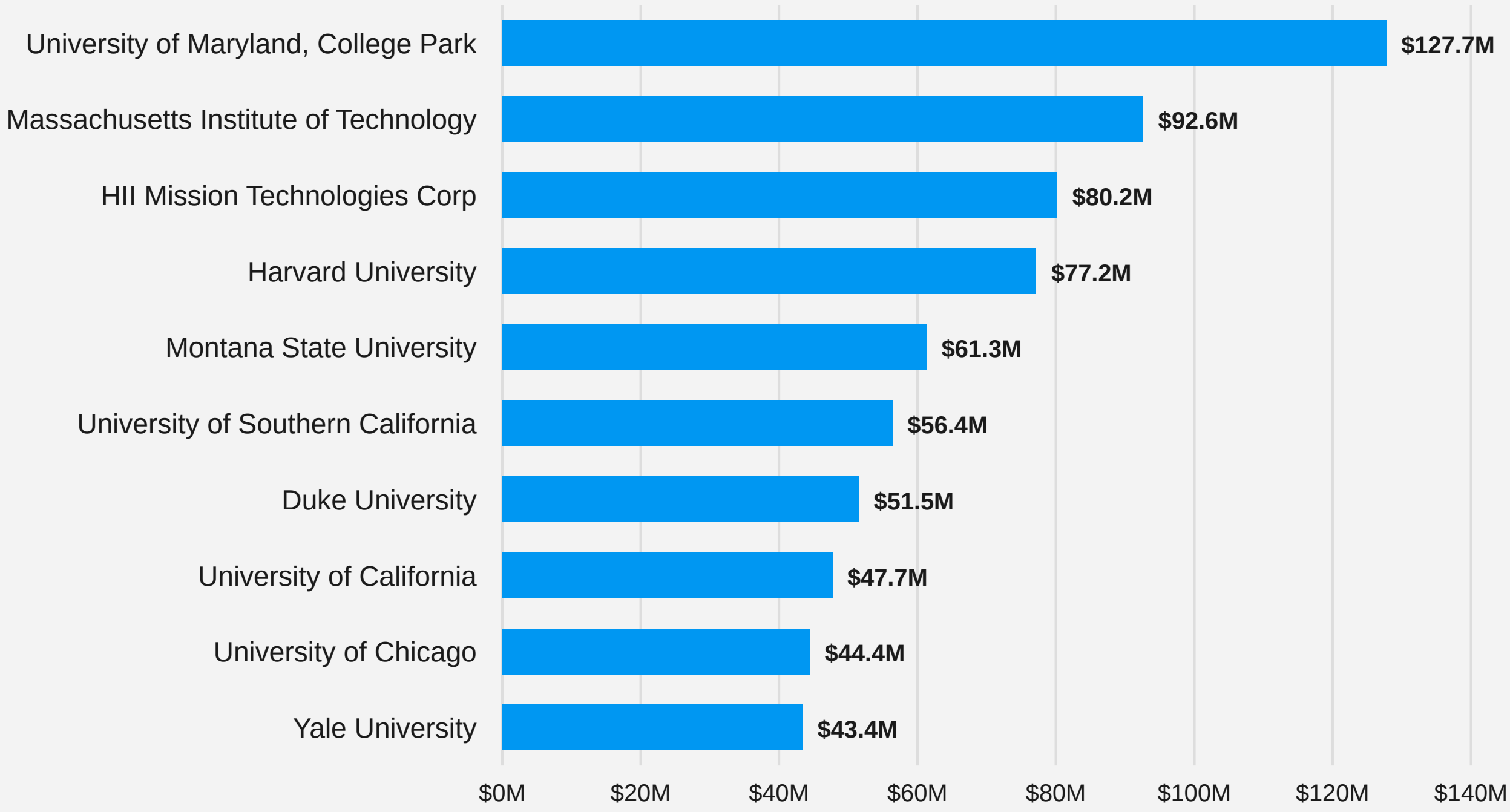

Among the top 10 recipients of DoD quantum technology awards, University of Maryland, College Park leads by a wide margin at $127.7 million, followed by MIT at $92.6 million and HII Mission Technologies Corp at $80.2 million, the latter standing out as the only non-university entity in the top tier, pointing to significant defense-industry involvement alongside academic research.

Harvard ($77.2 million) and Montana State University ($61.3 million) round out the upper half, with Montana State's strong showing notably consistent with Montana's outsized state-level ranking seen earlier in the data. The remaining institutions, University of Southern California ($56.4 million), Duke ($51.5 million), the University of California system ($47.7 million), University of Chicago ($44.4 million), and Yale ($43.4 million), cluster more closely together, each receiving between roughly $43 and $56 million. Together, these ten institutions account for just over $680 million in awards, illustrating that DoD quantum research funding, while spread across several elite universities and research centers, remains dominated by a small set of institutions with deep, established research capacity and, in at least one case, direct defense-sector expertise.

## 11.2 | Future research

To build a fuller picture in the future editions of this report, we welcome contributions from defense agencies, industry partners, and researchers who can share unclassified funding figures, program milestones, or technology-readiness assessments regarding quantum technologies. We are also interested in tracking how quantum defense programs for different applications move from field trials to operational deployment, and in developing clearer benchmarks for comparing defense quantum initiatives across countries and alliances.

You can reach us at contact@qir.mit.edu

## Footnotes

[1] European Commission, 'Quantum Technologies in Defence Factsheet' (2025).

[2] 'NATO Releases First Ever Quantum Strategy' (*North Atlantic Treaty Organization (NATO)*) <https://www.nato.int/en/news-and-events/articles/news/2024/01/17/nato-releases-first-ever-quantum-strategy> accessed 19 August 2026.

[3] 'Under Secretary of War for Research and Engineering Emil Michael Announces Six Critical Technology Areas' (*U.S. Department of War*) <https://www.war.gov/News/Releases/Release/Article/4333074/under-secretary-of-war-for-research-and-engineering-emil-michael-announces-six/> accessed 19 August 2026.

[4] Executive Order No 14413, 91 Fed Reg 38487 (22 June 2026).

[5] Sarah Willrich, 'White House Releases National Security Science & Technology Strategy' [2026] Lawfare <https://www.lawfaremedia.org/article/white-house-releases-national-security-science---technology-strategy> accessed 19 August 2026.

[6] The White House, 'National Security Science & Technology Strategy' (2026).

[7] 'QBI | DARPA' <https://www.darpa.mil/research/programs/quantum-benchmarking-initiative> accessed 19 August 2026.

[8] 'Stage B Selection | DARPA' <https://www.darpa.mil/research/programs/quantum-benchmarking-initiative/stage-b-selection> accessed 19 August 2026.

[9] 'Quantum Benchmarking Initiative Expands Quest to Separate Hype from Reality | DARPA' <https://www.darpa.mil/news/2026/qbi-stage-a-qbit> accessed 19 August 2026.

[10] 'DARPA, State of Maryland Sign Agreement to Propel Quantum Research | DARPA' <https://www.darpa.mil/news/2025/agreement-propel-quantum-research> accessed 19 August 2026.

[11] 'DARPA, State of New Mexico Establish Framework to Advance Quantum Computing | DARPA' <https://www.darpa.mil/news/2025/darpa-new-mexico-establish-framework-advance-quantum-computing> accessed 19 August 2026.

[12] 'Summary of NATO's Quantum Technologies Strategy' (*Official texts and resources | NATO*) <https://www.nato.int/en/about-us/official-texts-and-resources/official-texts/2024/01/16/summary-of-natos-quantum-technologies-strategy> accessed 19 August 2026.

[13] 'Quantum Mechanics, Classical Backbone: DARPA's QuANET Advances Practical Quantum Networking | DARPA' <https://www.darpa.mil/news/2025/quanet-advances-practical-quantum-networking> accessed 19 August 2026.

[14] European Commission, 'Commission Implementing Decision on the Financing of the European Defence Fund and the Adoption of the Work Programme for 2026 and Amending Commission Implementing Decision C(2025) 568 Final on the Financing of the European Defence Fund and the Adoption of the Work Programme for 2025 - Part 2' (2026).

[15] Adam Lewis and Martino Travagnin, 'A Secure Quantum Communications Infrastructure for Europe: Technical Background for a Policy Vision' (*JRC Publications Repository*, 2022) <https://doi.org/10.2760/180945> accessed 19 August 2026.

[16] 'Quantum Memory and Repeaters' [2022] NIST <https://www.nist.gov/pml/productsservices/quantum-networks-nist/technologies-quantum-networks/quantum-memory-and> accessed 19 August 2026.

[17] European Commission (n 1).

[18] 'NATO Releases First Ever Quantum Strategy' (n 2).

[19] 'Summary of NATO's Quantum Technologies Strategy' (n 12).

[20] 'The Netherlands Takes on Chair of NATO's Transatlantic Quantum Community' (*North Atlantic Treaty Organization (NATO)*) <https://www.nato.int/en/news-and-events/articles/news/2026/05/13/the-netherlands-takes-on-chair-of-natos-transatlantic-quantum-community> accessed 19 August 2026.

[21] 'AUKUS Defense Ministers Meeting Joint Statement' (*U.S. Department of War*) <https://www.war.gov/News/Releases/Release/Article/3604511/aukus-defense-ministers-meeting-joint-statement/> accessed 19 August 2026.

[22] ‘World-Leading Australian Quantum Clocks Successfully Trialled under AUKUS Pillar II’ (4 November 2025) <https://www.minister.defence.gov.au/media-releases/2025-11-04/world-leading-australian-quantum-clocks-successfully-trialled-under-aukus-pillar-ii> accessed 19 August 2026.

[23] National Defence, ‘Government of Canada Launches Call for Proposals to Establish Defence Innovations Secure Hubs for Quantum and Uncrewed System Technologies’ (18 February 2026) <https://www.canada.ca/en/department-national-defence/news/2026/02/government-of-canada-launches-call-for-proposals-to-establish-defence-innovations-secure-hubs-for-quantum-and-uncrewed-system-technologies.html> accessed 19 August 2026.

[24] ‘Top Secret Lab Develops Atomic Clock Using Quantum Technology’ (*GOV.UK*) <https://www.gov.uk/government/news/top-secret-lab-develops-atomic-clock-using-quantum-technology> accessed 19 August 2026.

[25] ‘Forum quantique défense : transformer l’innovation en capacités opérationnelles | Ministère des Armées et des Anciens combattants’ (16 April 2026) <http://www.defense.gouv.fr/actualites/forum-quantique-defense-transformer-linnovation-capacites-operationnelles> accessed 19 August 2026.

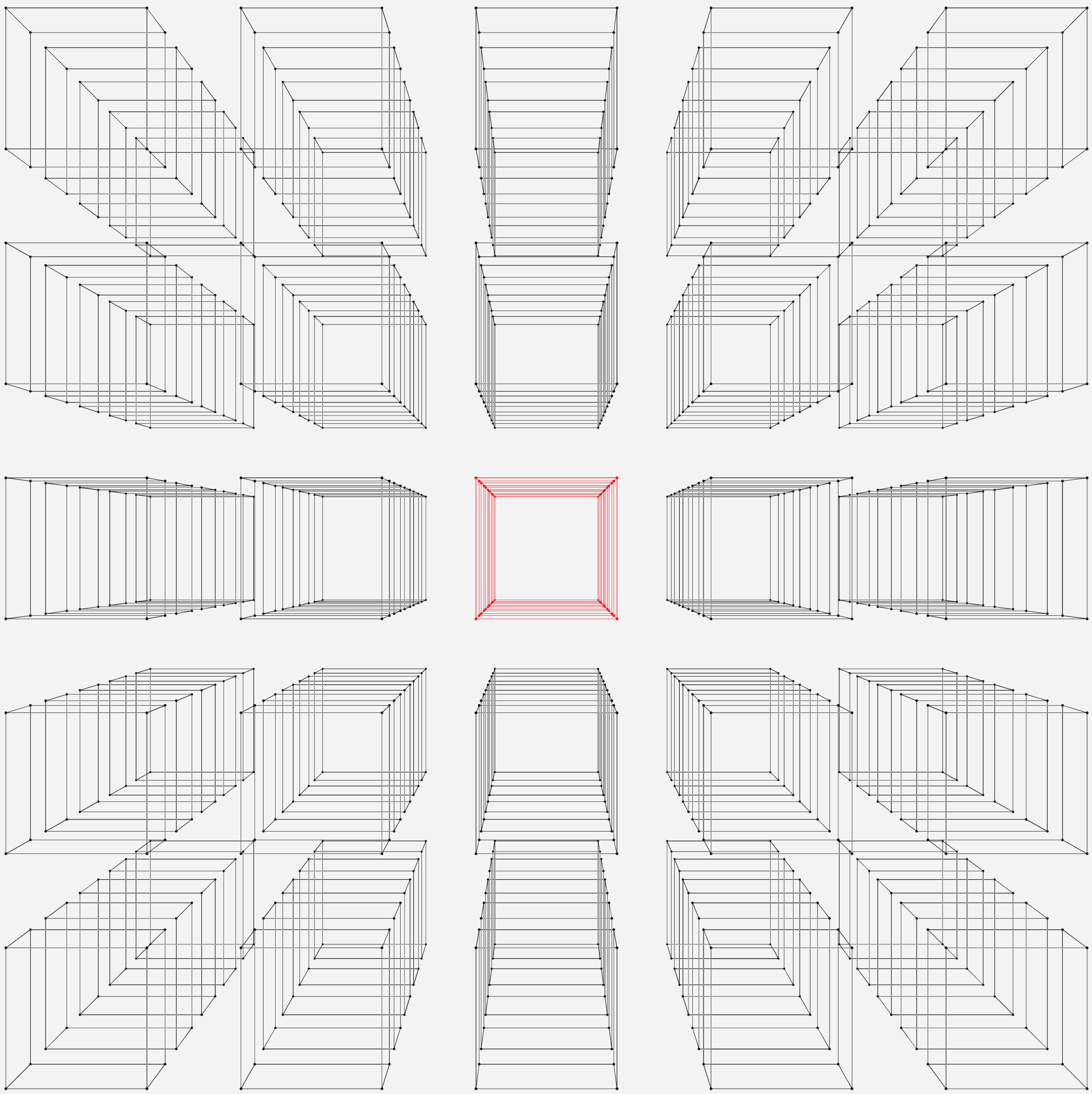

# 12 | Appendix

# Chapter 1 | Patents Methodology

The data in this section relies on patent families and was based on data acquired by Accenture Research from the LexisNexis patent database using select IPC codes relevant to quantum technologies. For patents without owner type, LLMs were used to classify owners by type. No screening was made regarding international patent families. There is an approximately 18-month lag in patent publication data; therefore, recent years' data may be subject to change.

The list of IPC codes used for this dataset can be found here:

| IPC Code | Description | Sub-category |
|---|---|---|
| G06N0010200000 | Models of quantum computing, e.g. quantum circuits or universal quantum computers [2022.01] | Quantum Computing |
| G06N0010600000 | Quantum algorithms, e.g. based on quantum optimization, or quantum Fourier or Hadamard transforms | Quantum Computing, Software |
| H04B0010700000 | Photonic quantum communication | Quantum Communications, Quantum Networking |
| G06N0010000000 | Quantum computing, i.e. information processing based on quantum-mechanical phenomena | Quantum Computing |
| B82Y0020000000 | Nanooptics, e.g. quantum optics or photonic crystals | Quantum Computing, Hardware |
| G02F0002020000 | Frequency-changing of light, e.g. by quantum counters | Quantum Computing, Hardware |
| G06N0010800000 | Quantum programming, e.g. interfaces, languages or software-development kits for creating or handling programs capable of running on quantum computers; Platforms for simulating or accessing quantum computers, e.g. cloud-based quantum computing | Quantum Computing, Software |
| H01L0033040000 | with a quantum effect structure or superlattice, e.g. tunnel junction | Quantum Computing, Hardware |
| H01S0005340000 | comprising quantum well or superlattice structures, e.g. single quantum well [SQW] lasers, multiple quantum well [MQW] lasers or graded index separate confinement heterostructure [GRINSCH] lasers (H01S 5/36 takes precedence) | Quantum Computing |
| G02F0001017000 | Structures with periodic or quasi periodic potential variation, e.g. superlattices, quantum wells | Quantum Computing |
| H01L0029775000 | with one-dimensional charge carrier gas channel, e.g. quantum wire FET | Quantum Computing |
| H01L0033060000 | within the light emitting region, e.g. quantum confinement structure or tunnel barrier | Quantum Computing |

| IPC Code | Description | Sub-category |
|---|---|---|
| H10K0050115000 | comprising active inorganic nanostructures, e.g. luminescent quantum dot | Quantum Computing |
| B82Y0010000000 | Nanotechnology for information processing, storage or transmission, e.g. quantum computing or single electron logic | Quantum Computing, Quantum Networking |
| B82Y0015000000 | Nanotechnology for interacting, sensing or actuating, e.g. quantum dots as markers in protein assays or molecular motors | Quantum Computing |
| G06N0010400000 | Physical realisations or architectures of quantum processors or components for manipulating qubits, e.g. qubit coupling or qubit | Quantum Computing |
| G06N0010700000 | Quantum error correction, detection or prevention, e.g. surface codes or magic state distillation | Quantum Computing |
| G16C0010000000 | Computational theoretical chemistry, i.e. ICT specially adapted for theoretical aspects of quantum chemistry, molecular mechanics, molecular dynamics or the like | Quantum Computing |
| H01L0029150000 | Structures with periodic or quasi periodic potential variation, e.g. multiple quantum wells, superlattices (such structures applied for the control of light G02F 1/017; applied in semiconductor lasers H01S 5/34) | Quantum Computing, Quantum Networking |
| H04B0010000000 | Transmission systems employing electromagnetic waves other than radio-waves, e.g. infrared, visible or ultraviolet light, or employing corpuscular radiation, e.g. quantum communication | Quantum Computing |

Regarding the categorization of the patent data presented by origin, Accenture researchers applied the following methodology:

Step 1: For applicants with name but without type (e.g. university, corporate, etc.), search was filtered using keywords such as specific origin entity or specific keywords (such as hospital, etc.) to categorize some of the applicants. The filtered results are stored as Result 1.

Step 2: Researchers identified those without keywords (over 1,000 names) using an LLM (Gemini-1.5-flash-001) to automatically detect their type. This result is stored as Result 2.

Step 3: Researchers merged the applicants with identified types (from both initial filtering and LLM results) with the applicants who originally had type information provided by the vendor.

Step 4: Merged data was manually checked to ensure non-duplication and accuracy.

The data presented in patent applications by country and leading organizations by country was sourced from The Quantum Insider (TQI).

TQI acquired the patent data directly from patent offices. For national patents, data was sourced from each country's national patent office (e.g. in the U.S. patents are sourced from the United States Patent and Trademark Office (USPTO); in China patents are sourced from the China National Intellectual Property Administration). WO = World Intellectual Property Organization (WIPO), EP = European Patent Office. For 'Other Countries' it is typically a longer tail of nations which have been aggregated. For the analysis in this report the TQI data category "simple family" was used which states that results grouped by patents that claim the exact same set of priority filings, giving one entry per unique set of priorities.

The TQI patent data was collected based on the following keywords:

Adiabatic Theorem, Bosonic Creutz Ladder, Dicke Model, Distributed quantum computation, Fault-tolerant quantum computation, gray zone assault, Hadamard Gate, Harrow Hassidim Lloyd, HHL algorithm, ion traps, Josephson junctions, neutral atoms, Noisy Intermediate-Scale Quantum era, Open Quantum Systems, Photonic Quantum Computing, QAOA, qbits, qbytes, QEC, QNLP, QSVM, qtrits, quantum annealing, Quantum Advantage, Quantum accelerators, quantum annealing, quantum algorithms, Quantum applications, quantum approaches, quantum approximate optimization, quantum approximate optimization algorithms, quantum arithmetic, Quantum artificial intelligence, quantum backtracking, quantum bits, Quantum Bosonic Systems, quantum bytes, quantum chaos, quantum chaos, quantum chemistry, Quantum circuits, quantum classifier, quantum communication, quantum compiler, Quantum complexity, Quantum component, Quantum computation, Quantum computational, Quantum computer, Quantum Computing Architectures, Quantum Control, quantum correlation, Quantum cryptanalysis, Quantum cryptoalgorithm, Quantum cryptog, Quantum cryptographic, Quantum cryptology, Quantum Cryptology, Quantum cryptosystem, Quantum decoding, quantum devices, quantum distillation, quantum dots, quantum dynamics, quantum dynamics, quantum eigensolver, quantum encryption, Quantum Entanglement, quantum entanglement distillation, quantum error correction, Quantum Error Detection, Quantum Field theory, quantum Fourier transform, Quantum gas sensors, Quantum gate, Quantum gate fidelity, Quantum gate-based, Quantum gates, quantum Grover, Quantum hardware,

Quantum hardware security, quantum image sensor, quantum imager array, quantum information, Quantum information processing, Quantum information science, Quantum information systems, Quantum information theory, Quantum interference, Quantum ions, quantum Jozsa, Quantum Kernel, Quantum key, Quantum key distribution, Quantum key distribution network, Quantum key distribution protocol, Quantum key distribution systems, quantum key exchange, Quantum LDPC Codes, Quantum Linear Optics, quantum logic, quantum machine learning, quantum machines, quantum magic states, Quantum Maps, Quantum Measurement, Quantum Memristors, Quantum metrological, Quantum metrology, Quantum metrology standards, Quantum Monte Carlo, Quantum Natural Language Processing, Quantum Networks, quantum neural networks, Quantum Oscillator, quantum phase amplifiers, Quantum precision measurement sensors, Quantum process, Quantum processing, Quantum Programs, Quantum proof, quantum public key, Quantum Quantile Mechanics, Quantum Quantizer, Quantum random number, Quantum random number generation, Quantum random number generation device, Quantum random number generator, Quantum random number sequences, Quantum safe network, Quantum sensing, Quantum sensing technology, Quantum sensor, Quantum sensor networks, quantum sensors, Quantum shared key, quantum Shor, Quantum Signal Processing, quantum simulation, Quantum Simulator, quantum single photon, quantum software, Quantum Speedup, quantum spin, quantum spintronics, quantum state, Quantum Subroutines, Quantum superposition, quantum supremacy, quantum supremacy, quantum switches, quantum systems, Quantum Technologies, quantum teleportation, quantum tensor, quantum toffoli gate, Quantum transmons, quantum variational, quantum video, Quantum Week, Quantum-enhanced, Quantum-enhanced, Quantum-resistance, qubits, qubytes, qudits, qumodes, qutrits, silicon qubits, VQE

## Chapter 2 | Academic Research Methodology

Accenture provided the data for this chapter, which was gathered from the Critical Technology Tracker Project by the Australian Strategic Policy Institute.[1]

### ASPI Methodology Summary

The Web of Science (WoS) Core Collection database was used as the primary source of research publication data.

WoS Core Collection was chosen because it has well-understood performance characteristics and is used extensively by researchers who study scientific trends. The dataset includes conference and journal publications and excludes bibliographic records that were deemed to not reflect research advances, such as book reviews, retracted publications and letters submitted to academic journals. In addition, data from the Research Organization Registry (ROR) was used to clean institution names, and data from the Open Researcher and Contributor ID (ORCID) database was used to build career profiles for the researchers plotted in the ASPI Talent Tracker.

Country-level quality metrics:

The methodology includes “both the top 10% and the H-index as neither is perfect and both add a unique insight”. In technologies in which 1st and 2nd place flip depending on which quality metric is used, the race really is too close to call. However, more often, the lead is large and unambiguous, and both metrics are consistent regarding who is leading”.

ASPI methodology states that “The top 10% of the most highly cited papers were analyzed to generate insights into which countries are publishing the greatest share of high-quality, innovative and high-impact research. Credit for each publication was divided among authors and their affiliations and not assigned only to the first author (for example, if there were two authors, they would each be assigned half the allocation). Fractional allocation of credit is a better prediction of individuals who go on to win Nobel Prizes or fellowship of prestigious societies. Fractional allocation of credit was used for all metrics.”

“The number of institutions that a country has in the world’s top 10 institutions is used to illustrate research concentration and dominance. This list is based on the number of papers that the institutions have in the top 10% of highly cited papers.”

Definitions as explained in the ASPI methodology:

Quality Metrics: “Distinguishing innovative and high-impact research papers from low-quality papers is critical when estimating the current and future technical capability of nations. Not all the millions of research papers published each year are high quality.”

Citation: “When a scientific paper references another paper, that’s known as a citation.

The number of times a paper is cited reflects the impact of the paper. As time goes by, there are more opportunities for a paper to be cited, so only papers of a similar age should be compared using citation counts.”

## Chapter 3 | Funding Methodology

This data was sourced from The Quantum Insider (TQI) using The Quantum Insider Funding Database. Limitations exist because the funding numbers are obtained from open media sources (press releases, articles, etc.).

The ’Other’ funding category encompasses a wide variety of investments that either did not fit neatly into the standard classification groups or were not precisely reported. Individual investments are sometimes complex and opaque as companies may secure funding from multiple sources simultaneously which have different terms and structures. There is also a wide variety of sources including government, institutional, traditional venture capital, grant aid, or quasi-debt. Due to the evolving nature of quantum funding structures, a number of transactions that might otherwise be categorized under ’Late Venture’ might have been included in ’Other’ to maintain consistency with earlier data and emerging patterns.

## Chapter 4 | Quantum in Corporate Communications Methodology

The data was collected by Accenture through AlphaSense on 17 March 2026 using the keyword search term “quantum computing”. The documents presented in this section include five categories, (1) Company Documents consisting of U.S. Filings, Global Filings, Company House Filings, Private Company Filings, Event Transcripts, ESG, Thought Leadership, Other Company Publications; (2) Research Documents consisting of Broker Research, IDC Research, Consultancy Research, Broker Feed; (3) Transcript Documents consisting of Event Transcripts; (4) News Documents consisting of Financial Times, Market News, General News, Trade Publications, RSS Feeds, LexisNexis, (5) Expert Call Documents consisting of Expert interviews.

# Chapter 5 | Policy Methodology

The policy research was completed through comprehensive desk research specifically designed to capture the rapidly evolving landscape of quantum technology initiatives across multiple countries, including detecting and analyzing the national strategy documents, implementation plans, and bilateral agreements, which involved cross-referencing multiple official sources, and analysis of policy implementation progress.

# Chapter 6 | Workforce Methodology

This data was gathered by Accenture and sourced from Lightcast. Lightcast integrates economic, labor market, demographic, education, profile, and job posting data from dozens of government and private-sector sources, creating a comprehensive and current dataset that includes both published data and detailed estimates with full United States coverage. Further information on Lightcast data sources available [here] (https://kb.lightcast.io/en/collections/3904183-data-methodology).

Please note that historical job postings data is continuously updated by Lightcast through reclassification, source additions, and data quality improvements. As a result, recent extracts may show higher volumes for past periods.

The following keywords were used in our Lightcast database searches:

Quantum skills keywords: Quantum Gates, Heteronuclear Single Quantum Coherence Spectroscopy,Heteronuclear Multiple Quantum Coherence,Quantum Link,Quantum Mechanics/Molecular Mechanics (QM/MM),Quantum Point Contact,Quantum Phase Transition,Quantum Dynamics,Quantum Imaging,Quantum Technology,Superconducting Quantum Interference Device (SQUID),Amazon Quantum Ledger Database (QLDB),Quantum Cryptography,Quantum GIS (QGIS),Quantum Scalar Servers,Quantum Chemistry,Quantum Mechanics,Quantum Physics,Quantum Information Sciences,Quantum Computing,Quantum Dots,Quantum Information,Quantum ESPRESSO.

Quantum mentions keywords: quantum computer quantum computing quantum supremacy quantum entanglement quantum superposition quantum bit qubit topological qubit silicon spin qubit quantum advantage quantum simulation quantum machine learning quantum optimization quantum chemistry optical quantum computing gate model quantum computing photonic quantum computer quantum dots superconducting quantum computer trapped ion quantum computer quantum annealing quantum as a service quantum cloud post-quantum cryptography PQC quantum algorithm Shor's algorithm Grover's algorithm quantum encryption quantum research quantum use case

The PQC keywords were excluded from the final research results in order to provide a more focused status report. The keyword search analysis was completed following the approach below:

$$\text{Quantum Job Share (12M rolling)} = \frac{\text{12-month rolling sum of job postings mentioning quantum-related keywords}}{\text{12-month rolling sum of all job postings}} = \frac{\Sigma Q_i}{\Sigma T_i}$$

$t$ = current month

$i$ = each month in the 12-month window (from $t-11$ to $t$)

$Q_i$ = number of job postings in month $i$ that mention quantum (in the job description/body)

$T_i$ = total number of job postings in month $i$

## Chapter 7 | Education Methodology

The education data represented in the "Postgraduate Education" section was collected from the StudyPortals resource in March 2026 and presents the master's degrees that make a specific reference to "quantum" in the degree name found in the named resource.

Education Enrollment data was collected from the publicly available data set "Current Term Enrollment Estimates" with the June 2026 updates of the NSC Research Center.

The NSC states in its methodology for compiling the relevant dataset that the data is based on administrative data directly derived from college and university registrars. NSC declares that since the fall of 2021, “institutions actively submitting enrollment data to the Clearinghouse account for 97 percent of all enrollments at Title IV, degree-granting institutions in the U.S.”

The 2021 U.S. Report “The Role of International Talent in Quantum Information Science” states that “the most QIST-relevant degree fields are physics, electrical engineering, and computer science” and explains that these domains were selected based on two criteria: preliminary search of keywords for online job postings and analysis of doctoral thesis titles, abstracts, and keywords. Therefore, we retrieved the student enrollment data for the relevant three degrees.

We used ’Electrical, Electronics, and Communications Engineering’ (141000 – Major Field Group CIP) enrollment data as a subcategory of Engineering, Physics degree (400800 – Major Field Group CIP) enrollment data as a subcategory of Physical Sciences, Computer Science degree (110700 – Major Field Group CIP) enrollment data as a subcategory of Computer and Information Sciences and Support Services. The data categorization had challenges as the major field groups at times had interconnected degrees such as “Computer and Information Science, general”, “Astronomy and Astrophysics” which are not included in the subject-level enrollment data. In order to provide a fuller picture, the report also presents the enrollment numbers for the three major field families Engineering, Physical Sciences, and Computer and Information Sciences and Support Services.

## Chapter 8 | Public Opinion Methodology

Survey on Public Opinion refers to a general population survey conducted online in December 2025 with U.S. participants. The survey instrument was administered to a representative panel of 1,309 U.S. residents, with demographic sampling aligned to U.S. Census Bureau distributions for both gender and age groups, ensuring population representativeness on the referred categories. Data collection procedures followed established survey research protocols, with items grouped by thematic content to enhance respondent engagement and reduce cognitive load.

The 5-point scale format was selected to balance response sensitivity with participant comprehension, avoiding the potential ambiguity associated with finer-grained scales while still capturing meaningful variations in opinion intensity.

Each item was measured using a symmetric response format anchored by “Strongly Disagree” and “Strongly Agree”, with intermediate positions of “Somewhat Disagree”, “Neither Agree nor Disagree”, and “Somewhat Agree”. Regarding emotional response questions, the response format was anchored by “Very Nervous” and “Very Excited”, with intermediate positions of “Somewhat Nervous”, “Neither Nervous nor Excited” and “Somewhat Excited”. This approach enabled precise quantification of attitudinal responses while maintaining respondent comprehension through clear, distinct categories. To better demonstrate the difference in responses, in the relevant chapter, graphs were prepared to represent the ratio of “Agree” and “Disagree” responses as well as “Excited” and “Nervous” emotional responses where the neutral answers were not represented. For the referred visualizations “Strongly Disagree” and “Somewhat Disagree” responses were grouped into the category “Disagree”, and “Strongly Agree” and “Somewhat Agree” responses were grouped into the category “Agree”. Utilizing the same approach, for the visualizations of positive and negative responses, “Very Nervous” and “Somewhat Nervous” responses were grouped into the category “Nervous”, and “Very Excited” and “Somewhat Excited” responses were grouped into the category “Excited”.

## Chapter 9 | Quantum Networking Testbeds Methodology

The 2025 dataset on quantum networking testbeds was created by merging input from the Center for Quantum Networks (CQN) researchers, QIR professional network and publicly available information by GQI Quantum Computing Report (accessed in June 2024). In the 2026 edition, updates were made through stakeholder input and desk research.

The quantum networking application dataset was retrieved from the CQN – QED-C Quantum Networking Applications Roadmap 2.0, published in April 2026.

# **Chapter 10** | Quantum Processor Benchmarking Methodology

The 2025 dataset of QPUs was compiled from a combination of a keyword-based online search and official announcements, references to QPU lists made available to us, and direct queries to QPU manufacturers. The previous report's data was collected from Jan 2024 to Apr 2025. In this edition, we merged our former data with the publicly accessible Quantum Insider QPU dataset. Official websites of manufacturers were also used until July 2026 to fill further gaps in the dataset.

For our 2025 dataset, a list of known manufacturers was created based on the sources of the Quantum Insider, Olivier Ezratty, and Wikipedia. For each manufacturer, the official website was interrogated to retrieve the indicated benchmarks. For data not on manufacturers' websites, we utilized web searches (Google) for official announcements from manufacturers and related news articles. Additionally, scholarly articles were identified via arXiv and Google Scholar using the following keywords for benchmarks: Quantum Volume, CLOPS, EPLG, Q-Score, benchmarking. During this process, additional manufacturers/QPUs were identified and added to the QPU list. Lastly, each manufacturer was contacted for verification of records - either to an existing contact of the QIR team, or via the communications address listed on manufacturer's website. This last step was not repeated in the 2026 edition and the data collection was limited to official websites and the community-updated TQI dataset. The final list was reviewed by the QIR team and experts in their professional network.

# **Chapter 11** | Quantum in Defense Methodology

The data used in this chapter was gathered from the USAspending.gov website, which is the official open data source of federal spending information, including information about federal awards such as contracts, grants, and loans. The dataset used was created after filtering for awarding agency (Department of Defense) and award description (keyword: "quantum") for the fiscal years 2008–2025. The approach might have presented the limitation of exclusion of quantum technology relevant awards that do not refer to the word "quantum" in the award description.

## Footnotes

[1] Gaida, J., Wong-Leung, J., & Robin, S. (2023). Critical technology tracker. Who Is Leading the Critical Technology Race. A Project by the Australian Strategic Policy Institute. https://techtracker.aspi.org.au. Data gathered April-May 2026.

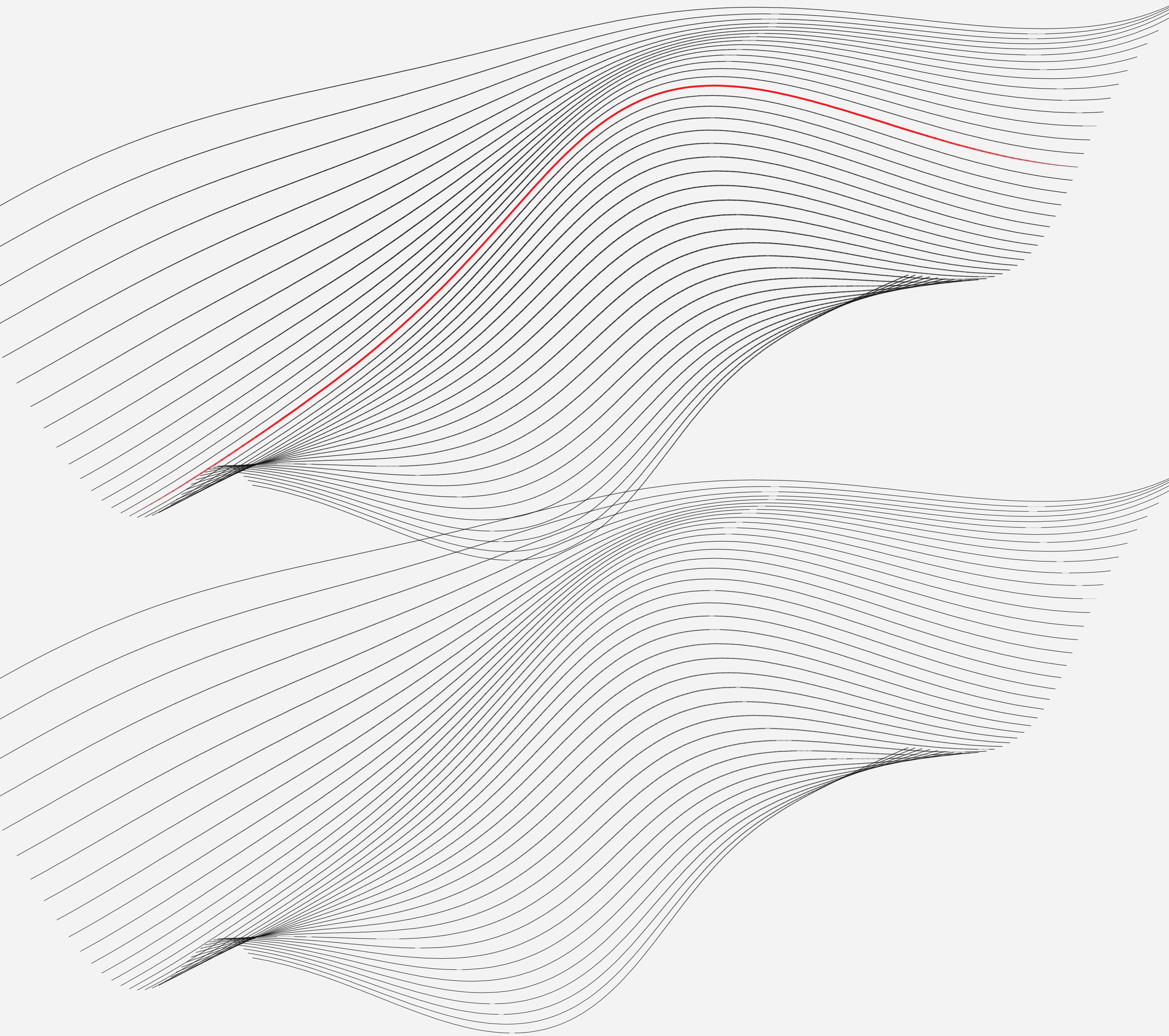

Quantum Index Report 2026

MIT Initiative on the Digital Economy

245 First Street, Room E94-1521, Cambridge, MA 02142 USA

qir.mit.edu | ide.mit.edu